\documentclass[aps,reprint,nofootinbib,nobibnotes,showpacs,superscriptaddress,longbibliography]{revtex4-1}
\usepackage[english]{babel}
	\addto\captionsenglish{%
	}
\usepackage{graphicx}
\usepackage{longtable}	% tables on more than one page
\usepackage{comment}

	\graphicspath{{graphics/}}

\usepackage{amsmath}
\usepackage{amssymb}
\usepackage[version=3]{mhchem}	% chemical formulas -> e.g. \ce{^{130}Te} instead of $^{130}$Te
\usepackage{bm}	% bold
\usepackage{nicefrac}	% fractions: a/b
\usepackage{mathrsfs}

\usepackage[hidelinks,linktocpage]{hyperref}
\hypersetup{
    colorlinks,
    linkcolor={red!70!black},
    citecolor={green!70!black},
    urlcolor={blue!70!black}
}
\usepackage[usenames,dvipsnames]{xcolor}	%colored text
\usepackage{upgreek}	% non italic μ symbol

\newcommand{\bb}{$0\nu \beta \beta$} % neutrinoless double beta decay
\newcommand{\bbvv}{$2\nu \beta \beta$} % 2 neutrino double beta decay

\newcommand{\mbb}{m_{\beta \beta}} % m_ee
\newcommand{\mbbM}{\mbb^{\max}} % m_ee max
\newcommand{\mbbm}{\mbb^{\min}} % m_ee min

\newcommand{\taubb}{\mbox{\normalsize $t^{\mbox{\tiny $\nicefrac{1}{2}$}}$}} % t_0v
\newcommand{\NH}{$\mathcal{NH}$} % Normal Hierarchy
\newcommand{\IH}{$\mathcal{IH}$} % Inverted Hierarchy

\newcommand{\ckky}{counts$/\keV/\kg/\yr$}

\newcommand{\meV}{\text{meV}}	\newcommand{\eV}{\text{eV}}	\newcommand{\keV}{\text{keV}}		
\newcommand{\MeV}{\text{MeV}}	\newcommand{\GeV}{\text{GeV}}	\newcommand{\TeV}{\text{TeV}}

\newcommand{\fm}{\text{fm}}	\newcommand{\cm}{\text{cm}}

\newcommand{\yr}{\text{yr}}	\newcommand{\kg}{\text{kg}}

\newcommand{\el}{\text{e}}	\newcommand{\pr}{\text{p}}

\newcommand{\xcCL}{\mbox{\tiny(90\% C.\,L.)}}		
\newcommand{\CL}{\mbox{C.\,L.}}		
\newcommand{\zb}{\mbox{\tiny(zero bkg)}}		
\newcommand{\Nev}{N_\mathrm{events}}		

\newcommand{\meas}{\mbox{\tiny meas}}		

\newcommand{\npartial}{\partial \hspace{-5.5pt}/}	% negation of \partial symbol
\newcommand{\notaA}{$^{\mbox{\scriptsize \,a}}$}
\newcommand{\notaB}{$^{\mbox{\scriptsize \,b}}$}
\newcommand{\notaC}{$^{\mbox{\scriptsize \,c}}$}

\begin{document}

%	\title{Expectations for neutrinoless double beta decay}
	\title{Neutrinoless double beta decay: 2015 review}

	\author{Stefano Dell'Oro}
		\email{stefano.delloro@gssi.infn.it}
	\author{Simone Marcocci}
		\email{simone.marcocci@gssi.infn.it}
		\affiliation{INFN, Gran Sasso Science Institute, Viale F.\ Crispi 7, 67100 L'Aquila, Italy \\}%	
	\author{Matteo Viel}
		\email{viel@oats.inaf.it}
		\affiliation{INAF, Osservatorio Astronomico di Trieste, Via G.\,B.\ Tiepolo 11, 34131 Trieste, Italy \\}%	
		\affiliation{INFN, Sezione di Trieste, Via Valerio 2, 34127 Trieste, Italy \\}%	
	\author{Francesco Vissani}
		\email{francesco.vissani@lngs.infn.it}
		\affiliation{INFN, Laboratori Nazionali del Gran Sasso, Via G.\ Acitelli 22, 67100 Assergi (AQ), Italy \\}%	
		\affiliation{INFN, Gran Sasso Science Institute, Viale F.\ Crispi 7, 67100 L'Aquila, Italy \\}%	

	\date{\today}

	\begin{abstract}
		
		The discovery of neutrino masses through the observation of oscillations boosted the importance of 
		neutrinoless double beta decay (\bb). In this paper, we review the main features of this process, underlining 
		its key role both from the experimental and theoretical point of view. In particular, we contextualize the 
		\bb~in the panorama 
		of lepton-number violating processes, also assessing some possible particle physics mechanisms mediating the process.
		Since the \bb~existence is correlated with neutrino masses, we also review the state-of-art of the 
		theoretical understanding of neutrino masses.
		In the final part, the status of current \bb~experiments is presented and the prospects for the future hunt 
		for \bb~are discussed. Also, experimental data coming from cosmological
		surveys are considered and their impact on \bb~expectations is examined.
		
	\end{abstract}

	\pacs{14.60.Pq,% Neutrino Mass and mixing
		23.40.-s% beta decay, double beta decay, e/mu capture
		\hspace{173pt}%
		DOI: \href{http://www.hindawi.com/journals/ahep/2016/2162659/}{10.1155/2016/2162659}
		}

	\maketitle

%	{\small \tableofcontents}

%------------------------------------------------
%------------------------------------------------
	%------------------------------------------------
%------------------------------------------------
\section{Introduction}	

\begin{figure}[bt]
		\centering
		\includegraphics[width=.6\columnwidth]{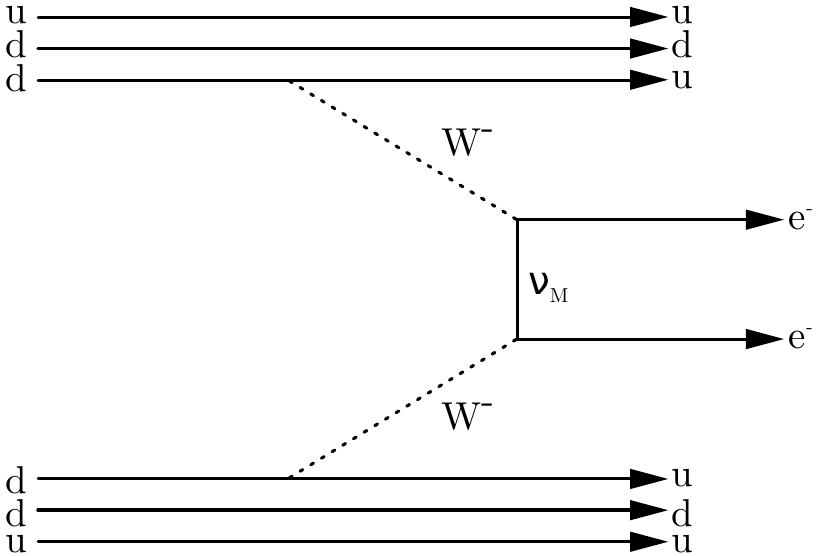}
		\caption{Diagram of the \bb~process due to the exchange of massive Majorana neutrinos, here denoted generically 
		by $\nu_{\mbox{\tiny{M}}}$.}
		\label{fig:DBD-diagram}
	\end{figure}

	In 1937, almost ten years after Paul Dirac's ``The quantum theory of electron''~\cite{Dirac:1928hu,Dirac:1928ej}, 
	Ettore Majorana proposed a new way to represent fermions in a relativistic quantum field theory~\cite{Majorana:1937vz}, 
	and remarked that this could be especially useful for neutral particles. 
	A single Majorana quantum field characterizes the situation in which particles and antiparticles coincide, 
	as it happens for the photon. 
	Giulio Racah stressed that such a field could fully describe massive neutrinos,
	noting that the theory by E.\ Majorana leads to physical predictions essentially 
	different from those coming from Dirac theory,~\cite{Racah:1937qq}.
	Two years later, Wendell Furry~\cite{Furry:1939qr} studied within this scenario a new process similar to the 
	``double beta disintegration'', introduced by Maria Goeppert-Mayer in 1935~\cite{GoeppertMayer:1935qp}. 
	It is the double beta decay without neutrino emission, or \emph{neutrinoless double beta decay} (\bb). 
	This process assumes a simple form, namely
	\begin{equation}
	\label{eq:DBD}
		(A,Z) \rightarrow (A,Z+2) + 2\el^-.
	\end{equation}
	The Feynman diagram of the \bb~process, written in terms of the particles we know today and of massive Majorana 
	neutrinos, is given Fig.\ \ref{fig:DBD-diagram}.
	 
	The main and evident feature of the \bb~transition is the explicit violation of the number of leptons and, 
	more precisely, the creation of a pair of electrons. 
	The discovery of \bb~would therefore demonstrate that lepton number is not a symmetry of nature.
	This, in turn, would support the exciting theoretical picture that leptons played a part in the creation of the 
	matter-antimatter asymmetry in the universe.
	
	In the attempt to investigate the nature of the \bb~process, various other theoretical possibilities were considered, 
	beginning by postulating new super-weak interactions~\cite{Feinberg&Goldhaber:1959,Pontecorvo:1968wp}. 
	However, the general interest has always remained focused on the neutrino mass mechanism. 
%	 that is supported by two important circumstances:
	In fact, this scenario is supported by two important facts: 
	\begin{enumerate} 
		\item On the theoretical side, the triumph of the Standard Model (SM) of electroweak interactions in the 
			1970s~\cite{Glashow:1961tr,Weinberg:1967tq,Salam:1968rm} led to formulate the discussion of new physics 
			signals using the language of effective operators, suppressed by powers of the new physics mass scale. 
		 	There is only one operator that is suppressed only by one power of the new mass scale and 
			violates the global symmetries of the SM or, more precisely, the lepton number: 
			it is the one that gives rise to Majorana neutrino masses \cite{Weinberg:1979sa} 
			(see also Refs.~\cite{Minkowski:1977sc,Yanagida:1979as,GellMann:1980vs,Mohapatra:1979ia}).
		\item On the experimental side, some anomalies in neutrino physics, which emerged in throughout 30 years, 
			found their natural explanation in terms of oscillations of massive neutrinos \cite{Pontecorvo:1957cp}. 
			This explanation was confirmed by several experiments 
			(see Refs.\ \cite{Strumia:2006db,GonzalezGarcia:2002dz} for reviews). 
			Thus, although oscillation phenomena are not sensitive to the Majorana nature of neutrinos~\cite{Bilenky:1980cx}, 
			the concept of neutrino mass has changed its status in physics, from the one of hypothesis to the one of fact. 
			This, of course, strengthened the case for light massive neutrinos to play a major role for the \bb~transition.
%		\footnote{%
%			The general operator analysis of Ref.\ \cite{Weinberg:1979sa} has been preceded by several works where the 
%			conclusion is obtained within specific models of new physics, e.\,g.\ in Refs.\ 
%			\cite{Minkowski:1977sc,Ramond:1979py,GellMann:1980vs,Yanagida:1979as,Glashow:1979nm,Mohapatra:1979ia}.}
	\end{enumerate}
	For these reasons, besides being an interesting nuclear process, \bb~is a also a key tool for studying neutrinos, 
	probing whether their nature is the one of Majorana 
	particles and providing us with precious information on the neutrino mass scale and ordering. 
	Even though the predictions of the \bb~lifetime still suffer from numerous uncertainties, great progresses in assessing 
	the expectations for this process have been and are being made. These will be discussed later in this review.

%------------------------------------------------
\vspace{10pt}
\paragraph*{\bf About the present review}

	In recent years, several review papers concerning neutrinoless double beta decay have been written. 
	They certainly witness the vivid interest of the scientific community in this topic.
	Each work emphasizes one or more relevant aspects such as 
	the experimental part~\cite{Elliott:2012sp,Giuliani:2012zu,GomezCadenas:2011it,Schwingenheuer:2012zs,Cremonesi:2013vla},
	the nuclear physics~\cite{Vergados:2012xy,Vogel:2012ja},
	the connection with neutrino masses~\cite{Petcov:2013poa,Bilenky:2014uka},
	other particle physics mechanisms~\cite{Rodejohann:2011a,Rodejohann:2012xd,Deppisch:2012nb,Pas:2015eia},
	\dots
	The present work is not an exception. We mostly focus on the first three aspects. 
	This choice is motivated by our intention to follow the theoretical ideas that describe the most plausible expectations 
	for the experiments. 
	In particular, after a general theoretical introduction (Secs.~\ref{sec:lepton} and \ref{sec:mechanism}), we examine 
	the present knowledge on neutrino masses (Sec.~\ref{sec:masses}) and the status of expectations from nuclear physics 
	in (Sec.~\ref{sec:nuclear}). Then we review the experimental situation (Sec.~\ref{sec:experimental}) and emphasize the link 
	between neutrinoless double beta decay and cosmology (Sec.\ref{sec:cosm_bounds}).

	A more peculiar aspect of this review is the effort to follow the historical arguments, without worrying too much about 
	covering once more well-known material or about presenting an exhaustive coverage of the huge recent 
	literature on the subject.
	Another specific characteristic is the way the information on the neutrino Majorana mass is dealt with.
	In order to pass from this quantity to the (potentially measurable) decay rate, we have to dispose of 
	quantitative information on the neutrino masses and on the matrix elements of the transition, which in turn requires the 
	description of the nuclear wave functions and of the operators that are implied. 
	Therefore, our approach is to consider the entire available information on neutrino masses and, in particular, 
	the one coming from cosmology. We argue that the recent progresses (especially those coming from the Planck satellite 
	data~\cite{Planck:2015xua}) play a very central role for the present discussion.
	On the other side, the matrix elements have to be calculated (rather than measured) and are thus subject to 
	uncertainties which are difficult to assess reliably. Moreover, the adopted methods of calculation do not precisely 
	reproduce other measurable quantities (single beta decay, two-neutrino double beta decay, \dots).
	We thus prefer to adopt a cautious/conservative assessment of the theoretical ranges of these matrix elements.	

	We would like to warn the Reader that other attitudes in the discussion are surely possible, 
	and it is indeed the case for some of the mentioned review works. Using less stringent limits from cosmology 
	and disregarding the uncertainties from nuclear physics is equivalent to assume the most favorable situation for 
	the experiments.
	This could be considered beneficial for the people involved in experimental search for the neutrinoless double beta decay. 
	However, we prefer to adhere to a more problematic view in the present work, simply because we 
	think that it more closely reflects the present status of facts.
	Considering the numerous experiments involved in the field, we deem that an updated discussion on these two issues has 
	now become quite urgent. 
	This will help us to assess and appreciate better the progresses expected in the close future, concerning the 
	cosmological measurements of neutrino masses and perhaps also the theoretical calculations of the relevant nuclear 
	matrix elements.

%------------------------------------------------
%------------------------------------------------
	%------------------------------------------------
%------------------------------------------------
\section{The total lepton number}
\label{sec:lepton}

	\begin{figure}[tb]
		\centering
		\includegraphics[width=1.\columnwidth]{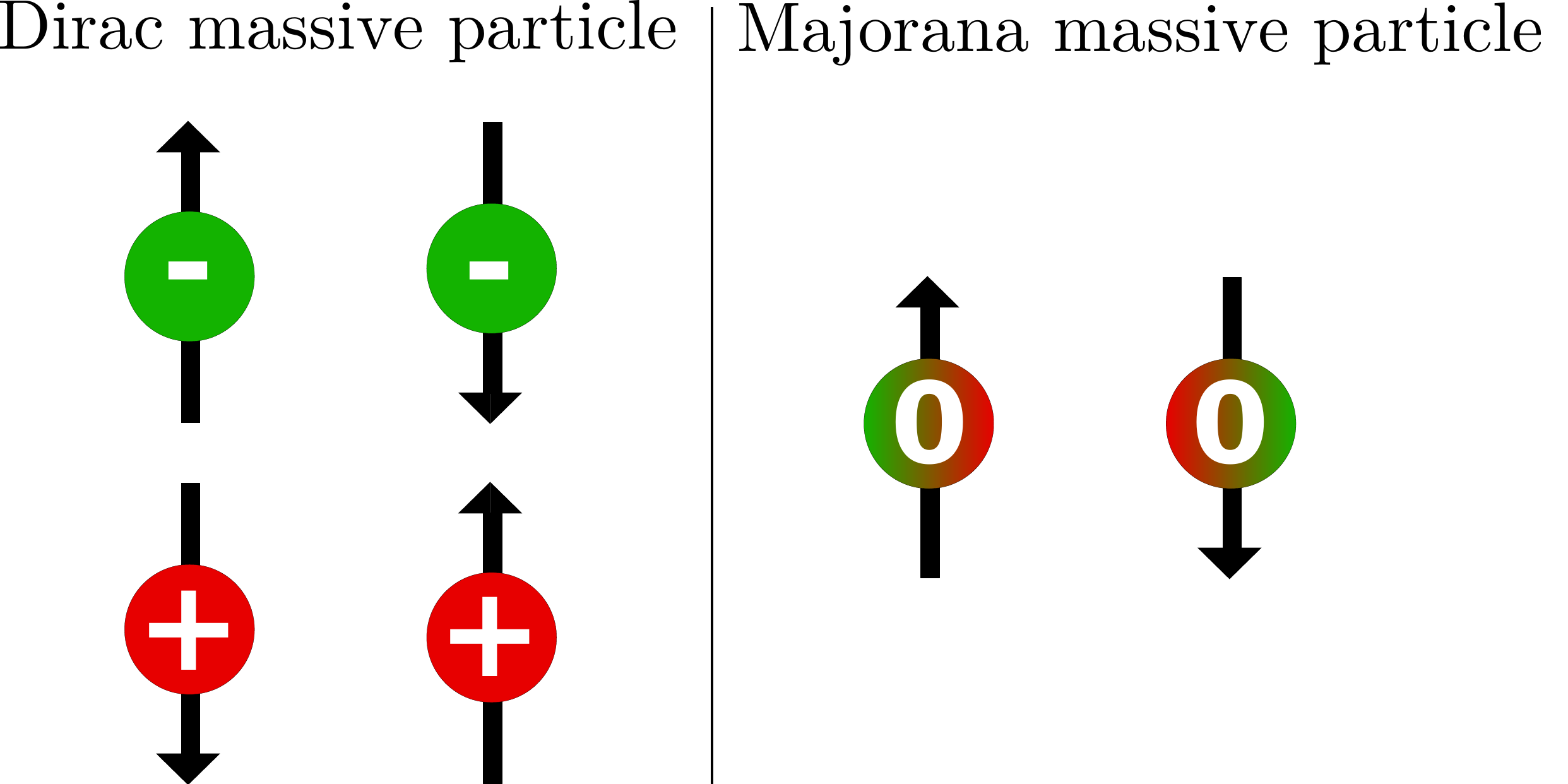}
		\caption{Massive fields in their rest frames. The arrows show the possible directions of the spin. 
			(Left) The 4 states of Dirac massive field. The signs indicate the charge that distinguishes particles and 
			antiparticles, e.\,g. the electric charge of an electron. 
			(Right) The 2 states of Majorana massive field. 
			The symbol ``zero'' indicates the absence of any $U(1)$ charge: particles and antiparticles coincide.}
		\label{fig:DiracMajoDiagr}
	\end{figure}
		
	No elementary process where the number of leptons or the number of hadrons varies has been observed yet. 
	This suggests the hypothesis that the lepton number $L$ and the baryon $B$ are subject to conservations laws. 
	However, we do not have any deep justification for which these laws should be exact. 
	In fact, it is possible to suspect that their validity is just approximate or circumstantial, since it is  
	related to the range of energies that we can explore in laboratories.%
	\footnote{Notice also that the fact that neutral leptons (i.\,e.\ neutrinos or antineutrinos) are very difficult to 
		observe restricts the experimental possibilities to test the total lepton number.}

	In this section, we discuss the status of the investigations on the total lepton number in the SM and in 
	a number of minimal extensions, focusing on theoretical considerations.
	In particular, we introduce the possibility that neutrinos are endowed with Majorana mass and consider a 
	few possible manifestations of lepton number violating phenomena. The case of the \bb~will instead 
	be addressed in the rest of this work.
	
%------------------------------------------------
\subsection{$B$ and $L$ symmetries in the SM}

	The SM in its minimal formulation has various global symmetries, including $B$ and $L$, which are called ``accidental''.
	This is due to the specific particle content of the model and to the hypothesis of renormalizability.
	Some combinations of these symmetries, like for example ``$B-L$'', are conserved also non-perturbatively.
	This is sufficient to forbid the \bb~transition completely in the SM. In other words, 
	a hypothetical evidence for such a transition would directly point out to physics beyond the SM.
	At the same time, the minimal formulation of the SM implies that neutrinos are massless, and this contradicts 
	the experimental findings. Therefore, the question of how to modify the SM arises, and this in turn poses the 
	related burning question concerning the nature of neutrino masses.
	
%	Before proceeding, we recall the existence of three families of quarks and leptons, each one including 15 elementary
%	fermions. These are listed for reference in Tab.~\ref{tab:fields} along with their gauge numbers and assignments of 
%	$L$ and $B$. 

%------------------------------------------------
\subsection{Majorana neutrinos}
\label{sec:Majorana_nu}

	In 1937, Majorana proposed a theory of massive and ``real'' fermions,~\cite{Majorana:1937vz}. This theory contains less 
	fields than the one used by Dirac for the description of the electron~\cite{Dirac:1928hu,Dirac:1928ej} and, 
	in this sense, it is simpler.
	Following the formalism introduced in 1933 by Fermi when describing the $\beta$ decay~\cite{Fermi:1933jpa}, the condition 
	of reality for a quantized fermionic field can be written as:
	\begin{equation}
		\chi = C \bar{\chi}^t
	\end{equation}
	where $C$ is the charge conjugation matrix, while $\bar{\chi}~\equiv~\chi^\dagger \gamma_0$ is the Dirac conjugate of the 
	field.
%	(similarly the transpose, conjugate and hermitian conjugate of $u$ are denoted by $u^t$,  $u^*$ and $u^\dagger$, 
%	respectively). 
	In particular, Majorana advocated a specific choice of the Dirac $\gamma$-matrices, such that 
	$C \gamma_0 ^t = 1$, which simplifies various equations.
	The free particle Lagrangian density formally coincides with the usual one:
	\begin{equation}
	\label{eq:lagrangian}
		\mathcal{L}_\mathrm{Majorana}=\frac{1}{2} 
		\bar{\chi} (i \npartial - m) \chi.
	\end{equation}

	Following Majorana's notations, the decomposition of the quantized 
	fields into oscillators is:
	\begin{equation}
		\chi(x) = \sum_{\mathbf{p},\lambda} \left[ {a}({\mathbf{p}\lambda})\ \psi(x; {\mathbf{p}\lambda}) + 
		{a}^*({\mathbf{p}\lambda})\ \psi^*(x; {\mathbf{p}\lambda}) \right]
	\end{equation}
	where $\lambda = \pm 1$ is the relative orientation between the spin and the momentum (helicity).
	We adopt the normalization for the wave functions: $\int d\mathbf{x} |\psi(t,\mathbf{x})|^2 = 1$, 
	and for the oscillators:
	${a}({\mathbf{p}\lambda})\,{a}^*({\mathbf{p'}\lambda'}) + 
	{a}^*({\mathbf{p'}\lambda'})\,{a}({\mathbf{p}\lambda}) = \delta_{\mathbf{p}\mathbf{p'}} \delta_{\lambda\lambda'}$.
	For any value of the momentum, there are 2 spin (or helicity) states:
	\begin{equation}
		{a}^*({\mathbf{p}\, +}) | \mbox{vac.} \rangle = |\mathbf{p}\uparrow\ \rangle \quad \mbox{and} \quad
		{a}^*({\mathbf{p}\, -}) | \mbox{vac.} \rangle = |\mathbf{p}\downarrow\ \rangle.
	\end{equation}
	Fig.~\ref{fig:DiracMajoDiagr} illustrates the comparison between the particle content both of a Dirac and a Majorana field 
	in the case $\mathbf{p}=0$ (rest frame).

	Evidently, a Majorana neutrino is incompatible with any $U(1)$ transformation, e.\,g.\ $L$ or the weak hypercharge 
	(that is however broken in the vacuum). 
%	In particular, the latter is a gauge symmetry, but it is subject to spontaneous symmetry breaking in the SM. 
%	The former instead will be in general violated by the presence of Majorana mass. 
	In general, $L$ will be violated by the presence of Majorana mass. 

	In the SM, the neutrino field appears only in the combination
	\begin{equation}
		\psi_L = P_L \ \psi
	\end{equation}
	where $P_L \equiv (1-\gamma_5)/2$ is the so-called chiral projector (Table \ref{tab:fields}). 
	It is then possible to implement the hypothesis of Majorana in the most direct way by defining the real field
	\begin{equation}\label{mimpl}
		\chi \equiv \psi_L + C \bar{\psi}_L^t.
	\end{equation}
	In fact, we can conversely obtain the SM field by a projection:
	\begin{equation}
	 \psi_L \equiv P_L \chi.
	\end{equation}

%------------------------------------------------
\subsection{Ultrarelativistic limit and massive neutrinos}
\label{sec:ultra_rel}
	
	\begin{figure}[tb]
		\centering
		\includegraphics[width=1.\columnwidth]{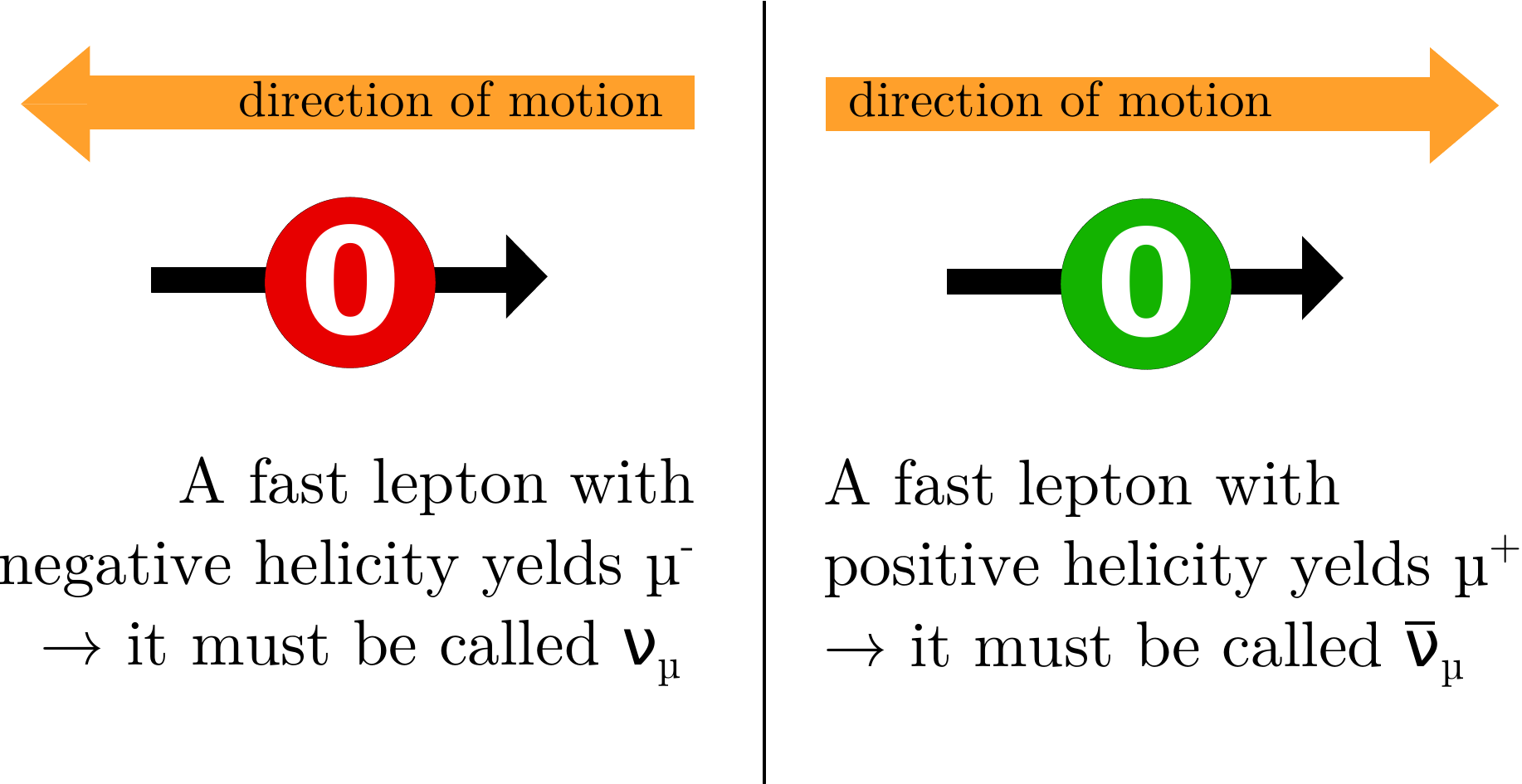}
		\caption{The chiral nature of weak interactions allows us to define what is a neutrino and what it an antineutrino 
			\emph{in the ultra-relativistic limit}, when chirality coincides with helicity and the value of the mass plays only 
			a minor role.}
		\label{fig:hel}
	\end{figure}

	The discovery that parity is a violated symmetry in weak interactions~\cite{Lee:1956qn,Wu:1957my} was soon followed 
	by the understanding that the charged current (which contains the neutrino field) always includes the left chiral 
	projector~\cite{Landau:1957tp,Lee:1957qr,Salam:1957st} (see Secs.\ \ref{sec:Majorana_nu} and \ref{sec:nu_exchange}).

	It is interesting to note the following implication. Within the hypothesis that neutrino are massless, the Dirac 
	equation becomes equivalent to two Weyl equations~\cite{Weyl:1929fm} corresponding to the Hamiltonian functions
	\begin{equation}
	\label{eq:Weyl}
		H_{\nu/\bar\nu}= \, \mp c \mathbf{p} \, \sigma
	\end{equation} 
	where $\sigma$ are the three Pauli matrices and the two signs apply to the neutral leptons that thanks to the 
	interaction produce 
	charged leptons of charge $\mp 1$, respectively. In other words, we can define these states as neutrinos and 
	antineutrinos, respectively. 
	Moreover, by looking at Eq.~(\ref{eq:Weyl}), one can see that the energy eigenstates are also helicity eigenstates.
	More precisely, the spin of the neutrino (antineutrino) is antiparallel (parallel) to its momentum.  
	See Fig.~\ref{fig:hel} for illustration.

	The one-to-one connection between chirality and helicity holds only in the ultrarelativistic limit,
	when the mass of the neutrinos is negligible. 
	This is typically the case that applies for detectable neutrinos, since the weak interaction cross sections are 
	bigger at larger energies. 
	However, these remarks do not imply in any way that neutrinos are massless. 
	On the contrary, we know that neutrinos are massive. 
	
	A consequence of the chiral nature of weak interactions is that, if we assume that neutrinos have the type of mass
	introduced by Dirac, we have a couple of states that are sterile under weak interactions in the ultrarelativistic
	limit. 
	Conversely, the fact that the left chiral state exists can be considered a motivation in favor of the hypothesis 
	of Majorana. In fact, this does not require the introduction of the right chiral state, as instead required by the 
	Dirac hypothesis.  
	Most importantly, it should be noticed that in the case of Majorana mass 
	\emph{it is not possible to define the difference between a neutrino and an antineutrino in a Lorentz invariant way.}	
	%\emph{it is not possible to give 
	%a definition of neutrino which is invariant under any Lorentz transformation}.

%------------------------------------------------
\subsection{Right-handed neutrinos and unified groups}
\label{sec:rh}
	
	The similarity between $L$ and $B$ is perceivable already within the SM. 
	The connection is even deeper within the so called Grand Unified Theories (GUT), i.\,e.\ gauge theories with a single 
	gauge coupling at a certain high energy scale. 
	The standard prototypes are $SU(5)$ \cite{Georgi:1974sy} and $SO(10)$ \cite{Georgi:1975qb,Fritzsch:1974nn}.
	GUTs undergo a series of symmetry-breaking stages at lower energies, eventually reproducing the SM.
	They lead to predictions on the couplings of the model and suggest the existence of new particles, even if theoretical 
	uncertainties make it difficult to obtain reliable predictions.
	The possibilities to test these theories are limited, and major manifestations could be violations of $L$ and $B$.
	
\begin{figure}[tb]
		\centering
		\includegraphics[width=.8\columnwidth]{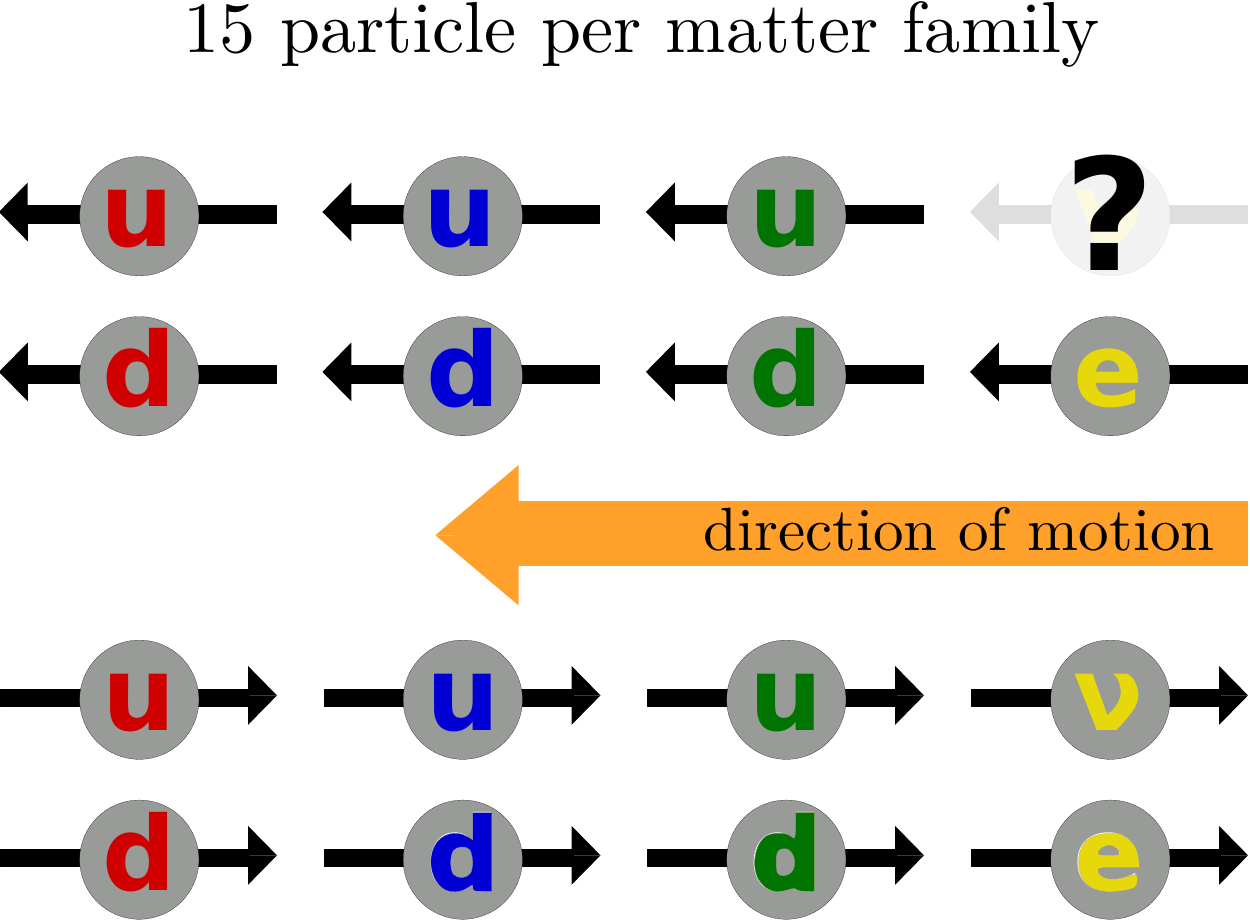}
		\caption{Helicity of the 15 massless matter particles contained in each family of the SM  
			(see Table~\ref{tab:fields}). The arrow gives the direction of the momentum.}
		\label{fig:matterpart}
	\end{figure}

	The matter content of GUT theories is particularly relevant for the discussion.
	In fact, the organization of each family of the SM  
	suggests the question whether right-handed neutrinos (RH) exist along with the other 7 RH particles
	(Fig.~\ref{fig:matterpart}). 
	This question is answered affirmatively in some extensions of the SM. 
	For example, this is true for gauge groups that also include a $SU(2)_{\mbox{\tiny R}}$ factor, 
	on top of the usual $SU(2)_{\mbox{\tiny L}}$ factor. 
	In the $SO(10)$ gauge group, which belongs to this class of models, each family of matter includes the 15 SM particles 
	plus 1 RH neutrino.
	
	It should be noted that RH neutrinos do not participate in SM interactions and can therefore be endowed 
	with a Majorana mass $M$, still respecting the SM gauge symmetries. 
	However, they do participate in the new interactions, and more importantly for the discussion, they can mix with 
	the ordinary neutrinos via the Dirac mass terms, $m^{\mbox{\tiny Dirac}}$. 
	Therefore, in presence of RH neutrinos, the SM Lagrangian (after spontaneous symmetry breaking) will 
	include the terms,
	\begin{equation}
		\mathcal{L}_\mathrm{mass} = 
		- \bar{\nu}_{\mbox{\tiny R}i }\ m^{\mbox{\tiny Dirac}}_{\ell i}\ \nu_{\mbox{\tiny L}\ell} + 
		\frac{1}{2}\overline{\nu}_{\mbox{\tiny R}i }\  M_i\, C\ \bar{\nu}_{\mbox{\tiny R}i }^t + h.\,c.
	\end{equation}
	where $\ell=e,\mu,\tau$ and $i=1,2,3$.
	It is easy to understand that, at least generically, this framework implies that the lepton number is broken. 

	Let us assume the existence of RH neutrinos, either embedded in an unified group or not, and let us suppose that 
	they are heavy (this happens, for example, if the scale of the new gauge bosons is large and the couplings of the RH
	neutrinos to the scalar bosons implementing spontaneous symmetry breaking of the new gauge group are not small).
	In this case, upon integrating away the heavy neutrinos from the theory, the light neutrinos will receive Majorana 
	mass, with size inversely proportional to the mass of the 
	RH ones,~\cite{Minkowski:1977sc,Yanagida:1979as,GellMann:1980vs,Mohapatra:1979ia}.
	This is the celebrated Type I Seesaw Model.
	In other words, the hypothesis of heavy RH neutrinos allows us to account for the observed small mass of the 
	neutrinos. Unfortunately, we cannot predict the size of the light neutrino mass precisely, unless we know both 
	$M$ and $m^{\mbox{\tiny Dirac}}$.
	
%	In principle, RH neutrinos could be also quite light.
%	An extreme possibility is that some of them have masses of the order of eV or less and give rise to new 
%	flavor oscillations, although this is not favored by the existing information from cosmology 
%	(see Sec.\ \ref{sec:nucosmology}).
	In principle, RH neutrinos could also be quite light.
	An extreme possibility is that some of them have masses of the order of eV or less and give rise to new 
	flavor oscillations observable in terrestrial laboratories~\cite{Borexino:2013xxa,An:2014bik,Ashenfelter:2015uxt}.
	This could help to address some experimental anomalies~\cite{Abazajian:2012ys,Gariazzo:2013gua}.
	However, as it has been known since long~\cite{Dolgov:2003sg,Cirelli:2004cz}, that the presence of eV neutrinos
	would also imply large effects in cosmology, both in the number of relativistic species and in the value of the 
	neutrino mass. 
	These effects are not in agreement with the existing information from cosmology (see Sec.~\ref{sec:nucosmology})
	and, for this reason, we will not investigate this hypothesis further 
	(we refer an interested Reader to the various discussion on the impact of eV neutrinos on the
	\bb, see e.\,g.\ Refs. \cite{Barry:2011wb,Girardi:2013zra,Gariazzo:2015rra}).
	
	In view of the evidences of neutrinos masses, theories like $SO(10)$ are particularly appealing, 
	since they offer a natural explanation of light Majorana neutrino masses. However, a complete theory able 
	to link in a convincing way fermion masses (including those of neutrinos) and to provide us reliable 
	predictions of new phenomena, such as \bb, does not exist yet.
	Despite many attempts were made in the past, it seems that this enterprise is still in its initial stages. 

%------------------------------------------------
\subsection{Leptogenesis}

	Although particles and antiparticles have the same importance in our understanding of particle physics, we know that the 
	Universe contains mostly baryons rather than antibaryons.%
	\footnote{The lepton number in the Universe is probed much less precisely. While we know that cosmic neutrinos 
		and antineutrinos are abundant, it is not easy to measure their asymmetry which, according to standard cosmology,
		should be very small. However, we expect to have the same number of electrons and protons to guarantee the overall 
		charge neutrality.}
	In 1967, Sakharov proposed a set of necessary conditions to generate the cosmic baryon asymmetry~\cite{Sakharov:1967dj}.
	This has been the beginning of many theoretical attempts to ``explain'' these observations in terms of new  physics. 
	
	In the SM, although $L$ and $B$ are not conserved separately at the non-perturbative 
	level~\cite{'tHooft:1976up,Kuzmin:1985mm,Harvey:1990qw}, the observed value of the Higgs mass is
	not big enough to account for the observed baryon asymmetry~\cite{Bochkarev:1987wf,Kajantie:1995kf}. 
	New violations of the global $B$ or $L$ are needed.
	
	An attractive theoretical possibility is that RH neutrinos not only enhance the SM endowing neutrinos with Majorana 
	mass, but also produce a certain amount of leptonic asymmetry in the Universe. This is subsequently converted 
	into a baryonic asymmetry thanks to $B+L$ violating effects, which are built-in in the SM. It is the 
	so called Leptogenesis mechanism, and it can be wittingly described by asking the following question: 
	\emph{do we all descend from neutrinos?} 
	The initial proposal of Leptogenesis dates back to 1980s~\cite{Fukugita:1986hr}, and there is a large 
	consensus that this type of idea is viable and attractive. 
	Subsequent investigators showed that the number of alternative theoretical possibilities is very large and, in
	particular, that there are other possible sources of $L$ violations besides RH neutrinos.
	Conversely, the number of testable possibilities is quite limited~\cite{Shaposhnikov:2012_COSMO}. 

	We believe that it is important to stay aware of the possibility of explaining the baryon number excess through
	Leptogenesis theories. However, at the same time, one should not overestimate the heuristic power of this theoretical 
	scheme, at least within the presently available information. 

%------------------------------------------------
\subsection{Neutrino nature and cosmic neutrino background}

	The Big Bang theory predicts that the present Universe is left with a residual population of $\sim 56$ 
	non-relativistic neutrinos and 
	antineutrinos per $\cm^3$ and per species. It constitutes a Cosmic Neutrino Background (C$\nu$B).
	Due to their very low energy, Eq.\ (\ref{eq:Weyl}) does not hold for these neutrinos. This happens because
	at least two species of neutrinos are non-relativistic. The detection of this C$\nu$B could 
	therefore allow to understand which hypothesis (Majorana or Dirac) applies for the neutrino description. 

	Let us assume to have a target of 100\,g of \ce{^3H}. Electron neutrinos can be detected through 
	the reaction~\cite{Weinberg:1962zza,Cocco:2007za}
	\begin{equation}
		\nu_\el + \ce{^3H} \to \ce{^3He} + \el^-.
	\end{equation}
	In the standard assumption of a homogeneous Fermi-Dirac distribution of the C$\nu$B, we 
	expect $\sim 8$ events per year if neutrinos are Majorana particles and about half if the Dirac hypothesis 
	applies~\cite{Long:2014zva}.
	Indeed, in the former case, the states with positive helicity (by definition, antineutrinos) will act just as 
	neutrinos, since they are almost at rest. 
	Instead, in the latter case, they will remain antineutrinos and thus they will not react.
	
	It can be noticed that the signal rate is not prohibitively small, but the major difficulty consists in attaining 
	a sufficient energy resolution to keep at a manageable level the background from beta decay.
	We will not discuss further the feasibility of such an experiment, 
	and refer to Refs.~\cite{Cocco:2007za,Long:2014zva} for more details.

%------------------------------------------------
%------------------------------------------------
	%------------------------------------------------
%------------------------------------------------
\section{Particle physics mechanisms for \bb}
\label{sec:mechanism}

	\begin{center}
	\begin{table*}[tb]
	\caption{List of the matter particles in the SM. The label ``singlet'' is often replaced with ``right'' and likewise 
		for ``doublet'' it can become ``left''. Hypercharge is assigned according to $Q=T_{\mbox{\tiny 3L}}+Y$.
		The chirality of a field (and all its U(1) numbers) can be exchanged by considering the charge 
		conjugate field; e.\,g.\ $\el_{\mbox{\tiny L}}^c\equiv C\bar{\el}_{\mbox{\tiny R}}^t $ has electric charge $+1$
			and leptonic charge $-1$.}
	\begin{ruledtabular}
	\begin{tabular}{l lll l l l}
		Name	&Field 	&SU(3)$_{\mbox{\tiny c}}$ 	&SU(2)$_{\mbox{\tiny L}}$	&U(1)$_{\mbox{\tiny Y}}$	&lepton	&baryon	\\ 
				& symbol &multiplicity 					&multiplicity 					&charge 					&number	$L$	&number $B$ \\
		\hline
		quark doublet 				&$q_{\mbox{\tiny L}}$		&3		&2		&$+1/6$	&0		&$1/3$	\\ 
		singlet up quark 			&$u_{\mbox{\tiny R}}$		&3		&1		&$+2/3$	&0		&$1/3$	\\ 
		singlet down quark 		&$d_{\mbox{\tiny R}}$		&3		&1		&$-1/3$	&0		&$1/3$	\\[+3pt]
		lepton doublet 			&$l_{\mbox{\tiny L}}$		&1		&2		&$-1/2$	&1		&$0$		\\ 
		singlet charged lepton	&$\el_{\mbox{\tiny R}}$		&1		&1		&$-1$		&1		&$0$		\\
%		higgs doublet			 	&$H$ 								&1		&2		&+1/2		&0		&$0$		\\ \hline
	\end{tabular}
	\end{ruledtabular}
	\label{tab:fields}
	\end{table*}
	\end{center}

	In this section, we focus on one of the most appealing lepton number violating process, the \bb.
	The exchange of light Majorana neutrinos is up to now the most appealing mechanism to eventually 
	explain the \bb. Some reasons justifying this statement were already mentioned, but here a more elaborate 
	discussion is proposed.
	In particular, we review the basic aspects of the light neutrino exchange mechanism for \bb~and compare it to 
	other ones. 
	Moreover, the possibility of inferring the size of neutrino masses from a hypothetical observation of \bb~and 
	of constraining (or proving the correctness) some alternative mechanisms with searches at the accelerators
	are also discussed.

%------------------------------------------------
\subsection{The neutrino exchange mechanism}
\label{sec:nu_exchange}

	The definition of a key quantity for the description of the neutrino exchange mechanism needs to be introduced.
	It is the propagator of virtual Majorana neutrinos.
	Due to the reality condition, Eq.~(\ref{eq:lagrangian}) can lead to new types of propagators 
	that do not exist within the Dirac theory. In fact, in this case we can use the antisymmetry of the charge 
	conjugation matrix and get:
	\begin{equation}
		\langle 0 | T[ \chi(x) \chi(y) ] | 0 \rangle = -\Delta(x-y)\,C
	\end{equation} 
	where $\Delta$ denotes the usual propagator,
	\begin{equation}
		\Delta(x) \equiv \int \frac{d^4q}{(2\pi)^4} \frac{i (\hat{q}+m) }{q^2-m^2+i 0} \el^{-iq x}.
	\end{equation}

	In the low energy limit (relevant for $\beta$ decay processes) the interaction of neutrinos are well described by the 
	current-current 4-fermion interactions, corresponding to the Hamiltonian density
	\begin{equation}
		\mathcal{H}_\mathrm{Fermi} = \frac{G_F}{\sqrt{2}} \ J^{a\dagger} J_a
	\end{equation}
	where $G_F$ is the Fermi coupling, and we introduced the current 
	$J^a = J^a_\mathrm{lept} + J^a_\mathrm{hadr}$ for $a=0,1,2,3$, that decreases 
	the charge of the system (its conjugate, $ J_a^\dagger$, does the contrary). 
	In particular, the leptonic current
	\begin{equation}
	\label{eq:lept_current}
		J^a_\mathrm{lept} = \sum_{\ell=\el,\mu,\tau}
		\bar{\psi}_\ell  \ \gamma^a (1-\gamma_5)\ \psi_{\nu_\ell}
	\end{equation} 
	defines the ordinary neutrino with ``flavor'' $\ell$. 
	In order to implement the Majorana hypothesis, one can use Eq.\ (\ref{mimpl}) and introduce the field 
	$\chi = \psi_L + C \bar{\psi}_L^t$.
	\begin{comment}
	It is worth noting that the neutrino appears only in the combination 
	\begin{equation}
		\psi_L = P_L \ \psi
	\end{equation}
	where $P_L =(1-\gamma_5)/2$ is the so-called chiral projector. 
	It is then possible to implement the hypothesis of Majorana most directly, just by defining the real field
	\begin{equation}
		\chi \equiv \psi_L + C \bar{\psi}_L^t
	\end{equation}
	and by replacing it in Eq.\ (\ref{eq:lept_current}).
	\end{comment}
	Nothing changes in the interactions if one substitutes the field $\psi_{\nu_\ell}$ 
	with the corresponding field $\chi_{\nu_\ell}$, since the chiral projector selects only the first piece, 
	$\psi_{\nu_\ell\, L}$.

	Let us assume that the field $\chi$ is a mass eigenstate. A contribution to the \bb~transition arises at the second 
	order of the Fermi interaction.
	Let us begin from the operator:
	\begin{multline}
%		-\frac{1}{2} \int d^4x \mathcal{L}_{\mbox{\tiny Fermi}}(x)  \int d^4y  \mathcal{L}_{\mbox{\tiny Fermi}}(y)
%		\end{equation}
%		\begin{equation}
		- {G_F^2} \int d^4x \, J^{a\dagger}_\mathrm{hadr}(x) \, \bar{\psi}_\el(x) \gamma_a  P_L \chi_{\nu_\el}\!(x) \\
					 \int d^4y \, J^{b\dagger}_\mathrm{hadr}(y) \, \bar{\psi}_\el(y) \gamma_b  P_L \chi_{\nu_\el}\!(y).
	\end{multline}
	By contracting the neutrino fields, the leptonic part of this operator becomes
	\begin{equation}
		\bar{\psi}_\el(x)\ \gamma^a \,  P_L \, \Delta(x-y) P_L \, \gamma^b \, C \bar{\psi}_\el^t(y)
	\end{equation}
	while the ordinary propagator, sandwiched between two chiral projectors, reduces to
	\begin{equation}
	\label{eq:pulp}
		P_L \, \Delta(x) P_L = P_L \int \frac{d^4q}{(2\pi)^4} \, \frac{i \, m}{q^2-m^2+i 0} \el^{-i q x}.
	\end{equation}
	The momentum $q$ represents the virtuality of the neutrino, whose value is connected to the momenta of the 
	final state electrons and to those of the intermediate virtual nucleons. 
	In particular, since the latter are confined in the nucleus, the typical 3-momenta are of the order of the 
	inverse of the nucleonic size, namely 
	\begin{equation}
	\label{eq:impulso_dbd}
		|\vec{q}| \sim \hbar c/\fm \sim \mbox{ few 100\,MeV}
	\end{equation} 
	whereas the energy ($q_0$) is small. 
	The comparison of this scale with the one of neutrino mass identifies and separates ``light'' from 
	``heavy'' neutrinos for what concerns \bb.
	
	The most interesting mechanism for \bb~is the one that sees light neutrinos as mediators.
	It is the one originally considered in Ref.~\cite{Furry:1939qr} 
	and it will be discussed with great details in the subsequent sections.
	In the rest of this section, instead, we examine various alternative possibilities. 

	We have some hints, mostly of theoretical nature, that the light neutrinos might have Majorana mass.
	However, the main reason for which the hypothesis that the \bb~receives its main contribution from 
	light Majorana neutrinos is the fact that experiments point out the existence of 3 light massive 
	neutrinos. 

%------------------------------------------------
\subsection{Alternative mechanisms to the light neutrino exchange}

%------------------------------------------------
\subsubsection{Historical proposals}

	A few years after the understanding of the $K^0-\bar{K}^0$ oscillation~\cite{GellMann:1955jx,Lande:1956pf,Jackson:1957zzb},
	which led Pontecorvo to conjecture that also neutrino oscillations could exist~\cite{Pontecorvo:1957cp},
	alternative theoretical mechanisms for the \bb~other than the neutrino exchange were firstly advocated.
	In 1959, Feinberg and Goldhaber~\cite{Feinberg&Goldhaber:1959} proposed the addition of the following term in the 
	effective Lagrangian density:
	\begin{equation}
	\label{eq:DBD_alt}
		\mathcal{H}_\mathrm{pion}=\frac{g}{m_\el} \pi^+ \pi^+ \el^t C^{-1} \el
	\end{equation}
	where $m_\el$ is the electron mass and $g$ an unspecified dimensionless coupling.
	Similarly, after the hypothesis of super-weak interactions in weak decays~\cite{Prentki:1965tt,Lee:1965hi}, 
	the importance for \bb~of operators like the one of Eq.~(\ref{eq:DBD_alt}) was stressed by 
	Pontecorvo~\cite{Pontecorvo:1968wp}.
	He also emphasized that the size and the origin of these operators could be quite independent from the neutrino masses.

%------------------------------------------------
\subsubsection{Higher dimensional operators}

	The SM offers a very convenient language to order the interesting operators leading to violation of $L$ and $B$. 
	It is possible to consider effective (non-renormalizable) operators that respect the gauge symmetry
%	\begin{equation}
%		\mbox{SU(3)}_{\mbox{\tiny c}}\times \mbox{SU(2)}_{\mbox{\tiny L}}\times \text{U(1)}_{\mbox{\tiny Y}}
%	\end{equation}
	$\text{SU(3)}_{\mbox{\tiny c}} \times \text{SU(2)}_{\mbox{\tiny L}} \times \text{U(1)}_{\mbox{\tiny Y}}$,
	but that violate $L$ and/or $B$~\cite{Weinberg:1979sa,Wilczek:1979hc}.
	Here, we consider a few representative cases (a more complete list can be found in Refs.\ \cite{Choi:2002bb,Bonnet:2012kh}), 
	corresponding to the following terms of the Lagrangian and Hamiltonian densities:
	\begin{equation}
	\label{eq:ggg}
		\mathcal{H}_\mathrm{Weinberg} = \frac{(l_{\mbox{\tiny L}} H)^2}{M} 
		+ \frac{l_{\mbox{\tiny L}} q_{\mbox{\tiny L}} q_{\mbox{\tiny L}} q_{\mbox{\tiny L}} }{M'^{\,2}}
		+ \frac{(l_{\mbox{\tiny L}} q_{\mbox{\tiny L}} d_{\mbox{\tiny R}}^c)^2}{M''^{\,5}}.
	\end{equation}
	The matter fields (fermions) in the equation are written in the standard notation of Table~\ref{tab:fields}, 
	$H$ is the Higgs field, while the constrains on the masses are: $M<10^{11}\,\TeV$, $M'>10^{12}\,\TeV$ 
	and $M''>5\,\TeV$. 
	In particular:
	\begin{itemize}
		\item	the first (dimension-5) operator generates Majorana neutrino masses, and the bound on $M$ derives from 
		neutrino masses $m_\nu<0.1$\,eV
	\item the dimension-6 operator leads to proton decay and this implies the tight bound on the mass $M'$
	\item the dimension-9 operator contributes to the \bb. 
		Its role in the transition can be relevant if the scale of lepton number violation is low.
	\end{itemize}
	
	Summarizing, if one assumes that the scale of new physics is much higher than the electroweak scale, 
	it is natural to expect that the leading mechanism behind the \bb~is the exchange of light neutrinos endowed with 
	Majorana masses.
	It is also worth to be noted that if light sterile neutrinos, dark matter or, generally, other light states are added, 
	more operators may be required.
	A large effective mass could also come from small adimensional couplings $y$, e.\,g.\ $1/M=y^2/\mu$.

	The number of possible mechanisms that eventually can lead to the above effective operators is also very large.  
	One possible (plausible) origin of the dimension-5 operator is discussed in Sec.\ \ref{sec:rh}. 
	However, other cases are possible and the same is true for the other operators. 
%	Two specific cases of particular interest are investigated in more detail.
	
%------------------------------------------------
\subsubsection{Heavy neutrino exchange}
	
	Let us now consider the case of heavy RH neutrino exchange mechanism. 
	The corresponding operator gives rise to the effective Hamiltonian density
	(for heavy neutrinos, the propagator of Eq.~(\ref{eq:pulp}) is proportional to $\delta(x-y)$\,):
	\begin{equation}
	\label{eq:RH}
		\mathcal{H}_{\nu_\mathrm{heavy}} = - \frac{G_F^2}{M_H}
		J^{a\dagger}_\mathrm{hadr} \bar{\psi}_{\el_L} \gamma^a  \gamma^b C 
		\bar{\psi}_{\el_L}^t J^{b\dagger}_\mathrm{hadr}.
	\end{equation}
	It is evident that this is a dimension-9 operator and it has in front a constant with mass 
	dimension $m^{-5}$, since $M_H$ indicates the relevant heavy neutrino mass.
	It has to be noted that such a definition can be used in an effective formula, but a gauge model requires to
	express $M_H$ in terms of the single RH neutrino masses $M_I$ and of the mixing between left handed 
	neutrinos $\nu_{\el_L}$ and heavy neutrinos:
	\begin{equation}
	\label{eq:restr}
		\frac{1}{M_H} = \frac{U_{\el I}^2}{M_I}.
	\end{equation}
	In particular, the mixings are small if $M_I$ is large since $U_{\el i}=m_{\el i}^{\mbox{\tiny Dirac}}/M_I$. 
	This suggests a suppression of the above effective operator with the cube of $M_I$, whereas the light neutrino 
	exchange mechanism leads to a milder suppression, linear in $M_I$
	(if the mixing matrices have specific flavor structures, deviations from this generic expectation are possible).
%	The heavy RH neutrinos are known to be compatible with an important contribution to the \bb, but it has been 
%	argued that their masses should not be much larger than about $10\,\GeV$~\cite{Mitra:2011qr} in order to avoid 
%	fine tunings on the light neutrino ones.
	However, it is still possible that RH neutrinos are heavy, but not ``very'' heavy. Actually, this was the first case to be 
	considered~\cite{Minkowski:1977sc}, and it could be of interest both for direct searches at accelerators 
	(see Sec.~\ref{sec:accelerators}) and for the \bb. In fact, in this case the mixing $U_{\el I}$ is not strongly 
	suppressed and RH neutrinos can give an important contribution to the transition~\cite{Atre:2009rg}.
	However, two remarks on this case are in order. As it was argued in Ref.~\cite{Mitra:2011qr}, 
	in order to avoid fine tunings on the light neutrinos, the masses of RH neutrinos should not be much larger than 
	about $10\,\GeV$.
	Moreover, in the extreme limit in which the mass becomes light (i.\,e.\ it is below the value in 
	Eq.~(\ref{eq:impulso_dbd})) and Type I Seesaw applies, the contribution of RH neutrinos cancels the one of 
	ordinary neutrinos~\cite{Blennow:2010th,LopezPavon:2012zg}.
	
%------------------------------------------------
\subsubsection{Models with RH currents}

	Another class of models of great interest are those that include RH currents and intermediate bosons. 
	In the language of SM, the neutrino exchange leads to a core operator 
	\begin{equation}
	\label{eq:ku}
		\mathcal{H}_{W\mathrm{bosons}}=\frac{1}{M} W^+ W^+ \el^t C^{-1} \el
	\end{equation}
	where $M$ is a mass scale and $W$ identify the fields of the usual $W$ bosons.
	When we consider virtual $W$ bosons, this may eventually lead to the usual case. 
	In principle, it is possible to replace the usual $W$ bosons with the corresponding $W_{\mbox{\tiny R}}$ bosons 
	of a new $SU(2)_{\mbox{\tiny R}}$ gauge group. 
	In this hypothesis, the RH neutrinos play a more important role and are no more subject to restrictions of the 
	mixing matrix, as those of Eq.\ (\ref{eq:restr}).
	However, the resulting dimension-9 operator is suppressed by 4 powers of the masses of the new gauge bosons.

	Evidently, new RH gauge bosons with masses accessible to direct experimental investigation are of special interest 
	(see Sec.~\ref{sec:accelerators}). 
	Since to date we do not have any experimental evidence, this possibility will not be emphasized in the following
	discussion. Anyway, investigations at the LHC are currently in progress and
	the interpretation of some anomalous events (among the collected data) as a hint in favor of relatively 
	light $W_{\mbox{\tiny R}}$ bosons has already been proposed~\cite{Dev:2015pga,Deppisch:2015cua,Deppisch:2016scs}.

%------------------------------------------------
\subsection{From \bb~to Majorana mass: a remark on ``natural'' gauge theories}

\begin{figure}[tb]
		\centering
		\includegraphics[width=.9\columnwidth]{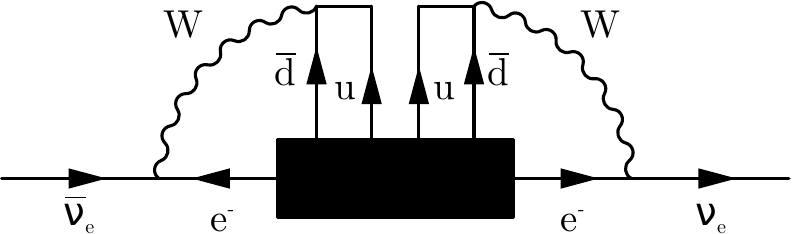}
		\caption{Diagram representing the contribution of the ``black box'' operator to the Majorana mass.
					Figure from Ref.~\cite{Duerr:2011zd}.}
		\label{fig:SV-diagram}
	\end{figure}

	In a well-known work, Schechter and Valle~\cite{Schechter:1981bd} employ the basic concepts of gauge theories to derive 
	some important considerations on the \bb. In particular, their argument proceeds as follows:
	\begin{enumerate}
		\item if the~\bb~is observed, there will be some process (among elementary particles) where the 
			electron-, up- and down-fields are taken twice. This ``black box'' process in Ref.\ \cite{Schechter:1981bd}
			(Fig.\ \ref{fig:SV-diagram}) effectively resembles the one caused by the dimension-9 operator in Eq.\ (\ref{eq:ggg})
		\item using $W$ bosons, it is possible to contract the two quark pairs and obtain something like the operator 
			in Eq.\ (\ref{eq:ku})
		\item finally, the electron- and the $W$-fields can be converted into neutrino-fields. 
			A contribution to the Majorana neutrino mass is therefore obtained
		\item the possibility that this contribution could be canceled by others is barred out as ``unnatural''.
	\end{enumerate}
	
	This argument works in the ``opposite direction'' with respect to ones presented so far. Instead of starting from 
	the Majorana mass to derive a contribution for the \bb, it shows that from the observation of the \bb, it is possible 
	to conclude the existence of the Majorana mass.
	The result could be seen as an application (or a generalization) of the Symanzik's rule as given by 
	Coleman~\cite{Coleman}: if a theory predicts $L$-violation, it will not be possible to screen it to forbid 
	\emph{only} a Majorana neutrino mass. 
	
	The size of the neutrino masses is not indicated in the original work, but a straightforward estimation of 
	the diagram of Fig.~\ref{fig:SV-diagram} shows that they are so small that they have no physical interest, 
	being of the order of $10^{-24}$\,eV~\cite{Duerr:2011zd}.
	However, what can be seen as a weak point of the argumentation, is the concept of ``natural theory'', whose definition 
	is not discussed in Ref.~\cite{Schechter:1981bd}, but simply proclaimed.
	In fact, it is possible to find examples of models where the \bb~exists but the Majorana neutrino mass contribution 
	is zero~\cite{Mitra:2011qr}, in accordance with the claim of Pontecorvo~\cite{Pontecorvo:1968wp}, but clashing with 
	the expectations deriving from that of Ref.~\cite{Schechter:1981bd}.
	
	We think that the (important) point made in Ref.~\cite{Schechter:1981bd} is valid not quite as a theorem (word that, 
	anyway, the authors never use to indicate their work). We rather believe it acts mostly as a reminder that 
	any specific theory that includes Majorana neutrino masses will have various specific links between these masses, 
	\bb~and possibly other manifestations of $L$-violation.
	We see as a risk the fact that, due to the impossibility of avoiding the issue of model dependence, 
	we will end up with the idea that we can accept ``petition of principles''.

%------------------------------------------------
\subsection{Role of the search at accelerators}
\label{sec:accelerators}

	There is the hope that the search for new particles at the accelerators might reveal new physics relevant for the 
	interpretation or in some way connected to the \bb. 
	This is a statement of wide validity. For example, the minimal supersymmetric extension of the SM is compatible with 
	new $L$-violating phenomena taking place already at the level of renormalizable operators~\cite{Weinberg:1981wj}.
	Also the hypothesized extra-dimensions at the TeV scale might be connected to new $L$-violating 
	operators~\cite{Mohapatra:2006gs}. Or even, models where the smallness of the neutrino mass is explained through 
	loop effects imply typically new particles that are not ultra-heavy~\cite{Mohapatra:1998rq}. 
	Notice that these are just a few among the many theoretical possibilities to select which, unfortunately, 
	lack of clear principles.
	
	The recent scientific literature tried at least to exploit some minimality criteria, and the theoretical models that 
	received the largest attention are indeed those discussed above.
	A specific subclass, named~$\nu$SM \cite{Alekhin:2015byh}, is found interesting enough to propose a dedicated search 
	at the CERN SPS~\cite{Alekhin:2015byh,Anelli:2015pba}, aiming to find rare decays of the ordinary mesons into heavy 
	neutrinos. 
	Other models, that foresee a new layer of gauge symmetry at accessible energies and, more specifically, those 
	connected to left-right gauge symmetry~\cite{Pati:1974yy} might instead lead to impressive $L$-violation at 
	accelerators~\cite{Keung:1983uu,Tello:2010am,Nemevsek:2011hz}. This should be quite analogous to the \bb~process itself 
	and that could be seen as manifestations of operators similar to those in Eq.~(\ref{eq:ku}).

	We would like just to point out that in both cases, in order to explain the smallness of neutrino masses, 
	very small adimensional couplings are required. Although this position is completely legitimate, in front 
	of the present understanding of particle physics, it seems fair to say that this leaves us with some 
	theoretical question to ponder.

%------------------------------------------------
%------------------------------------------------
	%------------------------------------------------
%------------------------------------------------
\section{Present knowledge of neutrino masses}
\label{sec:masses}

	In this section we discuss the crucial parameter describing the \bb~if the process is mediated by light Majorana 
	neutrinos (as defined in Sect.~\ref{sec:nu_exchange}).
	We take into account the present information coming from the oscillation parameters, cosmology 
	and other data. On the theoretical side, we motivate the interest for a minimal interpretation of the results. 

%------------------------------------------------
\subsection{The parameter $\mbb$}
\label{sec:mbb}

	We know three light neutrinos. They are identified by their charged current interactions i.\,e.\ 
	they have ``flavor'' $\ell=\rm e,\mu,\tau$. 
	The Majorana mass terms in the Lagrangian density is described by a symmetric matrix:
	\begin{equation}
	\label{eq:lagramajo3}
		\mathcal{L}_\mathrm{mass} = \frac{1}{2} 
		\sum_{\ell,\ell'=\rm e,\mu,\tau}
		\nu_\ell^t\, C^{-1} M_{\ell \ell'}\, \nu_{\ell'} + h.\,c.\,.
	\end{equation}
	The only term that violates the electronic number by two units is $M_{\el\el}$, and this simple consideration 
	motivates the fact that the amplitude of the \bb~decay has to be proportional to this parameters,
	while the width to its squared modulus. 
%	Thus, we expect that $\mbb=|M_{\mbox{\tiny ee}}|$. 
	We can diagonalize the neutrino mass matrix by mean of a unitary matrix
	\begin{equation}
	\label{eq:massamajo3}
		M = U^t \, \mbox{diag}(m_1,m_2,m_3) \, U^\dagger
	\end{equation}
	where the neutrino masses $m_i$ are real and non-negative. 
	Thus, we can define:
	\begin{equation}
	\label{eq:mbb3}
		\mbb \equiv \left| \sum_{i=1,2,3}   U_{\el i}^2 \ m_i \right|
	\end{equation}
	where the index $i$ runs on the 3 light neutrinos with given mass.
	This parameter is often called ``effective Majorana mass''
	(it can be thought of as the ``electron neutrino mass'' that rules the \bb~transition, but keeping in mind 
	that it is different from the ``electron neutrino mass'' that rules the $\beta$ decay transition).

	The previous intuitive argument in favor of this definition is corroborated by calculating the Feynman 
	diagram of Fig.\ \ref{fig:DBD-diagram}. Firstly, it has to be noted that the electronic neutrino 
	$\nu_\el$ is not a mass eigenstate in general. 
	Then, substituting Eq.\ (\ref{eq:massamajo3}) into Eq.\ (\ref{eq:lagramajo3}), we see that we go from the flavor basis 
	to the mass basis by setting
	\begin{equation}
	\label{eq:unit}
		\nu_\ell= \sum_{i=1,2,3}   U_{\ell i} \ \nu_i.
	\end{equation}
	Therefore, in the neutrino propagators of Fig.\ \ref{fig:DBD-diagram}, we will refer to the masses $m_i$ 
	(that in our case are ``light'') while, in the two leptonic vertices, we will have $U_{\el i}$. 
	Taking the product of these factors, we get the expression given in Eq.\ (\ref{eq:mbb3}).
 
	It should be noted that the leptonic mixing matrix $U$ as introduced above differs from the ordinary one used in 
	neutrino oscillation analyses. 
	Indeed, the latter is given after rotating away the phases of the neutrino fields and observing that oscillations 
	depend only upon the combination $M M^\dagger/(2 E)$. 
	This matrix contains only one complex phase which plays a role in oscillations (the ``CP-violating phase'').
	Instead, in the case of \bb~the observable is different. It is just $|M_{\el\el}|$.
	Here, there are new phases that cannot be rotated away and that play a physical role. 
	These are sometimes called ``Majorana phases''. Their contribution can be made explicit by rewriting 
	Eq.~(\ref{eq:mbb3}) as follows:
	\begin{equation}
	\label{eq:mbb4}
		\mbb = \left| \sum_{i=1,2,3}   \el^{i\xi_i}\ |U_{\el i}^2| \ m_i \right|.
	\end{equation}
	We can now identify $U_{\el i}$ of Eq.~(\ref{eq:mbb4}) with the mixing matrix used in neutrino oscillation analyses.%
	\footnote{Note that the specific choice and the symbols for these phases may differ among authors.}
	
	Before proceeding in the discussion, some remarks are in order:
	\begin{itemize}
		\item it is possible to adopt a convention for the neutrino mixing matrix such that the 3 mixing elements 
			$U_{\el i}$ are real and positive. However, in the most common convention 
			$U_{\el 3}$ is defined to be complex
		\item only two Majorana phases play a physical role, the third one just being matter of convention
		\item it is not possible even in principle to reconstruct the Majorana mass matrix simply on experimental bases, 
			unless we find another observable which depends Majorana phases.
	\end{itemize}
%	The discussion of the maximum and minimum values of $\mbb$, varying the unknown Majorana phases, is given in the 
%	appendix. 
%	The actual chances to probe Majorana phases experimentally by measuring $\mbb$ will be assessed later.
	Furthermore, a specific observation on the Type I Seesaw model is useful. 
	Let us consider the simplest case with only $\nu_{\el}$ and one heavy neutrino $\nu_H$ that mix with this state. 
	The Majorana mass matrix is of the form:
	\begin{equation}
		\begin{pmatrix} 
		0 									&m^{\mbox{\tiny Dirac}}	 \\ 
		m^{\mbox{\tiny Dirac}} 		&M_H 
		\end{pmatrix}.
	\end{equation}
	One should not be mislead, concluding that in this case (and, generally, in the Type I Seesaw) $\mbb$ is zero. 
	In fact, as it is well-known, the masses of the light neutrinos (in this case, of $\nu_{\el}$) 
	arise when one integrates away the heavy neutrino state, getting 
	\begin{equation}
		m_{\nu_\el} = - \frac{(m^{\mbox{\tiny Dirac}})^2}{M_H}.
	\end{equation}
	As discussed in Ref.\ \cite{Mitra:2011qr}, we obtain in this one-flavor case the non-zero contribution
	\begin{equation}
		\label{eq:moment}
		\mbb = \left| m_{\nu_\el} \right| \, \left (1+\frac{\langle q^2\rangle}{M_H^2} \right).
	\end{equation}
	The second factor is the direct contribution of the heavy neutrino.%
	\footnote{This formula agrees with the naive scaling expected from the heavy neutrino contribution. 
		But in specific three-flavor models it is possible, at least in principle, that heavy neutrinos give a large and 
		even dominating contribution to the \bb~decay rate~\cite{Mitra:2011qr}.}
	The quantity $\langle q^2 \rangle$ depends on the nuclear structure and it is of the order of $(100\,\MeV)^2$
	and thus Eq.~(\ref{eq:moment}) is valid if we assume $|m_{\nu_{\mbox{\tiny \el}}}| \ll 100\,\MeV \ll M_H$.
	
	In the above discussion, we have emphasized the three flavor case. 
	The main reason for this is evidently that we know about the existence of only 3 light neutrinos. 
	It is possible to test this hypothesis by searching for new oscillation phenomena, by testing the universality of the 
	weak leptonic couplings and/or the unitarity of the matrix in Eq.~(\ref{eq:unit}), by searching directly at 
	accelerators new and (not too) light neutrino states, etc.\,. 
	However, we believe that it is fair to state that, to date, we have no conclusive experimental evidence or 
	strong theoretical reason to deviate from this minimal theoretical scheme.
	We will adopt it in the proceeding of the discussion. In this way, we can take advantage of the precious information 
	that was collected on the neutrino masses to constrain the parameter $\mbb$ and to clarify the various expectations.

%------------------------------------------------
\subsection{Oscillations}
\label{sec:osc_data}

	\begin{figure*}[tb]
		\centering
		\includegraphics[width=\columnwidth]{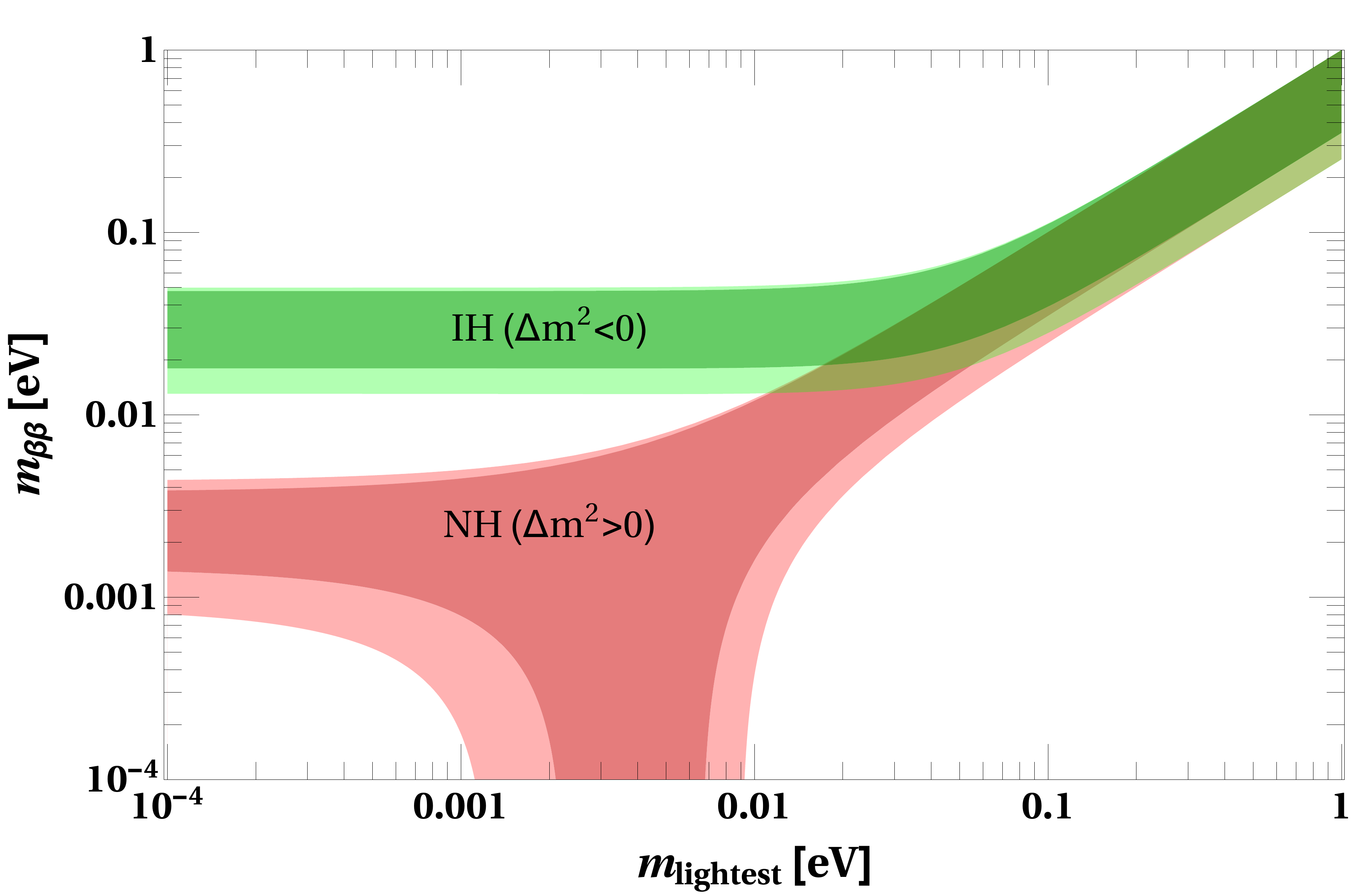}
		\includegraphics[width=\columnwidth]{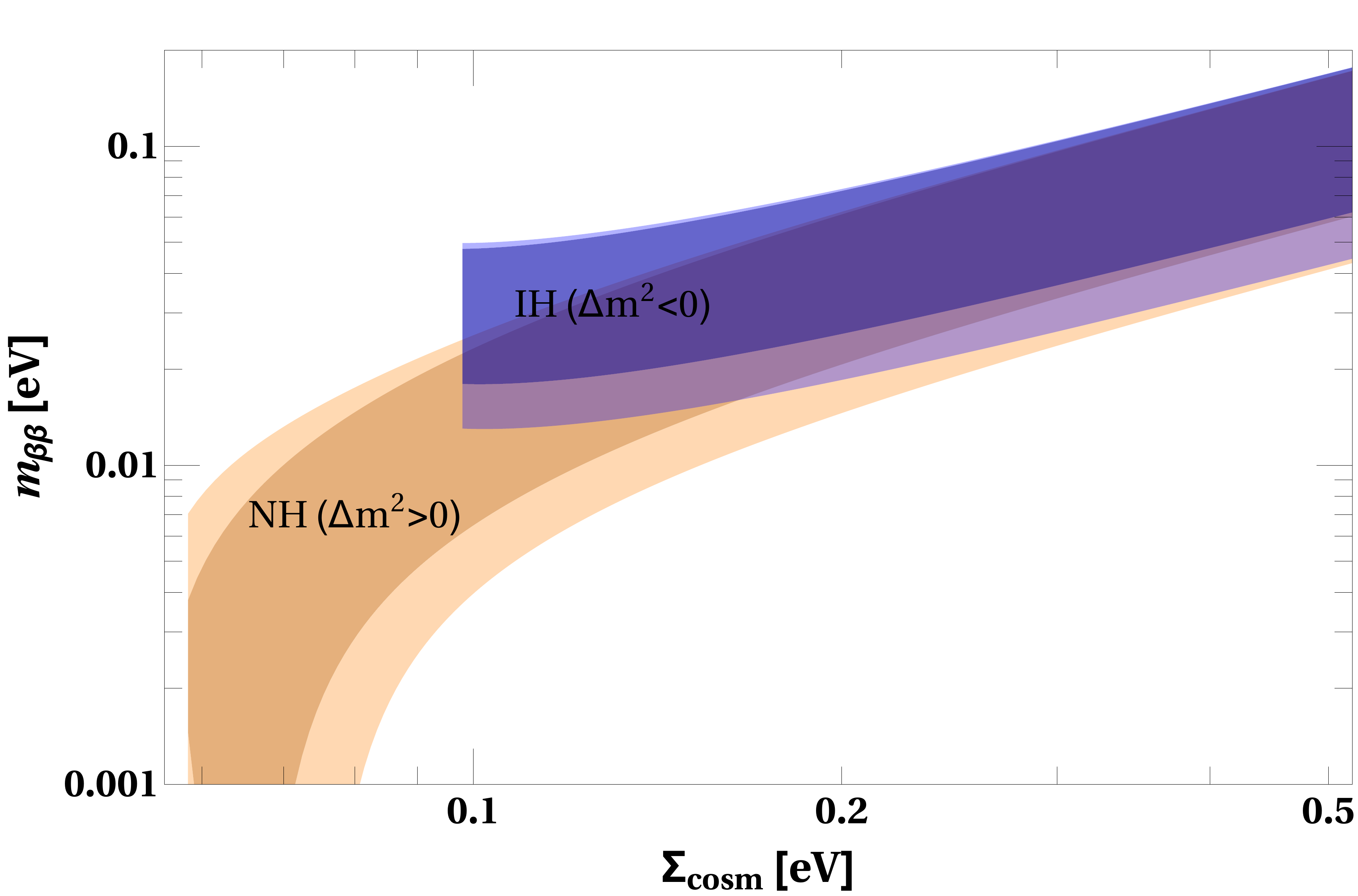}
		\caption{Updated predictions on $\mbb$ from oscillations as a function of the lightest neutrino mass (left) 
			and of the cosmological mass (right) in the two cases of \NH~and \IH.
			The shaded areas correspond to the $3\sigma$ regions due to error propagation of the uncertainties on the 
			oscillation parameters.
			Figure from Ref.\ \cite{Dell'Oro:2014yca}.}
		\label{fig:DBD_graph}
	\end{figure*}

%	\begin{figure}[tb]
%		\centering
%		\includegraphics[width=1.\columnwidth, angle=0]{R_mbb_vs_ml_VS}
%		\caption{Updated predictions on $\mbb$ from oscillations as a function of the lightest neutrino mass in the two 
%			cases of \NH~and \IH.
%			The shaded areas correspond to the $3\sigma$ regions due to error propagation of the uncertainties on the 
%			oscillation parameters. Figure from Ref.\ \cite{Dell'Oro:2014yca}.}
%		\label{fig:DBD_graphVS}
%	\end{figure}

	\begin{center}
	\begin{table}[t]
	\caption{Results of the global 3$\nu$ oscillation analysis, in terms of best-fit values and allowed 1$\sigma$ range 
		for the 3$\nu$ mass-mixing parameters relevant for our analysis as reported in Ref.~\cite{Capozzi:2013csa}. 
		The last column is our estimate of the $\sigma$ while assuming symmetric uncertainties.}
	\small{
	\begin{ruledtabular}
	\begin{tabular}{l c c c}
		Parameter						&Best fit						&$1\sigma$ range 			&$\sigma_{\mbox{\tiny symmetric}}$	\\
		\hline \\[-10pt]
		{\NH}				&									&	 										&									\\
		\cline{1-1}
		\\[-8pt]
		$\sin^2(\theta_{12})$		&$3.08 \cdot 10^{-1}$		&$(2.91-3.25) \cdot 10^{-1}$		&$0.17 \cdot 10^{-1}$		\\
		$\sin^2(\theta_{13})$		&$2.34 \cdot 10^{-2}$		&$(2.16-2.56) \cdot 10^{-2}$		&$0.22 \cdot 10^{-2}$		\\
		$\sin^2(\theta_{23})$		&$4.37 \cdot 10^{-1}$		&$(4.14-4.70) \cdot 10^{-1}$		&$0.33 \cdot 10^{-1}$		\\
		\\[-8pt]
		$\delta m^2\,\,[\eV^2]$		&$7.54 \cdot 10^{-5}$		&$(7.32-7.80) \cdot 10^{-5}$		&$0.26 \cdot 10^{-5}$		\\
		$\Delta m^2\,[\eV^2]$		&$2.44 \cdot 10^{-3}$		&$(2.38-2.52) \cdot 10^{-3}$		&$0.08 \cdot 10^{-3}$		\\[+2pt]
		{\IH}				&									&											&									\\
		\cline{1-1}
		\\[-8pt]
		$\sin^2(\theta_{12})$		&$3.08 \cdot 10^{-1}$		&$(2.91-3.25) \cdot 10^{-1}$		&$0.17 \cdot 10^{-1}$		\\
		$\sin^2(\theta_{13})$		&$2.39 \cdot 10^{-2}$		&$(2.18-2.60) \cdot 10^{-2}$		&$0.21 \cdot 10^{-2}$ 		\\
		$\sin^2(\theta_{23})$		&$4.55 \cdot 10^{-1}$		&$(4.24-5.94) \cdot 10^{-1}$		&$1.39 \cdot 10^{-1}$		\\
		\\[-8pt]
		$\delta m^2\,\,[\eV^2]$		&$7.54 \cdot 10^{-5}$		&$(7.32-7.80) \cdot 10^{-5}$		&$0.26 \cdot 10^{-5}$		\\
		$\Delta m^2\,[\eV^2]$		&$2.40 \cdot 10^{-3}$		&$(2.33-2.47) \cdot 10^{-3}$		&$0.07 \cdot 10^{-3}$		\\
		\end{tabular}
		\end{ruledtabular}
		}
	\label{tab:osc_parameters}
	\end{table}
	\end{center}
	
	In Ref.~\cite{Capozzi:2013csa}, a complete analysis of the current knowledge of the oscillation parameters and 
	of neutrino masses can be found.
%	\footnote{This analysis has just been updated. The new results are reported in Ref.~\cite{Capozzi:2016rtj}.}
	Although the absolute neutrino mass scale is still unknown, it has been possible to measure, through oscillation 
	experiments, the squared mass splittings between the three active neutrinos. 
	In Table \ref{tab:osc_parameters}, the parameters relevant for 
	our analysis are reported. The mass splittings are labeled by $\delta m^2$ and $\Delta m^2$. The former is 
	measured through the observation of solar neutrino oscillations, while the latter comes from atmospheric neutrino data. 
	The definitions of these two parameters are the following: 
	\begin{equation}
		\delta m^2 \equiv m_2^2 - m_1^2 \quad \mbox{and} \quad
		\Delta m^2 \equiv m_3^2 - \frac{m_1^2 + m_2^2}{2}.
	\end{equation} 
	Practically, $\delta m^2$ regards the splitting between $\nu_1$ and $\nu_2$, while $\Delta m^2$ refers to the distance 
	between the $\nu_3$ mass and the mid-point of $\nu_1$ and $\nu_2$ masses.
	
	The sign of $\delta m^2$ can be determined by observing matter 
	enhanced oscillations as explained within the MSW 
	theory~\cite{Wolfenstein:1977ue,Mikheev:1986gs}. It turns out, after comparing with 
	experimental data, that $\delta m^2>0$~\cite{Abe:2008aa}.
	Unfortunately, determining the sign of $\Delta m^2$ is still unknown and it is not simple to measure it. 
	However, it has been argued (see e.\,g. Ref.~\cite{Ghoshal:2010wt}) 
	that by carefully measuring the oscillation pattern, it could possible to distinguish between 
	the two possibilities, $\Delta m^2>0$ and $\Delta m^2<0$. 
	This is a very promising perspective in order to solve this ambiguity, which is sometimes called the 
	``mass hierarchy problem''. In fact, standard names for the two mentioned possibilities for the neutrino mass
	spectra are ``Normal Hierarchy'' (\NH) for $\Delta m^2>0$ and ``Inverted Hierarchy'' (\IH) for $\Delta m^2<0$.

	The oscillation data are analyzed in Ref.~\cite{Capozzi:2013csa} by writing the leptonic (PMNS) mixing matrix 
	$U|_\mathrm{osc.}$ in terms of the mixing angles $\theta_{12}$, 
	$\theta_{13}$ and $\theta_{23}$ and of the CP-violating phase $\phi$ according to the (usual) representation
	{\small
	\begin{align}
		&U|_\mathrm{osc.} = \\
		&\left(  
			\begin{matrix} 
				c_{12}c_{13} && s_{12}c_{13} && s_{13}\el^{-i\phi} \\ 
				-s_{12}c_{23}-c_{12}s_{13}s_{23}\el^{i\phi} && c_{12}c_{23}-s_{12}s_{13}s_{23}\el^{i\phi} && c_{13}s_{23} \\ 
				s_{12}s_{23}-c_{12}s_{13}c_{23}\el^{i\phi} && -c_{12}s_{23}-s_{12}s_{13}c_{23}\el^{i\phi} && c_{13}c_{23}
			\end{matrix}  
		\right) \nonumber
	\end{align}
	}%
	where $s_{ij},c_{ij} \equiv \sin\theta_{ij},\cos\theta_{ij}$.
%	Note that this are the same phase convention and parameterization that are used for the quark (CKM) mixing matrix but, 
%	of course, the values of the parameters are different.
	Note the usage of the same phase convention and parameterization of the quark (CKM) mixing matrix even if, of course,
	the values of the parameters are different.
	With this convention, it is possible to obtain Eq.~(\ref{eq:mbb4}) by defining
	\begin{equation} 
		U \equiv U|_\mathrm{osc.} \cdot \mbox{diag}\left( \el^{-i\xi_1/2},\el^{-i\xi_2/2},\el^{i\phi-i\xi_3/2} \right).
	\end{equation}

	Table~\ref{tab:osc_parameters} shows the result of the best fit and of the $1\sigma$ range for the different 
	oscillation parameters. It can be noted that the values are slightly different depending on the mass hierarchy. 
	This comes from the different analysis procedures used during the evaluation, as explained in Ref.~\cite{Capozzi:2013csa}. 
	Therefore, throughout this work the two neutrino mass spectra are treated differently one from the other, 
	since we used these hierarchy-dependent parameters.
	The uncertainties are not completely symmetric around the best fit point, but the deviations are quite small, 
	as claimed by the authors themselves in the reference. 
	In particular, the plots in the paper show Gaussian likelihoods for the parameters 
	determining $\mbb$. In order to later propagate the errors, we decided to neglect the asymmetry, which 
	has no relevant effects on the presented results. 
	We computed the maximum between the distances of the best fit values and the borders of the $1\sigma$ range 
	(fourth column of Tab. \ref{tab:osc_parameters}) and we assumed that the 
	parameters fluctuate according to a Gaussian distribution around the best fit value, with a standard deviation given 
	by that maximum.
	
	Thanks to the knowledge of the oscillation parameters, it is possible to put a first series of constraints on $\mbb$.
	However, as already recalled, since the complex phases of the mixing parameters in Eq.\ (\ref{eq:mbb4}) cannot be 
	probed by oscillations, the allowed region for $\mbb$ is obtained letting them vary freely. 
	The expressions for the resulting extremes (i.\,e.\ the $\mbb$ maximum and minimum values
	due to the phase variation) can be found in App.~\ref{app:mbb_extr}.
	We adopt the graphical representation of $\mbb$ introduced in Ref.~\cite{Vissani:1999tu} and refined 
	in Refs.~\cite{Feruglio:2002af,Strumia:2006db}. 
	It consists in plotting $\mbb$~in bi-logarithmic scale as a function of the mass of the lightest neutrino, 
	both for the cases of \NH~and of \IH. 
	The resulting plot is shown in the left panel of Fig.~\ref{fig:DBD_graph}.
	The uncertainties on the various parameters are propagated using the procedures described in App.\ \ref{app:stat}.
	This results in a wider allowed region, which corresponds to the shaded parts in the picture.
	
%------------------------------------------------
\subsubsection{Mass eigenstates composition}

	\begin{figure}[tb]
		\centering
		\includegraphics[width=1.\columnwidth, angle=0]{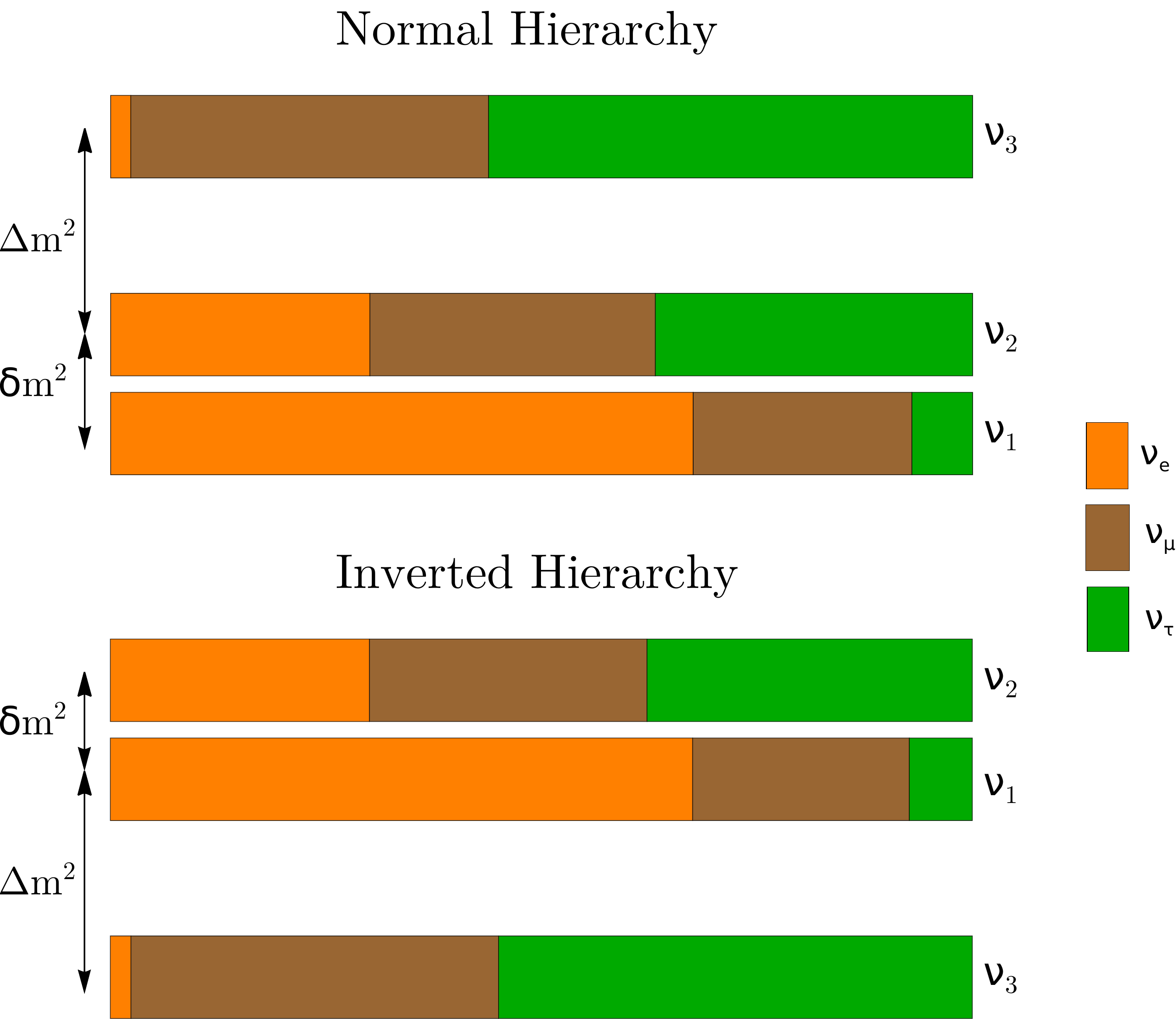}
		\caption{Graphic view of the probability of finding one of the flavor eigenstates if the neutrino is in a certain
			mass eigenstate. The value $\phi=0$ for the CP-violating phase is assumed.}
		\label{fig:hierarchies}
	\end{figure}
	
%	It can be interesting to have a view graph to monitor the situation regarding neutrino masses and mixing. 
%	Eq.~(\ref{eq:unit}) implies the following relation between ultra relativistic neutrinos:
	The standard three flavor oscillations involve three massive states
	that, consistently with Eq.~(\ref{eq:unit}), are given by\,%
	\footnote{Note that in this case we are in the ultra-relativistic limit. See Sec.~\ref{sec:ultra_rel}.}
	\begin{equation}
		\left|\nu_i\right\rangle=\sum_{\ell=\el,\mu,\tau} U_{\ell i}\left|\nu_\ell\right\rangle.
	\end{equation}
	Thus, it is possible to estimate the probability of finding the component $\nu_\ell$ of each 
	mass eigenstate $\nu_i$.
	This probability is just the squared module of the matrix element $U_{\ell i}$, 
	since the matrix is unitary. 
	The result is graphically shown in Fig.~\ref{fig:hierarchies}. 
	As already mentioned, since hierarchy-dependent parameters were used, the flavor composition of the 
	various eigenstates slightly depends on the mass hierarchy.
	It is worth noting that the results also depend on the possible choices of $\phi$, 
	while they do not depend on the eventual Majorana phases.
	Table \ref{tab:eigenstates} reports the calculation for the cases $\phi=0$ and $\phi=1.39\pi \,\,(1.31\pi)$, best 
	fit value for the \NH~(\IH) according to Ref.~\cite{Capozzi:2013csa}.

%------------------------------------------------
\subsection{Cosmology and neutrino masses}
\label{sec:nucosmology}

%------------------------------------------------
\subsubsection{The parameter $\Sigma$}

	The three light neutrino scenario is consistent with all known facts in particle physics including the new measurements 
	by Planck~\cite{Planck:2015xua}. In this assumption, the physical quantity probed by cosmological surveys, $\Sigma$,
	is the sum of the masses of the three light neutrinos: 
	\begin{equation}
	\label{eq:sigma}
		\Sigma \equiv m_1 + m_2 + m_3.
	\end{equation} 
	Depending on the mass hierarchy, is it possible to express $\Sigma$ as a function of the lightest neutrino mass $m$
	and of the oscillation mass splittings.
	In particular, in the case of \NH~one gets:
	\begin{equation}
	\label{eq:sigmaNH}
		\left\{
		\begin{aligned}
			m_1 &= m \\
			m_2 &= \sqrt{m^2 + \delta m^2} \\
			m_3 &= \sqrt{m^2 + \Delta m^2 + \delta m^2/2}
		\end{aligned}
		\right.
	\end{equation}
	while, in the case of \IH:
	\begin{equation}
	\label{eq:sigmaIH}
		\left\{
		\begin{aligned}
			m_1 &= \sqrt{m^2 + \Delta m^2 - \delta m^2/2} \\
			m_2 &= \sqrt{m^2 + \Delta m^2 + \delta m^2/2} \\
			m_3 &= m.
		\end{aligned}
		\right.
	\end{equation}
	
	It can be useful to compute the mass of the lightest neutrino, given a value of $\Sigma$. 
	This can be convenient in order to compute $\mbb$ as a function of $\Sigma$ instead of $m$.%
	\footnote{In App.\ \ref{app:sigma}, an approximate (but accurate) alternative method for the numerical calculation 
		needed to make this conversion is given.}
	In this way, $\mbb$ is expressed as a function of a directly observable parameter.
	
	The close connection between the neutrino mass measurements obtained in the 
	laboratory and those probed by cosmological observations was outlined long ago~\cite{Zeldovich:1981wf}.
	Furthermore, the measurements of $\Sigma$ have recently reached important sensitivities, as 
	discussed in Sec.~\ref{sec:cosm_bounds}.%
	
	In the right panel of Fig.\ \ref{fig:DBD_graph}, an updated version of the plot ($\mbb$ vs.\ $\Sigma$)
	originally introduced in Ref.~\cite{Fogli:2004as} is shown.
	Concerning the treatment of the uncertainties, we use again the assumption 
	of Gaussian fluctuations and the prescription reported in App.\ \ref{app:stat}.

	\begin{center}
	\begin{table}[b]
	\caption{Flavor composition of the neutrino mass eigenstates. The two cases refer to the values for the 
		CP-violating phase $\phi=0$ and $\phi=1.39\pi \,\,(1.31\pi)$, best fit value in case of \NH~(\IH) 
		according to Ref.~\cite{Capozzi:2013csa}}
	\small{
	\begin{ruledtabular}
	\begin{tabular}{l l l l l}
%		Eigenstate						&\NH~($\phi=0$)	&\IH~($\phi=0$)	&\NH~($\phi=1.39\pi$)		&\IH~($\phi=1.31$)	\\
		Eigenstate						&\multicolumn{2}{c}{\NH}	&\multicolumn{2}{c}{\IH}	\\
		\cline{2-3}	 \cline{4-5} \\[-10pt]
											&($\phi=0$)		&($\phi=1.39\pi$)		&($\phi=0$)		&($\phi=1.31\pi$)	\\[+2pt]
		\hline
		$\nu_1$	\\
		\cline{1-1} \\[-10pt]
		$\nu_\el$						&.676			&.676							&.675				&.675		\\
		$\nu_\mu$						&.254			&.160							&.252				&.141		\\
		$\nu_\tau$						&.070			&.164							&.073				&.184		\\[+3pt]
		$\nu_2$	\\
		\cline{1-1} \\[-10pt]
		$\nu_\el$						&.301			&.301							&.301				&.301		\\
		$\nu_\mu$						&.331			&.425							&.322				&.432		\\
		$\nu_\tau$						&.368			&.274							&.378				&.267		\\[+3pt]
		$\nu_3$	\\
		\cline{1-1} \\[-10pt]
		$\nu_\el$						&.023			&.023							&.024				&.024		\\
		$\nu_\mu$						&.415			&.415							&.426				&.426		\\
		$\nu_\tau$						&.562			&.562							&.550				&.550		\\[+3pt]
		\end{tabular}
		\end{ruledtabular}
		}
	\label{tab:eigenstates}
	\end{table}
	\end{center}
	
%------------------------------------------------
\subsubsection{Constraints from cosmological surveys}
\label{sec:sigma_bounds}

	\begin{figure}[t]
		\centering
		\includegraphics[width=1.\columnwidth]{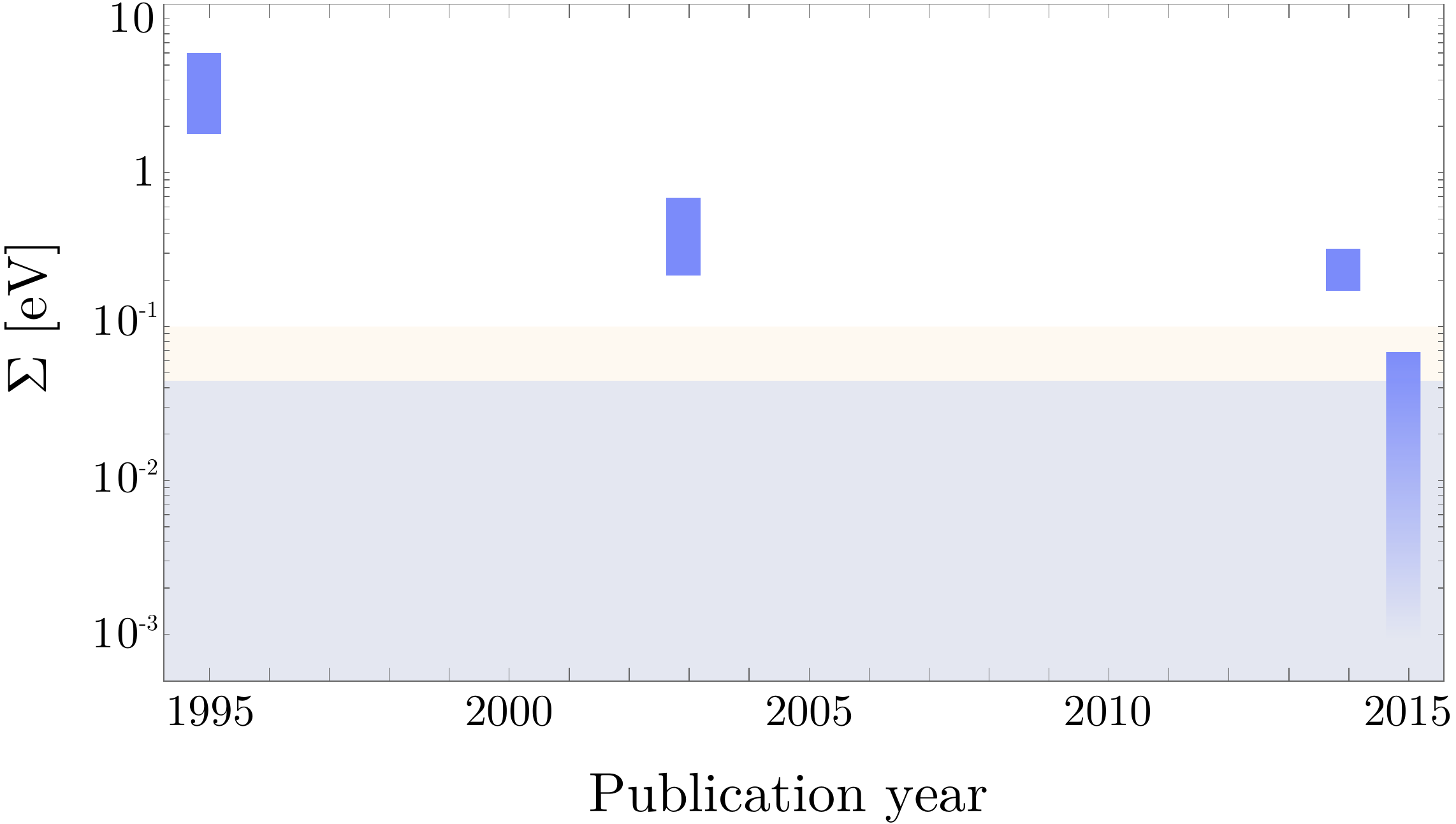}
		\caption{Evolution of some significant values for $\Sigma$ as indicated by cosmology, based on well-known 
			works~\cite{Primack:1994pe,Allen:2003pta,Battye:2013xqa,Palanque-Delabrouille:2014jca}.
%			The first value~\cite{Primack:1994pe} was used to argue in favor of the ``cold + hot dark matter cosmology'' and 
%			motivated the CHORUS~\REF and NOMAD~\REF experiments. 
			Since the error for the first value is not reported in the reference, we assumed an error of 50\% 
			for the purpose of illustration. 
			The yellow region includes values of $\Sigma$ compatible with the \NH~spectrum, but not with the \IH~one.
			The gray band includes values of $\Sigma$ incompatible with the standard cosmology and with oscillation 
			experiments.}
		\label{fig:limit_sigma}
	\end{figure}
	
	The indications for neutrino masses from cosmology has kept changing for the last 20 years.
	A comprehensive review on the topic can be found in Ref.~\cite{Lesgourgues&al:2013}.
	In Fig.~\ref{fig:limit_sigma} the values for $\Sigma$ given in 
	Refs.~\cite{Primack:1994pe,Allen:2003pta,Battye:2013xqa,Palanque-Delabrouille:2014jca} are shown.
	The scientific literature contains several authoritative claims for a non-zero value for $\Sigma$ but,
	being different among each others, these values cannot be all correct (at least) and this calls us for a cautious 
	attitude in the interpretation. 
%	Moreover, it has been shown in Ref.~\cite{Kovalenko:2013eba} that the presence in the nuclear medium of 
%	$L$-violating four-fermion interactions of neutrinos with quarks from a decaying nucleus could account for an apparent 
%	incompatibility between the \bb~searches in the laboratory and the cosmological data. 
%	In fact, the net effect of these interactions (not present in the latter case) would be the generation of an effective 
%	``in-medium'' Majorana neutrino mass matrix with a corresponding enhancement of the \bb~rate.
	Referring to the most recent years, two different positions emerge: on one side, we find claims that cosmology provides 
	us a hint for non-zero neutrino masses. On the other, we have very tight limits on $\Sigma$.

	In the former case, it has been suggested~\cite{Wyman:2013lza,Battye:2013xqa} that a total non zero neutrino 
	mass around $0.3\,\eV$ could alleviate some tensions present between cluster number counts 
	(selected both in X-ray and by Sunyaev-Zeldovich effect) and weak lensing data.
	A sterile neutrino particle with mass in a similar range is sometimes also
	advocated~\cite{Hamann:2013iba,Ade:2013zuv}.
	However, evidence for non-zero neutrino masses either in the active or sterile sectors seems to be claimed 
	in order to fix the significant tensions between different data sets (cosmic microwave background (CMB) and 
	baryonic acoustic oscillations (BAOs) on one side and weak lensing, 
	cluster number counts and high values of the Hubble parameter on the other).  

	In the latter case, the limit on $\Sigma$ is so stringent, that it better agrees with the \NH~spectrum, rather than with 
	\IH~one (see the discussion in Sec.~\ref{sec:upper_bound}).%
	\footnote{Actually, it has been shown in Ref.~\cite{Kovalenko:2013eba} that the presence in the nuclear medium of 
		$L$-violating four-fermion interactions of neutrinos with quarks from a decaying nucleus could account for an apparent 
		incompatibility between the \bb~searches in the laboratory and the cosmological data. 
		In fact, the net effect of these interactions (not present in the latter case) would be the generation of an effective 
		``in-medium'' Majorana neutrino mass matrix with a corresponding enhancement of the \bb~rate.}
	The tightest experimental limits on $\Sigma$ are usually obtained by combining CMB data with the ones probing smaller 
	scales. In this way, their combination allows a more effective investigation of the 
	neutrino induced suppression in terms of matter power spectrum, both in scale and redshift. 
	Quite recently, a very stringent limit, $\Sigma < 146\,\meV~(2\sigma \,\mbox{C.\,L.})$, was set by 
	Palanque-Delabrouille and collaborators~\cite{Palanque-Delabrouille:2014jca}.
%	by using the one-dimensional Lyman-$\alpha$ forest power spectrum~\cite{Palanque-Delabrouille:2013gaa}.
	New tight limits were presented after the data release by the Planck Collaboration in 2015~\cite{Planck:2015xua}.
	Some of the most significant results are reported in Table~\ref{tab:postPlanck}.
	The bounds on $\Sigma$ indicated by these post-Planck studies are quite small, but they are still larger than the 
	final sensitivities expected, especially thanks to the inclusion of other cosmological data sets probing smaller scales
	(see e.\,g.\ Refs.~\cite{Wong:2011ip,Lesgourgues:2014zoa} for review works). 
	Therefore, these small values cannot be considered surprising and, conversely, margins of further progress are present.

	In our view, this situation should be considered as favorable
	since more proponents are forced to carefully examine and discuss all the available hypotheses.
	In view of this discussion, in Sec.~\ref{sec:cosm_bounds} we consider two possible scenarios and discuss 
	the implications from the cosmological investigations for the \bb~in both cases.

	\begin{center}
	\begin{table}[b]
	\caption{Tight constraints on $\Sigma$ obtained in 2015, by analyzing the data on the CMB by the Planck 
		Collaboration~\cite{Planck:2015xua}, polarization included, along with other relevant cosmological data 
		probing smaller scales.}
	\small{
	\begin{ruledtabular}
	\begin{tabular}{l l}
		upper bound 				&included dataset	\\
		on $\Sigma$ ($2\sigma$\,C.\,L.)								\\[+2pt]
		\hline
		\\[-10pt]
		$153\,\meV$,~\cite{Planck:2015xua}\notaA					&SNe, BAO, $H_0$ prior \\	
		$120\,\meV$,~\cite{Palanque-Delabrouille:2015pga}		&Lyman-$\alpha$ \\	
		$126\,\meV$,~\cite{DiValentino:2015sam}					&BAO, $H_0, \tau$ priors , Planck SZ clusters \\	
		$177\,\meV$,~\cite{Zhang:2015uhk}							&BAO \\	
		$110\,\meV$,~\cite{Cuesta:2015iho}							&BAO, galaxy clustering, lensing \\	
		\end{tabular}
		\end{ruledtabular}
		\begin{flushleft}
		\notaA {\scriptsize Results as reported in 
			\href{http://wiki.cosmos.esa.int/planckpla2015/images/0/07/Params\_table\_2015\_limit95.pdf}%
			{wiki.cosmos.esa.int/planckpla2015}, page 311.} \\
		\end{flushleft}
		}
	\label{tab:postPlanck}
	\end{table}
	\end{center}

%------------------------------------------------
\subsection{Other non-oscillations data}

	For the sake of completeness, we mention other two potential sources of information on neutrinos masses. They are:
	\begin{itemize}
		\item the study of kinematic effects (in particular of supernova neutrinos)
		\item the investigation of the effect of mass in single beta decay processes.
	\end{itemize}
	The first type of investigations, applied to SN1987A, produced a limit of about 
	$6\,\eV$ on the electron antineutrino mass~\cite{Loredo:2001rx,Pagliaroli:2010ik}. 
	The perspectives for the future are connected to new detectors, or to the existence of antineutrino pulses in the 
	first instants of a supernova emission. 
	The second approach, instead, is presently limited to about $2\,\eV$~\cite{Kraus:2004zw,Aseev:2011dq}, 
	even having the advantage of being obtained in controlled conditions -- i.\,e.\ in laboratory. 
	Its future is currently in the hands of new experiments based on a \ce{^3H} source~\cite{Osipowicz:2001sq} and
	on the electron capture of \ce{^{163}Ho}~\cite{Alpert:2014lfa,Doe:2013jfe,Ranitzsch:2012}, 
	which have the potential to go below the eV in sensitivity.

%------------------------------------------------
%------------------------------------------------
\subsection{Theoretical understanding}

	Theorists have not been very successful in anticipating the discoveries 
	on neutrino masses obtained by means of oscillations. The discussion within gauge models clarified that it is possible 
	or even likely to have neutrino masses in gauge models (compare with Sec.\ \ref{sec:rh}). 
	However, a large part of the theoretical community focused for a long time on models such as ``minimal $SU(5)$'', 
	where the neutrino masses are zero, emphasizing the interest in proton decay search rather than in neutrino mass search. 
	On top of that, we had many models that aimed to predict e.\,g., the correct solar neutrino solution or the size of 
	$\theta_{13}$ before the measurements, but none of them were particularly convincing. More specifically, 
	a lot of attention was given to the ``small mixing angle solution'' and the ``very small $\theta_{13}$ scenario'', 
	that are now excluded from the data. 

	Moreover, it is not easy to justify the theoretical position where neutrino masses are not considered along the masses 
	of other fermions. 
	This remark alone explains the difficulty of the theoretical enterprise that theorists have to face.
	For the reasons commented in Sec.~\ref{sec:rh}, the $SO(10)$ models are quite attractive to address a discussion of 
	neutrino masses. However, even considering this specific class of well-motivated Grand Unified groups, 
	it remains difficult to claim that we have a complete and convincing formulation of the theory.
	In particular, this holds for the arbitrariness in the choice of the representations (especially that of the Higgs 
	bosons), for the large number of unknown parameters (especially the scalar potential), for the possible role 
	of non-renormalizable operators, for the uncertainties in the assumption concerning low scale supersymmetry, for the 
	lack of experimental tests, etc.\,.
	Note that, incidentally, preliminary investigations on the size of $\mbb$ in $SO(10)$ did not provide a clear evidence 
	for a significant lower bound~\cite{Buccella:2004cd}.
	Anyway, even the case of an exactly null effective Majorana mass does not increase the symmetry of the Lagrangian, 
	and thus does not forbid the \bb, as remarked in Ref.~\cite{Pascoli:2007qh}.

	Here, we just consider one specific theoretical scheme, for illustration purposes. 
	This should not be considered a full fledged theory, but rather it attempts to account for the theoretical uncertainties 
	in the predictions. 
	The hierarchy of the masses and of the mixing angles has suggested the hypothesis that the elements of the Yukawa 
	couplings and thus of the mass matrices are subject to some selection rule. 
	The possibility of a $U(1)$ selection rule has been proposed in Ref.~\cite{Froggatt:1978nt} and, since then, it 
	has become very popular. 
	
	Immediately after the first strong evidences of atmospheric neutrino oscillations (1998) specific realizations 
	for neutrinos have been discussed in various works (see Ref.~\cite{Vissani:2001im} for references). 
	These correspond to the neutrino mass matrix 
	\begin{equation}
		M_\mathrm{neutrino} = m \times \mbox{diag}(\varepsilon,1,1)\ C\ \mbox{diag}(\varepsilon,1,1)
%		h\ C \ h\mbox{ with }h=
	\end{equation}
	where the flavor structure is dictated by a diagonal matrix that acts only on the electronic flavor and suppresses 
	the matrix elements $M_{\el\mu},M_{\el\tau}$ and $M_{\el\el}$ (twice).
	The dimensionful parameter (the overall mass scale) is given by 
	$\Delta \equiv \sqrt{\Delta m^2_\mathrm{atm}} \approx 50\,\meV$.
	We thus have a matrix of coefficients $C$ with elements $C_{\ell\ell'}=\mathcal{O}(1)$ 
	that are usually treated as random numbers of the order of 1 in the absence of a theory.
	A choice of $\varepsilon$ that suggested values of $\theta_{12}$ and $\theta_{13}$ in the correct region (before their 
	measurement) is $\varepsilon = \theta_C$ or $\sqrt{m_\mu/m_\tau}$~\cite{Vissani:2001im}. 
	Within these assumptions, the matrix element in which we are interested is 
	\begin{equation}
		\mbb = \left| m\ \varepsilon^2\ \mathcal{O}(1) \right| \approx (2-4)\,\meV.
	\end{equation}
	Finally, we note that the SM renormalization of the elements of the neutrino mass matrix is multiplicative. 
	The effect of renormalization is therefore particularly small for $\mbb$ 
	(see e.\,g.\ Eq.~(17) of Ref.~\cite{Vissani:2003aj} and the discussion therein). 
	In other words, the value $\mbb=0$ (or values close to this one) should be regarded as a stable point of the 
	renormalization flow. 
	
	Let us conclude repeating that, anyway, there are many reasons to consider the theoretical expectations with 
	detachment, and the above theoretical scheme is not an exception to this rule. 
	It is very important to keep in mind this fact in order to properly assess the value of the search for the \bb~and 
	to proceed accordingly in the investigations.

%------------------------------------------------
%------------------------------------------------
	¡ø%------------------------------------------------
%------------------------------------------------
\section{The role of nuclear physics}
\label{sec:nuclear}

	\begin{figure}[tb]
		\centering
		\includegraphics[width=1.\columnwidth, angle=0]{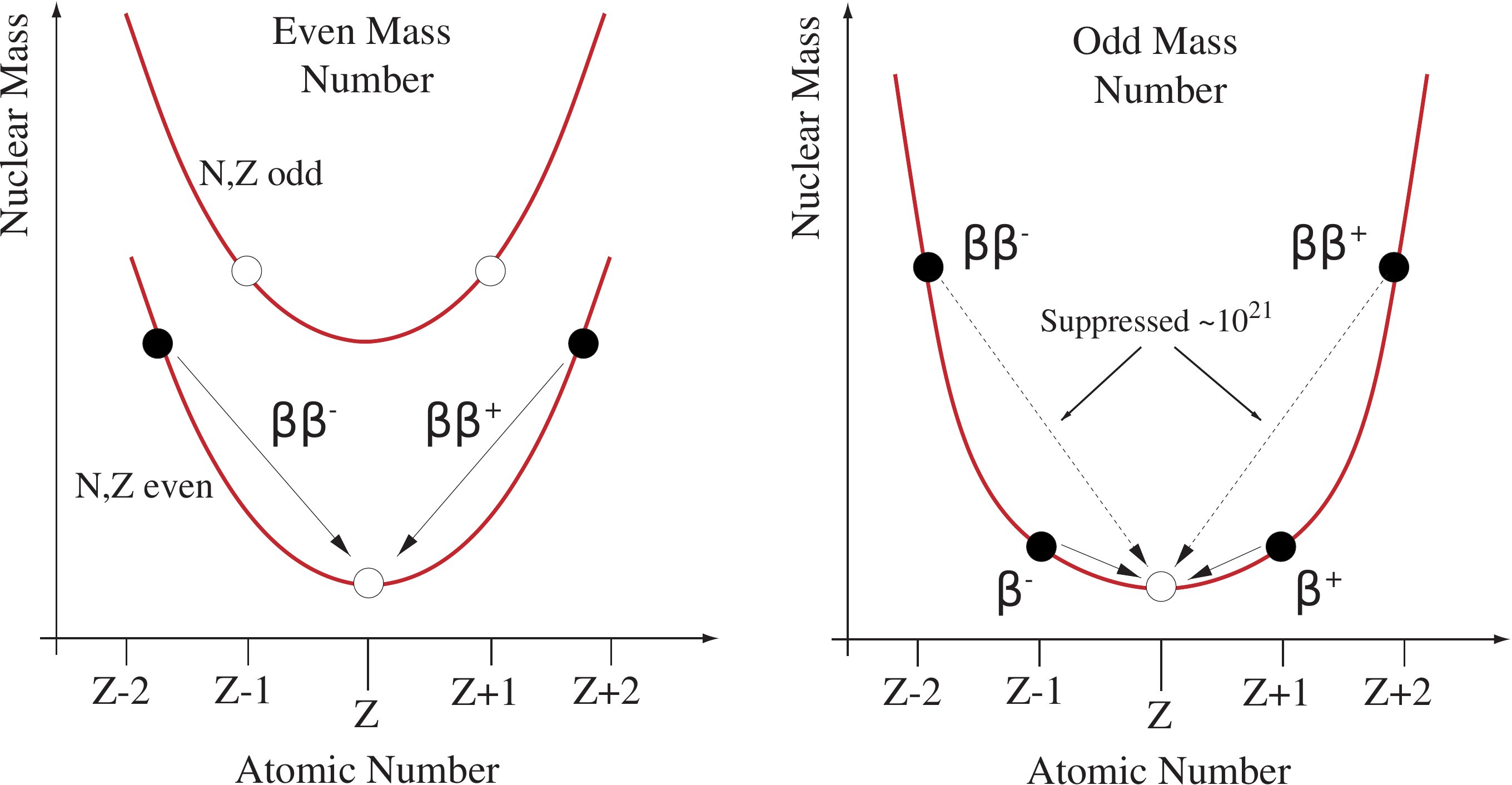}
		\caption{Nuclear mass as a function of the atomic number Z in the case of an isobar candidate with $A$ even 
			(left) and $A$ odd (right).}
		\label{fig:parabola}
	\end{figure}
	
	\bb~is first of all a nuclear process. Therefore, the transition has to be described 
	properly, taking into account the relevant aspects that concern nuclear structure and dynamics.
	In particular, it is a second order nuclear weak process and it corresponds to the transition from a nucleus 
	$(A, Z)$ to its isobar $(A, Z + 2)$ with the emission of two electrons.
	In principle, a nucleus $(A, Z)$ can decay via double beta decay as long as the nucleus $(A, Z + 2)$ is lighter. 
	However, if the nucleus can also decay by single beta decay, $(A, Z + 1)$, the branching 
	ratio for the \bb~will be too difficult to be observed due to the overwhelming background rate from the single beta decay. 
	Therefore, candidate isotopes for detecting the \bb~are even-even nuclei that, due to the nuclear pairing force, 
	are lighter than the odd-odd $(A, Z + 1)$ nucleus, making single beta decay kinematically forbidden 
	(Fig.~\ref{fig:parabola}). 
	It is worth noting that, since the \bb~candidates are even-even nuclei, 
	it follows immediately that their spin is always zero.

	The theoretical expression of the half-life of the process in a certain nuclear species can be factorized as:
	\begin{equation}
		\left[\taubb \right]^{-1}=G_{0\nu}\left|\mathcal{M}\right|^2 \left|f(m_i,U_{\el i})\right|^2
		\label{eq:tau}
	\end{equation}
	where $G_{0\nu}$ is the phase space factor (PSF), $\mathcal{M}$ is the nuclear matrix element (NME) and 
	$f(m_i, U_{\el i})$ is an adimensional function containing the particle physics beyond the SM that could explain 
	the decay through the neutrino masses $m_i$ and the mixing matrix elements $U_{\el i}$. 
	
%	$\mathcal{M}$ is an adimensional quantity of essential importance in the subsequent discussion.
	In this section, we review the crucial role of nuclear physics in the expectations, predictions and eventual 
	understanding of the \bb, also assessing the present knowledge and uncertainties.
	We mainly restrict to the discussion of the light neutrino exchange as the candidate process for mediating the 
	\bb~transition, but the mechanism of heavy neutrino exchange is also considered.
	
	In the former case ($m\lesssim 100\,\mbox{MeV}$, see Eq.~(\ref{eq:impulso_dbd})), the factor $f$ is proportional 
	$\mbb$:
%	\begin{align}
%		f(m_i, U_{ei} ) &= \frac{\mbb}{m_\el} = \frac{1}{m_\el} 
%		\Biggl| \sum_{k=\underset{(e,\mu,\tau)}{\mbox{\tiny light}}} U_{ek}^2 m_k \Biggr|   \\
%		&= \Biggl| e^{i\alpha_1}|U_{ei}^2|m_1 + e^{i\alpha_2}|U_{e2}^2|m_2 + e^{-2i\phi}|U_{e3}^2|m_3 \Biggr| \nonumber
%	\label{eq:mbb_def}
%	\end{align}
	\begin{equation}
		f(m_i, U_{\el i} ) \equiv \frac{\mbb}{m_\el} 
		= \frac{1}{m_\el} \left| \sum_{k=1,2,3} U_{\el k}^2 m_k \right|
	\label{eq:mbb_def}
	\end{equation}
	where the electron mass $m_\el$ is taken as a reference value.
	In the scheme of the heavy neutrino exchange ($m \gtrsim 100\,\mbox{MeV}$), the effective parameter is instead: 
	\begin{equation}
		f(m_i,U_{\el i}) \equiv m_\pr \, \left\langle M_H^{-1}\right\rangle 
		=m_\pr \, \left| \sum_{I=\mathrm{heavy}}U_{\el I}^2 \frac{1}{M_I} \right|
	\end{equation}
	where the proton mass $m_\pr$ is now used, according to the tradition, as the reference value.

%------------------------------------------------
\subsection{Recent developments on the phase space factor calculations}

%	The PSF, $G_{0\nu}$, accounts for the atomic physics behind the \bb~process.
	The first calculations of PSFs date back to the late 1950s~\cite{Primakoff:1959} and used a simplified
	description of the wave functions. 
	The improvements in the evaluation of the PSFs are due to always more accurate descriptions and less 
	approximations~\cite{Doi:1981mj,Doi:1982dn,Tomoda:1990rs}.
	
	Recent developments in the numerical evaluation of Dirac wave functions and in the solution of the Thomas-Fermi 
	equation allowed to calculate accurately the PSFs both for single and double beta decay.
	The key ingredients are the scattering electron wave functions. 
	The new calculations take into account relativistic corrections, the finite nuclear size and the effect of the 
	atomic screening on the emitted electrons. 
	The main difference between these calculations and the older 
	ones is of the order of a few percent for light nuclei ($Z=20$), about $30\%$ for \ce{Nd} ($Z=60$), and a rather 
	large $90\%$ for \ce{U} ($Z=92$).

	In Refs.~\cite{Kotila:2012zza,Stoica:2013lka,Stefanik:2015twa}, the most up to date calculations of the PSFs 
	for \bb~can be found.
	The results obtained in these works are quite similar. Throughout this paper, we use the values from the first 
	reference.
	
%------------------------------------------------
\subsection{Models for the NMEs}

	\begin{figure}[tb]
		\centering
		\includegraphics[width=1.\columnwidth, angle=0]{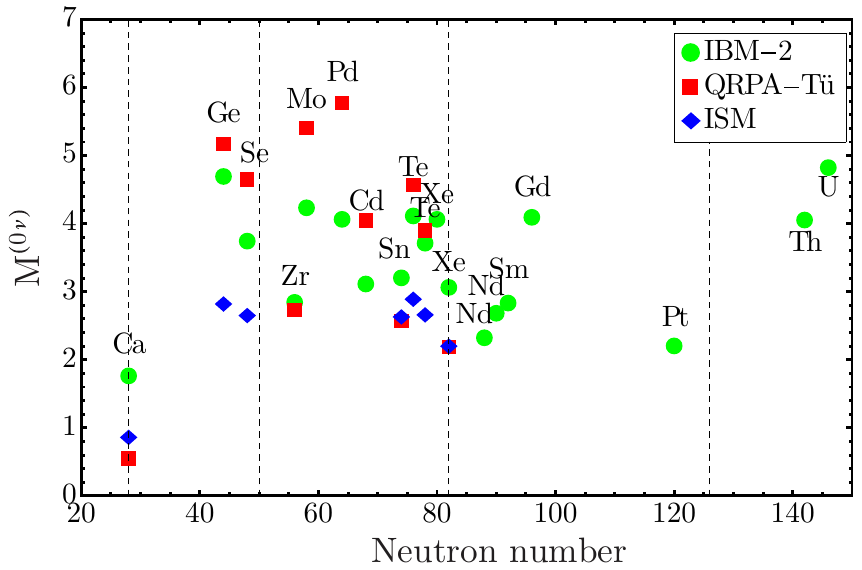}
		\caption{Most updated NMEs calculations for the \bb~with the IBM-2~\cite{Barea:2015kwa}, 
			QRPA-T\"{u}~\cite{Simkovic:2013qiy} and ISM~\cite{Menendez:2008jp} models. 
			The results somehow differ among the models, but are not too far away.
%			It is clear that this is not the main issue about theoretical uncertainties in the NME, 
%			as discussed in Sec.\ \ref{sec:theo_uncertainties}.
			Figure from Ref.~\cite{Barea:2015kwa}.}
		\label{fig:NME_all}
	\end{figure}

	Let us suppose that the decay proceeds through an $s$-wave. Since we have just two electrons 
	in the final state, we cannot form an angular momentum greater than one. Therefore, usually only \bb~matrix elements to 
	final $0^+$ states are considered.  These can be the ground state, $0^+_1$, or the first excited state, $0^+_2$. Of course, 
	we consider as a starting state just a $0^+$ state, since the double beta decay is possible only for $(Z,A)$ even-even 
	isobar nuclei. 
	
	The calculation of the NMEs for the \bb~is a difficult task because the ground and many excited states of open-shell 
	nuclei with complicated nuclear structure have to be considered. 
	The problem is faced by using different approaches and, especially in the last few years, the reliability of 
	the calculations improved a lot. 
	Here, a list of the main theoretical models is presented. The most relevant features for each of them are highlighted.
	\begin{itemize}
		\item \emph{Interacting Shell Model} (ISM), \cite{Caurier:2004gf,Menendez:2008jp}. 
			In the ISM only a limited number of orbits around the Fermi level is considered, 
			but all the possible correlations within the space are included and the pairing correlations in the valence space are 
			treated exactly. Proton and neutron numbers are conserved and angular momentum conservation is preserved. 
			A good spectroscopy for parent and daughter nuclei is achieved
%			The calculations performed with ISM are available for several nuclei
		\item \emph{Quasiparticle Random Phase Approximation} (QRPA), \cite{Simkovic:2013qiy,Hyvarinen:2015bda}. 
%			Rodin:2003eb
			The QRPA uses a large valence space and thus it cannot comprise all the possible configurations. 
			Typically, single particle states in a Woods-Saxon potential are considered. 
			The proton-proton and neutron-neutron pairings are taken into account and treated in the BCS approximation
			(proton and neutron numbers are not exactly conserved) 
		\item \emph{Interacting Boson Model} (IBM-2), \cite{Barea:2015kwa}
%			Barea:2009zza
			In the IBM, the low-lying states of the nucleus are modeled in terms of bosons. The bosons are in either 
			$s$ boson ($L=0$) or $d$ boson ($L=2$) states. Therefore, one is restricted to $0^+$ and $2^+$ neutron 
			pairs transferring into two protons. 
			The bosons interact through one- and two-body forces giving rise to bosonic wave functions.
		\item \emph{Projected Hartree-Fock Bogoliubov Method} (PHFB), \cite{Rath:2013fma} 
%			Rath:2010zz
			In the PHFB, the NME are calculated using the projected-Hartree-Fock-Bogoliubov wave functions, which 
			are eigenvectors of four different parameterizations of a Hamiltonian with pairing plus multipolar 
			effective two-body interaction. In real applications, the nuclear Hamiltonian is restricted only to 
			quadrupole interactions
		\item \emph{Energy Density Functional Method} (EDF), \cite{Rodriguez:2010mn}. 
			The EDF is considered to be an improvement with respect to the PHFB. The state-of-the-art density functional 
			methods based on the well-established Gogny D1S functional and a large single particle basis are used.
	\end{itemize}
	
	The most common methods are ISM, QRPA and IBM-2. 
	In Fig.~\ref{fig:NME_all}, a comparison among the most recent NME calculations computed with these three models is shown.
	It can be seen that the disagreement can be generally quantified in some tens of percents, instead of the factors 
	$2-4$ of the past. This can be quite satisfactory. 
	As it will be discussed in Sec.~\ref{sec:theo_uncertainties}, the main source of uncertainty in the inference does not rely 
	in the NME calculations anymore, but in the determination of the quenching of the axial vector coupling constant.
	For this reason, in the subsequent discussion we will restrict to one of the considered models, namely the 
	IBM-2 \cite{Barea:2015kwa}, without significant loss of generality.
		
%------------------------------------------------
\subsection{Theoretical uncertainties}
\label{sec:theo_uncertainties}	

%------------------------------------------------
\subsubsection{Generality}

	Following Eq.~(\ref{eq:tau}), an experimental limit on the \bb~half-life translates in a limit on the effective Majorana 
	mass:
	\begin{equation}
		\mbb \le \frac{m_\el}{\mathcal{M}\  \sqrt{G_{0\nu} \, \taubb}}.
	\label{eq:mbb_bound}
	\end{equation}
	From the theoretical point of view, in order to constrain $\mbb$, the estimation of the uncertainties both on 
	$G_{0\nu}$ and $\mathcal{M}$ is crucial.
	Actually, the PSFs can be assumed quite well known, being the error on their most 
	recent calculations around $~7\%$~\cite{Kotila:2012zza}.
	
	A convenient parametrization for the NMEs is the following~\cite{Simkovic:1999re}:
	\begin{equation}
		\mathcal{M} \equiv g_A^2 \, \mathcal{M}_{0\nu} =
		g_A^2\left(M^{(0\nu)}_{GT}-\left(\frac{g_V}{g_A}\right)^2M^{(0\nu)}_{F}+M^{(0\nu)}_T\right)
	\label{eq:matrix_elements}
	\end{equation}
	where $g_V$ and $g_A$ are the axial and vector coupling constants of the nucleon, $M^{(0\nu)}_{GT}$ is the Gamow-Teller 
	(GT) operator matrix element between initial and final states (spin-spin interaction), $M^{(0\nu)}_{F}$ is the Fermi 
	contribution (spin independent interaction) and $M^{(0\nu)}_T$ is the tensor operator matrix element. 
	The form of Eq.~(\ref{eq:matrix_elements}) emphasizes the role of $g_A$.
	Indeed, $\mathcal{M}_{0\nu}$ mildly depends on $g_A$ and can be evaluated 
	by modeling theoretically the nucleus. Actually, it is independent on $g_A$ if the same quenching is assumed both for 
	the vector and axial coupling constants, as we do here for definiteness, following Ref.~\cite{Barea:2013bz}.%
%	\footnote{Some residual dependence upon $g_A$ could be attributed to a different renormalization of the two 
%		coupling constants.} 

%------------------------------------------------
\subsubsection{Is the uncertainty large or small?}
	
	The main sources of uncertainties in the inference on $\mbb$ are the NMEs.
	A comparison of the calculations from 1984 to 1998 revealed an uncertainty of more than a 
	factor 4~\cite{Feruglio:2002af}. 
	A similar point of view comes out from the investigation of Ref.~\cite{Bahcall:2004ip}, where the results of the 
	various calculations were used to attempt a statistical inference.

	An important step forward was made with the first calculations of $\mathcal{M}_{0\nu}$ that estimated also 
	the errors, see Refs.~\cite{Rodin:2003eb,Rodin:2006yk}. 
	These works, based on the QRPA model, assessed a relatively small intrinsic error of 
	$\sim 20\%$. The validity of these conclusions have been recently supported 
	by the (independent) calculation based on the IBM-2 description of the 
	nuclei~\cite{Kotila:2012zza,Barea:2015kwa}, which assesses an intrinsic error of $~15\%$ on $\mathcal{M}_{0\nu}$.
	However, the problem in assessing the uncertainties in the NMEs is far from being solved. 
	Each scheme of calculation can estimate its own uncertainty, but it is still hard to understand the differences in 
	the results among the models (Fig.~\ref{fig:NME_all}) and thus give an overall error.
%	Also, the computation of other already measured processes like single beta decay and \bbvv~is not accurate.
	Notice also that when a process ``similar'' to the \bb~is considered (single beta decay, electron capture, \bbvv) and 
	the calculations are compared with the measured rates, the actual differences are much larger 
	than $20\%$~\cite{Barea:2013bz}. 
	This suggests that it is not cautious to assume that the uncertainties on the \bb~are instead subject to such a level 
	of theoretical control.
%	In addition, the papers~\cite{Barea:2013bz,Barea:2015kwa}, have also emphasized a more important role of axial 
%	coupling $g_A$ than originally assumed. 
%	Furthermore, a more important role than originally assumed is played by the axial coupling constant.
%	In fact, the issue concerning the value to adopt for $g_A$ translates into a big uncertainty in the overall 
%	evaluation of $\mathcal{M}$.

	Recently, there has been a lively interest in a specific and important reason of uncertainty, namely the value 
	of the axial coupling constant $g_A$. 
	This has a direct implication on the issue that we are discussing, since any uncertainty on the value of $g_A$ 
	reflects itself into a (larger) uncertainty factor on the value of the matrix element $\mathcal{M}$. 
	We will examine these arguments in greater details in the rest of this section.
	 
	It is important to appreciate the relevance of these considerations for the experimental 
	searches. If the value of the axial coupling in the nuclear medium is decreased by a factor $\delta$, namely 
	$g_A~\rightarrow~g_A\cdot(1 - \delta)$, the expected decay rate and therefore the number of signal events $S$ will also 
	decrease, approximatively as $S\cdot(1 -  \delta)^4$. 
	This change can be compensated by increasing the time of data taking or the mass of the experiment. 
	However, the figure of merit, namely $S/\sqrt{B}$, which quantifies 
	the statistical significance of the measurement, changes only with the square root of the time or of the mass, 
	in the typical case in which there are also background events $B$.
	For instance, if we have a decrease by $\delta= 10$\,($20$)\% of the axial coupling, we will obtain the same 
	measurement after a time that is larger by a factor of $1/(1-\delta)^8=2.3$\,($6$). In other words, an effect 
	that could be naively considered small has instead a big impact for the experimental search for the \bb.
	
%------------------------------------------------
\subsubsection{The size of the axial coupling}
	
	\begin{figure}[t]
		\centering
		\includegraphics[width=1.\columnwidth, angle=0]{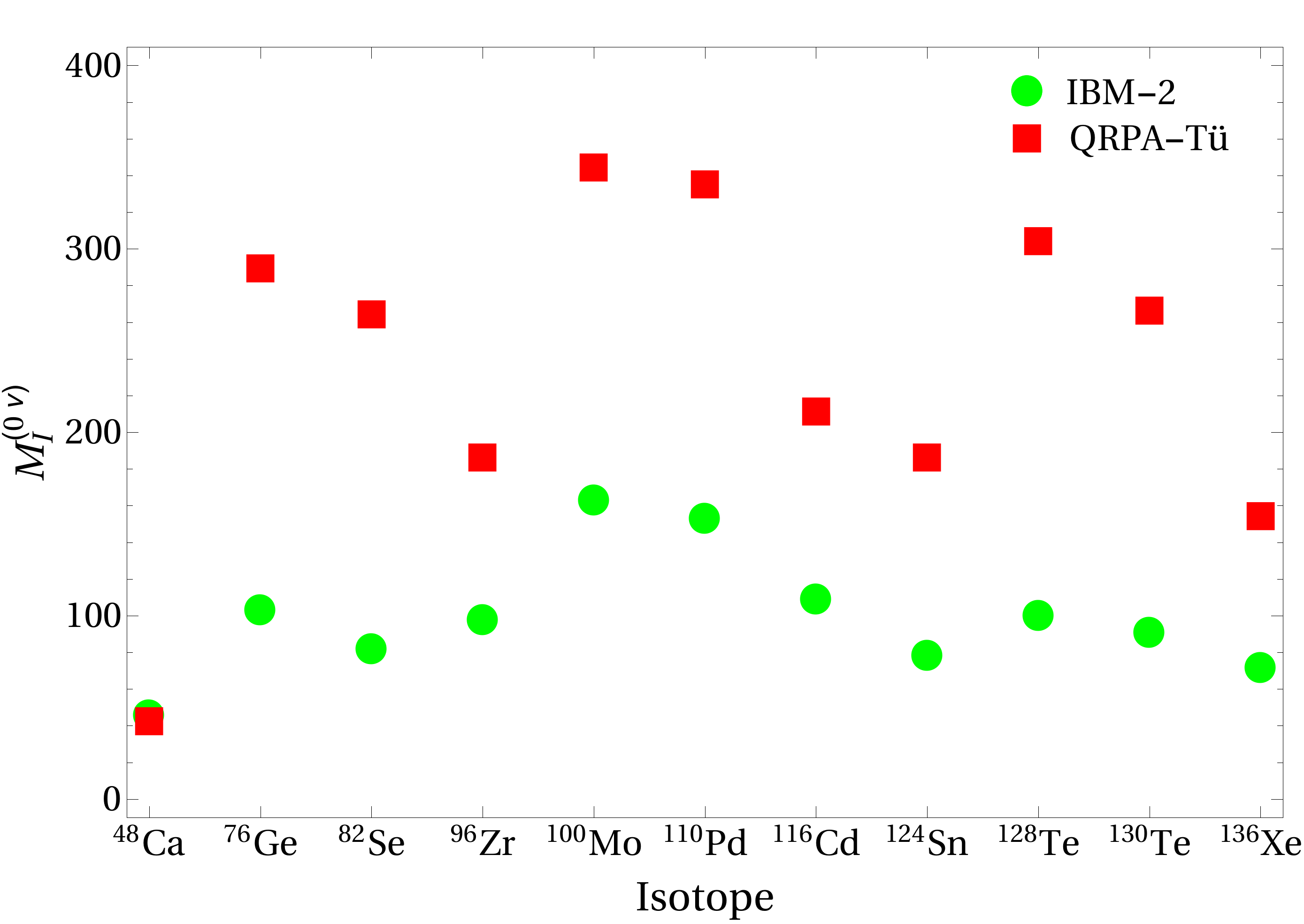}
		\caption{Most updated NMEs calculations for the \bb~via heavy neutrino exchange with the 
			IBM-2~\cite{Barea:2015kwa} and QRPA-T\"{u}~\cite{Faessler:2014kka} models. 
			In both cases, the value $g_A=g_\mathrm{nucleon}$ for the axial coupling constant and the Argonne parametrization 
			for the short-range correlations are assumed. 
%			The QRPA results are systematically overestimating the IBM-2 ones.
			The results show a continuous overestimation of the QRPA estimations over the IBM-2 ones.}
		\label{fig:NMEheavy}
	\end{figure}

	It is commonly expected that the value $g_A\simeq 1.269$ measured in the weak 
	interactions and decays of nucleons is ``renormalized'' in the nuclear medium towards the value appropriate for 
	quarks~\cite{Rodin:2003eb,Rodin:2006yk,Simkovic:2012hq}. 
	It was argued in Ref.~\cite{Barea:2013bz} that a further modification (reduction) is rather plausible. 
	This is in agreement with what was stated some years before in Ref.~\cite{Faessler:2007hu},
	where the possibility of a ``strong quenching'' of $g_A$ (i.\,e.\ $g_A<1$) is actually favored.
	The same was also confirmed by recent study on single beta decay and \bbvv~\cite{Suhonen:2013laa}.
	It has to be noticed that within the QRPA framework, the dependence of $\mathcal{M}$ upon $g_A$ is actually milder than 
	quadratic, because the model is calibrated through the experimental \bbvv~decay rates using also another parameter,
	the particle-particle strength $g_{pp}$~\cite{Lisi:2015yma}.
	
%	Following the discussion of Ref.~\cite{Barea:2013bz}, the quenching of $g_A$ can mainly be attributed to two issues:
	There could be different causes for the quenching of $g_A$. It was found that it can be 
	attributed mainly to the following issues~\cite{Barea:2013bz,Konieczka:2015ela}:
	\begin{itemize}
		\item the limited model space (i.\,e.\ the size of the basis of the eigenstates) in which the calculation is done. 
			This problem is by definition model dependent and it was extensively investigated in light nuclei in the 
			1970s~\cite{Fujita:1965,Wilkinson:1973zz,Wilkinson:1973_2,Wilkinson:1974}, 
			when it was argued that $g_A \sim1$.
			In heavy nuclei, the question of quenching was first discussed in Ref.\ \cite{Fujita:1965}. 
			In this case, $g_A$ was found to be even lower than $1$, thus stimulating the statement that massive 
			renormalization of $g_A$ occurs;
		\item the contribution of non-nucleonic degrees of freedom. This effect does not depend much on the nuclear 
			model adopted, but rather on the mechanism of coupling to non-nucleonic degrees of freedom. 
			It was extensively investigated theoretically in the 
			1970s~\cite{Ericson:1971,Barshay:1974fh,Shimizu:1974}.
			Recently, it has been investigated again within the framework of the chiral Effective Field 
			Theory (EFT)~\cite{Menendez:2011qq}. It turns out that it may depend on momentum transfer and that it may lead in 
			some cases to an enhancement rather than a quenching;
		\item the renormalization of the GT operator due to two-body currents.
			The first calculations for GT transitions for the \bb~operator based on the chiral 
			EFT~\cite{Menendez:2011qq} showed the importance of two-body currents for the effective quenching of $g_A$.
			This was later confirmed in independent works~\cite{Ekstrom:2014iya,Engel:2014pha} and, more recently, by the use 
			of a no-core-configuration-interaction formalism within the density functional theory~\cite{Konieczka:2015ela}.
	\end{itemize}
	
	It is still not clear if the quenching in both the transitions (\bb~and \bbvv) is the same. 
	One argument which suggests that this is not unreasonable, consists in noting that the \bbvv~can occur only through a 
	GT ($1^+$) transition. Instead, the \bb~could happen through all the possible intermediate states, 
	so it is possible to argue that the transitions through states with spin parity different from $1^+$ can be unquenched 
	or even enhanced. Incidentally, it turns out that the dominant multipole in the \bb~transition 
	is the GT one, thus making the hypothesis that the quenching in \bbvv~and \bb~is the same quite solid.
%	This will be our working hypothesis for the following part of the work, but it is worth it to notice that within 
%	the QRPA model and the ISM model there are some indications for which this might not be 
%	true~\cite{Engel:2014pha, Menendez:2008jp}.
	Following Ref.~\cite{Barea:2015kwa}, we adopt this as a working hypothesis in our discussion, however keeping in 
	mind that some indications that the quenching might be different in the \bb~and \bbvv~transitions are present 
	in other models~\cite{Engel:2014pha,Menendez:2008jp}.
	
	It would be extremely precious if these theoretical questions could be answered by some experimental data. 
	It has been argued that the experimental study of nuclear transitions where the nuclear charge is changed by two units
	leaving the mass number unvaried, in analogy to the \bb~decay, could give important information. 
	Despite the Double Charge Exchange reactions and \bb~processes are mediated by different interactions, 
	some similarities between the two cases are presents. These could be exploited to assess effectively the NME for 
	the \bb~(and, more specifically, the entity of the quenching of $g_A$). 
%	In the near future, a new project called NUMEN \cite{Cappuzzello:2015oza} will start in Catania (Italy) to try to get 
%	some inputs for the improvement of the theoretical understanding of this nuclear process.
	In the near future, a new project will be started at the Laboratori Nazionali del Sud (Italy)~\cite{Cappuzzello:2015oza} 
	with the aim of getting some inputs to deepen our theoretical understanding of this nuclear process.
	
%------------------------------------------------
\subsubsection{Quenching as a major cause of uncertainty}
	
	In view of the above considerations, we think that currently the value of $g_A$ in the nuclear medium cannot be 
	regarded as a quantity that is known reliably.
	It is rather an \emph{important reason of uncertainty in the predictions}. 
	In a conservative treatment, we should consider at least the following three cases,
	\begin{equation}
	g_A~=~
		\begin{cases} \label{eq:cazzes}
			g_\mathrm{nucleon} &= 1.269 \\
			g_\mathrm{quark}   &= 1 \\
			g_\mathrm{phen.}   &= g_\mathrm{nucleon}\cdot A^{-0.18}
		\end{cases}
	\end{equation}
	where the last formula
%	is called ``maximal quenching'' ($g_A<1$) and it 
	includes phenomenologically the effect of the atomic number $A$.
	It represents the worst possible scenario for the \bb~search.
%	(see Secs.~\ref{sec:present_sensitivity} and \ref{sec:fut_exp})
	The $g_\mathrm{phen.}$ parametrization as a function of $A$ comes directly from 
	the comparison between the theoretical half-life for \bbvv~and its observation in different nuclei, 
	as reported in Ref.~\cite{Barea:2013bz}. 
	From the comparison between the theoretical half-life for the process and the experimental value it was 
	possible to extract an effective value for $g_A$, thus determining its quenching.  
	The assumption that $g_A$ depends only upon the atomic number $A$ is rather convenient for a cursory exploration of 
	the potential impact of unaccounted nuclear physics effects on \bb, but most likely it is also an oversimplification 
	of the truth, as suggested by the residual difference between the calculated \bbvv~rates. 
	Surely, it cannot replace an adequate theoretical modeling, that in the light of the following discussion has become 
	rather urgent. Anyway, we stress that this is just a phenomenological 
	description of the quenching, since the specific behavior is different in each nucleus and it somewhat differs from 
	this parametrization~\cite{Barea:2013bz}. 
		
	The question which is the ``true value'' of $g_A$ is still open and introduces a considerable uncertainty in the 
	inferences concerning massive neutrinos. The implications are discussed in Secs.~\ref{sec:present_sensitivity}
	and \ref{sec:fut_exp}.

%------------------------------------------------
\subsection{The case of heavy neutrino exchange}
	
	\begin{figure}[t]
		\centering
		\includegraphics[width=1.\columnwidth, angle=0]{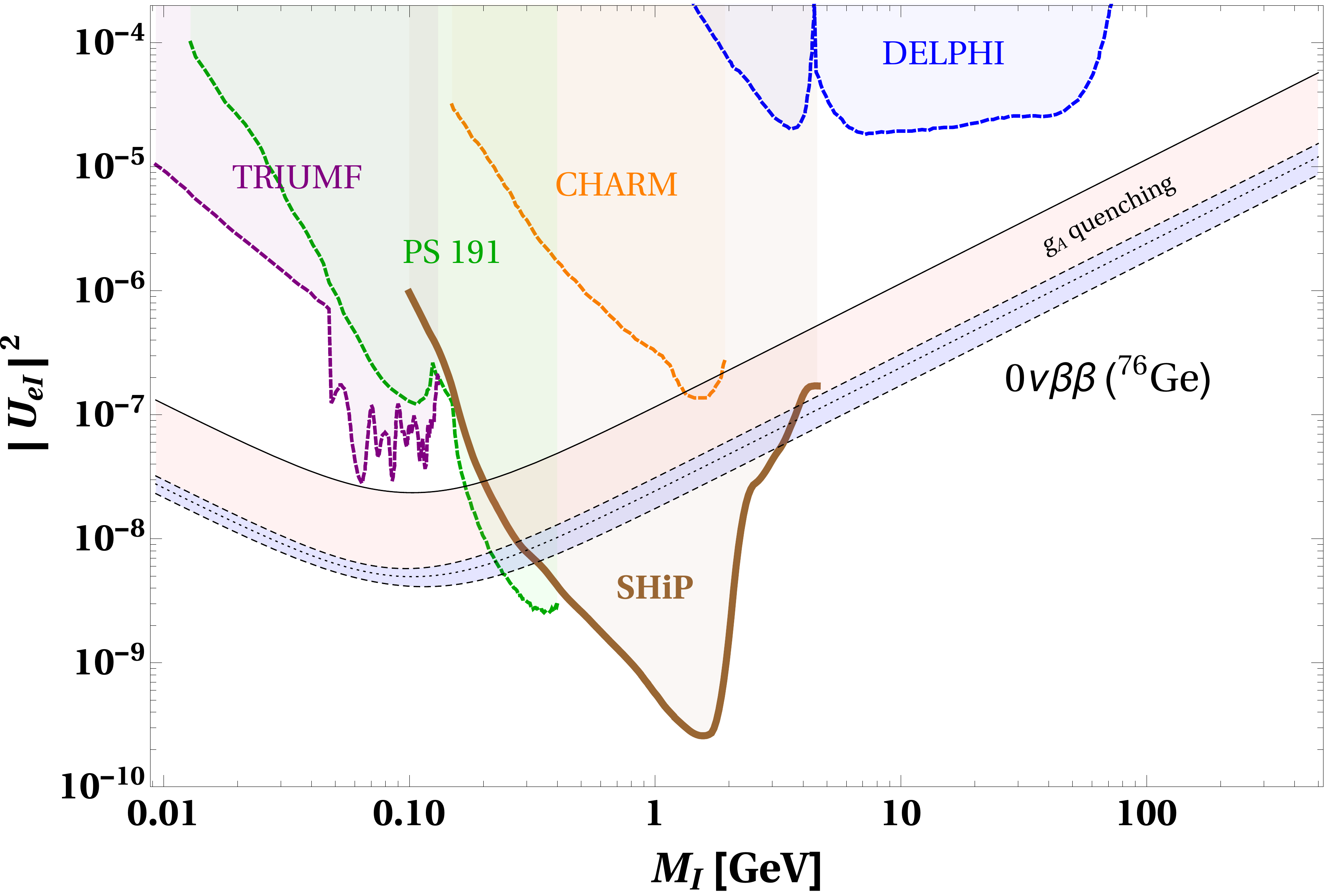}
		\caption{Bounds on the mixing between the electron neutrino and a single heavy neutrino from the 
			combination of bounds obtained with Ge~\bb~experiments~\cite{Agostini:2013mzu} using the representation 
			introduced in Ref.~\cite{Kovalenko:2009td}. 
			The bands correspond to the uncertainties discussed in the text.
			The dashed contours indicate the mass regions excluded by some of the accelerator experiments considered in 
			Ref.~\cite{Atre:2009rg}: CHARM (90\%\,C.\,L., \cite{Bergsma:1985is}), 
			DELPHI (95\%\,C.\,L., \cite{Abreu:1996pa}), 
			PS\,191 (90\%\,C.\,L., \cite{Bernardi:1987ek}), 
			TRIUMF (90\%\,C.\,L., \cite{Britton:1992pg,Britton:1992xv}).
			The continuous contour indicate the expected probed region by the new proposed SHiP experiment 
			at the CERN SPS~\cite{Alekhin:2015byh}. Figure from Ref.~\cite{Alekhin:2015byh}.}
		\label{fig:plut}
	\end{figure}

	As already discussed in Sec.~\ref{sec:mechanism}, it is possible to attribute the \bb~decay rate to the same 
	particles that are added to the SM spectrum to explain oscillations, e.\,g.\ heavy neutrinos. 
	In this context one can assume that the exchange of $M_H>100\,\MeV$ saturates the \bb~decay rate, also 
	reproducing the ordinary neutrino masses. 
	Heavy neutrino masses and mixing angles, compatible with the rate of \bb, depend on the NMEs of 
	the transition (compare e.\,g.~\cite{Atre:2009rg} and~\cite{Mitra:2011qr}). 
	Thus, nuclear physics has an impact also on the limits that are relevant for a direct search for heavy neutrinos 
	with accelerators. Each scheme of nuclear physics calculation can estimate its intrinsic uncertainty. 
	This is usually found to be small in modern computations (about $28\%$ for heavy neutrino exchange~\cite{Barea:2015kwa}).
	In a conservative treatment, this uncertainty plus the already discussed unknown value of $g_A$ should be taken 
	into account. 
	It has to be noticed that if the \bb~is due to a point-like (dimension-9) operator, as for heavy neutrino exchange, 
	two nucleons are in the same point. 
	Therefore, the effect of a hard core repulsion, estimated modeling the ``short-range correlations'', plays
	an important role in the determination of the uncertainties.
	A significant step forward has been recently made, pushing down this source of theoretical error of about an order 
	of magnitude~\cite{Barea:2015kwa}. 
%	The inaccuracy of 28\% previously mentioned takes it already into account.
			
%	Also in this case, we use the NMEs evaluated with the IBM-2 model~\cite{Barea:2015kwa}.
%	Considering, for example, the case of \ce{^{76}Ge}, we get:
%	\begin{equation}
%		\mathcal{M}(\ce{Ge}) = 
%		\begin{cases}
%			104\pm 29 						&g_A=g_\mathrm{nucleon}	\\
%			22 \pm 6							&g_A=g_\mathrm{phen.}
%		\end{cases}.
%		\label{eq:Mge}
%	\end{equation}
%	
%	It has to be noted that the NMEs evaluated within the QRPA model~\cite{Faessler:2014kka} are larger in comparison 
%	with those obtained with the IBM-2 model. 
%	In Fig.~\ref{fig:NMEheavy}, we show a comparison between the NMEs obtained with the IBM-2 and the QRPA models. 
%	The difference is quite big for many of the nuclei and might be due to the different treatment of the intermediate 
%	states. The IBM-2 values can be assumed in a conservative approach.

	The most updated NMEs for the \bb~via heavy neutrino exchange are evaluated within the frames of the 
	IBM-2~\cite{Barea:2015kwa} and QRPA~\cite{Faessler:2014kka} models.
	A comparison between these results is shown in Fig.~\ref{fig:NMEheavy}.
	It can be seen that the values obtained within the QRPA model are always larger than those obtained with the IBM-2.
	The difference is quite big for many of the nuclei and might be due to the different treatment of the intermediate 
	states.	
	Also in this case, we use the NMEs evaluated with the IBM-2 model.
	This allows us to keep a more conservative approach by getting less stringent limits.
	Considering, for example, the case of \ce{^{76}Ge}, we have:
	\begin{equation}
		\mathcal{M}_{0\nu}(\ce{Ge}) = 
		\begin{cases}
			104\pm 29 						&g_A=g_\mathrm{nucleon}	\\
			22 \pm 6							&g_A=g_\mathrm{phen.}
		\end{cases}.
		\label{eq:Mge}
	\end{equation}
	
	From the experimental point of view, the limits on \bb~indicate that the mixings of heavy neutrinos 
	$|U_{\el I}|^2$ are small. 
	Using the current values for the PSF, NME and sensitivity for the isotope~\cite{Brugnera:2013xma}, we get:
	\begin{equation}
		\left|\sum_I \frac{U^2_{\el I}}{M_I} \right| < \frac{7.8 \cdot  10^{-8} }{m_\pr}  
		\cdot \biggl[\frac{104}{\mathcal{M}_{0\nu}(\mbox{Ge})}\biggr]
		\cdot \biggl[\frac{3\cdot  10^{25}\mbox{ yr}}{\tau_{1/2}^{0\nu}}\biggr]^{\frac{1}{2}}
	\label{eq:formula}
	\end{equation}
	where $m_\pr$ is the proton mass and the heavy neutrino masses $M_I$ are assumed to be $\gtrsim$\,GeV.
	
	Fig.\ \ref{fig:plut} illustrates the case of a single heavy neutrino mixing with the light ones and mediating 
	the \bb~transition.
	In particular, the plot shows the case of the mixing for $^{76}$Ge assuming that a single heavy neutrino dominates 
	the amplitude.
	The two regimes of heavy and light neutrino exchange are matched as proposed in Ref.~\cite{Kovalenko:2009td}.
	The colored bands reflect the different sources of theoretical uncertainty.

	As it is clear from Fig.\ \ref{fig:plut}, the bound coming from \bb~searches is still uncertain. 
	It weakens \emph{by one order of magnitude} if 
	the axial vector coupling constant is strongly quenched in 	the nuclear medium.
	
	The potential of the \bb~sensitivity to heavy neutrinos is therefore weakened and very sensitive to theoretical 
	nuclear physics uncertainties.
	For some regions of the parameter space, even the limits obtained more than 15 years ago with accelerators are 
	more restrictive than the current limits coming from \bb~search.

%------------------------------------------------
%------------------------------------------------
	%------------------------------------------------
%------------------------------------------------
\section{Experimental search for the \bb}
\label{sec:experimental}

	The process described by Eq.~(\ref{eq:DBD}) is actually just one of the forms that \bb~can assume. 
	In fact, depending on the relative numbers of the nucleus protons and neutrons, four different mechanisms are possible:
	\begin{equation}
	\label{eq:DBD_all}
		\begin{aligned}
			(A,Z) 			&\rightarrow (A,Z+2) + 2\el^-		\\
			(A,Z)				&\rightarrow (A,Z+2) + 2\el^+		\\
			(A,Z) + 2\el^- &\rightarrow (A,Z-2)					\\
			(A,Z) + \el^- 	&\rightarrow (A,Z-2) + \el^+
		\end{aligned}
		\qquad
		\begin{aligned}
			&(\beta^- \beta^-)	\\
			&(\beta^+ \beta^+)	\\
			&(EC\,EC)				\\
			&(EC\,\beta^+).
		\end{aligned}
	\end{equation}
	Here, $\beta^-$ ($\beta^+$) indicate the emission of an electron (positron) and $EC$ stands for electron capture 
	(usually a K-shell electron is captured).

	The explicit violation of the number of electronic leptons $\el$, $\bar{\el}$, $\nu_e$ or $\bar\nu_\el$ appears evident 
	in each process in Eq.~(\ref{eq:DBD_all}).
%	Since none of these and, more in general, no process involving lepton number violation has been observed yet
%	(see Sec.~ref{sec:lepton}), a large number of experiments has been and presently is involved in the 
%	search for the \bb.
	A large number of experiments has been and is presently involved in the search for these processes, 
	especially of the first one.

	In this section, we introduce the experimental aspects relevant for the \bb~searches and we present an 
	overview of the various techniques. 
	We review the status of the past and present experiments, highlighting the main features and the sensitivities.
	The expectations take into account the uncertainties coming from the theoretical side and, in particular, 
	those from nuclear physics.
	The requirements for future experiments are estimated and finally, the new constraints from cosmology are used 
	as a complementary information to that coming from the \bb~experiments.
		
%------------------------------------------------
\subsection{The \bb~signature}
\label{sec:signature}

	\begin{figure}[tb]
		\centering
		\includegraphics[width=.9\columnwidth]{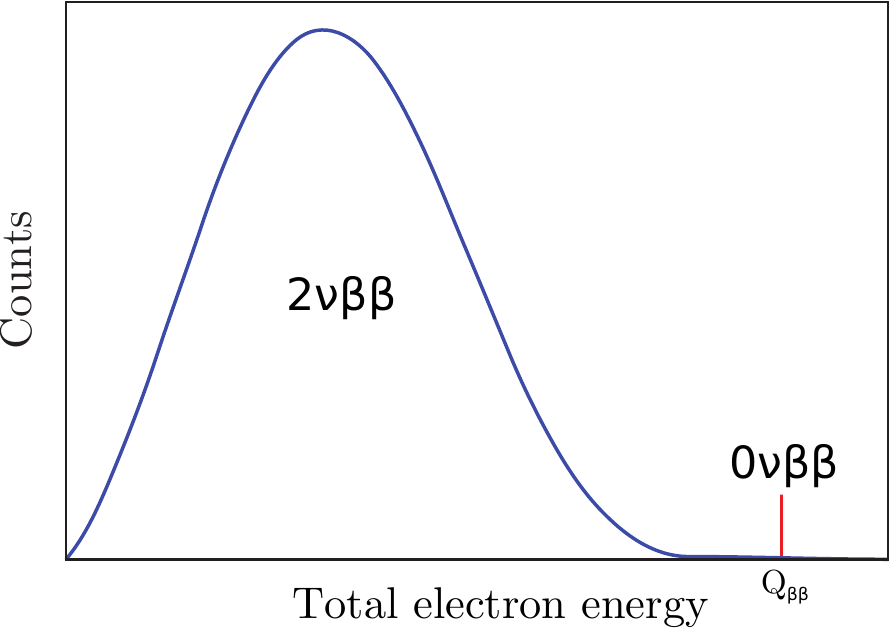}
		\caption{Schematic view of the \bbvv~and the \bb~spectra.}
		\label{fig:DBD_spectrum}
	\end{figure}
	
	From the experimental point of view, the searches for a \bb~signal rely on the detection of the two emitted electrons. 
	In fact, being the energy of the recoiling nucleus negligible, the sum of kinetic energy of the two electrons is equal 
	to the Q-value of the transition. Therefore, if we consider these as a single body, we expect to observe a monochromatic 
	peak at the Q-value (Fig.\ \ref{fig:DBD_spectrum}).
	
	Despite this very clear signature, because of the rarity of the process, the detection of the two electrons is 
	complicated by 
	the presence of background events in the same energy region, which can mask the \bb~signal.
	The main contributions to the background come from the environmental radioactivity, the cosmic rays, and the \bbvv~itself. 
	In particular, the last contribution has the problematic feature of being unavoidable in presence of finite 
	energy resolution, since it is originated by the 
	same isotope which is expected to undergo \bb.

	In principle, any event producing an energy deposition similar to that of the \bb~decay increases the 
	background level, and hence spoils the experiment sensitivity. 
	The capability of discriminate the background events is thus of great important for this kind of search. 

%------------------------------------------------
\subsection{The choice of the isotope}

	\begin{center}
	\begin{table}[t]
	\caption{Isotopic abundance and Q-value for the known \bbvv~emitters~\cite{Saakyan:2013yna}.}
	\small{
	\begin{ruledtabular}
	\begin{tabular}{r l l}
		Isotope				&isotopic abundance  (\%)			&$Q_{\beta\beta}$ [MeV]	\\[+2pt]
		\hline
		\\[-10pt]
		\ce{^{48}Ca}		&\!$0.187$				&$4.263$ \\	
		\ce{^{76}Ge}		&$7.8$					&$2.039$ \\	
		\ce{^{82}Se}		&$9.2$					&$2.998$ \\	
		\ce{^{96}Zr}		&$2.8$					&$3.348$ \\	
		\ce{^{100}Mo}		&$9.6$					&$3.035$ \\	
		\ce{^{116}Cd}		&$7.6$					&$2.813$ \\	
		\ce{^{130}Te}		&\!\!\!$34.08$			&$2.527$ \\	
		\ce{^{136}Xe}		&$8.9$					&$2.459$ \\	
		\ce{^{150}Nd}		&$5.6$					&$3.371$ \\	
		\end{tabular}
		\end{ruledtabular}
		}
	\label{tab:isotopes}
	\end{table}
	\end{center}

	The choice for the best isotope to look for \bb~is the first issue to deal with. 
	From one side, the background level and the energy resolution need
	to be optimized. From the other, since the live-time of the experiment cannot exceed some years, the scalability 
	of the technique, i.\,e.\ the possibility to build a similar experiment with enlarged mass and higher exposure, is 
	also fundamental.
	This translates in a series of criteria for the choice of the isotope:
	\begin{itemize}
		\item \emph{high Q-value ($Q_{\beta\beta}$)}. 
			This requirement is probably the most important, since it directly influences the 
			background. The $2615\,\keV$ line of \ce{^{208}Tl}, which represents the end-point of the natural gamma 
			radioactivity, constitutes an important limit in terms of background level. $Q_{\beta\beta}$ 
			should not be lower than $\sim 2.4\,\MeV$ (the only exception is the \ce{^{76}Ge}, due to the extremely 
			powerful detection technique, see Sec.~\ref{sec:detection}).
			The ideal condition would be to have it even larger than $3270\,\keV$, 
			the highest energy beta among the \ce{^{222}Rn} daughters (\ce{^{238}U} chain), coming from \ce{^{214}Bi};
		\item \emph{high isotopic abundance}. This is a fundamental requirement to have
			experiments with sufficiently large mass. With the only exception of the \ce{^{130}Te}, 
			all the relevant isotopes have a natural isotopic abundance $< 10\%$. 
			This practically means that the condition translates into \emph{ease of enrichment} for the material;
		\item \emph{compatibility with a suitable detection technique}. It has to be possible to integrate the isotope
			of interest in a working detector. The source can either be separated from the detector or coincide with it. 
			Furthermore, the detector has to be competitive in providing results and has to guarantee the 
			potential for the mass scalability.
	\end{itemize}
	
	This results in a group of ``commonly'' studied isotopes among all the possible candidate \bb~emitters.
	It includes:
	\ce{^{48}Ca}, \ce{^{76}Ge}, \ce{^{82}Se}, \ce{^{96}Zr}, \ce{^{100}Mo}, \ce{^{116}Cd}, \ce{^{130}Te}, \ce{^{136}Xe} and 
	\ce{^{150}Nd}.
	Table \ref{tab:isotopes} reports the Q-value and the isotopic abundance for the mentioned isotopes.
	
\begin{figure}[b]
		\centering
		\includegraphics[width=1.\columnwidth]{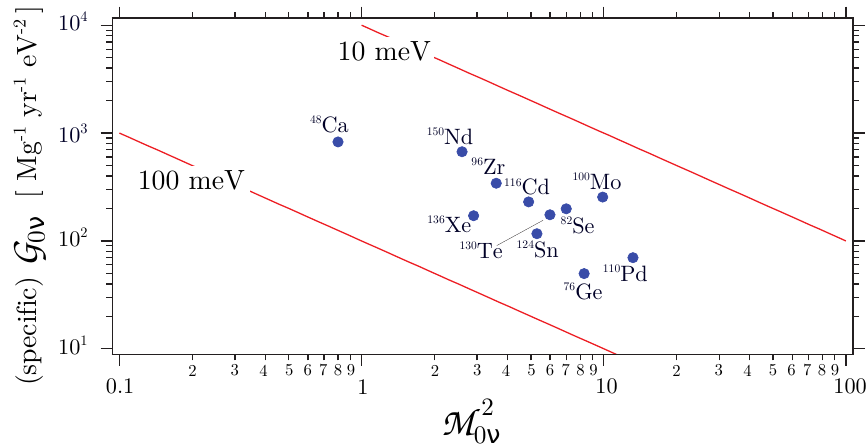}
		\caption{Geometric mean of the squared $\mathcal{M}_{0\nu}$ considered in Ref.~\cite{Robertson:2013cy} 
			vs. the specific $G_{0\nu}$. The case $g_A=g_\mathrm{quark}$ is assumed.
			Adapted from  Ref.~\cite{Robertson:2013cy}.}
		\label{fig:Robertson}
	\end{figure}

	From the theoretical side, referring to Eq.~(\ref{eq:tau}), one should also try to maximize both the PSF and the 
	NME in order to get more strict bounds on $\mbb$ with the same sensitivity in terms of half-life time. 
	However, as recently discussed in Ref.~\cite{Robertson:2013cy}, a uniform inverse correlation between the PSF and 
	the square of the NME emerges in all nuclei (Fig.~\ref{fig:Robertson}). 
	This happens to be more a coincidence than something physically motivated and, as a consequence, no isotope is 
	either favored or disfavored for the search for the \bb.
	It turns out that all isotopes have qualitatively the same decay rate per unit mass for any given value of $\mbb$.
	
	In recent time, also another criterion is becoming more and more relevant. 
	This is simply the availability of the isotope itself in view of the next generations of \bb~experiments, 
	which will have a very large mass. 
	In fact, once the \bb~isotope mass for an experiment will 
	be of the order of some tons, a non negligible fraction of the annual world production of the isotope of interest 
	could be needed. 
	This is e.\,g.\ the case of \ce{^{136}Xe}, where the requests from the \bb~experiments also ``compete'' with those 
	from the new proposed dark matter ones. The consequences are a probable price increase and a long storage for the 
	isotope that needs to be taken into account.
	
%------------------------------------------------
\subsection{Sensitivity}
\label{sec:sensitivity}

	In the fortunate event of a \bb~peak showing up in the energy spectrum, starting from the law of radioactive decay, 
	the decay half-life can be evaluated as 
	\begin{equation}
		\taubb = \ln 2 \cdot T \cdot \varepsilon \cdot \frac{N_{\beta \beta}}{N_\mathrm{peak}}
		\label{eq:exp_tau}
	\end{equation}
	where $T$ is the measuring time, $\varepsilon$ is the detection efficiency, $N_{\beta \beta}$ is the number of 
	$\beta \beta$ decaying nuclei under observation, and $N_\mathrm{peak}$ is the number of observed 
	decays in the region of interest. 
	If we assume to know exactly the detector features (i.\,e.\ the number of decaying nuclei, the efficiency and the 
	time of measurement), the uncertainty on $\taubb$ is only due to the statistical fluctuations of the counts:
	\begin{equation}
	\label{eq:fluctuations}
	\frac{\delta \taubb}{\taubb}=\frac{\delta N_\mathrm{peak}}{N_\mathrm{peak}}.
	\end{equation}
	It seems reasonable to suppose Poisson fluctuations on $N_\mathrm{peak}$. Since the expected number of events is 
	``small'', the Poisson distribution differs in a non negligible way from the Gaussian. 
	In order to quantify this discrepancy, we consider two values for $N_\mathrm{peak}$, namely $N_\mathrm{peak}=5$ 
	and $N_\mathrm{peak}=20$. In Tab.\ \ref{tab:gaus_pois} we show the confidence intervals at $1\sigma$ for the 
	counts both considering a purely Poisson distribution (with mean equal to $N_\mathrm{peak}$) and a Gaussian one 
	(with mean $N_\mathrm{peak}$ and standard deviation $\sqrt{N_\mathrm{peak}}$). 
	Notice that, even if the number of counts is just 5, the Poisson and Gaussian distributions give almost the same 
	relative uncertainties. 

	If no peak is detected, the sensitivity of a given \bb~experiment is usually expressed in
	terms of  ``detector factor of merit'', $S^{0\nu}$~\cite{Cremonesi:2013vla}. 
	This can be defined as the process half-life corresponding to 
	the maximum signal that could be hidden by the background fluctuations $n_B$ (at a given statistical C.\,L.). 
	To obtain an estimation for $S^{0\nu}$ as a function of the experiment parameters, it is sufficient to require that 
	the \bb~signal exceeds the standard deviation of the total detected counts in the interesting energy window. 
	At the confidence level $n_{\sigma}$, this means that we can write:
	\begin{equation}
		n_{\beta \beta}\geq n_{\sigma} \sqrt{n_{\beta \beta}+n_B}	
		\label{eq:def_sensib}
	\end{equation} 
	where $n_{\beta \beta}$ is the number of \bb~events and 
	Poisson statistics for counts is assumed. If one now states that the background counts scale linearly with the mass of 
	the detector,%
	\footnote{This is reasonable since, a priori, impurities are uniform inside the detector but, of course, 
		this might not be always the case e.\ g.\ if the main source of background is removed with volume fiducialization.}
	from Eq.~(\ref{eq:exp_tau}) it is easy to find an expression for $S^{0\nu}$:
	\begin{equation}
	\label{eq:fiorini}
		S^{0\nu} = \ln 2 \cdot T \cdot \varepsilon \cdot \frac{n_{\beta \beta}}{n_\sigma \cdot n_B} =
		\ln 2\cdot \varepsilon \cdot \frac{1}{n_\sigma} \cdot \frac{x\,\eta\,N_A}{\mathscr{M}_A} \cdot 
		\sqrt{\frac{M\cdot T}{B \cdot \Delta}}
	\end{equation}
	where $B$ is the background level per unit mass, energy, and time, $M$ is the detector mass, $\Delta$ is the 
	FWHM energy resolution, $x$ is the stoichiometric multiplicity of the element containing the $\beta \beta$ candidate, 
	$\eta$ is the $\beta \beta$ candidate isotopic abundance, $N_A$ is the Avogadro number and, finally, $\mathscr{M}_A$ is the 
	compound molecular mass.
	Despite its simplicity, Eq.~(\ref{eq:fiorini}) has the advantage of emphasizing the role of the essential experimental 
	parameters.
	
	Of particular interest is the case in which the background level $B$ is so low that the expected number of background 
	events in the region of interest along the experiment life is of order of unity:
	\begin{equation}
		M\cdot T\cdot B\cdot \Delta \lesssim 1.
	\label{eq:zb}
	\end{equation}
	This is called as the ``zero background'' experimental condition and it is likely the experimental condition 
	that next generation experiments will face.
	Practically, it means that the goal is a great mass and a long time of data taking, keeping the background level and 
	the energy resolution as little as possible. 
	
	In this case, $n_B$ is a constant, Eq.~(\ref{eq:fiorini}) is no more valid and the 
	sensitivity is given by:
	\begin{equation}
		S^{0\nu} _{0B} = \ln 2 \cdot T \cdot \varepsilon \cdot \frac{N_{\beta \beta}}{n_\sigma \cdot n_B} =
		\ln 2\cdot \varepsilon \cdot \frac{x\,\eta\,N_A}{\mathscr{M}_A} \cdot \frac{M\,T}{N_{\mbox{\tiny S}}}.
	\label{eq:0bkg_sens}
	\end{equation}
	The constant $N_{\mbox{\tiny S}}$ is now the number of observed events in the region of interest.

	\begin{center}
	\begin{table}[t]
	\caption{$1\sigma$ ranges both for Gaussian and Poisson distributions for two different values of $N_\mathrm{peak}$. 
	In the former case, we assumed a standard deviation equal to $\sqrt{N_\mathrm{peak}}$. To compute the error 
	columns, we halved the total width of the range and divided it by $N_\mathrm{peak}$.}
	\small{
	\begin{ruledtabular}
	\begin{tabular}{l r l l}
	Distribution	&$N_\mathrm{peak}$	&range 		&relative error (\%)	\\
  	\hline
  	Gauss				&5									&2.8 - 7.2 		&44.7	\\
  						&20								&15.5 - 24.5 	&22.4	\\[+3pt]
  	Poisson			&5									&3.1 - 7.6 		&45.0	\\
  						&20								&15.8 - 24.8 	&22.5	\\
 	\end{tabular}
 	\end{ruledtabular}
 	}
	\label{tab:gaus_pois}
	\end{table}
	\end{center}
	
%------------------------------------------------
\subsection{Experimental techniques}
\label{sec:detection}

	\begin{figure*}[tb]
		\centering
		\includegraphics[width=1.8\columnwidth]{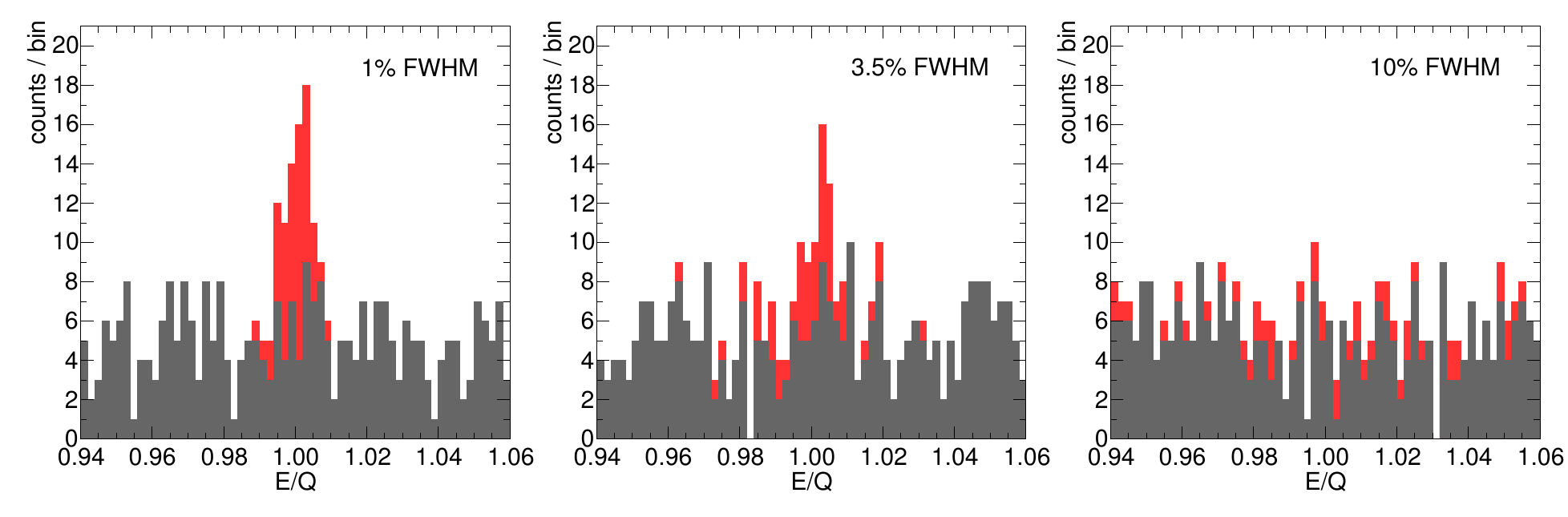}
		\caption{Signal and background (red and grey stacked histograms, respectively) in the region of interest 
			around $Q_{\beta\beta}$ for 3 Monte Carlo experiments with the same signal strength (50 counts) and 
			background rate (1 count$\,\keV^{-1}$), but different energy resolution:
			top: 1\% FWHM, centre: 3.5\% FWHM, bottom: 10\% FWHM. 
			The signal is distributed normally around $Q_{\beta\beta}$, while the background is assumed flat.
			Figure from Ref.\ \cite{Gomez-Cadenas:2015twa}.}
		\label{fig:JJC}
	\end{figure*}
	
	The experimental approach to search for the \bb~consists in the development of a proper detector, able to reveal the two 
	emitted electrons and to collect their sum energy spectrum (see Sec.\ \ref{sec:signature}).%
	\footnote{Additional information (e.\,g.\ the single electron energy or the initial momentum) 
		can also be provided sometimes.}
	The desirable features for such a detector are thus:
	\begin{itemize}
		\item \emph{good energy resolution}. This is a fundamental requirement to identify the sharp \bb~peak over an almost 
			flat background, as shown in Fig.\ \ref{fig:JJC}, and 
			it is also the only protection against the (intrinsic) background induced by the tail of the \bbvv~spectrum.
			Indeed, it can be shown that the ratio $R_{0\nu/2\nu}$ of counts due to \bb~and those due to \bbvv~in the peak 
			region can be approximated by~\cite{Elliott:2002xe}:
			\begin{equation}
			\label{eq:bbvv_bkg}
				R_{0\nu/2\nu} \propto \left( \frac{Q_{\beta\beta}}{\Delta} \right)^6 \,
				\frac{t^{\mbox{\tiny $\nicefrac{1}{2}$}}_{2\nu}}{t^{\mbox{\tiny $\nicefrac{1}{2}$}}_{0\nu}}.
			\end{equation}
			This expression clearly indicates that a good energy resolution is critical. But it also shows that the minimum 
			required value actually depends on the chosen isotope, considered a strong dependence of Eq.~(\ref{eq:bbvv_bkg})
			upon the \bbvv~half-life $t^{\mbox{\tiny $\nicefrac{1}{2}$}}_{2\nu}$;
		\item \emph{very low background}. Of course \bb~experiments have to be located underground in order to be protected from 
			cosmic rays. Moreover, radio-pure materials for the detector and the surrounding parts, as well as 
			proper passive and/or active shielding are mandatory to protect against environmental radioactivity;%
			\footnote{The longest natural radioactivity decay competing to the \bb~are of the order of $(10^9-10^{10})\,\yr$ 
			versus lifetimes $\gtrsim 10^{25}\,\yr$.}
		\item \emph{large isotope mass}. Present experiments have masses of the order of some tens of kg up to a few hundreds kg.
			Tons will be required for experiments aiming to cover the \IH~region (see Sec.\ \ref{sec:fut_exp})
	\end{itemize}
	It has to be noted that it is impossible to optimize the listed features simultaneously in a single detector.
	Therefore, it is up to the experimentalists to choose which one to privilege in order to get the best sensitivity.
	
\begin{figure}[tb]
		\centering
		\includegraphics[width=.6\columnwidth]{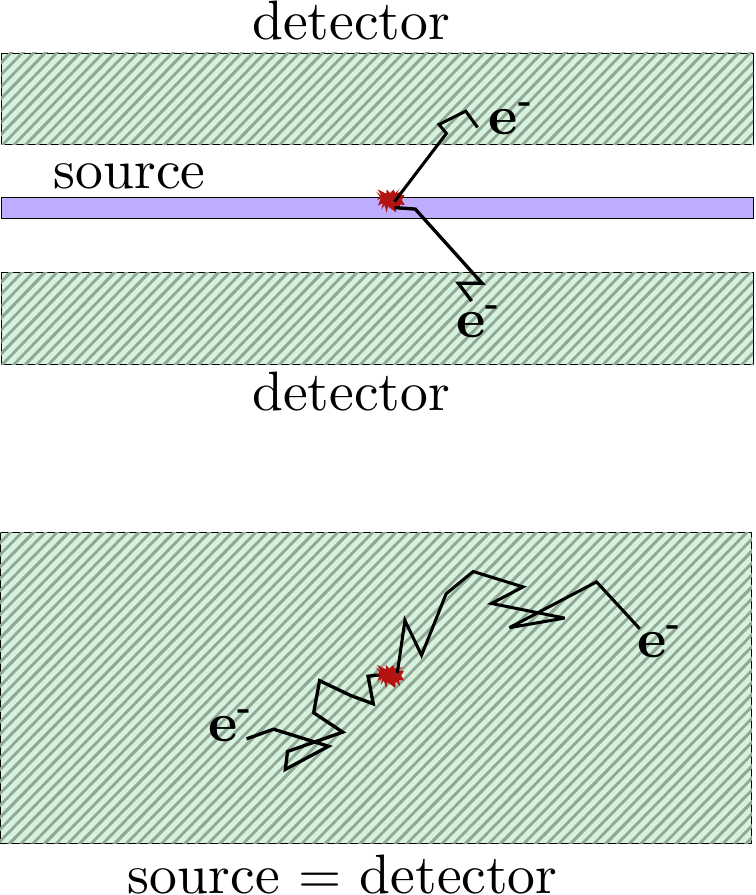}
		\caption{Schematic representation of the two main experimental categories for the \bb~search: calorimetric technique 
			(source $\equiv$ detector) and external source approach (source $\ne$ detector).}
		\label{fig:detector-source}
	\end{figure}

	The experiments searching for the \bb~of a certain isotope can be classified into two main categories: 
	detectors based on a calorimetric technique, in which the source is embedded in the detector itself, and 
	detector using an external source approach, in which source and detector are two separate systems 
	(Fig.\ \ref{fig:detector-source}).
	
%------------------------------------------------
\subsubsection{Calorimetric technique}

	The calorimetric technique has already been implemented in various types of detectors. 
	The main advantages and limitations for this technique can be summarized as follows~\cite{Cremonesi:2013vla}:
	\begin{itemize}
		\item[$(+)$] large source masses are achievable thanks to the intrinsically high efficiency of the method.
			Experiments with masses up to $\sim 200\,\mbox{kg}$ have already proved to work and ton-scale detectors seem possible
		\item[$(+)$] very high resolution is achievable with the proper type of detector ($\sim 0.1\%$ FWHM with \ce{Ge} diodes 
			and bolometers)
		\item[$(-)$] severe constraints on detector material (and thus on the isotope that can be investigated) arise from 
			the request that the source material has to be embedded in the structure of the detector. However, this is not 
			the case for some techniques (e.\,g.\ for bolometers and loaded liquid scintillators)
		\item[$(-)$] the event topology reconstruction is usually difficult, with the exception of liquid or 
			gaseous \ce{Xe} TPC. However, the cost is paid in terms of a lower energy resolution.
	\end{itemize}

	Among the most successful examples of detectors using the calorimetric technique, we find:
	\begin{itemize}
		\item Ge-diodes. The large volume, high-purity, and high-energy resolution achievable make this kind 
			of detector suitable for the \bb~search, despite the low $Q_{\beta\beta}$ of \ce{^{76}Ge}
		\item bolometers. Macro-calorimeters with masses close to 1\,kg very good energy resolution (close to that of 
			Ge-diodes) are now available for many compounds including \bb~emitters. The most significant case is 
			the search for the \bb~of \ce{^{130}Te} with \ce{TeO_2} bolometers
		\item Xe liquid and gaseous TPC. The lower energy resolution is ``compensated'' by the capability of 
			reconstructing the event topology
		\item liquid scintillators loaded with the \bb~isotope. These detectors have a poor energy resolution.
			However, a huge amount of material can be dissolved and, thanks to the purification processes, 
			very low backgrounds are achievable. They are ideal detectors to set very stringent limits on the 
			decay half-life.
	\end{itemize}

%------------------------------------------------
\subsubsection{External source approach}

	Also in the case of the external source approach, different detection techniques have been adopted, namely 
	scintillators, solid state detectors, and gas chambers. 
	The main advantages and limitations for this technique can be summarized as follows:
	\begin{itemize}
		\item[$(+)$] the reconstruction of the event topology is possible, thus making in principle easier the achievement of 
			the zero background condition. However, the poor energy resolution does not allow to distinguish between 
			\bb~events and \bbvv~events with total electron energy around $Q_{\beta\beta}$. Therefore the \bbvv~represent 
			an important background source
		\item[$(-)$] the energy resolutions are low (of the order of 10\%). The limit is intrinsic and it mainly due 
			to the electron energy deposition in the source itself
		\item[$(-)$] large isotope masses are hardly achievable due to self-absorption in the source.
			Up to now, only masses of the order of some tens of kg have been possible, but an increase to about $100\,\kg$ target 
			seems feasible
		\item[$(-)$] the detection efficiencies are low (of the order of 30\%).
	\end{itemize}
	
	So far, the most stringent bounds come from the calorimetric approach which, anyway, remains the one promising the 
	best sensitivities and it is therefore the chosen technique for most of the future projects.
	However, the external source detector type has provided excellent results on the studies of the \bbvv. 
	Moreover, in case of discovery of a \bb~signal, the event topology reconstruction could represent a fundamental tool 
	for the understanding of the mechanism behind the \bb. 

%------------------------------------------------
\subsection{Experiments: a brief review}

	\begin{center}
	\begin{table*}[tb]
	\caption{In this table, the main features and performances of some past, present and future \bb~experiments are listed.}
	\small{
	\begin{ruledtabular}
	\begin{tabular}{l l l l l l l l}
		Experiment	&Isotope		&Techinique	&Total mass	&Exposure	&FWHM @\,$Q_{\beta\beta}$	&Background	&$S^{0\nu} \xcCL$ \\
						&				&					&[kg]		&[kg\,yr]	&[keV]		&[\ckky]			&[$10^{25}$\,yr]\\[+2pt]
		\hline
		\\[-4pt]
		\emph{Past} \\
		\cline{1-1} \\[-9pt]

		Cuoricino,\,\cite{Andreotti:2010vj}								&\ce{^{130}Te}	&bolometers
					&40.7 \footnotesize{(\ce{TeO_2})}		&19.75	&$5.8\pm2.1$		&$0.153\pm0.006$		&$0.24$	\\

		CUORE-0,\,\cite{Alfonso:2015wka}								&\ce{^{130}Te}		&bolometers
					&39 \footnotesize{(\ce{TeO_2})}			&9.8	&$5.1\pm0.3$			&$0.058\pm0.006$		&$0.29$	\\

		Heidelberg-Moscow,\,\cite{KlapdorKleingrothaus:2000sn}	&\ce{^{76}Ge}	&Ge diodes
					&11 \footnotesize{(\ce{^{enr}Ge})}		&35.5		&$4.23\pm0.14$		&$0.06\pm0.01$			&$1.9$	\\

		IGEX,\,\cite{Aalseth:1999ji,Aalseth:2002rf}					&\ce{^{76}Ge}	&Ge diodes		
					&8.1 \footnotesize{(\ce{^{enr}Ge})}		&8.9		&$\sim 4$			&$\lesssim 0.06$		&$1.57$	\\

		GERDA-I,\,\cite{Ackermann:2012xja,Agostini:2013mzu}		&\ce{^{76}Ge}	&Ge diodes
					&17.7 \footnotesize{(\ce{^{enr}Ge})}	&21.64	&$3.2\pm0.2$		&$\sim 0.01$			&$2.1$	\\

		NEMO-3,\,\cite{Arnold:2013dha}									&\ce{^{100}Mo}	&\footnotesize{tracker +}	
					&6.9 \footnotesize{(\ce{^{100}Mo})}	&34.7		&$350$				&$0.013$					&$0.11$	\\[-3pt]
					&					&\footnotesize{calorimeter}	&	&	&							&							&	\\

		\emph{Present} \\
		\cline{1-1} \\[-9pt]

		EXO-200,\,\cite{Albert:2014awa}									&\ce{^{136}Xe}	&LXe TPC
					&175 \footnotesize{(\ce{^{enr}Xe})}		&100		&$89\pm3$			&$(1.7\pm0.2)\cdot10^{-3}$	&$1.1$	\\

		KamLAND-Zen,\,\cite{Gando:2012zm,Asakura:2014lma}			&\ce{^{136}Xe}	&\footnotesize{loaded liquid}		
					&348 \footnotesize{(\ce{^{enr}Xe})}		&89.5			&$244\pm11$		&$\sim 0.01$			&$1.9$	\\[-3pt]
					&					&\footnotesize{scintillator}	&	&	&							&							&			\\

		\emph{Future} \\
		\cline{1-1} \\[-9pt]

		CUORE,\,\cite{Artusa:2014lgv}								&\ce{^{130}Te}		&bolometers
					&741 \footnotesize{(\ce{TeO_2})}			&1030		&$5$				&$0.01$					&$9.5$	\\

		GERDA-II,\,\cite{Brugnera:2013xma}						&\ce{^{76}Ge}	&Ge diodes
					&37.8 \footnotesize{(\ce{^{enr}Ge})}	&100		&3					&$0.001$					&$15$		\\

		LUCIFER,\,\cite{Pattavina:2015_LUCIFER}			&\ce{^{82}Se}		&bolometers
					&17 \footnotesize{(\ce{Zn^{82}Se})}	&18		&$10$				&$0.001$					&$1.8$	\\

		MAJORANA D.,\,\cite{Abgrall:2013rze}					&\ce{^{76}Ge}	&Ge diodes
					&44.8 \footnotesize{(\ce{^{enr/nat}Ge})}	&100\notaA		&4					&$0.003$		&12	\\

		NEXT,\,\cite{Gomez-Cadenas:2013lta,Laing:2015_NEXT}	&\ce{^{136}Xe}	&Xe TPC
					&100 \footnotesize{(\ce{^{enr}Xe})}		&300	&$12.3-17.2$		&$5\cdot10^{-4}$		&5		\\[+12pt]

		AMoRE,\,\cite{Kim:2015_AMoRE}			&\ce{^{100}Mo}		&bolometers
					&200 \footnotesize{(\ce{Ca^{enr}MoO_4})}	&295		&$9$	&$1\cdot10^{-4}$					&$5$	\\

		nEXO,\,\cite{Ostrovskiy:2015_nEXO}						&\ce{^{136}Xe}	&LXe TPC
					&4780 \footnotesize{(\ce{^{enr}Xe})}	&12150\notaB	&58		&$1.7\cdot10^{-5}$\notaB	&66	\\

		PandaX-III,\,\cite{Ji:2015_PandaX}							&\ce{^{136}Xe}	&Xe TPC
					&1000 \footnotesize{(\ce{^{enr}Xe})}	&3000\notaC	&$12-76$		&$0.001$			&11\notaC		\\

		SNO+,\,\cite{Andringa:2015tza}			&\ce{^{130}Te}	&\footnotesize{loaded liquid}		
					&2340 \footnotesize{(\ce{^{nat}Te})}		&3980			&270		&$2\cdot10^{-4}$			&$9$	\\[-3pt]
					&					&\footnotesize{scintillator}	&	&	&							&							&			\\

		SuperNEMO,\,\cite{Arnold:2015wpy,Blot:2015_SuperNEMO}	&\ce{^{82}Se}	&\footnotesize{tracker +}	
					&100 \footnotesize{(\ce{^{82}Se})}	&500			&120				&$0.01$			&$10$	\\[-3pt]
					&					&\footnotesize{calorimeter}	&	&	&							&							&	\\

		\end{tabular}
		\end{ruledtabular}
		\begin{flushleft}
		\notaA {\scriptsize our assumption (corresponding sensitivity from Fig.\ 14 of Ref.\ \cite{Abgrall:2013rze}).} \\
		\notaB {\scriptsize we assume 3\,tons fiducial volume.} \\
		\notaC {\scriptsize our assumption by rescaling NEXT.} 
		\end{flushleft}
		}
	\label{tab:experiments}
	\end{table*}
	\end{center}

	The first attempt to observe the \bb~process dates back to 1948~\cite{Fireman:1948,Fireman:1949qs}.
	Actually, the old experiments aiming to set a limit on the double beta decay half-lives did not distinguish 
	between \bbvv~and \bb. 
	In the case of indirect investigations through geochemical observation, this was not possible even in principle.

	However, the importance that the \bb~was acquiring in particle physics provided a 
	valid motivation to continuously enhance the efforts in the search for this decay.
	On the experimental side, the considerable technological improvements allowed to increase the half-life sensitivity 
	of several orders of magnitude.%
	\footnote{The \bbvv~was first observed in the laboratory in \ce{^{82}Se} in 1987~\cite{Elliott:1987kp}, and 
		in many other isotopes in the subsequent years. See Ref.\ \cite{Saakyan:2013yna} for a review on \bbvv.}
	The long history of \bb~measurements up to about the year 2000 can be found 
	in Refs.~\cite{KlapdorKleingrothaus:2010zza,Tretyak:2002dx,Barabash:2011mf}.
	Here, we concentrate only on a few experiments starting from the late 1990s.
%	Much more detailed reviews on the experimental searches can be found e.\,g.\ in 
%	Refs.~\cite{Cremonesi:2013vla,Avignone:2007fu}.
	
	Table \ref{tab:experiments} summarizes the main characteristics and performances of the selected experiments. 
	It has to be noticed that, due to their different specific features, the actual 
	comparison among all the values is not always possible. 
	We tried to overcome this problem by choosing a common set of units of measurement.
%	It is thus possible that some mistake has been introduced during the conversions, but we believe that the values we report 
%	are quite reliable.

%------------------------------------------------
\subsubsection{The claimed observation}

	In 2001, after the publication of the experiment final results~\cite{KlapdorKleingrothaus:2000sn}, 
	a fraction of the Heidelberg-Moscow Collaboration claimed to observe a peak in the spectrum, whose energy corresponded
	to the \ce{^{76}Ge} \bb~transition Q-value~\cite{KlapdorKleingrothaus:2001ke}.
	After successive re-analysis (by fewer and fewer people), the final value for the half-life was found to be:
	$\taubb = (2.23^{+0.44}_{-0.31})\cdot10^{25}\,\yr$~\cite{KlapdorKleingrothaus:2006ff}.
	This claim and the subsequent papers by the same authors aroused a number of critical replies 
	(see e.\,g. Refs.\cite{Aalseth:2002dt,Feruglio:2002af,Zdesenko:2002kz,Schwingenheuer:2012zs}).
	Many of the questions and doubts still remain unanswered.
	To summarize, caution suggests that we disregard the claim that the transition was observed.

	Anyway, to date, the limit on the \ce{^{76}Ge} \bb~half-life is more stringent than the reported 
	value~\cite{Brugnera:2013xma}. 

%------------------------------------------------
\subsection{Present sensitivity on $\mbb$}
\label{sec:present_sensitivity}

	\begin{figure*}[tb]
		\centering
		\includegraphics[width=\columnwidth]{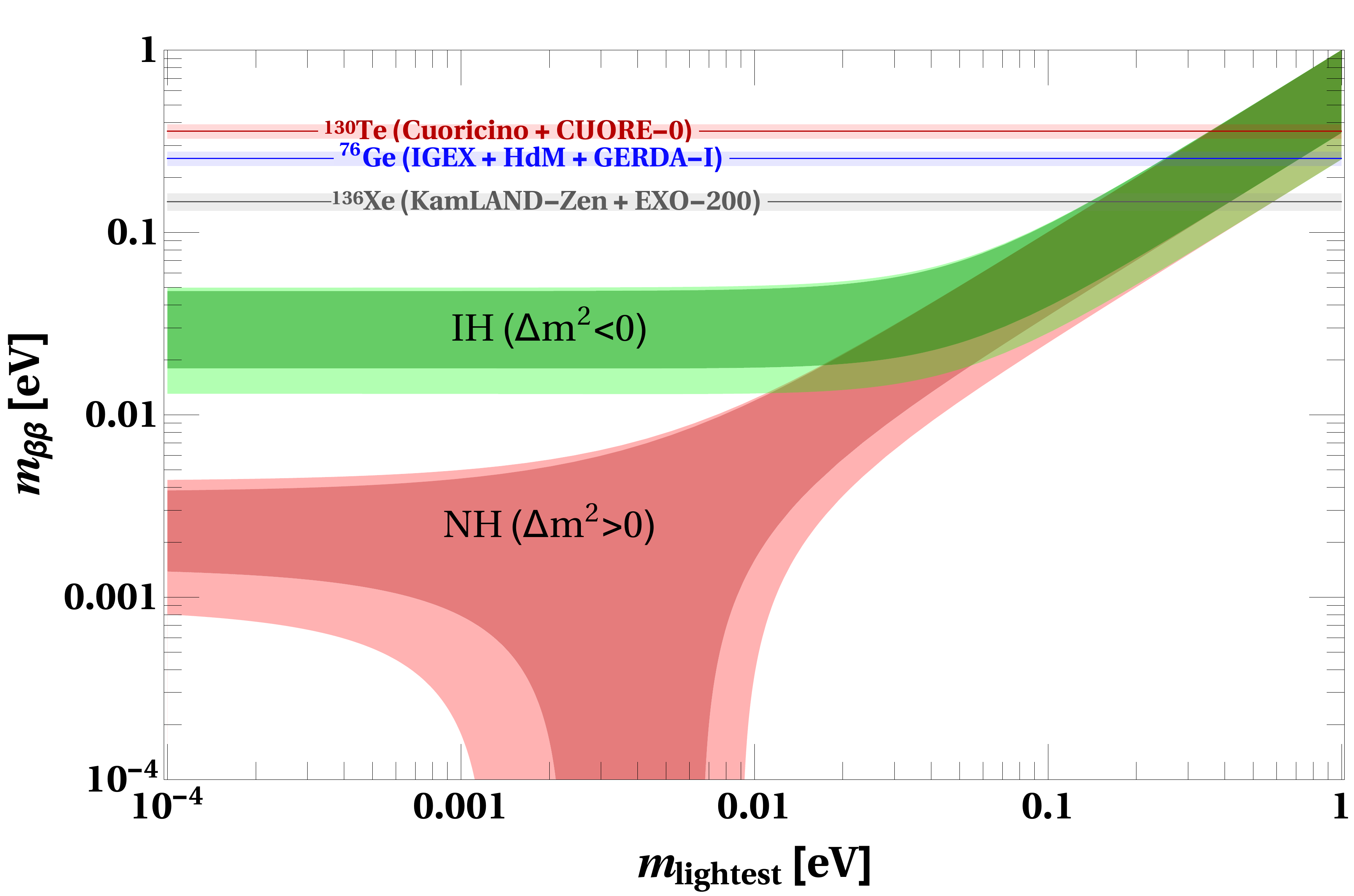}
		\includegraphics[width=\columnwidth]{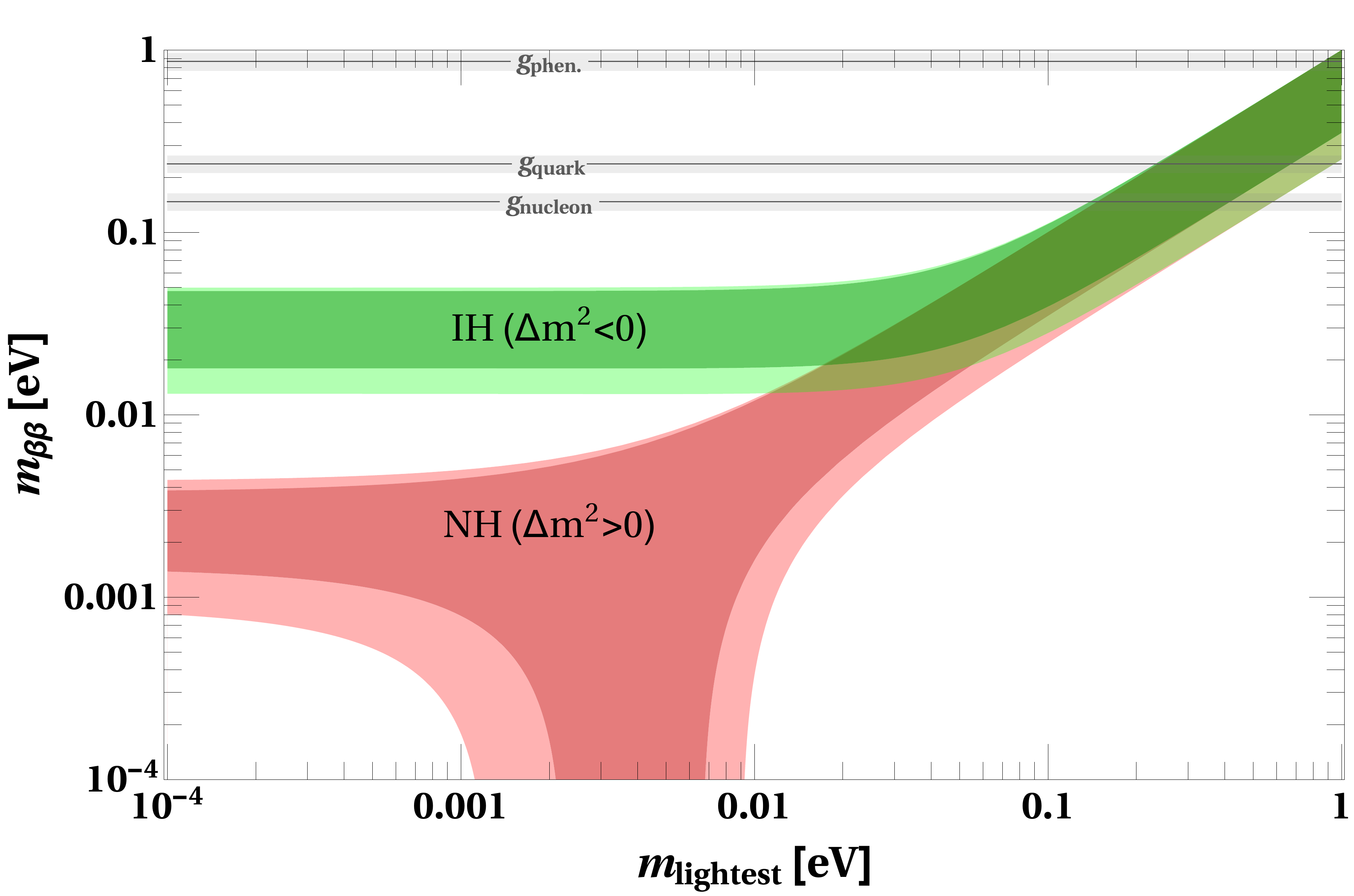}
		\caption{The colored regions show the predictions on $\mbb$ from oscillations as a function of the lightest 
			neutrino mass with the relative the $3\sigma$ regions.
			The horizontal bands show the experimental limits with the spread due to the theoretical uncertainties on the 
			NME~\cite{Barea:2015kwa} and PSF~\cite{Kotila:2012zza}~\cite{Gando:2012zm}.
			(Left) Combined experimental limits for the three isotopes: $\ce{^{76}Ge}$~\cite{Brugnera:2013xma},
			$\ce{^{130}Te}$~\cite{Alfonso:2015wka} and $\ce{^{136}Xe}$. The case $g_A=g_\mathrm{nucleon}$.
			(Right) Combined experimental limit on $\ce{^{136}Xe}$ for the three different values for $g_A$, according to 
			Eq.~(\ref{eq:cazzes}).
		}
		\label{fig:mbb_present_bounds}
	\end{figure*}

	Once the experimental sensitivities are known in terms of $S^{0\nu}$, by using Eq.~(\ref{eq:mbb_bound}),
	it is possible to correspondingly find the lower bounds on $\mbb$.

	Fig.~\ref{fig:mbb_present_bounds} shows the most stringent limits up to date. 
	They come from \ce{^{76}Ge}~\cite{Brugnera:2013xma}, \ce{^{130}Te}~\cite{Alfonso:2015wka} 
	and \ce{^{136}Xe}~\cite{Gando:2012zm}. 
	In particular, the combined sensitivity from the single experimental limits is taken from the corresponding references. 

	In the left panel of the figure, the case $g_A=g_\mathrm{nucleon}$ (unquenched value) is assumed. 
	The uncertainties on NME and PSF are taken into account according to the procedure shown in App.~\ref{app:stat}, 
	and they result in the broadening of the lines describing the limits. 
	As the plot shows, the current generation of experiments is probing the quasi degenerate part of the neutrino 
	mass spectrum.
	
	The effect of the quenching of $g_A$ appears evident in the right panel: the sensitivity for the 
	same combined \ce{^{136}Xe} experiment in the two cases of $g_\mathrm{nucleon}$ and $g_\mathrm{phen.}$ 
	differs of a factor $\gtrsim 5$. It is clear from the figure that this is the biggest uncertainty, with respect to all 
	the other theoretical ones.
	
	The single values for the examined cases are reported in Table~\ref{tab:mbb_present_bounds}.
	
	\begin{center}
	\begin{table*}[tb]
	\caption{Lower bounds for $\mbb$ for \ce{^{76}Ge}, \ce{^{130}Te} and \ce{^{136}Xe}. The sensitivities 
		were obtained by combining the most stringent limits from the experiments studying the isotopes.
		Refs.\ \cite{Kotila:2012zza} and~\cite{Barea:2015kwa} were used for the PSFs and for the NME, respectively.
		The different results correspond to different values of $g_A$ according to Eq.~(\ref{eq:cazzes}).
		%\lam{Non mi tornano le incertezze: per esempio lo xenon, 7\% di PSF e 15\% di NME dovrebbe fare circa il 16\% 
		%(sommiamo in quadratura no?), mentre viene 26\%.}
		}
	\small{
	\begin{ruledtabular}
	\begin{tabular}{llllll}
		Experiment		&Isotope				&$S^{0\nu} \xcCL$	&\multicolumn{3}{c}{Lower bound for $\mbb\, [\eV]$}	\\
		\cline{4-6}
			&	&$[10^{25}\,\yr]$		&$g_\mathrm{nucleon}$	&$g_\mathrm{quark}$	&$g_\mathrm{phen.}$	\\[+2pt]
		\hline
%		StatisticalFactor = 1
%		IGEX + HdM + GERDA-I,~\cite{Brugnera:2013xma}
%							&\ce{^{76}Ge}		&$3.0$			&$0.25\pm0.05$		&$0.40\pm0.08$		&$1.19\pm0.23$	\\
%		Cuoricino + CUORE-0,~\cite{Alfonso:2015wka}	
%							&\ce{^{130}Te}		&$0.4$			&$0.36\pm0.07$		&$0.58\pm0.11$		&$2.07\pm0.39$	\\
%		EXO-200 + KamLAND-ZEN,~\cite{Gando:2012zm}
%							&\ce{^{136}Xe}		&$3.4$			&$0.15\pm0.04$		&$0.24\pm0.06$		&$0.88\pm0.21$	\\
%		StatisticalFactor = 1/sqrt[3]
		IGEX + HdM + GERDA-I,~\cite{Brugnera:2013xma}
							&\ce{^{76}Ge}		&$3.0$			&$0.25\pm0.02$		&$0.40\pm0.04$		&$1.21\pm0.11$	\\
		Cuoricino + CUORE-0,~\cite{Alfonso:2015wka}	
							&\ce{^{130}Te}		&$0.4$			&$0.36\pm0.03$		&$0.58\pm0.05$		&$2.07\pm1.05$ \\
		EXO-200 + KamLAND-ZEN,~\cite{Gando:2012zm}
							&\ce{^{136}Xe}		&$3.4$			&$0.15\pm0.02$		&$0.24\pm0.03$		&$0.87\pm0.10$	\\
		\end{tabular}
		\end{ruledtabular}
		}
	\label{tab:mbb_present_bounds}
	\end{table*}
	\end{center}

%------------------------------------------------
\subsection{Near and far future experiments}
\label{sec:fut_exp}

	\begin{center}
	\begin{table*}[tb]
	\caption{Lower bounds for $\mbb$ for the more (upper group) and less (lower group) near future \bb~experiments.
		Refs.\ \cite{Kotila:2012zza} and~\cite{Barea:2015kwa} were used for the PSFs and for the NME, respectively.
		The different results correspond to different values of $g_A$ according to Eq.~(\ref{eq:cazzes}).
		%\lam{Non mi tornano le incertezze: per esempio lo xenon, 7\% di PSF e 15\% di NME dovrebbe fare circa il 16\% 
		%(sommiamo in quadratura no?), mentre viene 26\%.}
		}
	\small{
	\begin{ruledtabular}
	\begin{tabular}{llllll}
		Experiment		&Isotope				&$S^{0\nu} \xcCL$	&\multicolumn{3}{c}{Lower bound for $\mbb\, [\eV]$}	\\
		\cline{4-6}
			&	&$[10^{25}\,\yr]$		&$g_\mathrm{nucleon}$	&$g_\mathrm{quark}$	&$g_\mathrm{phen.}$	\\[+2pt]
		\hline
%		StatisticalFactor = 1/sqrt[3]
		CUORE,\,\cite{Artusa:2014lgv}
							&\ce{^{130}Te}		&$9.5$			&$0.073\pm0.008$		&$0.14\pm0.01$		&$0.44\pm0.04$	\\
		GERDA-II,\,\cite{Brugnera:2013xma}
							&\ce{^{76}Ge}		&$15$			&$0.11\pm0.01$		&$0.18\pm0.02$		&$0.54\pm0.05$	\\
		LUCIFER,\,\cite{Pattavina:2015_LUCIFER}
							&\ce{^{82}Se}		&$1.8$			&$0.20\pm0.02$		&$0.32\pm0.03$		&$0.97\pm0.09$	\\
		MAJORANA D.,\,\cite{Abgrall:2013rze}
							&\ce{^{76}Ge}		&$12$			&$0.13\pm0.01$		&$0.20\pm0.02$		&$0.61\pm0.06$	\\
		NEXT,\,\cite{Laing:2015_NEXT}
							&\ce{^{136}Xe}		&$5$			&$0.12\pm0.01$		&$0.20\pm0.02$		&$0.71\pm0.08$	\\[+8pt]

		AMoRE,\,\cite{Kim:2015_AMoRE}
							&\ce{^{100}Mo}		&$5$			&$0.084\pm0.008$		&$0.14\pm0.01$		&$0.44\pm0.04$	\\
		nEXO,\,\cite{Ostrovskiy:2015_nEXO}
							&\ce{^{136}Xe}		&$660$			&$0.011\pm0.001$		&$0.017\pm0.002$	&$0.062\pm0.007$	\\
		PandaX-III,\,\cite{Ji:2015_PandaX}
							&\ce{^{136}Xe}		&$11$			&$0.082\pm0.009$		&$0.13\pm0.01$		&$0.48\pm0.05$	\\
		SNO+,\,\cite{Andringa:2015tza}
							&\ce{^{130}Te}		&$9$			&$0.076\pm0.007$		&$0.12\pm0.01$		&$0.44\pm0.04$	\\
		SuperNEMO,\,\cite{Arnold:2015wpy}
							&\ce{^{82}Se}		&$10$			&$0.084\pm0.008$		&$0.14\pm0.01$		&$0.41\pm0.04$	\\
		\end{tabular}
		\end{ruledtabular}
		}
	\label{tab:mbb_future_bounds}
	\end{table*}
	\end{center}

	It is also possible to extract the bounds on $\mbb$ coming from the near future experiments starting from the 
	expected sensitivities and using Eq.~(\ref{eq:mbb_bound}). The results are shown in Table~\ref{tab:mbb_future_bounds}.
	It can be seen the mass region below $100\,\meV$ will begin to be probed in case of unqueched value for $g_A$.
	But still we will not enter the \IH~region.
	In case $g_A$ is maximally quenched, instead, the situation is much worse. Indeed, the expected sensitivity 
	would correspond to values of $\mbb$ which we already consider probed by the past experiments. 
	
	Let us now consider a next generation experiment (call it a ``mega'' experiment) and a next-to-next 
	generation one (an ``ultimate'' experiment) with enhanced sensitivity. 
	To define the physics goal we want to achieve, we refer to Ref.~\cite{Dell'Oro:2014yca}.
	
	The most honest way to talk of the sensitivity is in terms of exposure or of half-life time that can be 
	probed. From the point of view of the physical interest, however, besides the hope of 
	discovering the \bb, the most exciting investigation 
	that can be imagined at present is the exclusion of the \IH~case. 
	This is the goal that most of the experimentalists are trying to reach with future \bb~experiments
	(see e.\,g.\ Ref.~\cite{Artusa:2014wnl}).
	For this reason, we require a sensitivity $\mbb = 8\,\meV$.
	The mega experiment is the one that satisfies this requirement in the most favorable case, namely, 
	when the quenching of $g_A$ is absent. Instead, the ultimate experiment assumes that $g_A$ is maximally 
	quenched. We chose the $8\,\meV$ value because, even taking into account the residual uncertainties on the NME and 
	on the PSF, the overlap with the allowed band for $\mbb$ in the \IH~is excluded at more than $3\sigma$.
	Notice that we are assuming that at some point the issue of the quenching will be sorted out.
	Through Eq.~(\ref{eq:mbb_bound}), 
	we obtain the corresponding value of $\taubb$ and thus we calculate the needed exposure to accomplish the task.

	Referring to Eq.~(\ref{eq:0bkg_sens}), if we suppose $\varepsilon \simeq 1$ (detector efficiency of $100\%$ and no 
	fiducial volume cuts), 
	$x\simeq\eta\simeq1$ (all the mass is given by the candidate nuclei), and we assume one observed event 
	(i.\,e.\ $N_{\mbox{\tiny S}} = 1$) in the region of interest, we get the simplified equation:
	\begin{equation} 
		M \cdot T =\frac{\mathscr{M}_A \cdot S^{0\nu}}{\ln 2 \cdot N_{\mbox{\tiny A}}} . 
	\label{eq:our_sensitivity}
	\end{equation}
	This is the equation we used to estimate the product $M\cdot T$ (exposure), 
	and thus to assess the sensitivity of the mega and ultimate scenarios.
	The key input is, of course, the theoretical expression of $\taubb$.
	The calculated values of the exposure 
	are shown in Table \ref{tab:exposure} for the 
	three considered nuclei: \ce{^{76}Ge}, \ce{^{130}Te} and \ce{^{136}Xe}.
	The last column of the table gives 
	the maximum allowed value of the product $B\cdot \Delta$ that satisfies  Eq.~(\ref{eq:zb}).
	
\begin{figure}[b]
		\centering
		\includegraphics[width=.4\columnwidth]{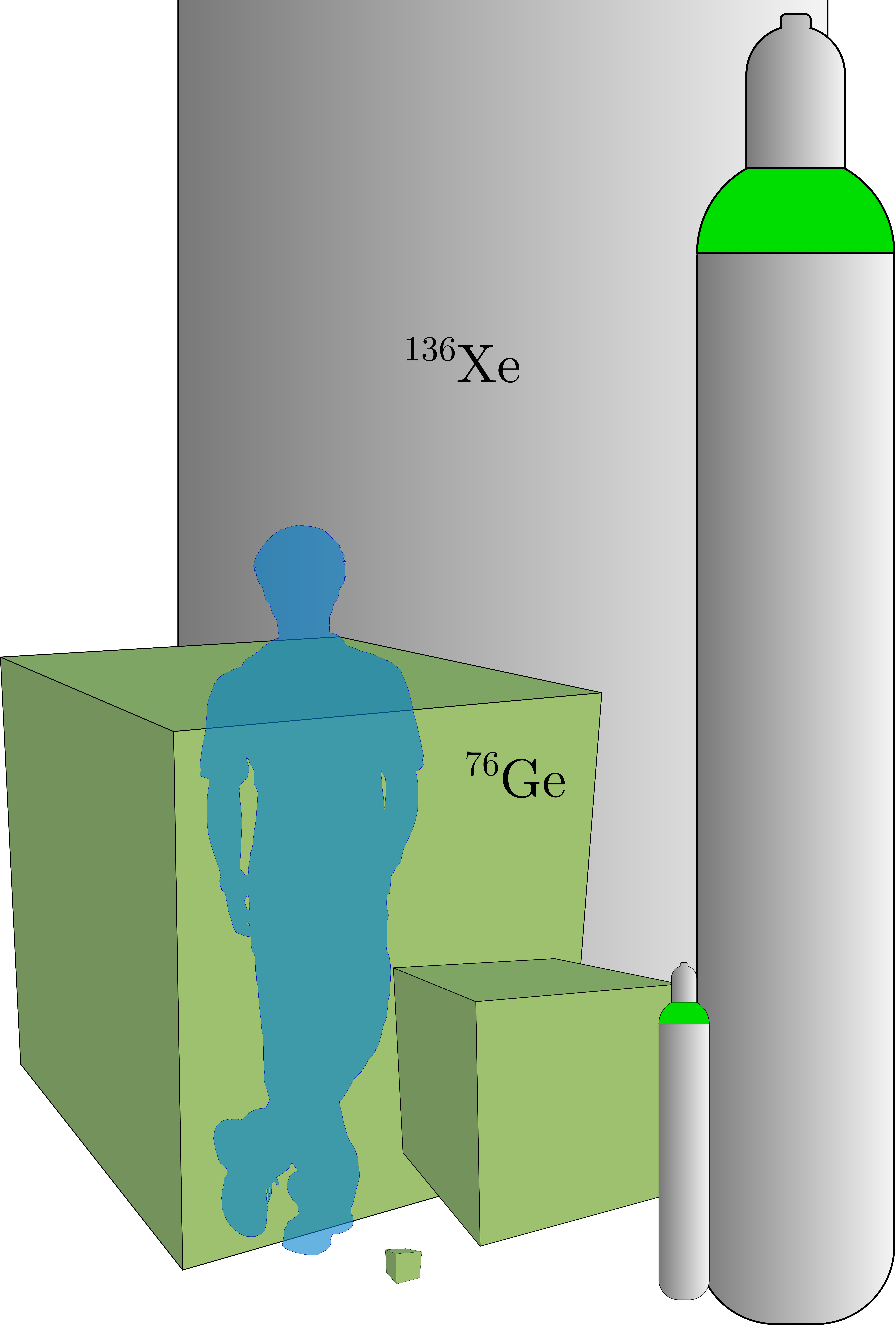}
		\caption{Masses corresponding to present, mega and ultimate exposures, assuming zero background condition and 5 
			years of data acquisition. The cubes represent the amount of \ce{^{76}Ge}, the (150 bar) bottles the one of 
			\ce{^{136}Xe}. The smallest masses depict the present exposure, while the biggest bottle is out of scale.}
		\label{fig:future_masses}
	\end{figure}

	Fig.\ \ref{fig:future_masses} compares (in a schematic view) the masses of \ce{^{76}Ge} and \ce{^{136}Xe} corresponding 
	to the present sensitivity~\cite{Brugnera:2013xma,Gando:2012zm} assuming zero background condition and 5 years of 
	data acquisition to those of the ``mega'' and ``ultimate'' experiments with the same assumptions. 

	\begin{center}
	\begin{table*}[tb]
	\caption{Sensitivity and exposure necessary to discriminate between \NH~and \IH: the goal is $\mbb=8\,\meV$. The two cases 
		refer to the unquenched value of $g_A=g_\mathrm{nucleon}$ (mega) and $g_A=g_\mathrm{phen.}$ 
		(ultimate). The calculations are performed assuming \emph{zero background} experiments with $100\%$ detection 
		efficiency and no fiducial volume cuts. The last column shows the maximum value of the product $B \cdot \Delta$ in 
		order to actually comply with the zero background condition.}
	\small{
	\begin{ruledtabular}
	\begin{tabular}{l c c c c}
		Experiment		&Isotope			&$S^{0\nu} _{0B} \, [\yr] \quad$		&\multicolumn{2}{c}{Exposure (estimate)} \\
		\cline{4-5}
		&	&	&$M \cdot T$ \,[ton$\cdot$yr]		 &$B \cdot \Delta \, \zb$\,[counts$\,\kg^{-1}\,\yr^{-1}$]					\\
		\hline
		mega Ge			&\ce{^{76}Ge}		&$3.0 \cdot 10^{28}$	&$5.5$	&$1.8\cdot10^{-4}$				\\
		mega Te			&\ce{^{130}Te}		&$8.1 \cdot 10^{27}$	&$2.5$	&$4.0\cdot10^{-4}$				\\
		mega Xe			&\ce{^{136}Xe}		&$1.2 \cdot 10^{28}$	&$3.8$ 	&$2.7\cdot10^{-4}$				\\
		\\[-2.0mm]
		ultimate Ge		&\ce{^{76}Ge}		&$6.9 \cdot 10^{29}$	&$125$	&$8.0\cdot10^{-6}$				\\
		ultimate Te		&\ce{^{130}Te}		&$2.7 \cdot 10^{29}$	&$84$		&$1.2\cdot10^{-5}$				\\
		ultimate Xe		&\ce{^{136}Xe}		&$4.0 \cdot 10^{29}$	&$130$	&$7.7\cdot10^{-6}$				\\
		\end{tabular}
		\end{ruledtabular}
		}
	\label{tab:exposure}
	\end{table*}
	\end{center}

%%%	Figura Mega-Ultimate sizes spostata in CosmologyDBD %%%

%------------------------------------------------
%------------------------------------------------
	%------------------------------------------------
%------------------------------------------------
\section{Interplay with cosmology}
\label{sec:cosm_bounds}

	Here, we want to assess the possibility of taking advantage of the knowledge about the neutrino cosmological mass 
	to make inferences on some \bb~experiment results (or expected ones). 
	In particular, we follow Ref.~\cite{Dell'Oro:2015tia}.
	As already discussed in Sec.~\ref{sec:sigma_bounds}, we consider two possible scenarios. 
	Firstly, we assume only upper limits both on $\Sigma$ and $\mbb$, without any observation of \bb.
	Later, we imagine an observation of \bb~together with a non zero measurement of $\Sigma$
	(in both cases, we consider the unquenched value $g_A=g_\mathrm{phen.}$ for the axial vector coupling constant).
	
%------------------------------------------------
\subsection{Upper bounds scenario}
\label{sec:upper_bound}

	The tight limit on $\Sigma$ in Ref.~\cite{Palanque-Delabrouille:2014jca} was obtained by combining the 
	Planck 2013 results~\cite{Ade:2013zuv} with the one-dimensional flux power spectrum measurement of the Lyman-$\alpha$ 
	forest extracted from the BAO Spectroscopic Survey of the Sloan Digital Sky Survey~\cite{Palanque-Delabrouille:2013gaa}.
	In particular, the data from a new sample of quasar spectra were analyzed and a novel theoretical framework 
	which incorporates neutrino non-linearities self consistently was employed.

	The authors of Ref.~\cite{Palanque-Delabrouille:2014jca} computed a probability for $\Sigma$ that can be summarized to 
	a very a good approximation by:
	\begin{equation} \label{eq:chiCosm}
		\Delta \chi^2(\Sigma) = \frac{(\Sigma- 22\,\meV)^2}{(62\,\meV)^2}.
	\end{equation}
	Starting from the likelihood function $\mathcal{L}~\propto~\exp{-(\Delta \chi^2 /2)}$ with $\Delta \chi^2$ 
	as derived from Fig.~7 in the reference, one can obtain the following limits:
	\begin{equation} \label{eq:clCosm}
		\begin{aligned}
			\Sigma < \hphantom{1}84\,\meV		&\qquad (1\sigma \,\mbox{C.\,L.}) \\
			\Sigma < 146\,\meV					&\qquad (2\sigma \,\mbox{C.\,L.}) \\
			\Sigma < 208\,\meV					&\qquad (3\sigma \,\mbox{C.\,L.})
		\end{aligned}
	\end{equation}
	which are very close to those predicted by the Gaussian $\Delta \chi^2$ of Eq.~(\ref{eq:chiCosm}). 
	In particular, it is worth noting that, even if this measurement is compatible with zero at less than $1\sigma$, 
	the best fit value is different from zero, as expected from the oscillation data and as evidenced by Eq.~(\ref{eq:chiCosm}).
	We want to remark that, despite the impact relative impact of systematic versus statistical errors on the estimated 
	flux power is considered and discussed~\cite{Palanque-Delabrouille:2013gaa}, 
	it is anyway advisable to take these results from cosmology with the due caution.

\begin{figure*}[tb]
		\centering
		\includegraphics[width=\columnwidth]{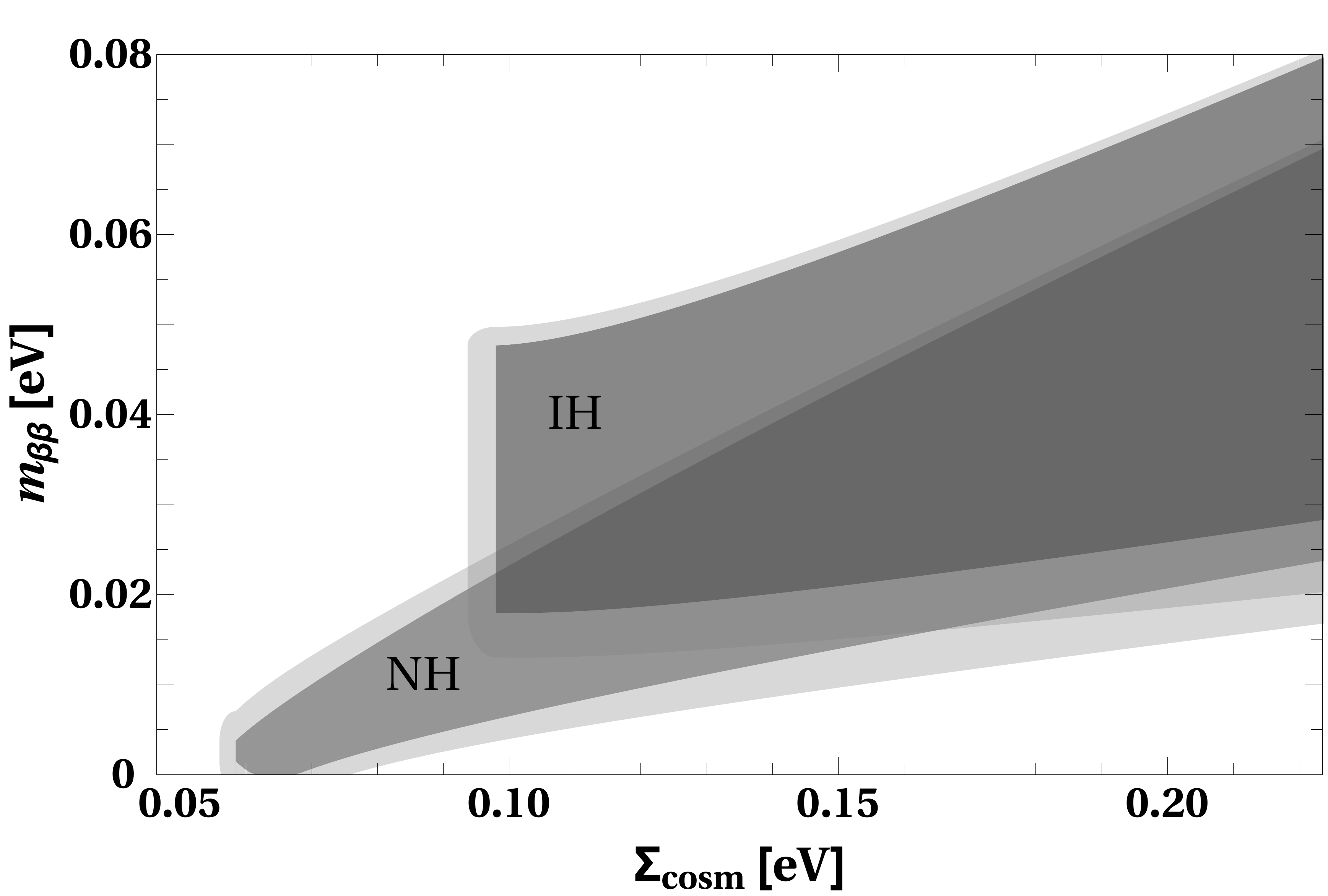}
		\includegraphics[width=\columnwidth]{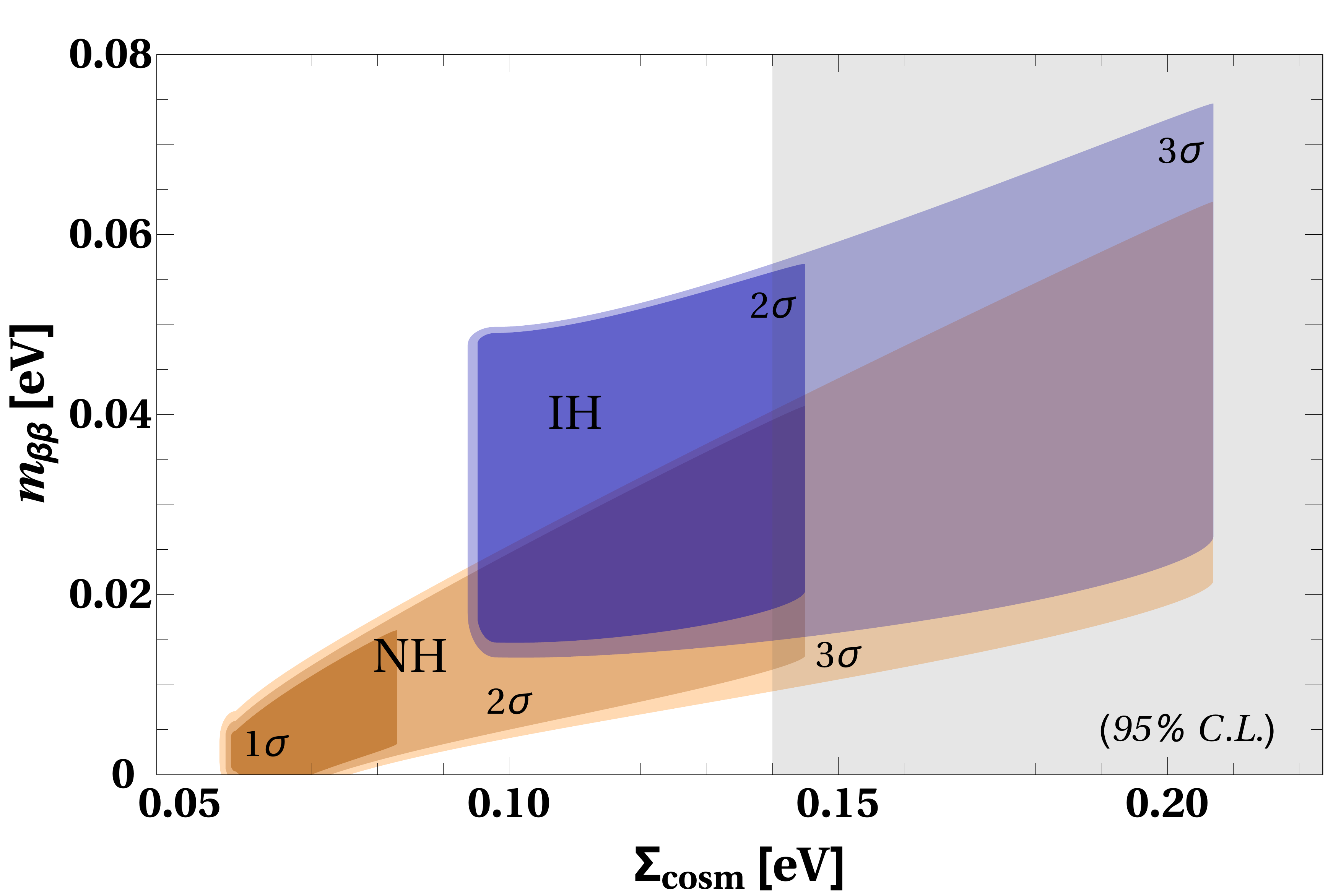}
		\caption{(Left) Allowed regions for $\mbb$ as a function of $\Sigma$ with constraints given by 
			the oscillation parameters. The darker regions show the spread induced by Majorana phase variations, 
			while the light shaded areas correspond to the $3\sigma$ regions due to error propagation of the uncertainties on 
			the oscillation parameters.
			(Right) Constraints from cosmological surveys are added to those from oscillations. 
			Different C.\,L.\ contours are shown for both hierarchies. Notice that the 1$\sigma$ region for the 
			\IH~case is not present, being the scenario disfavored at this confidence level. 
			The dashed band signifies the 95\% C.\,L.\ excluded region coming from Ref.\ \cite{Palanque-Delabrouille:2014jca}.
			Figure from Ref.~\cite{Dell'Oro:2015tia}.
			}
		\label{fig:blobs}
	\end{figure*}

	The plot showing $\mbb$ as a function of $\Sigma$ which was already shown in the right panel of Fig.~\ref{fig:DBD_graph}, 
	is again useful for the discussion.
	A zoomed version of that plot (with linear instead of logarithmic scales for the axis) is presented in the left panel of 
	Fig.~\ref{fig:blobs}. 
	As already mentioned, the extreme values for $\mbb$ after variation of the Majorana phases can be easily calculated 
	(see App.~\ref{app:mbb_extr}). 
	This variation, together with the uncertainties on the oscillation parameters, results in a widening of the allowed regions. 
	It is also worth noting that the error on $\Sigma$ contributes to the total uncertainty.
	Its effect is a broadening of the light shaded area on the left side of the minimum allowed value $\Sigma(m=0)$ for 
	each hierarchy. 
	In order to compute this uncertainty, we considered Gaussian errors on the oscillation parameters, namely
	\begin{equation}
		\delta \Sigma = \sqrt{	
				\left(\frac{\partial\,\Sigma \hspace{11pt}}{\partial\,\delta m^2}\,\sigma(\delta m^2) \right)^2 + 
				\left(\frac{\partial\,\Sigma \hspace{11pt}}{\partial\,\Delta m^2}\,\sigma(\Delta m^2) \right )^2}\,.
	\end{equation}

% --- %
% --- %

	It is possible to include the new cosmological constraints on~$\Sigma$ from Ref.~\cite{Palanque-Delabrouille:2014jca}
	considering the following inequality:
	\begin{equation}
		\frac{(y-\mbb(\Sigma))^2}{ (n\,\sigma[ \mbb(\Sigma) ])^2}+\frac{(\Sigma-\Sigma(0))^2}{(\Sigma_n-\Sigma(0)) ^2}<1
		\label{eq:chiblob}
	\end{equation}
	where $\mbb(\Sigma)$ is the Majorana Effective Mass as a function of $\Sigma$ and
	$\sigma[\mbb(\Sigma)]$ is the 1$\sigma$ associated error, computed as discussed in Ref.~\cite{Dell'Oro:2014yca}.
	$\Sigma_n$ is the limit on $\Sigma$ derived from Eq.~(\ref{eq:chiCosm}) for the C.\,L.\ $n=1,2,3,\dots$
	By solving Eq.~(\ref{eq:chiblob}) for $y$, it is thus possible to get the allowed contour for $\mbb$ considering both the 
	constraints from oscillations and from cosmology. In particular, the Majorana phases are taken into account by 
	computing $y$ along the two extremes of $\mbb(\Sigma)$, namely $\mbb^{max}(\Sigma)$ and $\mbb^{min}(\Sigma)$, 
	and then connecting the two contours. The resulting plot is shown in the right panel of Fig.~\ref{fig:blobs}.
	
	The most evident feature of Fig.~\ref{fig:blobs} is the clear difference in terms of expectations for both $\mbb$ 
	and $\Sigma$ in the two hierarchy cases. 
	The relevant oscillation parameters (mixing angles and mass splittings) are well known and they induce only 
	minor uncertainties on the expected value of $\mbb$. These uncertainties widen the allowed contours in the upper, 
	lower and left sides of the picture.
	The boundaries in the rightmost regions are due to the new information from cosmology and are cut at various 
	confidence levels. It is notable that at 1$\sigma$, due to the exclusion of the 
	\IH, the set of plausible values of $\mbb$ is highly restricted.
	
\begin{figure}[tb]
		\centering
		\includegraphics[width=\columnwidth]{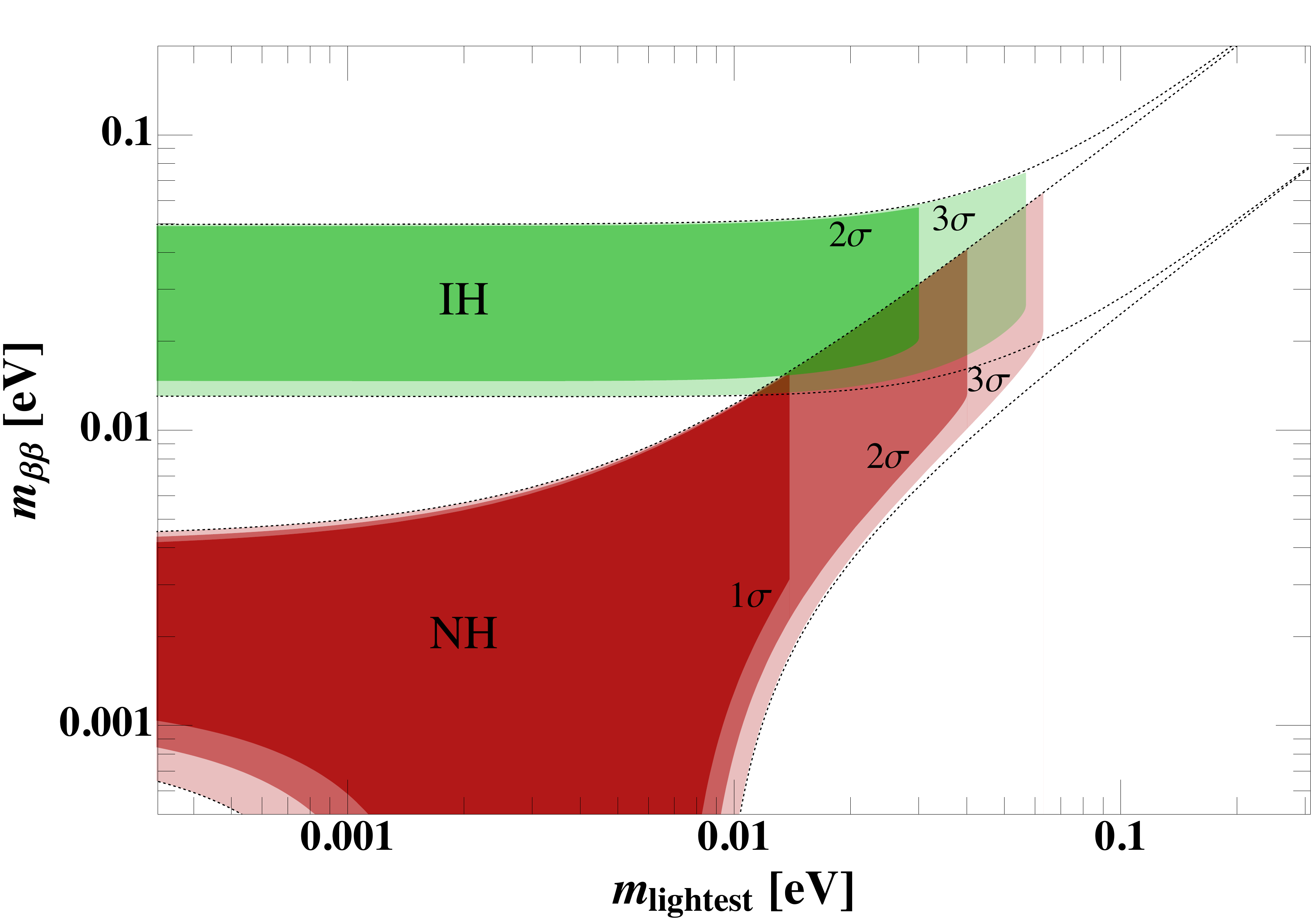}
		\caption{Constraints from cosmological surveys are added to those from oscillations in the representation
			of $\mbb$ as a function of the lightest neutrino mass. 
			The dotted contours represent the $3\sigma$ regions allowed considering oscillations only.
			The shaded areas show the effect of the inclusion of cosmological constraints at different C.\,L.\,. 
			The horizontal bands correspond to the expected sensitivity for future experiments. 
			Figure from Ref.~\cite{Dell'Oro:2015tia}.}
		\label{fig:VS_cosm}
	\end{figure}

	The impact of the new constraints on $\Sigma$ appears even more evident by plotting $\mbb$ as a function of the mass of 
	the lightest neutrino.
	In this case, Eq.~(\ref{eq:chiblob}) becomes:
	\begin{equation}
		\frac{(y-\mbb(m))^2}{ (n\,\sigma[ \mbb(m) ])^2}+\frac{m^2}{m(\Sigma_n) ^2}<1.
		\label{eq:chiblobVS}
	\end{equation}
	The plot in Fig.~\ref{fig:VS_cosm} globally shows that the next generation of experiments will have small possibilities of 
	detecting a signal of \bb~due to light Majorana neutrino exchange.
	Therefore, if the new results from cosmology are confirmed or improved, ton or even multi-ton scale detectors 
	will be needed~\cite{Dell'Oro:2014yca}.

	On the other hand, a \bb~signal in the near future could either disprove some assumptions of the present cosmological 
	models, or suggest that a different mechanism other than the light neutrino exchange mediates the transition.
	New experiments are interested in testing the latter possibility by probing scenarios beyond the 
	SM~\cite{Cirigliano:2004tc,Tello:2010am,Alekhin:2015byh}.  
	
%------------------------------------------------
\subsection{Measurements scenario}
	\label{sec:measurement}
	
	Here we consider the implications of the following non-zero value of $\Sigma$~\cite{Battye:2013xqa}:
	\begin{equation}
		\Sigma = (0.320 \pm 0.081)\, \mbox{eV}.
	\end{equation}
	We focus on the light neutrino exchange scenario and assume that \bb~is observed with a rate compatible with:
	\begin{enumerate}
		\item the present sensitivity on $\mbb$. In particular, we use the limit coming from the combined 
		\ce{^{136}Xe}-based experiments~\cite{Gando:2012zm}. We refer to this as to the ``present'' case
		\item a value of $\mbb$ that will be likely probed in the next few years. In particular, we use the 
		CUORE experiment sensitivity~\cite{Artusa:2014lgv}, as an example of next generation of \bb~experiments.
		We refer to this as to the ``near future'' case.
	\end{enumerate}
	
	For the sake of completeness, it is useful to recall a few definitions and relations. The likelihood of a 
	simultaneous observation of some values for $\Sigma$ and $\mbb$ (respectively with uncertainties $\sigma(\Sigma^{\meas})$ 
	and $\sigma(\mbb^{\meas})$ and distributed according to Gaussian distributions) can be written as following:
	\begin{equation}
		\mathcal{L} \propto \exp \left[ -\frac{(\Sigma-\Sigma^{\meas})^2}{2\sigma(\Sigma^{\meas})^2} \right]
		\exp \left[ -\frac{(\mbb - \mbb^{\meas})^2}{2\sigma(\mbb^{\meas})^2} \right].
	\label{eq:likelihood}
	\end{equation}
	Recalling the relation between the $\chi^2$ and the likelihood, namely $\mathcal{L} \propto \el^{-\chi^2/2}$, we obtain:
	\begin{equation}
		\chi^2=\frac{(\Sigma-\Sigma^{\meas})^2}{\sigma(\Sigma^{\meas})^2}+\frac{(\mbb - \mbb^{\meas})^2}{\sigma(\mbb^{\meas})^2}
	\end{equation}
	which represents an elliptic paraboloid. Since we are dealing with a two parameter $\chi^2$, we need to find 
	the appropriate prescription to define the confidence intervals. 
	At the desired confidence level, we get:
	\begin{equation}
		\CL = \iint_{\chi^2 < \chi_0^2}  dx\,dy \,\frac{1}{2\pi \sigma_x \sigma_y} \,
		\el^{-\frac{x^2}{2\sigma^2_x} - \frac{y^2}{2\sigma^2_y}}
	\end{equation}
	and thus 
	\begin{equation}
		\chi^2_0 = -2\,\ln\,(1-\CL).
	\end{equation}	
	This defines the value for $\chi^2$ correspondent to the confidence level $\CL$\,.
	
	In order to write down the likelihood we need to evaluate the standard deviations both on $\Sigma$ and on $\mbb$. 
	While the error on $\Sigma$ 
	comes directly from the cosmological measurement, the one on $\mbb$ has to be determined. 
	It has two different contributions: one is statistical and comes from the Poisson fluctuations on the 
	observed number of events (see Sec.~\ref{sec:sensitivity}), while the other comes from the 
	uncertainties on the nuclear physics (see Sec.~\ref{sec:theo_uncertainties}). 
	Actually, a greater effect would rise if we took into account the error on $g_A$, but 
	here we assume that the quenching is absent.
%	Refer to App.\ \ref{app:stat} for technical details on the error propagation.
	
	For a few observed events, let us say less 
	than 10 events, the global error is dominated by the statistical fluctuations. The error on the nuclear physics
	becomes the main contribution only if many events (more than a few tens) are detected.
	Using the described procedure and for the present case, we find an uncertainty on $\mbb$ of about $31\,\meV$ for 
	5 observed events, which reduces to $24\,\meV$ for 10 events. If we neglect the statistical uncertainty, e.\,g.\ 
	we put $\Nev$, the uncertainty becomes $14\,\meV$. This means that the Poisson fluctuations effect 
	is not negligible at all. 
	Similarly, repeating the same work for the near future case, we obtain an uncertainty of $17\,\meV$ 
	for 5 events, $13\,\meV$ for 10 events and $8\,\meV$ for $\Nev$.
	
\begin{figure}[tb]
		\centering
		\includegraphics[width=\columnwidth]{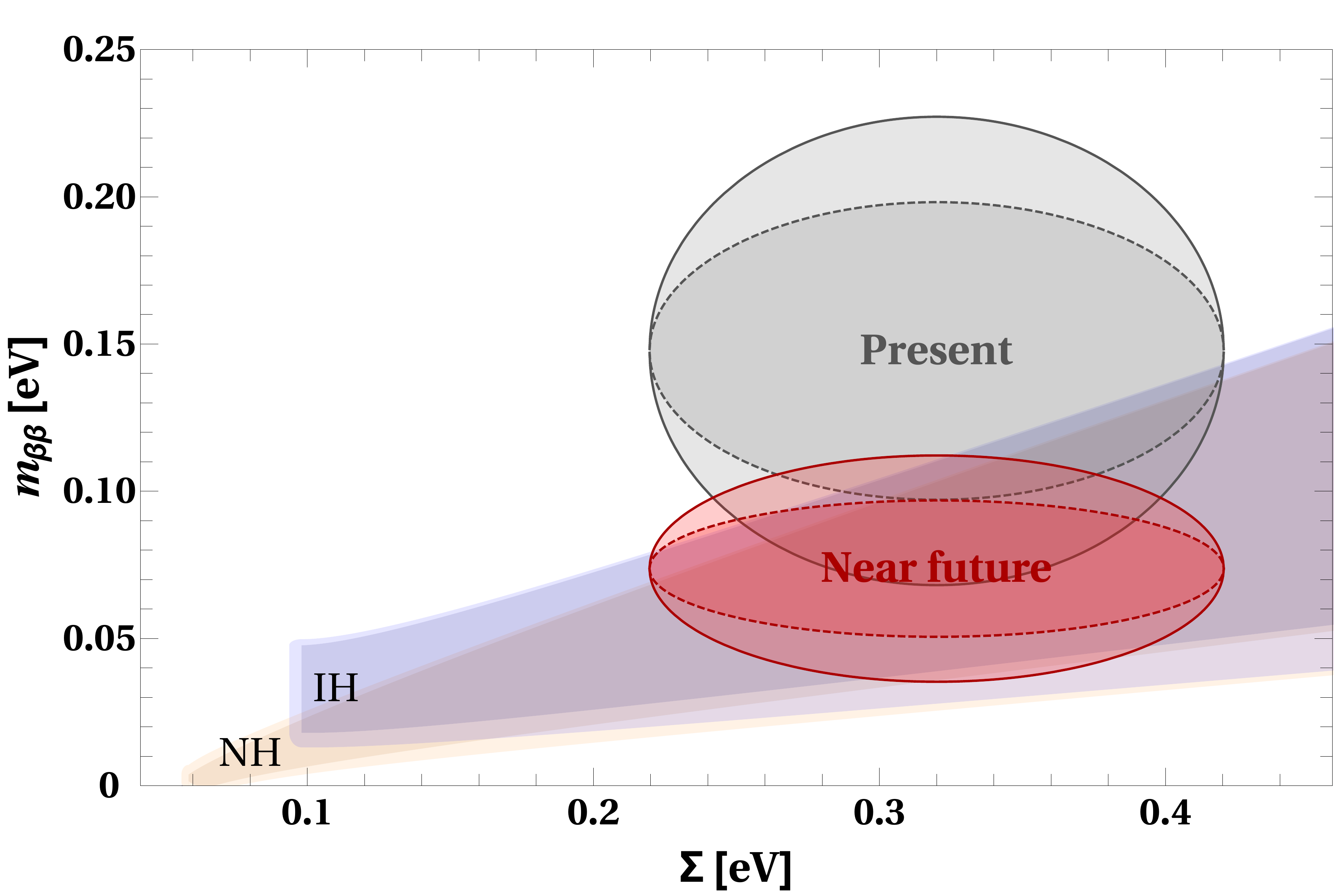}
		\caption{The plots show the allowed regions for $\mbb$ as a function of the neutrino cosmological mass $\Sigma$. 
			The ellipses show the 90\%\,C.\,L.\ regions in which a positive observation of \bb~could be contained, 
			according to the experimental uncertainties and 5 (solid) and 20 (dashed) actually observed events. 
			In particular, the upper ellipse assumes the present limit from the combined \ce{^{136}Xe}
			experiments~\cite{Gando:2012zm}. The lower one assumes the sensitivity of CUORE~\cite{Artusa:2014lgv}.}
		\label{fig:ell1}
	\end{figure}

	Let us now concentrate on the case of 5 \bb~observed events. If we cut the $\chi^2$ at the $90\%\,\CL$ and we consider 
	the data previously mentioned, we obtain the bigger, solid ellipses drawn in Fig.~\ref{fig:ell1}. 
	This shows that in the near future case, a detection of \bb~would allow to say nothing neither about the mass 
	hierarchy nor about the Majorana phases.
	Interestingly, if \bb~were actually discovered with a $\mbb$ a little bit lower than the one probed in the present case, 
	some conclusions about the Majorana phases could be carried out. In any case, in order to state anything precise about 
	$\mbb$ and the Majorana phases, even assuming the discovery of \bb, the uncertainty on the quenching of the axial 
	vector coupling constant has to be dramatically decreased.
	
	If we repeat the same exercise assuming an observed number of events of 20, we obtain the smaller, dashed ellipses of 
	Fig.~\ref{fig:ell1}. In this case, an hypothetical observation coming from the present case is highly 
	disfavored while in the future case, even if nothing can be said about the hierarchy, some conclusions could be 
	carried out regarding the Majorana phases. 
	
	This simple analysis shows that, thanks to the great efforts done in the NME and PSF 
	calculations, it is most likely that the biggest contribution to the error will come from the statistical fluctuations 
	of the counts.
	However, the theoretical uncertainty from the nuclear physics could make the picture really hard to 
	understand because, up to now, it is a source of uncertainty of a factor $4-8$ on $\mbb$.

%------------------------------------------------
\subsection{Considerations on the information from cosmological surveys}

	The newest results reported in Table~\ref{tab:postPlanck} confirms and strengthens the cosmological indications 
	of upper limits on $\Sigma$, and it is likely that we will have soon other 
	substantial progress. Moreover, the present theoretical understanding of neutrino
	masses does not contradict these cosmological indications. 
	These considerations emphasize the importance of exploring the issue of mass hierarchy in laboratory experiments and 
	with cosmological surveys.
	However, as already stated, a cautious approach in dealing with the results from cosmological surveys is highly advisable.
	
	From the point of view of \bb, these results show that ton or multi-ton scale detectors will be needed in order to 
	probe the range of $\mbb$ now allowed by cosmology. 
	Nevertheless, if next generation experiments see a signal, it will likely be a \bb~signal of new physics different from 
	the light Majorana neutrino exchange.

%------------------------------------------------
%------------------------------------------------
	%------------------------------------------------
%------------------------------------------------
\section{Summary}
   
	In this review, we analyzed the \bb~process under many different aspects.
	We assessed its importance to test lepton number, to determine the nature of neutrino mass and to probe its values.
	Various particle physics mechanisms that could contribute to the \bb~were examined, although with the conclusion that
	from the theoretical point of view the most interesting and promising remains the light Majorana neutrino exchange.
	We studied the current experimental sensitivity, focussing on the critical point of determining the uncertainties
	in the theoretical calculations and predictions. In view of all these considerations, the prospects for the near 
	future experimental sensitivity were presented and the main features of present, past and future \bb~experiments 
	were discussed.
	Finally, we stressed the huge power of cosmological surveys in constraining neutrino masses and consequently 
	the \bb~process.

%------------------------------------------------
%------------------------------------------------

	\begin{acknowledgments}
		We wish to acknowledge extensive discussions with Professor F.\ Iachello and thank him for stimulating this study.
		F.\ V. also thanks E.\ Lisi.
	\end{acknowledgments}

%------------------------------------------------
\appendix

%------------------------------------------------
%------------------------------------------------
	%------------------------------------------------
%------------------------------------------------
\section{Extremal values of $\mbb$}	
\label{app:mbb_extr}

	Recalling the definition of Eq.~(\ref{eq:mbb3}) for the Majorana effective mass:
	\begin{equation}
		\mbb = \Biggl| \sum_{i=1}^3 U_{\el i}^2 m_i \Biggr|
	\end{equation}
	it is possible to demonstrate
	that the extreme values assumed by this parameter due to free variations of the phases are:%
	\footnote{\,The proof shown here is based on the work reported in Ref.~\cite{Vissani:1999tu}.} 
	\begin{align}
			&\mbbM = \sum_{i=1}^3 \bigl| U_{\el i}^2 \bigr| m_i \\
			&\mbbm = \max \Bigl\{ 2 \bigl| U_{\el i}^2 \bigr| m_i - \mbbM, 0 \Bigr\} \quad i=1,2,3.
			\label{eq:mee_min_app}
	\end{align}
	
%------------------------------------------------
\subsection{Formal proof}

	Regarding the first assertion, it is obvious that the sum of $n$ complex numbers has the biggest allowed module when 
	those numbers have aligned phases. 
	Since the physical quantities depend on $m_{\beta\beta}^2$, without any loss of generality it is possible to choose 
	the first term ($U_{\el 1}^2 m_1$) to be real. It thus follows that also 
	the other two terms must be real: this is equivalent to considering the sum of the modules of the single terms.
	
	To prove the second statement, let us consider the general case $\mbb \sim |z_1 + z_2 + z_3| \equiv r$ where $z_i$ are 
	complex numbers. We want to minimize $r$, by keeping fixed the $|z_i|$.
	Let us define:
	\begin{equation}
		\left\{
		\begin{aligned}
			&r_1 = |z_1| - |z_2| - |z_3| \\
			&r_2 = |z_2| - |z_1| - |z_3| \\
			&r_3 = |z_3| - |z_1| - |z_2|
		\end{aligned}
		\right.
	\end{equation}
	and
	\begin{equation}
		\left\{
		\begin{aligned}
			&q_1 = |z_1| - |z_2 + z_3| \\
			&q_2 = |z_2| - |z_1 + z_3| \\
			&q_3 = |z_3| - |z_1 + z_2|.
		\end{aligned}
		\right.
	\end{equation}
	It is worth noting that only one of the $r_i$ can be positive, at most. Therefore, it is possible to distinguish 4 cases:
	\begin{enumerate}
		\item[$i)$] $r_1>0$;
		\item[$ii)$] $r_2>0$;
		\item[$iii)$] $r_3>0$;
		\item[$iv)$] $r_i \le 0$ for $i=1,2,3$.
	\end{enumerate}
	In the first one, it is possible to show that $r^{\min}=r_1$. 
	In fact, 
	we can write:
	\begin{equation}
	\begin{split}
		r	&= |z_1 + z_2 + z_3| = |z_1 - (- z_2 - z_3)| \\
			&\ge \bigl| |z_1| - |- z_2 - z_3| \bigr| \\
			&= \bigl| |z_1| - |z_2 + z_3| \bigr|= |q_1|
	\end{split}
	\end{equation}
	and, since 
	\begin{equation}
		q_1 = |z_1| - |z_2 + z_3| \ge |z_1| - |z_2| - |z_3| = r_1 > 0
	\end{equation}
	we obtain:
	\begin{equation}
		r \ge |q_1| \ge q_1 \ge r_1.
	\end{equation}
	Similarly, $r_2>0 \, \Rightarrow \, r^{\min}=r_2$ and $r_3>0 \, \Rightarrow \, r^{\min}=r_3$ 
	in the second and in the third cases, respectively.
	In the last case, it is necessary to observe that, if one of the $r_i = 0$, then $r^{\min}=0$.
	Therefore, only the case in which $r_i<0 \, \forall i$ must be considered. 
	In this case, $q_1$ goes from negative when $arg(z_2) = arg(z_3)$, to positive, when $arg(z_2) = - arg(z_3)$.
	By continuity, this implies that a proper phase choice such that $q_1=0$ must exist.
	Thus, one can conclude also in this case that $r^{\min}=0$ (by choosing $r=|q_1|$).
	
	In synthesis, the single case analysis leads to:
	\begin{equation}
		r^{\min} = \max \{r_i,0 \}.
	\end{equation}
	This proves the original statement, since $r_i = |z_i| - |z_j| - |z_k| + |z_i| - |z_i| = 2|z_i| - \sum_{l=1}^3 |z_l|$ 
	for $i \ne j \ne k$, $\{i,j,k\} = \{1,2,3\}$.

%------------------------------------------------
\subsection{Remarks on the case $\mbbm=0$}

	\begin{figure}[t]
		\centering
		\includegraphics[width=\columnwidth]{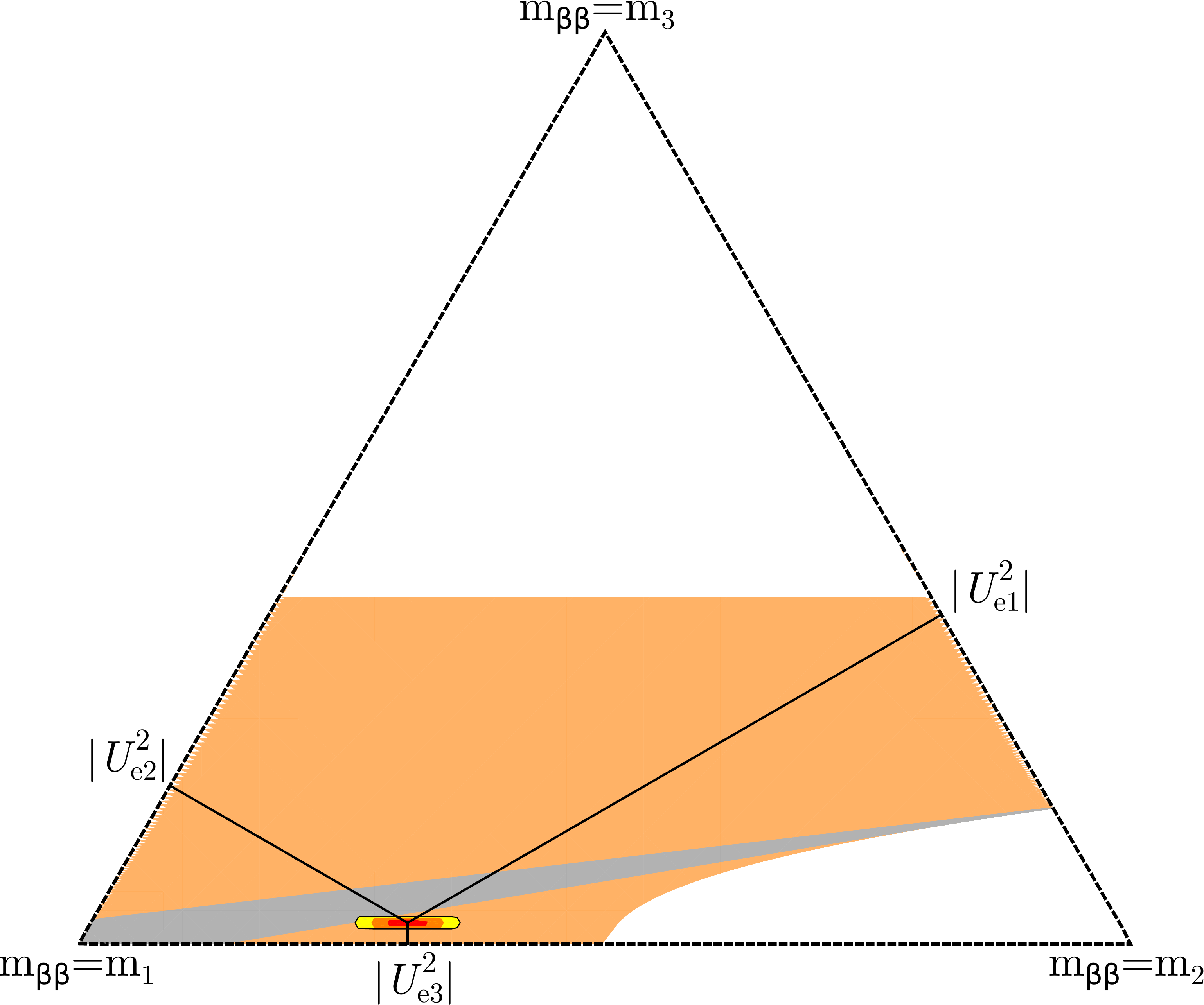}
		\caption{Representation of $\mbbm$ in the unitarity triangle.
			The internal point in the middle of the small colored bar is identified by the constraints from the oscillation 
			parameters. The colored regions correspond to the $1\sigma$ (red), $2\sigma$ (orange) and $3\sigma$ (yellow). 
			The distance from a side represent the size of the corresponding mixing element $\left| U_{\el i}^2 \right|$. 
			The inner shaded regions of the triangle enclose the areas where $\mbbm = 0$ for a lightest neutrino mass that 
			can vary from $10^{-5}\,\mbox{eV}$ to the value which corresponds to a cosmological mass 
			$\Sigma=0.14\,\mbox{eV}$ (orange, $90\%\,\CL$ current bound) and $\Sigma=0.06\,\eV$ 
			(gray, for purpose of illustration).}
		\label{fig:triangle}
	\end{figure}

	The three mixing elements $\left| U_{\el i}^2 \right|$ are constrained by the unitarity: 
	$\sum_i \left| U_{\el i}^2 \right| = 1$. This condition can be graphically pictured by using the inner region of 
	an equilateral triangle with unitary height, where the distance from the $i$-th side corresponds to the value of 
	$\left| U_{\el i}^2 \right|$ (see Ref.~\cite{Vissani:1999tu} for details). The result is displayed in 
	Fig.~\ref{fig:triangle}. 
	
	The experimental constraints on the oscillation parameters make it possible to evaluate the elements 
	$\left| U_{\el i}^2 \right|$
	and, therefore, to identify a point inside the triangle, which is placed at the center of the colored bar in 
	Fig.~\ref{fig:triangle}. The different colors of the bar correspond to the $1\sigma$, $2\sigma$ and $3\sigma$ regions.
	
	At each vertex, the value of $\mbb$ coincides with $\mbbm$ and with one of the mass eigenstates 
	($\nu_\el  \equiv \nu_i$). Then, the value of $\mbbm$ decreases moving from one vertex towards the inner part of 
	the triangle, until it becomes zero inside the region delimited by the vertices defined by the conditions:
	\begin{align}
			&\left| U_{\el 1}^2 \right| m_1 = \left| U_{\el 2}^2 \right| m_2 \mbox{ when } \left| U_{\el 3}^2 \right| = 0 
		\label{eq:Ue3} \\
			&\left| U_{\el 1}^2 \right| m_1 = \left| U_{\el 3}^2 \right| m_3 \mbox{ when } \left| U_{\el 2}^2 \right| = 0 \\
			&\left| U_{\el 2}^2 \right| m_2 = \left| U_{\el 3}^2 \right| m_3 \mbox{ when } \left| U_{\el 1}^2 \right| = 0.
	\end{align}
	In fact, if we consider, for example, the first condition, from Eq.~(\ref{eq:Ue3}) we have:
	\begin{multline}
		2 \bigl| U_{\el i}^2 \bigr| m_i - \mbbM \\ = 
		2 \bigl| U_{\el i}^2 \bigr| m_i - \bigl| U_{\el 1}^2 \bigr| m_1 - \bigl| 
			U_{\el 2}^2 \bigr| m_2 - \bigl| U_{\el 3}^2 \bigr| m_3 \\ =
		2 \bigl| U_{\el i}^2 \bigr| m_i - 2 \bigl| U_{\el 1}^2 \bigr| m_1.
	\end{multline}
	Substituting the possible values $i=1,2,3$, and recalling that the condition to get $\mbbm$ is expressed by 
	Eq.~(\ref{eq:mee_min_app}), we obtain:
	\begin{equation}
		\mbbm = \max \Bigl\{ -2 \bigl| U_{\el 1}^2 \bigr| m_1, 0 \Bigr\} = 0.
	\end{equation}
	The same argument can be applied also for the other two conditions.
	It is therefore possible to identify a region inside the triangle where $\mbbm$ is zero. The experimental constraints on 
	the oscillation parameters limit the possibility of $\mbbm=0$, only to the case of \NH. Of course, the position and the 
	extension of this region depends on the lightest neutrino mass. 
	
	Instead of choosing one particular value for the lightest neutrino mass, it is more convenient to plot the superposition 
	of the regions obtained for increasing values of this parameter. 
	In Fig.~\ref{fig:triangle}, in orange we show the region obtained varying $m_1$ from 
	$10^{-5}\,\eV$, up to the $90\%\,\CL$ maximum value it can have considering the limit on $\Sigma$ from 
	Ref.~\cite{Palanque-Delabrouille:2014jca}, according to Eq.~(\ref{eq:chiCosm}). 
	The gray region shows the superposition obtained when $m_1 \sim 0$, 
	namely we show what happens if it turns out that the cosmological mass is close to its lower limit 
	($\lesssim 0.06\,\eV$ for the \NH~case).
		
	The existence of a $\mbbm=0$ region implies that, in principle, \bb~could be forbidden just by particular combinations 
	of the phases, even if the neutrino is a Majorana particle.

%------------------------------------------------
%------------------------------------------------
	%------------------------------------------------
%------------------------------------------------
%------------------------------------------------
\section{Error propagation}
\label{app:stat}

	It is convenient and usually appropriate to adopt statistical procedures that are as direct and as practical as possible. 
	We are interested in the following situation. For any choice of the Majorana phases, the massive parameter that 
	regulates the \bb~can be thought as $M(m,\bm{x})$. 
	It is a function of the parameters that are determined by oscillation experiments up to their experimental error, 
	$x_i\pm \Delta x_i$, and of another massive parameter $m$.
	Here a remark is necessary. When in the literature we found 
	maximal or systematic uncertainties, in order to 
	propagate their effects in our calculations, we decided to interpret them as the semi-widths of flat distributions and 
	thus, dividing these numbers by $\sqrt{3}$ we could get the standard deviations of those distributions. 
	Then, we considered those values as standard deviations for Gaussian fluctuations of the parameters around the given values.

	For any fixed value of $m$, and for the other parameters set to their best fit values $x_i$, we can attach the 
	following error to $M$:
	\begin{equation}
		\left. \Delta M \right|_m =\sqrt{\sum_i \left( \frac{\partial M}{\partial x_i} \right)^2 \Delta x_i^2}.
	\label{eq:err_prop}
	\end{equation}
	When we want to consider the prediction and the error for a fixed value of another massive parameter
	$\Sigma(m,\bm{x})$, we have to vary also $m$, keeping
	$\delta \Sigma = \partial \Sigma/\partial m\, \delta m + \partial \Sigma/\partial x_i\, \delta x_i =0$. 
	Therefore, in this case we find:
	\begin{equation}
		\left. \Delta M \right|_\Sigma =
		\sqrt{\sum_i \left( \frac{\partial M}{\partial x_i} -
		\frac{{\partial \Sigma}/{\partial x_i}}{{\partial \Sigma}/{\partial m} } \,
		\frac{\partial M}{\partial m} \right)^2 \Delta x_i^2}.
	\end{equation}
	Of course, we will calculate $m$ by inverting the $\Sigma(m,\bm{x})=\Sigma$ 
	(here, the symbol $\Sigma$ denotes the function and also its value; however, this abuse of notation is harmless 
	in practice).

%------------------------------------------------
%------------------------------------------------
	%------------------------------------------------
%------------------------------------------------
%\section{Lightest neutrino mass and cosmological mass}
\section{$\Sigma = f(m_\mathrm{lightest})$, analitical solution}
\label{app:sigma}

	Let us write in full generality the three flavor relation for the mass probed in cosmology as 
	\begin{equation}
		\Sigma = m + \sqrt{m^2 + a^2} + \sqrt{m^2 + b^2}
	\label{eq:imhotep}
	\end{equation}
	where $m$, $\Sigma$, $a$ and $b$ are masses, i.\,e.\ non-negative parameters. It is possible to obtain $m$ as a 
	function of $\Sigma$ in the physical range 
	\begin{equation}
		\Sigma \ge a+b
	\notag
	\end{equation}
	simply by solving a quartic equation. Since we are interested in certain specific cases (\NH~or \IH) we specify 
	the discussion further.
	
	When $a \ll b$, corresponding to the \NH~case, it is convenient to write the quartic equation as 
	\begin{equation}
		(3 m^2-4 m \Sigma + \lambda^2) (m^2 - \lambda^2) + 4 a^2 b^2 = 0
	\end{equation}
	where
	\begin{equation}
		\lambda^2 \equiv \Sigma^2 - (a^2 + b^2).
	\end{equation}
	Indeed, we see that this quartic equation has spurious solutions in this limit, e.\,g. those for $m \approx \pm \lambda$. 
	Instead, we are interested in the one that (for $a = 0$) reads
	\begin{equation}
		m = m_{\mbox{\tiny \NH}}(\Sigma,b) \equiv \frac{2 \Sigma - \sqrt{\Sigma^2+3 b^2}}{3} 
	\label{eq:nhh}
	\end{equation}
	with $\Sigma \ge b$.
	In the case when $a\approx b$, instead, which corresponds to the \IH~case, it is convenient to write the quartic equation as
	\begin{equation}
		(3 m^2+2 m \Sigma - \lambda^2) (m - \Sigma)^2 - (a^2- b^2)^2=0
	\end{equation}
	where
	\begin{equation}
		\lambda^2 \equiv \Sigma^2-2 (a^2+b^2).
	\end{equation}
	Again, we see that this quartic equation has spurious solutions in the limit $a\approx b$, e.\,g. $m \approx \Sigma$. 
	We are interested to the one that in the case $a=b$ reads
	\begin{equation}
		m=m_{\mbox{\tiny \IH}}(\Sigma,b) \equiv \frac{- \Sigma + 2 \sqrt{\Sigma^2-3 b^2}}{3}
	\label{eq:ihh}
	\end{equation}
	with $\Sigma \ge 2 b$.
	
	Finally, we discuss useful approximate formulae for the specific parameterization suggested 
	in Ref.~\cite{Capozzi:2013csa}, namely
	\begin{equation}
		\left\{
			\begin{aligned}
				a &= \delta m^2 \\
				b &= \Delta m^2+\delta m^2/2
			\end{aligned}
		\right.
	\end{equation}
	for the \NH~case and
	\begin{equation}
		\left\{
			\begin{aligned}
				a &= \Delta m^2-\delta m^2/2 \\
				b &= \Delta m^2+\delta m^2/2
			\end{aligned}
		\right.
	\end{equation}
	for the \IH~one.
	
	In the latter case, the approximation obtained by Eq.~(\ref{eq:ihh}), namely,
	\begin{equation}
		m = m_{\mbox{\tiny \IH}}(\Sigma,\Delta m^2)  %\mbox{ with }\mu\ge 2 \sqrt{\Delta m^2} 
	\label{eq:ihh2}
	\end{equation}
	is already excellent, being better than $3\,\upmu\eV$ in the whole range of masses.
	Instead, Eq.~(\ref{eq:nhh}) implies a 
	maximum error that can reach 5\,meV for \NH. Although this is quite adequate for the present and near future 
	sensitivity, it is possible to improve the approximation also in the case of \NH~by using 
	\begin{equation}
		m = m_{\mbox{\tiny \NH}}(\Sigma,\Delta m^2) - \frac{\delta m^2}{4\ m_{\mbox{\tiny \NH}}(\Sigma,\Delta m^2)}.
	\label{eq:nhh2}
	\end{equation}
	This formula is obtained by linearly expanding in $\delta m^2$ the relation that links $\Sigma$ and $m$,
	Eq.~(\ref{eq:imhotep}), around the point $m = m_{\mbox{\tiny \NH}}(\Sigma,\Delta m^2)$.
	The error is remarkably small error and more than adequate for the present sensitivity: less than 0.2\,meV.

\bibliography{ref}

%merlin.mbs apsrev4-1.bst 2010-07-25 4.21a (PWD, AO, DPC) hacked
%Control: key (0)
%Control: author (0) dotless jnrlst
%Control: editor formatted (1) identically to author
%Control: production of article title (0) allowed
%Control: page (1) range
%Control: year (0) verbatim
%Control: production of eprint (0) enabled
\begin{thebibliography}{213}%
\makeatletter
\providecommand \@ifxundefined [1]{%
 \@ifx{#1\undefined}
}%
\providecommand \@ifnum [1]{%
 \ifnum #1\expandafter \@firstoftwo
 \else \expandafter \@secondoftwo
 \fi
}%
\providecommand \@ifx [1]{%
 \ifx #1\expandafter \@firstoftwo
 \else \expandafter \@secondoftwo
 \fi
}%
\providecommand \natexlab [1]{#1}%
\providecommand \enquote  [1]{``#1''}%
\providecommand \bibnamefont  [1]{#1}%
\providecommand \bibfnamefont [1]{#1}%
\providecommand \citenamefont [1]{#1}%
\providecommand \href@noop [0]{\@secondoftwo}%
\providecommand \href [0]{\begingroup \@sanitize@url \@href}%
\providecommand \@href[1]{\@@startlink{#1}\@@href}%
\providecommand \@@href[1]{\endgroup#1\@@endlink}%
\providecommand \@sanitize@url [0]{\catcode `\\12\catcode `\$12\catcode
  `\&12\catcode `\#12\catcode `\^12\catcode `\_12\catcode `\%12\relax}%
\providecommand \@@startlink[1]{}%
\providecommand \@@endlink[0]{}%
\providecommand \url  [0]{\begingroup\@sanitize@url \@url }%
\providecommand \@url [1]{\endgroup\@href {#1}{\urlprefix }}%
\providecommand \urlprefix  [0]{URL }%
\providecommand \Eprint [0]{\href }%
\providecommand \doibase [0]{http://dx.doi.org/}%
\providecommand \selectlanguage [0]{\@gobble}%
\providecommand \bibinfo  [0]{\@secondoftwo}%
\providecommand \bibfield  [0]{\@secondoftwo}%
\providecommand \translation [1]{[#1]}%
\providecommand \BibitemOpen [0]{}%
\providecommand \bibitemStop [0]{}%
\providecommand \bibitemNoStop [0]{.\EOS\space}%
\providecommand \EOS [0]{\spacefactor3000\relax}%
\providecommand \BibitemShut  [1]{\csname bibitem#1\endcsname}%
\let\auto@bib@innerbib\@empty
%</preamble>
\bibitem [{\citenamefont {Dirac}(1928{\natexlab{a}})}]{Dirac:1928hu}%
  \BibitemOpen
  \bibfield  {author} {\bibinfo {author} {\bibfnamefont {P.~A.~M.}\
  \bibnamefont {Dirac}},\ }\bibfield  {title} {\enquote {\bibinfo {title} {{The
  Quantum theory of electron}},}\ }\href {\doibase 10.1098/rspa.1928.0023}
  {\bibfield  {journal} {\bibinfo  {journal} {Proc.\ Roy.\ Soc.\ Lond.\ A}\
  }\textbf {\bibinfo {volume} {117}},\ \bibinfo {pages} {610--624} (\bibinfo
  {year} {1928}{\natexlab{a}})}\BibitemShut {NoStop}%
%%CITATION = PRSLA,A117,610;%%
\bibitem [{\citenamefont {Dirac}(1928{\natexlab{b}})}]{Dirac:1928ej}%
  \BibitemOpen
  \bibfield  {author} {\bibinfo {author} {\bibfnamefont {P.~A.~M.}\
  \bibnamefont {Dirac}},\ }\bibfield  {title} {\enquote {\bibinfo {title} {{The
  Quantum theory of electron. Part II.}}}\ }\href {\doibase
  10.1098/rspa.1928.0056} {\bibfield  {journal} {\bibinfo  {journal} {Proc.\
  Roy.\ Soc.\ Lond.\ A}\ }\textbf {\bibinfo {volume} {118}},\ \bibinfo {pages}
  {351} (\bibinfo {year} {1928}{\natexlab{b}})}\BibitemShut {NoStop}%
%%CITATION = PRSLA,A118,351;%%
\bibitem [{\citenamefont {Majorana}(1937)}]{Majorana:1937vz}%
  \BibitemOpen
  \bibfield  {author} {\bibinfo {author} {\bibfnamefont {E.}~\bibnamefont
  {Majorana}},\ }\bibfield  {title} {\enquote {\bibinfo {title} {{Theory of the
  Symmetry of Electrons and Positrons}},}\ }\href {\doibase 10.1007/BF02961314}
  {\bibfield  {journal} {\bibinfo  {journal} {Nuovo Cim.}\ }\textbf {\bibinfo
  {volume} {14}},\ \bibinfo {pages} {171--184} (\bibinfo {year}
  {1937})}\BibitemShut {NoStop}%
%%CITATION = NUCIA,14,171;%%
\bibitem [{\citenamefont {Racah}(1937)}]{Racah:1937qq}%
  \BibitemOpen
  \bibfield  {author} {\bibinfo {author} {\bibfnamefont {G.}~\bibnamefont
  {Racah}},\ }\bibfield  {title} {\enquote {\bibinfo {title} {{On the symmetry
  of particle and antiparticle}},}\ }\href {\doibase 10.1007/BF02961321}
  {\bibfield  {journal} {\bibinfo  {journal} {Nuovo Cim.}\ }\textbf {\bibinfo
  {volume} {14}},\ \bibinfo {pages} {322--328} (\bibinfo {year}
  {1937})}\BibitemShut {NoStop}%
%%CITATION = NUCIA,14,322;%%
\bibitem [{\citenamefont {Furry}(1939)}]{Furry:1939qr}%
  \BibitemOpen
  \bibfield  {author} {\bibinfo {author} {\bibfnamefont {W.~H.}\ \bibnamefont
  {Furry}},\ }\bibfield  {title} {\enquote {\bibinfo {title} {{On transition
  probabilities in double beta-disintegration}},}\ }\href {\doibase
  10.1103/PhysRev.56.1184} {\bibfield  {journal} {\bibinfo  {journal} {Phys.\
  Rev.}\ }\textbf {\bibinfo {volume} {56}},\ \bibinfo {pages} {1184--1193}
  (\bibinfo {year} {1939})}\BibitemShut {NoStop}%
%%CITATION = PHRVA,56,1184;%%
\bibitem [{\citenamefont {Goeppert-Mayer}(1935)}]{GoeppertMayer:1935qp}%
  \BibitemOpen
  \bibfield  {author} {\bibinfo {author} {\bibfnamefont {M.}~\bibnamefont
  {Goeppert-Mayer}},\ }\bibfield  {title} {\enquote {\bibinfo {title} {{Double
  beta-disintegration}},}\ }\href {\doibase 10.1103/PhysRev.48.512} {\bibfield
  {journal} {\bibinfo  {journal} {Phys.\ Rev.}\ }\textbf {\bibinfo {volume}
  {48}},\ \bibinfo {pages} {512--516} (\bibinfo {year} {1935})}\BibitemShut
  {NoStop}%
%%CITATION = PHRVA,48,512;%%
\bibitem [{\citenamefont {Feinberg}\ and\ \citenamefont
  {Goldhaber}(1959)}]{Feinberg&Goldhaber:1959}%
  \BibitemOpen
  \bibfield  {author} {\bibinfo {author} {\bibfnamefont {G.}~\bibnamefont
  {Feinberg}}\ and\ \bibinfo {author} {\bibfnamefont {M.}~\bibnamefont
  {Goldhaber}},\ }\bibfield  {title} {\enquote {\bibinfo {title} {{Microscopic
  tests of symmetry principles}},}\ }\href
  {http://www.pnas.org/content/45/8.toc} {\bibfield  {journal} {\bibinfo
  {journal} {Proc.\ Nat.\ Ac.\ Sci.}\ }\textbf {\bibinfo {volume} {45}},\
  \bibinfo {pages} {1301--1312} (\bibinfo {year} {1959})}\BibitemShut {NoStop}%
\bibitem [{\citenamefont {Pontecorvo}(1968)}]{Pontecorvo:1968wp}%
  \BibitemOpen
  \bibfield  {author} {\bibinfo {author} {\bibfnamefont {B.}~\bibnamefont
  {Pontecorvo}},\ }\bibfield  {title} {\enquote {\bibinfo {title} {{Superweak
  interactions and double beta decay}},}\ }\href {\doibase
  10.1016/0370-2693(68)90437-1} {\bibfield  {journal} {\bibinfo  {journal}
  {Phys.\ Lett.\ B}\ }\textbf {\bibinfo {volume} {26}},\ \bibinfo {pages}
  {630--632} (\bibinfo {year} {1968})}\BibitemShut {NoStop}%
%%CITATION = PHLTA,B26,630;%%
\bibitem [{\citenamefont {Glashow}(1961)}]{Glashow:1961tr}%
  \BibitemOpen
  \bibfield  {author} {\bibinfo {author} {\bibfnamefont {S.~L.}\ \bibnamefont
  {Glashow}},\ }\bibfield  {title} {\enquote {\bibinfo {title} {{Partial
  Symmetries of Weak Interactions}},}\ }\href {\doibase
  10.1016/0029-5582(61)90469-2} {\bibfield  {journal} {\bibinfo  {journal}
  {Nucl.\ Phys.}\ }\textbf {\bibinfo {volume} {22}},\ \bibinfo {pages}
  {579--588} (\bibinfo {year} {1961})}\BibitemShut {NoStop}%
%%CITATION = NUPHA,22,579;%%
\bibitem [{\citenamefont {Weinberg}(1967)}]{Weinberg:1967tq}%
  \BibitemOpen
  \bibfield  {author} {\bibinfo {author} {\bibfnamefont {S.}~\bibnamefont
  {Weinberg}},\ }\bibfield  {title} {\enquote {\bibinfo {title} {{A Model of
  Leptons}},}\ }\href {\doibase 10.1103/PhysRevLett.19.1264} {\bibfield
  {journal} {\bibinfo  {journal} {Phys.\ Rev.\ Lett.}\ }\textbf {\bibinfo
  {volume} {19}},\ \bibinfo {pages} {1264--1266} (\bibinfo {year}
  {1967})}\BibitemShut {NoStop}%
%%CITATION = PRLTA,19,1264;%%
\bibitem [{\citenamefont {Salam}(1968)}]{Salam:1968rm}%
  \BibitemOpen
  \bibfield  {author} {\bibinfo {author} {\bibfnamefont {A.}~\bibnamefont
  {Salam}},\ }\bibfield  {title} {\enquote {\bibinfo {title} {{Weak and
  Electromagnetic Interactions}},}\ }\href@noop {} {\bibfield  {journal}
  {\bibinfo  {journal} {8th Nobel Symposium, Lerum, Sweden, May 1968}\ }\textbf
  {\bibinfo {volume} {\!\!}},\ \bibinfo {pages} {367--377} (\bibinfo {year}
  {1968})}\BibitemShut {NoStop}%
%%CITATION = CONFP,C680519,367;%%
\bibitem [{\citenamefont {Weinberg}(1979)}]{Weinberg:1979sa}%
  \BibitemOpen
  \bibfield  {author} {\bibinfo {author} {\bibfnamefont {S.}~\bibnamefont
  {Weinberg}},\ }\bibfield  {title} {\enquote {\bibinfo {title} {{Baryon and
  Lepton Nonconserving Processes}},}\ }\href {\doibase
  10.1103/PhysRevLett.43.1566} {\bibfield  {journal} {\bibinfo  {journal}
  {Phys.\ Rev.\ Lett.}\ }\textbf {\bibinfo {volume} {43}},\ \bibinfo {pages}
  {1566--1570} (\bibinfo {year} {1979})}\BibitemShut {NoStop}%
%%CITATION = PRLTA,43,1566;%%
\bibitem [{\citenamefont {Minkowski}(1977)}]{Minkowski:1977sc}%
  \BibitemOpen
  \bibfield  {author} {\bibinfo {author} {\bibfnamefont {P.}~\bibnamefont
  {Minkowski}},\ }\bibfield  {title} {\enquote {\bibinfo {title} {{$\mu \to
  \text{e} \gamma$ at a Rate of One Out of $10^{9}$ Muon Decays?}}}\ }\href
  {\doibase 10.1016/0370-2693(77)90435-X} {\bibfield  {journal} {\bibinfo
  {journal} {Phys.\ Lett.\ B}\ }\textbf {\bibinfo {volume} {67}},\ \bibinfo
  {pages} {421--428} (\bibinfo {year} {1977})}\BibitemShut {NoStop}%
%%CITATION = PHLTA,B67,421;%%
\bibitem [{\citenamefont {Yanagida}(1979)}]{Yanagida:1979as}%
  \BibitemOpen
  \bibfield  {author} {\bibinfo {author} {\bibfnamefont {T.}~\bibnamefont
  {Yanagida}},\ }\bibfield  {title} {\enquote {\bibinfo {title} {{Horizontal
  symmetry and masses of neutrinos}},}\ }\href@noop {} {\bibfield  {journal}
  {\bibinfo  {journal} {Proceedings of the Workshop on the Baryon Number of the
  Universe and Unified Theories, Tsukuba, Japan, February 1979}\ }\textbf
  {\bibinfo {volume} {\!\!}},\ \bibinfo {pages} {95--99} (\bibinfo {year}
  {1979})}\BibitemShut {NoStop}%
%%CITATION = CONFP,C7902131,95;%%
\bibitem [{\citenamefont {Gell-Mann}\ \emph {et~al.}(1979)\citenamefont
  {Gell-Mann}, \citenamefont {Ramond},\ and\ \citenamefont
  {Slansky}}]{GellMann:1980vs}%
  \BibitemOpen
  \bibfield  {author} {\bibinfo {author} {\bibfnamefont {M.}~\bibnamefont
  {Gell-Mann}}, \bibinfo {author} {\bibfnamefont {P.}~\bibnamefont {Ramond}}, \
  and\ \bibinfo {author} {\bibfnamefont {R.}~\bibnamefont {Slansky}},\
  }\bibfield  {title} {\enquote {\bibinfo {title} {{Complex Spinors and Unified
  Theories}},}\ }\href@noop {} {\bibfield  {journal} {\bibinfo  {journal}
  {Proceedings of the Supergravity Workshop, Stony Brook, New York, USA,
  September 1979}\ }\textbf {\bibinfo {volume} {\!\!}},\ \bibinfo {pages}
  {315--321} (\bibinfo {year} {1979})}\BibitemShut {NoStop}%
%%CITATION = ARXIV:1306.4669;%%
\bibitem [{\citenamefont {Mohapatra}\ and\ \citenamefont
  {Senjanovic}(1980)}]{Mohapatra:1979ia}%
  \BibitemOpen
  \bibfield  {author} {\bibinfo {author} {\bibfnamefont {R.~N.}\ \bibnamefont
  {Mohapatra}}\ and\ \bibinfo {author} {\bibfnamefont {G.}~\bibnamefont
  {Senjanovic}},\ }\bibfield  {title} {\enquote {\bibinfo {title} {{Neutrino
  Mass and Spontaneous Parity Violation}},}\ }\href {\doibase
  10.1103/PhysRevLett.44.912} {\bibfield  {journal} {\bibinfo  {journal}
  {Phys.\ Rev.\ Lett.}\ }\textbf {\bibinfo {volume} {44}},\ \bibinfo {pages}
  {912} (\bibinfo {year} {1980})}\BibitemShut {NoStop}%
%%CITATION = PRLTA,44,912;%%
\bibitem [{\citenamefont {Pontecorvo}(1957)}]{Pontecorvo:1957cp}%
  \BibitemOpen
  \bibfield  {author} {\bibinfo {author} {\bibfnamefont {B.}~\bibnamefont
  {Pontecorvo}},\ }\bibfield  {title} {\enquote {\bibinfo {title} {{Mesonium
  and anti-mesonium}},}\ }\href@noop {} {\bibfield  {journal} {\bibinfo
  {journal} {Sov.\ Phys.\ JETP}\ }\textbf {\bibinfo {volume} {6}},\ \bibinfo
  {pages} {429} (\bibinfo {year} {1957})},\ \bibinfo {note} {[Zh.\ Eksp.\
  Teor.\ Fiz.\ {\bf 33}, 549 (1957)]}\BibitemShut {NoStop}%
%%CITATION = SPHJA,6,429;%%
\bibitem [{\citenamefont {Strumia}\ and\ \citenamefont
  {Vissani}(2006)}]{Strumia:2006db}%
  \BibitemOpen
  \bibfield  {author} {\bibinfo {author} {\bibfnamefont {A.}~\bibnamefont
  {Strumia}}\ and\ \bibinfo {author} {\bibfnamefont {F.}~\bibnamefont
  {Vissani}},\ }\bibfield  {title} {\enquote {\bibinfo {title} {{Neutrino
  masses and mixings and \dots}},}\ }\href@noop {} {\  (\bibinfo {year}
  {2006})},\ \Eprint {http://arxiv.org/abs/hep-ph/0606054}
  {arXiv:hep-ph/0606054 [hep-ph]} \BibitemShut {NoStop}%
%%CITATION = HEP-PH/0606054;%%
\bibitem [{\citenamefont {Gonzalez-Garcia}\ and\ \citenamefont
  {Nir}(2003)}]{GonzalezGarcia:2002dz}%
  \BibitemOpen
  \bibfield  {author} {\bibinfo {author} {\bibfnamefont {M.~C.}\ \bibnamefont
  {Gonzalez-Garcia}}\ and\ \bibinfo {author} {\bibfnamefont {Y.}~\bibnamefont
  {Nir}},\ }\bibfield  {title} {\enquote {\bibinfo {title} {{Neutrino masses
  and mixing: Evidence and implications}},}\ }\href {\doibase
  10.1103/RevModPhys.75.345} {\bibfield  {journal} {\bibinfo  {journal} {Rev.\
  Mod.\ Phys.}\ }\textbf {\bibinfo {volume} {75}},\ \bibinfo {pages} {345--402}
  (\bibinfo {year} {2003})}\BibitemShut {NoStop}%
%%CITATION = HEP-PH/0202058;%%
\bibitem [{\citenamefont {Bilenky}\ \emph {et~al.}(1980)\citenamefont
  {Bilenky}, \citenamefont {Hosek},\ and\ \citenamefont
  {Petcov}}]{Bilenky:1980cx}%
  \BibitemOpen
  \bibfield  {author} {\bibinfo {author} {\bibfnamefont {Samoil~M.}\
  \bibnamefont {Bilenky}}, \bibinfo {author} {\bibfnamefont {J.}~\bibnamefont
  {Hosek}}, \ and\ \bibinfo {author} {\bibfnamefont {S.T.}\ \bibnamefont
  {Petcov}},\ }\bibfield  {title} {\enquote {\bibinfo {title} {{On Oscillations
  of Neutrinos with Dirac and Majorana Masses}},}\ }\href {\doibase
  10.1016/0370-2693(80)90927-2} {\bibfield  {journal} {\bibinfo  {journal}
  {Phys.\ Lett.\ B}\ }\textbf {\bibinfo {volume} {94}},\ \bibinfo {pages} {495}
  (\bibinfo {year} {1980})}\BibitemShut {NoStop}%
%%CITATION = PHLTA,B94,495;%%
\bibitem [{\citenamefont {Elliott}(2012)}]{Elliott:2012sp}%
  \BibitemOpen
  \bibfield  {author} {\bibinfo {author} {\bibfnamefont {S.~R.}\ \bibnamefont
  {Elliott}},\ }\bibfield  {title} {\enquote {\bibinfo {title} {{Recent
  Progress in Double Beta Decay}},}\ }\href {\doibase
  10.1142/S0217732312300091} {\bibfield  {journal} {\bibinfo  {journal} {Mod.\
  Phys.\ Lett.\ A}\ }\textbf {\bibinfo {volume} {27}},\ \bibinfo {pages}
  {1230009} (\bibinfo {year} {2012})}\BibitemShut {NoStop}%
%%CITATION = ARXIV:1203.1070;%%
\bibitem [{\citenamefont {Giuliani}\ and\ \citenamefont
  {Poves}(2012)}]{Giuliani:2012zu}%
  \BibitemOpen
  \bibfield  {author} {\bibinfo {author} {\bibfnamefont {A.}~\bibnamefont
  {Giuliani}}\ and\ \bibinfo {author} {\bibfnamefont {A.}~\bibnamefont
  {Poves}},\ }\bibfield  {title} {\enquote {\bibinfo {title} {{Neutrinoless
  Double-Beta Decay}},}\ }\href {\doibase 10.1155/2012/857016} {\bibfield
  {journal} {\bibinfo  {journal} {Adv.\ High Energy Phys.}\ }\textbf {\bibinfo
  {volume} {2012}},\ \bibinfo {pages} {857016} (\bibinfo {year}
  {2012})}\BibitemShut {NoStop}%
%%CITATION = 00642,2012,857016;%%
\bibitem [{\citenamefont {G\'{o}mez-Cadenas}\ \emph {et~al.}(2012)\citenamefont
  {G\'{o}mez-Cadenas}, \citenamefont {Mart\`in-Albo}, \citenamefont {Mezzetto},
  \citenamefont {Monrabal},\ and\ \citenamefont {Sorel}}]{GomezCadenas:2011it}%
  \BibitemOpen
  \bibfield  {author} {\bibinfo {author} {\bibfnamefont {J.~J.}\ \bibnamefont
  {G\'{o}mez-Cadenas}}, \bibinfo {author} {\bibfnamefont {J.}~\bibnamefont
  {Mart\`in-Albo}}, \bibinfo {author} {\bibfnamefont {M.}~\bibnamefont
  {Mezzetto}}, \bibinfo {author} {\bibfnamefont {F.}~\bibnamefont {Monrabal}},
  \ and\ \bibinfo {author} {\bibfnamefont {M.}~\bibnamefont {Sorel}},\
  }\bibfield  {title} {\enquote {\bibinfo {title} {{The Search for neutrinoless
  double beta decay}},}\ }\href {\doibase 10.1393/ncr/i2012-10074-9} {\bibfield
   {journal} {\bibinfo  {journal} {Riv.\ Nuovo Cim.}\ }\textbf {\bibinfo
  {volume} {35}},\ \bibinfo {pages} {29--98} (\bibinfo {year}
  {2012})}\BibitemShut {NoStop}%
%%CITATION = ARXIV:1109.5515;%%
\bibitem [{\citenamefont {Schwingenheuer}(2013)}]{Schwingenheuer:2012zs}%
  \BibitemOpen
  \bibfield  {author} {\bibinfo {author} {\bibfnamefont {B.}~\bibnamefont
  {Schwingenheuer}},\ }\bibfield  {title} {\enquote {\bibinfo {title} {{Status
  and prospects of searches for neutrinoless double beta decay}},}\ }\href
  {\doibase 10.1002/andp.201200222} {\bibfield  {journal} {\bibinfo  {journal}
  {Ann.\ Phys.}\ }\textbf {\bibinfo {volume} {525}},\ \bibinfo {pages}
  {269--280} (\bibinfo {year} {2013})}\BibitemShut {NoStop}%
%%CITATION = ARXIV:1210.7432;%%
\bibitem [{\citenamefont {Cremonesi}\ and\ \citenamefont
  {Pavan}(2013)}]{Cremonesi:2013vla}%
  \BibitemOpen
  \bibfield  {author} {\bibinfo {author} {\bibfnamefont {O.}~\bibnamefont
  {Cremonesi}}\ and\ \bibinfo {author} {\bibfnamefont {M.}~\bibnamefont
  {Pavan}},\ }\bibfield  {title} {\enquote {\bibinfo {title} {{Challenges in
  Double Beta Decay}},}\ }\href {\doibase 10.1155/2014/951432} {\bibfield
  {journal} {\bibinfo  {journal} {Adv.\ High Energy Phys.}\ }\textbf {\bibinfo
  {volume} {2014}},\ \bibinfo {pages} {951432} (\bibinfo {year}
  {2013})}\BibitemShut {NoStop}%
%%CITATION = ARXIV:1310.4692;%%
\bibitem [{\citenamefont {Vergados}\ \emph {et~al.}(2012)\citenamefont
  {Vergados}, \citenamefont {Ejiri},\ and\ \citenamefont
  {\u{S}imkovic}}]{Vergados:2012xy}%
  \BibitemOpen
  \bibfield  {author} {\bibinfo {author} {\bibfnamefont {J.~D.}\ \bibnamefont
  {Vergados}}, \bibinfo {author} {\bibfnamefont {H.}~\bibnamefont {Ejiri}}, \
  and\ \bibinfo {author} {\bibfnamefont {F.}~\bibnamefont {\u{S}imkovic}},\
  }\bibfield  {title} {\enquote {\bibinfo {title} {{Theory of Neutrinoless
  Double Beta Decay}},}\ }\href {\doibase 10.1088/0034-4885/75/10/106301}
  {\bibfield  {journal} {\bibinfo  {journal} {Rep.\ Prog.\ Phys.}\ }\textbf
  {\bibinfo {volume} {75}},\ \bibinfo {pages} {106301} (\bibinfo {year}
  {2012})}\BibitemShut {NoStop}%
%%CITATION = ARXIV:1205.0649;%%
\bibitem [{\citenamefont {Vogel}(2012)}]{Vogel:2012ja}%
  \BibitemOpen
  \bibfield  {author} {\bibinfo {author} {\bibfnamefont {P.}~\bibnamefont
  {Vogel}},\ }\bibfield  {title} {\enquote {\bibinfo {title} {{Nuclear
  structure and double beta decay}},}\ }\href {\doibase
  10.1088/0954-3899/39/12/124002} {\bibfield  {journal} {\bibinfo  {journal}
  {J.\ Phys.\ G}\ }\textbf {\bibinfo {volume} {39}},\ \bibinfo {pages} {124002}
  (\bibinfo {year} {2012})}\BibitemShut {NoStop}%
%%CITATION = ARXIV:1208.1992;%%
\bibitem [{\citenamefont {Petcov}(2013)}]{Petcov:2013poa}%
  \BibitemOpen
  \bibfield  {author} {\bibinfo {author} {\bibfnamefont {S.~T.}\ \bibnamefont
  {Petcov}},\ }\bibfield  {title} {\enquote {\bibinfo {title} {{The Nature of
  Massive Neutrinos}},}\ }\href {\doibase 10.1155/2013/852987} {\bibfield
  {journal} {\bibinfo  {journal} {Adv.\ High Energy Phys.}\ }\textbf {\bibinfo
  {volume} {2013}},\ \bibinfo {pages} {852987} (\bibinfo {year}
  {2013})}\BibitemShut {NoStop}%
%%CITATION = ARXIV:1303.5819;%%
\bibitem [{\citenamefont {Bilenky}\ and\ \citenamefont
  {Giunti}(2015)}]{Bilenky:2014uka}%
  \BibitemOpen
  \bibfield  {author} {\bibinfo {author} {\bibfnamefont {S.~M.}\ \bibnamefont
  {Bilenky}}\ and\ \bibinfo {author} {\bibfnamefont {C.}~\bibnamefont
  {Giunti}},\ }\bibfield  {title} {\enquote {\bibinfo {title} {{Neutrinoless
  Double-Beta Decay: a Probe of Physics Beyond the Standard Model}},}\ }\href
  {\doibase 10.1142/S0217751X1530001X} {\bibfield  {journal} {\bibinfo
  {journal} {Int.\ J.\ Mod.\ Phys.\ A}\ }\textbf {\bibinfo {volume} {30}},\
  \bibinfo {pages} {1530001} (\bibinfo {year} {2015})}\BibitemShut {NoStop}%
%%CITATION = ARXIV:1411.4791;%%
\bibitem [{\citenamefont {Rodejohann}(2011)}]{Rodejohann:2011a}%
  \BibitemOpen
  \bibfield  {author} {\bibinfo {author} {\bibfnamefont {W.}~\bibnamefont
  {Rodejohann}},\ }\bibfield  {title} {\enquote {\bibinfo {title}
  {{Neutrinoless-double beta decay and particle physics}},}\ }\href {\doibase
  10.1142/S0218301311020186} {\bibfield  {journal} {\bibinfo  {journal} {Int.\
  J.\ Mod.\ Phys.\ E}\ }\textbf {\bibinfo {volume} {20}},\ \bibinfo {pages}
  {1833--1930} (\bibinfo {year} {2011})}\BibitemShut {NoStop}%
\bibitem [{\citenamefont {Rodejohann}(2012)}]{Rodejohann:2012xd}%
  \BibitemOpen
  \bibfield  {author} {\bibinfo {author} {\bibfnamefont {W.}~\bibnamefont
  {Rodejohann}},\ }\bibfield  {title} {\enquote {\bibinfo {title}
  {{Neutrinoless double beta decay and neutrino physics}},}\ }\href {\doibase
  10.1088/0954-3899/39/12/124008} {\bibfield  {journal} {\bibinfo  {journal}
  {J.\ Phys.\ G}\ }\textbf {\bibinfo {volume} {39}},\ \bibinfo {pages} {124008}
  (\bibinfo {year} {2012})}\BibitemShut {NoStop}%
%%CITATION = ARXIV:1206.2560;%%
\bibitem [{\citenamefont {Deppisch}\ \emph {et~al.}(2012)\citenamefont
  {Deppisch}, \citenamefont {Hirsch},\ and\ \citenamefont
  {P$\ddot{\text{a}}$s}}]{Deppisch:2012nb}%
  \BibitemOpen
  \bibfield  {author} {\bibinfo {author} {\bibfnamefont {F.~F.}\ \bibnamefont
  {Deppisch}}, \bibinfo {author} {\bibfnamefont {M.}~\bibnamefont {Hirsch}}, \
  and\ \bibinfo {author} {\bibfnamefont {H.}~\bibnamefont
  {P$\ddot{\text{a}}$s}},\ }\bibfield  {title} {\enquote {\bibinfo {title}
  {{Neutrinoless Double Beta Decay and Physics Beyond the Standard Model}},}\
  }\href {\doibase 10.1088/0954-3899/39/12/124007} {\bibfield  {journal}
  {\bibinfo  {journal} {J.\ Phys.\ G}\ }\textbf {\bibinfo {volume} {39}},\
  \bibinfo {pages} {124007} (\bibinfo {year} {2012})}\BibitemShut {NoStop}%
%%CITATION = ARXIV:1208.0727;%%
\bibitem [{\citenamefont {P$\ddot{\text{a}}$s}\ and\ \citenamefont
  {Rodejohann}(2015)}]{Pas:2015eia}%
  \BibitemOpen
  \bibfield  {author} {\bibinfo {author} {\bibfnamefont {H.}~\bibnamefont
  {P$\ddot{\text{a}}$s}}\ and\ \bibinfo {author} {\bibfnamefont
  {W.}~\bibnamefont {Rodejohann}},\ }\bibfield  {title} {\enquote {\bibinfo
  {title} {{Neutrinoless Double Beta Decay}},}\ }\href {\doibase
  10.1088/1367-2630/17/11/115010} {\bibfield  {journal} {\bibinfo  {journal}
  {New J.\ Phys.}\ }\textbf {\bibinfo {volume} {17}},\ \bibinfo {pages}
  {115010} (\bibinfo {year} {2015})}\BibitemShut {NoStop}%
%%CITATION = ARXIV:1507.00170;%%
\bibitem [{\citenamefont {Ade}\ \emph {et~al.}(2015)\citenamefont {Ade} \emph
  {et~al.}}]{Planck:2015xua}%
  \BibitemOpen
  \bibfield  {author} {\bibinfo {author} {\bibfnamefont {P.~A.~R.}\
  \bibnamefont {Ade}} \emph {et~al.} (\bibinfo {collaboration} {Planck
  Collaboration}),\ }\bibfield  {title} {\enquote {\bibinfo {title} {{Planck
  2015 results. XIII. Cosmological parameters}},}\ }\href@noop {} {\  (\bibinfo
  {year} {2015})},\ \Eprint {http://arxiv.org/abs/1502.01589} {arXiv:1502.01589
  [astro-ph.CO]} \BibitemShut {NoStop}%
%%CITATION = ARXIV:1502.01589;%%
\bibitem [{\citenamefont {Fermi}(1933)}]{Fermi:1933jpa}%
  \BibitemOpen
  \bibfield  {author} {\bibinfo {author} {\bibfnamefont {E.}~\bibnamefont
  {Fermi}},\ }\bibfield  {title} {\enquote {\bibinfo {title} {{Trends to a
  theory of $\beta$ ray emission (in Italian)}},}\ }\href@noop {} {\bibfield
  {journal} {\bibinfo  {journal} {Ric.\ Sci.}\ }\textbf {\bibinfo {volume}
  {4}},\ \bibinfo {pages} {491--495} (\bibinfo {year} {1933})}\BibitemShut
  {NoStop}%
%%CITATION = RISCA,4,491;%%
\bibitem [{\citenamefont {Lee}\ and\ \citenamefont {Yang}(1956)}]{Lee:1956qn}%
  \BibitemOpen
  \bibfield  {author} {\bibinfo {author} {\bibfnamefont {T.~D.}\ \bibnamefont
  {Lee}}\ and\ \bibinfo {author} {\bibfnamefont {C.~N.}\ \bibnamefont {Yang}},\
  }\bibfield  {title} {\enquote {\bibinfo {title} {{Question of Parity
  Conservation in Weak Interactions}},}\ }\href {\doibase
  10.1103/PhysRev.104.254} {\bibfield  {journal} {\bibinfo  {journal} {Phys.\
  Rev.}\ }\textbf {\bibinfo {volume} {104}},\ \bibinfo {pages} {254--258}
  (\bibinfo {year} {1956})}\BibitemShut {NoStop}%
%%CITATION = PHRVA,104,254;%%
\bibitem [{\citenamefont {Wu}\ \emph {et~al.}(1957)\citenamefont {Wu},
  \citenamefont {Ambler}, \citenamefont {Hayward}, \citenamefont {Hoppes},\
  and\ \citenamefont {Hudson}}]{Wu:1957my}%
  \BibitemOpen
  \bibfield  {author} {\bibinfo {author} {\bibfnamefont {C.~S.}\ \bibnamefont
  {Wu}}, \bibinfo {author} {\bibfnamefont {E.}~\bibnamefont {Ambler}}, \bibinfo
  {author} {\bibfnamefont {R.~W.}\ \bibnamefont {Hayward}}, \bibinfo {author}
  {\bibfnamefont {D.~D.}\ \bibnamefont {Hoppes}}, \ and\ \bibinfo {author}
  {\bibfnamefont {R.~P.}\ \bibnamefont {Hudson}},\ }\bibfield  {title}
  {\enquote {\bibinfo {title} {{Experimental Test of Parity Conservation in
  Beta Decay}},}\ }\href {\doibase 10.1103/PhysRev.105.1413} {\bibfield
  {journal} {\bibinfo  {journal} {Phys.\ Rev.}\ }\textbf {\bibinfo {volume}
  {105}},\ \bibinfo {pages} {1413--1414} (\bibinfo {year} {1957})}\BibitemShut
  {NoStop}%
%%CITATION = PHRVA,105,1413;%%
\bibitem [{\citenamefont {Landau}(1957)}]{Landau:1957tp}%
  \BibitemOpen
  \bibfield  {author} {\bibinfo {author} {\bibfnamefont {L.~D.}\ \bibnamefont
  {Landau}},\ }\bibfield  {title} {\enquote {\bibinfo {title} {{On the
  conservation laws for weak interactions}},}\ }\href {\doibase
  10.1016/0029-5582(57)90061-5} {\bibfield  {journal} {\bibinfo  {journal}
  {Nucl.\ Phys.}\ }\textbf {\bibinfo {volume} {3}},\ \bibinfo {pages}
  {127--131} (\bibinfo {year} {1957})}\BibitemShut {NoStop}%
%%CITATION = NUPHA,3,127;%%
\bibitem [{\citenamefont {Lee}\ and\ \citenamefont {Yang}(1957)}]{Lee:1957qr}%
  \BibitemOpen
  \bibfield  {author} {\bibinfo {author} {\bibfnamefont {T.~D.}\ \bibnamefont
  {Lee}}\ and\ \bibinfo {author} {\bibfnamefont {C.~N.}\ \bibnamefont {Yang}},\
  }\bibfield  {title} {\enquote {\bibinfo {title} {{Parity Nonconservation and
  a Two Component Theory of the Neutrino}},}\ }\href {\doibase
  10.1103/PhysRev.105.1671} {\bibfield  {journal} {\bibinfo  {journal} {Phys.\
  Rev.}\ }\textbf {\bibinfo {volume} {105}},\ \bibinfo {pages} {1671--1675}
  (\bibinfo {year} {1957})}\BibitemShut {NoStop}%
%%CITATION = PHRVA,105,1671;%%
\bibitem [{\citenamefont {Salam}(1957)}]{Salam:1957st}%
  \BibitemOpen
  \bibfield  {author} {\bibinfo {author} {\bibfnamefont {A.}~\bibnamefont
  {Salam}},\ }\bibfield  {title} {\enquote {\bibinfo {title} {{On parity
  conservation and neutrino mass}},}\ }\href {\doibase 10.1007/BF02812841}
  {\bibfield  {journal} {\bibinfo  {journal} {Nuovo Cim.}\ }\textbf {\bibinfo
  {volume} {5}},\ \bibinfo {pages} {299--301} (\bibinfo {year}
  {1957})}\BibitemShut {NoStop}%
%%CITATION = NUCIA,5,299;%%
\bibitem [{\citenamefont {Weyl}(1929)}]{Weyl:1929fm}%
  \BibitemOpen
  \bibfield  {author} {\bibinfo {author} {\bibfnamefont {H.}~\bibnamefont
  {Weyl}},\ }\bibfield  {title} {\enquote {\bibinfo {title} {{Electron and
  Gravitation. 1. (In German)}},}\ }\href {\doibase 10.1007/BF01339504}
  {\bibfield  {journal} {\bibinfo  {journal} {Z.\ Phys.}\ }\textbf {\bibinfo
  {volume} {56}},\ \bibinfo {pages} {330--352} (\bibinfo {year}
  {1929})}\BibitemShut {NoStop}%
%%CITATION = ZEPYA,56,330;%%
\bibitem [{\citenamefont {Georgi}\ and\ \citenamefont
  {Glashow}(1974)}]{Georgi:1974sy}%
  \BibitemOpen
  \bibfield  {author} {\bibinfo {author} {\bibfnamefont {H.}~\bibnamefont
  {Georgi}}\ and\ \bibinfo {author} {\bibfnamefont {S.~L.}\ \bibnamefont
  {Glashow}},\ }\bibfield  {title} {\enquote {\bibinfo {title} {{Unity of All
  Elementary Particle Forces}},}\ }\href {\doibase 10.1103/PhysRevLett.32.438}
  {\bibfield  {journal} {\bibinfo  {journal} {Phys.\ Rev.\ Lett.}\ }\textbf
  {\bibinfo {volume} {32}},\ \bibinfo {pages} {438--441} (\bibinfo {year}
  {1974})}\BibitemShut {NoStop}%
%%CITATION = PRLTA,32,438;%%
\bibitem [{\citenamefont {Georgi}(1975)}]{Georgi:1975qb}%
  \BibitemOpen
  \bibfield  {author} {\bibinfo {author} {\bibfnamefont {H.}~\bibnamefont
  {Georgi}},\ }\bibfield  {title} {\enquote {\bibinfo {title} {{Unified Gauge
  Theories}},}\ }\href@noop {} {\bibfield  {journal} {\bibinfo  {journal}
  {Conf.\ Proc.\ C75-01-20}\ }\textbf {\bibinfo {volume} {\!\!}} (\bibinfo
  {year} {1975})}\BibitemShut {NoStop}%
%%CITATION = INSPIRE-106900;%%
\bibitem [{\citenamefont {Fritzsch}\ and\ \citenamefont
  {Minkowski}(1975)}]{Fritzsch:1974nn}%
  \BibitemOpen
  \bibfield  {author} {\bibinfo {author} {\bibfnamefont {H.}~\bibnamefont
  {Fritzsch}}\ and\ \bibinfo {author} {\bibfnamefont {P.}~\bibnamefont
  {Minkowski}},\ }\bibfield  {title} {\enquote {\bibinfo {title} {{Unified
  Interactions of Leptons and Hadrons}},}\ }\href {\doibase
  10.1016/0003-4916(75)90211-0} {\bibfield  {journal} {\bibinfo  {journal}
  {Annals Phys.}\ }\textbf {\bibinfo {volume} {93}},\ \bibinfo {pages}
  {193--266} (\bibinfo {year} {1975})}\BibitemShut {NoStop}%
%%CITATION = APNYA,93,193;%%
\bibitem [{\citenamefont {Bellini}\ \emph {et~al.}(2013)\citenamefont {Bellini}
  \emph {et~al.}}]{Borexino:2013xxa}%
  \BibitemOpen
  \bibfield  {author} {\bibinfo {author} {\bibfnamefont {G.}~\bibnamefont
  {Bellini}} \emph {et~al.} (\bibinfo {collaboration} {Borexino
  Collaboration}),\ }\bibfield  {title} {\enquote {\bibinfo {title} {{SOX:
  Short distance neutrino Oscillations with BoreXino}},}\ }\href {\doibase
  10.1007/JHEP08(2013)038} {\bibfield  {journal} {\bibinfo  {journal} {J.\ High
  Energy Phys.}\ }\textbf {\bibinfo {volume} {1308}},\ \bibinfo {pages} {038}
  (\bibinfo {year} {2013})}\BibitemShut {NoStop}%
%%CITATION = ARXIV:1304.7721;%%
\bibitem [{\citenamefont {An}\ \emph {et~al.}(2014)\citenamefont {An} \emph
  {et~al.}}]{An:2014bik}%
  \BibitemOpen
  \bibfield  {author} {\bibinfo {author} {\bibfnamefont {F.~P.}\ \bibnamefont
  {An}} \emph {et~al.} (\bibinfo {collaboration} {Daya Bay Collaboration}),\
  }\bibfield  {title} {\enquote {\bibinfo {title} {{Search for a Light Sterile
  Neutrino at Daya Bay}},}\ }\href {\doibase 10.1103/PhysRevLett.113.141802}
  {\bibfield  {journal} {\bibinfo  {journal} {Phys.\ Rev.\ Lett.}\ }\textbf
  {\bibinfo {volume} {113}},\ \bibinfo {pages} {141802} (\bibinfo {year}
  {2014})}\BibitemShut {NoStop}%
%%CITATION = ARXIV:1407.7259;%%
\bibitem [{\citenamefont {Ashenfelter}\ \emph {et~al.}(2015)\citenamefont
  {Ashenfelter} \emph {et~al.}}]{Ashenfelter:2015uxt}%
  \BibitemOpen
  \bibfield  {author} {\bibinfo {author} {\bibfnamefont {J.}~\bibnamefont
  {Ashenfelter}} \emph {et~al.} (\bibinfo {collaboration} {PROSPECT
  Collaboration}),\ }\bibfield  {title} {\enquote {\bibinfo {title} {{The
  PROSPECT Physics Program}},}\ }\href@noop {} {\  (\bibinfo {year} {2015})},\
  \Eprint {http://arxiv.org/abs/1512.02202} {arXiv:1512.02202
  [physics.ins-det]} \BibitemShut {NoStop}%
%%CITATION = ARXIV:1512.02202;%%
\bibitem [{\citenamefont {Abazajian}\ \emph {et~al.}(2012)\citenamefont
  {Abazajian} \emph {et~al.}}]{Abazajian:2012ys}%
  \BibitemOpen
  \bibfield  {author} {\bibinfo {author} {\bibfnamefont {K.~N.}\ \bibnamefont
  {Abazajian}} \emph {et~al.},\ }\bibfield  {title} {\enquote {\bibinfo {title}
  {{Light Sterile Neutrinos: A White Paper}},}\ }\href@noop {} {\  (\bibinfo
  {year} {2012})},\ \Eprint {http://arxiv.org/abs/1204.5379} {arXiv:1204.5379
  [hep-ph]} \BibitemShut {NoStop}%
%%CITATION = ARXIV:1204.5379;%%
\bibitem [{\citenamefont {Gariazzo}\ \emph {et~al.}(2013)\citenamefont
  {Gariazzo}, \citenamefont {Giunti},\ and\ \citenamefont
  {Laveder}}]{Gariazzo:2013gua}%
  \BibitemOpen
  \bibfield  {author} {\bibinfo {author} {\bibfnamefont {S.}~\bibnamefont
  {Gariazzo}}, \bibinfo {author} {\bibfnamefont {C.}~\bibnamefont {Giunti}}, \
  and\ \bibinfo {author} {\bibfnamefont {M.}~\bibnamefont {Laveder}},\
  }\bibfield  {title} {\enquote {\bibinfo {title} {{Light Sterile Neutrinos in
  Cosmology and Short-Baseline Oscillation Experiments}},}\ }\href {\doibase
  10.1007/JHEP11(2013)211} {\bibfield  {journal} {\bibinfo  {journal} {J.\ High
  Energy Phys.}\ }\textbf {\bibinfo {volume} {1311}},\ \bibinfo {pages} {211}
  (\bibinfo {year} {2013})}\BibitemShut {NoStop}%
%%CITATION = ARXIV:1309.3192;%%
\bibitem [{\citenamefont {Dolgov}\ and\ \citenamefont
  {Villante}(2004)}]{Dolgov:2003sg}%
  \BibitemOpen
  \bibfield  {author} {\bibinfo {author} {\bibfnamefont {A.~D.}\ \bibnamefont
  {Dolgov}}\ and\ \bibinfo {author} {\bibfnamefont {F.~L.}\ \bibnamefont
  {Villante}},\ }\bibfield  {title} {\enquote {\bibinfo {title} {{BBN bounds on
  active sterile neutrino mixing}},}\ }\href {\doibase
  10.1016/j.nuclphysb.2003.11.031} {\bibfield  {journal} {\bibinfo  {journal}
  {Nucl.\ Phys.\ B}\ }\textbf {\bibinfo {volume} {679}},\ \bibinfo {pages}
  {261--298} (\bibinfo {year} {2004})}\BibitemShut {NoStop}%
%%CITATION = HEP-PH/0308083;%%
\bibitem [{\citenamefont {Cirelli}\ \emph {et~al.}(2005)\citenamefont
  {Cirelli}, \citenamefont {Marandella}, \citenamefont {Strumia},\ and\
  \citenamefont {Vissani}}]{Cirelli:2004cz}%
  \BibitemOpen
  \bibfield  {author} {\bibinfo {author} {\bibfnamefont {M.}~\bibnamefont
  {Cirelli}}, \bibinfo {author} {\bibfnamefont {G.}~\bibnamefont {Marandella}},
  \bibinfo {author} {\bibfnamefont {A.}~\bibnamefont {Strumia}}, \ and\
  \bibinfo {author} {\bibfnamefont {F.}~\bibnamefont {Vissani}},\ }\bibfield
  {title} {\enquote {\bibinfo {title} {{Probing oscillations into sterile
  neutrinos with cosmology, astrophysics and experiments}},}\ }\href {\doibase
  10.1016/j.nuclphysb.2004.11.056} {\bibfield  {journal} {\bibinfo  {journal}
  {Nucl.\ Phys.\ B}\ }\textbf {\bibinfo {volume} {708}},\ \bibinfo {pages}
  {215--267} (\bibinfo {year} {2005})}\BibitemShut {NoStop}%
%%CITATION = HEP-PH/0403158;%%
\bibitem [{\citenamefont {Barry}\ \emph {et~al.}(2011)\citenamefont {Barry},
  \citenamefont {Rodejohann},\ and\ \citenamefont {Zhang}}]{Barry:2011wb}%
  \BibitemOpen
  \bibfield  {author} {\bibinfo {author} {\bibfnamefont {J.}~\bibnamefont
  {Barry}}, \bibinfo {author} {\bibfnamefont {W.}~\bibnamefont {Rodejohann}}, \
  and\ \bibinfo {author} {\bibfnamefont {H.}~\bibnamefont {Zhang}},\ }\bibfield
   {title} {\enquote {\bibinfo {title} {{Light Sterile Neutrinos: Models and
  Phenomenology}},}\ }\href {\doibase 10.1007/JHEP07(2011)091} {\bibfield
  {journal} {\bibinfo  {journal} {J.\ High Energy Phys.}\ }\textbf {\bibinfo
  {volume} {1107}},\ \bibinfo {pages} {091} (\bibinfo {year}
  {2011})}\BibitemShut {NoStop}%
%%CITATION = ARXIV:1105.3911;%%
\bibitem [{\citenamefont {Girardi}\ \emph {et~al.}(2013)\citenamefont
  {Girardi}, \citenamefont {Meroni},\ and\ \citenamefont
  {Petcov}}]{Girardi:2013zra}%
  \BibitemOpen
  \bibfield  {author} {\bibinfo {author} {\bibfnamefont {I.}~\bibnamefont
  {Girardi}}, \bibinfo {author} {\bibfnamefont {A.}~\bibnamefont {Meroni}}, \
  and\ \bibinfo {author} {\bibfnamefont {S.~T.}\ \bibnamefont {Petcov}},\
  }\bibfield  {title} {\enquote {\bibinfo {title} {{Neutrinoless Double Beta
  Decay in the Presence of Light Sterile Neutrinos}},}\ }\href {\doibase
  10.1007/JHEP11(2013)146} {\bibfield  {journal} {\bibinfo  {journal} {J.\ High
  Energy Phys.}\ }\textbf {\bibinfo {volume} {1311}},\ \bibinfo {pages} {146}
  (\bibinfo {year} {2013})}\BibitemShut {NoStop}%
%%CITATION = ARXIV:1308.5802;%%
\bibitem [{\citenamefont {Gariazzo}\ \emph {et~al.}(2016)\citenamefont
  {Gariazzo}, \citenamefont {Giunti}, \citenamefont {Laveder}, \citenamefont
  {Li},\ and\ \citenamefont {Zavanin}}]{Gariazzo:2015rra}%
  \BibitemOpen
  \bibfield  {author} {\bibinfo {author} {\bibfnamefont {S.}~\bibnamefont
  {Gariazzo}}, \bibinfo {author} {\bibfnamefont {C.}~\bibnamefont {Giunti}},
  \bibinfo {author} {\bibfnamefont {M.}~\bibnamefont {Laveder}}, \bibinfo
  {author} {\bibfnamefont {Y.~F.}\ \bibnamefont {Li}}, \ and\ \bibinfo {author}
  {\bibfnamefont {E.~M.}\ \bibnamefont {Zavanin}},\ }\bibfield  {title}
  {\enquote {\bibinfo {title} {{Light sterile neutrinos}},}\ }\href {\doibase
  10.1088/0954-3899/43/3/033001} {\bibfield  {journal} {\bibinfo  {journal}
  {J.\ Phys.\ G}\ }\textbf {\bibinfo {volume} {43}},\ \bibinfo {pages} {033001}
  (\bibinfo {year} {2016})}\BibitemShut {NoStop}%
%%CITATION = ARXIV:1507.08204;%%
\bibitem [{\citenamefont {Sakharov}(1967)}]{Sakharov:1967dj}%
  \BibitemOpen
  \bibfield  {author} {\bibinfo {author} {\bibfnamefont {A.~D.}\ \bibnamefont
  {Sakharov}},\ }\bibfield  {title} {\enquote {\bibinfo {title} {{Violation of
  CP Invariance, c Asymmetry, and Baryon Asymmetry of the Universe}},}\ }\href
  {http://iopscience.iop.org/article/10.1070/PU1991v034n05ABEH002497/meta;
  jsessionid=66AF62A9E1943AE2989F9A561A276B1D.c4.iopscience.cld.iop.org}
  {\bibfield  {journal} {\bibinfo  {journal} {Pisma Zh.\ Eksp.\ Teor.\ Fiz.}\
  }\textbf {\bibinfo {volume} {5}},\ \bibinfo {pages} {32--35} (\bibinfo {year}
  {1967})}\BibitemShut {NoStop}%
%%CITATION = ZFPRA,5,32;%%
\bibitem [{\citenamefont {'t~Hooft}(1976)}]{'tHooft:1976up}%
  \BibitemOpen
  \bibfield  {author} {\bibinfo {author} {\bibfnamefont {G.}~\bibnamefont
  {'t~Hooft}},\ }\bibfield  {title} {\enquote {\bibinfo {title} {{Symmetry
  Breaking Through Bell-Jackiw Anomalies}},}\ }\href {\doibase
  10.1103/PhysRevLett.37.8} {\bibfield  {journal} {\bibinfo  {journal} {Phys.\
  Rev.\ Lett.}\ }\textbf {\bibinfo {volume} {37}},\ \bibinfo {pages} {8--11}
  (\bibinfo {year} {1976})}\BibitemShut {NoStop}%
%%CITATION = PRLTA,37,8;%%
\bibitem [{\citenamefont {Kuzmin}\ \emph {et~al.}(1985)\citenamefont {Kuzmin},
  \citenamefont {Rubakov},\ and\ \citenamefont {Shaposhnikov}}]{Kuzmin:1985mm}%
  \BibitemOpen
  \bibfield  {author} {\bibinfo {author} {\bibfnamefont {V.~A.}\ \bibnamefont
  {Kuzmin}}, \bibinfo {author} {\bibfnamefont {V.~A.}\ \bibnamefont {Rubakov}},
  \ and\ \bibinfo {author} {\bibfnamefont {M.~E.}\ \bibnamefont
  {Shaposhnikov}},\ }\bibfield  {title} {\enquote {\bibinfo {title} {{On the
  Anomalous Electroweak Baryon Number Nonconservation in the Early
  Universe}},}\ }\href {\doibase 10.1016/0370-2693(85)91028-7} {\bibfield
  {journal} {\bibinfo  {journal} {Phys.\ Lett.\ B}\ }\textbf {\bibinfo {volume}
  {155}},\ \bibinfo {pages} {36} (\bibinfo {year} {1985})}\BibitemShut
  {NoStop}%
%%CITATION = PHLTA,B155,36;%%
\bibitem [{\citenamefont {Harvey}\ and\ \citenamefont
  {Turner}(1990)}]{Harvey:1990qw}%
  \BibitemOpen
  \bibfield  {author} {\bibinfo {author} {\bibfnamefont {J.A.}\ \bibnamefont
  {Harvey}}\ and\ \bibinfo {author} {\bibfnamefont {M.~S.}\ \bibnamefont
  {Turner}},\ }\bibfield  {title} {\enquote {\bibinfo {title} {{Cosmological
  baryon and lepton number in the presence of electroweak fermion number
  violation}},}\ }\href {\doibase 10.1103/PhysRevD.42.3344} {\bibfield
  {journal} {\bibinfo  {journal} {Phys.\ Rev.\ D}\ }\textbf {\bibinfo {volume}
  {42}},\ \bibinfo {pages} {3344--3349} (\bibinfo {year} {1990})}\BibitemShut
  {NoStop}%
%%CITATION = PHRVA,D42,3344;%%
\bibitem [{\citenamefont {Bochkarev}\ and\ \citenamefont
  {Shaposhnikov}(1987)}]{Bochkarev:1987wf}%
  \BibitemOpen
  \bibfield  {author} {\bibinfo {author} {\bibfnamefont {A.I.}\ \bibnamefont
  {Bochkarev}}\ and\ \bibinfo {author} {\bibfnamefont {M.E.}\ \bibnamefont
  {Shaposhnikov}},\ }\bibfield  {title} {\enquote {\bibinfo {title}
  {{Electroweak Production of Baryon Asymmetry and Upper Bounds on the Higgs
  and Top Masses}},}\ }\href {\doibase 10.1142/S0217732387000537} {\bibfield
  {journal} {\bibinfo  {journal} {Mod.\ Phys.\ Lett.\ A}\ }\textbf {\bibinfo
  {volume} {2}},\ \bibinfo {pages} {417} (\bibinfo {year} {1987})}\BibitemShut
  {NoStop}%
%%CITATION = MPLAE,A2,417;%%
\bibitem [{\citenamefont {Kajantie}\ \emph {et~al.}(1996)\citenamefont
  {Kajantie}, \citenamefont {Laine}, \citenamefont {Rummukainen},\ and\
  \citenamefont {Shaposhnikov}}]{Kajantie:1995kf}%
  \BibitemOpen
  \bibfield  {author} {\bibinfo {author} {\bibfnamefont {K.}~\bibnamefont
  {Kajantie}}, \bibinfo {author} {\bibfnamefont {M.}~\bibnamefont {Laine}},
  \bibinfo {author} {\bibfnamefont {K.}~\bibnamefont {Rummukainen}}, \ and\
  \bibinfo {author} {\bibfnamefont {M.~E.}\ \bibnamefont {Shaposhnikov}},\
  }\bibfield  {title} {\enquote {\bibinfo {title} {{The Electroweak phase
  transition: A Nonperturbative analysis}},}\ }\href {\doibase
  10.1016/0550-3213(96)00052-1} {\bibfield  {journal} {\bibinfo  {journal}
  {Nucl.\ Phys.\ B}\ }\textbf {\bibinfo {volume} {466}},\ \bibinfo {pages}
  {189--258} (\bibinfo {year} {1996})}\BibitemShut {NoStop}%
%%CITATION = HEP-LAT/9510020;%%
\bibitem [{\citenamefont {Fukugita}\ and\ \citenamefont
  {Yanagida}(1986)}]{Fukugita:1986hr}%
  \BibitemOpen
  \bibfield  {author} {\bibinfo {author} {\bibfnamefont {M.}~\bibnamefont
  {Fukugita}}\ and\ \bibinfo {author} {\bibfnamefont {T.}~\bibnamefont
  {Yanagida}},\ }\bibfield  {title} {\enquote {\bibinfo {title} {{Baryogenesis
  Without Grand Unification}},}\ }\href {\doibase 10.1016/0370-2693(86)91126-3}
  {\bibfield  {journal} {\bibinfo  {journal} {Phys.\ Lett.\ B}\ }\textbf
  {\bibinfo {volume} {174}},\ \bibinfo {pages} {45} (\bibinfo {year}
  {1986})}\BibitemShut {NoStop}%
%%CITATION = PHLTA,B174,45;%%
\bibitem [{\citenamefont {Shaposhnikov}(2012)}]{Shaposhnikov:2012_COSMO}%
  \BibitemOpen
  \bibfield  {author} {\bibinfo {author} {\bibfnamefont {M.}~\bibnamefont
  {Shaposhnikov}},\ }\bibfield  {title} {\enquote {\bibinfo {title}
  {{Baryo/Leptogenesis}},}\ }\href@noop {} {\  (\bibinfo {year} {2012})},\
  \bibinfo {note} {[Presentation at
  \href{http://lss.bao.ac.cn/cosmo12/index.php?code=5}{Cosmo 2012}, Beijing,
  China, September 2012]}\BibitemShut {NoStop}%
\bibitem [{\citenamefont {Weinberg}(1962)}]{Weinberg:1962zza}%
  \BibitemOpen
  \bibfield  {author} {\bibinfo {author} {\bibfnamefont {S.}~\bibnamefont
  {Weinberg}},\ }\bibfield  {title} {\enquote {\bibinfo {title} {{Universal
  Neutrino Degeneracy}},}\ }\href {\doibase 10.1103/PhysRev.128.1457}
  {\bibfield  {journal} {\bibinfo  {journal} {Phys.\ Rev.}\ }\textbf {\bibinfo
  {volume} {128}},\ \bibinfo {pages} {1457--1473} (\bibinfo {year}
  {1962})}\BibitemShut {NoStop}%
%%CITATION = PHRVA,128,1457;%%
\bibitem [{\citenamefont {Cocco}\ \emph {et~al.}(2007)\citenamefont {Cocco},
  \citenamefont {Mangano},\ and\ \citenamefont {Messina}}]{Cocco:2007za}%
  \BibitemOpen
  \bibfield  {author} {\bibinfo {author} {\bibfnamefont {A.G.}\ \bibnamefont
  {Cocco}}, \bibinfo {author} {\bibfnamefont {G.}~\bibnamefont {Mangano}}, \
  and\ \bibinfo {author} {\bibfnamefont {M.}~\bibnamefont {Messina}},\
  }\bibfield  {title} {\enquote {\bibinfo {title} {{Probing low energy neutrino
  backgrounds with neutrino capture on beta decaying nuclei}},}\ }\href
  {\doibase 10.1088/1475-7516/2007/06/015} {\bibfield  {journal} {\bibinfo
  {journal} {J.\ Cosm.\ Astropart.\ Phys.}\ }\textbf {\bibinfo {volume}
  {0706}},\ \bibinfo {pages} {015} (\bibinfo {year} {2007})}\BibitemShut
  {NoStop}%
%%CITATION = HEP-PH/0703075;%%
\bibitem [{\citenamefont {Long}\ \emph {et~al.}(2014)\citenamefont {Long},
  \citenamefont {Lunardini},\ and\ \citenamefont {Sabancilar}}]{Long:2014zva}%
  \BibitemOpen
  \bibfield  {author} {\bibinfo {author} {\bibfnamefont {A.~J.}\ \bibnamefont
  {Long}}, \bibinfo {author} {\bibfnamefont {C.}~\bibnamefont {Lunardini}}, \
  and\ \bibinfo {author} {\bibfnamefont {E.}~\bibnamefont {Sabancilar}},\
  }\bibfield  {title} {\enquote {\bibinfo {title} {{Detecting non-relativistic
  cosmic neutrinos by capture on tritium: phenomenology and physics
  potential}},}\ }\href {\doibase 10.1088/1475-7516/2014/08/038} {\bibfield
  {journal} {\bibinfo  {journal} {J.\ Cosm.\ Astropart.\ Phys.}\ }\textbf
  {\bibinfo {volume} {1408}},\ \bibinfo {pages} {038} (\bibinfo {year}
  {2014})}\BibitemShut {NoStop}%
%%CITATION = ARXIV:1405.7654;%%
\bibitem [{\citenamefont {Gell-Mann}\ and\ \citenamefont
  {Pais}(1955)}]{GellMann:1955jx}%
  \BibitemOpen
  \bibfield  {author} {\bibinfo {author} {\bibfnamefont {M.}~\bibnamefont
  {Gell-Mann}}\ and\ \bibinfo {author} {\bibfnamefont {A.}~\bibnamefont
  {Pais}},\ }\bibfield  {title} {\enquote {\bibinfo {title} {{Behavior of
  neutral particles under charge conjugation}},}\ }\href {\doibase
  10.1103/PhysRev.97.1387} {\bibfield  {journal} {\bibinfo  {journal} {Phys.\
  Rev.}\ }\textbf {\bibinfo {volume} {97}},\ \bibinfo {pages} {1387--1389}
  (\bibinfo {year} {1955})}\BibitemShut {NoStop}%
%%CITATION = PHRVA,97,1387;%%
\bibitem [{\citenamefont {Lande}\ \emph {et~al.}(1956)\citenamefont {Lande},
  \citenamefont {Booth}, \citenamefont {Impeduglia}, \citenamefont {Lederman},\
  and\ \citenamefont {Chinowsky}}]{Lande:1956pf}%
  \BibitemOpen
  \bibfield  {author} {\bibinfo {author} {\bibfnamefont {K.}~\bibnamefont
  {Lande}}, \bibinfo {author} {\bibfnamefont {E.~T.}\ \bibnamefont {Booth}},
  \bibinfo {author} {\bibfnamefont {J.}~\bibnamefont {Impeduglia}}, \bibinfo
  {author} {\bibfnamefont {L.~M.}\ \bibnamefont {Lederman}}, \ and\ \bibinfo
  {author} {\bibfnamefont {W.}~\bibnamefont {Chinowsky}},\ }\bibfield  {title}
  {\enquote {\bibinfo {title} {{Observation of Long-Lived Neutral V
  Particles}},}\ }\href {\doibase 10.1103/PhysRev.103.1901} {\bibfield
  {journal} {\bibinfo  {journal} {Phys.\ Rev.}\ }\textbf {\bibinfo {volume}
  {103}},\ \bibinfo {pages} {1901--1904} (\bibinfo {year} {1956})}\BibitemShut
  {NoStop}%
%%CITATION = PHRVA,103,1901;%%
\bibitem [{\citenamefont {Lande}\ \emph {et~al.}(1957)\citenamefont {Lande},
  \citenamefont {Lederman},\ and\ \citenamefont {Chinowsky}}]{Jackson:1957zzb}%
  \BibitemOpen
  \bibfield  {author} {\bibinfo {author} {\bibfnamefont {K.}~\bibnamefont
  {Lande}}, \bibinfo {author} {\bibfnamefont {L.~M.}\ \bibnamefont {Lederman}},
  \ and\ \bibinfo {author} {\bibfnamefont {W.}~\bibnamefont {Chinowsky}},\
  }\bibfield  {title} {\enquote {\bibinfo {title} {{Report on Long-Lived K0
  Mesons}},}\ }\href {\doibase 10.1103/PhysRev.105.1925.2} {\bibfield
  {journal} {\bibinfo  {journal} {Phys.\ Rev.}\ }\textbf {\bibinfo {volume}
  {105}},\ \bibinfo {pages} {1925--1927} (\bibinfo {year} {1957})}\BibitemShut
  {NoStop}%
%%CITATION = PHRVA,105,1925;%%
\bibitem [{\citenamefont {Prentki}\ and\ \citenamefont
  {Veltman}(1965)}]{Prentki:1965tt}%
  \BibitemOpen
  \bibfield  {author} {\bibinfo {author} {\bibfnamefont {J.}~\bibnamefont
  {Prentki}}\ and\ \bibinfo {author} {\bibfnamefont {M.~J.~G.}\ \bibnamefont
  {Veltman}},\ }\bibfield  {title} {\enquote {\bibinfo {title} {{Possibility of
  CP violation in semistrong interactions}},}\ }\href {\doibase
  10.1016/0031-9163(65)91141-8} {\bibfield  {journal} {\bibinfo  {journal}
  {Phys.\ Lett.}\ }\textbf {\bibinfo {volume} {15}},\ \bibinfo {pages} {88--90}
  (\bibinfo {year} {1965})}\BibitemShut {NoStop}%
%%CITATION = PHLTA,15,88;%%
\bibitem [{\citenamefont {Lee}\ and\ \citenamefont
  {Wolfenstein}(1965)}]{Lee:1965hi}%
  \BibitemOpen
  \bibfield  {author} {\bibinfo {author} {\bibfnamefont {T.~D.}\ \bibnamefont
  {Lee}}\ and\ \bibinfo {author} {\bibfnamefont {L.}~\bibnamefont
  {Wolfenstein}},\ }\bibfield  {title} {\enquote {\bibinfo {title} {{Analysis
  of CP Noninvariant Interactions and the $K_1^0$, $K_2^0$ System}},}\ }\href
  {\doibase 10.1103/PhysRev.138.B1490} {\bibfield  {journal} {\bibinfo
  {journal} {Phys.\ Rev.}\ }\textbf {\bibinfo {volume} {138}},\ \bibinfo
  {pages} {B1490--B1496} (\bibinfo {year} {1965})}\BibitemShut {NoStop}%
%%CITATION = PHRVA,138,B1490;%%
\bibitem [{\citenamefont {Wilczek}\ and\ \citenamefont
  {Zee}(1979)}]{Wilczek:1979hc}%
  \BibitemOpen
  \bibfield  {author} {\bibinfo {author} {\bibfnamefont {F.}~\bibnamefont
  {Wilczek}}\ and\ \bibinfo {author} {\bibfnamefont {A.}~\bibnamefont {Zee}},\
  }\bibfield  {title} {\enquote {\bibinfo {title} {{Operator Analysis of
  Nucleon Decay}},}\ }\href {\doibase 10.1103/PhysRevLett.43.1571} {\bibfield
  {journal} {\bibinfo  {journal} {Phys.\ Rev.\ Lett.}\ }\textbf {\bibinfo
  {volume} {43}},\ \bibinfo {pages} {1571--1573} (\bibinfo {year}
  {1979})}\BibitemShut {NoStop}%
%%CITATION = PRLTA,43,1571;%%
\bibitem [{\citenamefont {Choi}\ \emph {et~al.}(2002)\citenamefont {Choi},
  \citenamefont {Jeong},\ and\ \citenamefont {Song}}]{Choi:2002bb}%
  \BibitemOpen
  \bibfield  {author} {\bibinfo {author} {\bibfnamefont {K.}~\bibnamefont
  {Choi}}, \bibinfo {author} {\bibfnamefont {K.S.}\ \bibnamefont {Jeong}}, \
  and\ \bibinfo {author} {\bibfnamefont {W.Y.}\ \bibnamefont {Song}},\
  }\bibfield  {title} {\enquote {\bibinfo {title} {{Operator analysis of
  neutrinoless double beta decay}},}\ }\href {\doibase
  10.1103/PhysRevD.66.093007} {\bibfield  {journal} {\bibinfo  {journal}
  {Phys.\ Rev.\ D}\ }\textbf {\bibinfo {volume} {66}},\ \bibinfo {pages}
  {093007} (\bibinfo {year} {2002})}\BibitemShut {NoStop}%
%%CITATION = HEP-PH/0207180;%%
\bibitem [{\citenamefont {Bonnet}\ \emph {et~al.}(2013)\citenamefont {Bonnet},
  \citenamefont {Hirsch}, \citenamefont {Ota},\ and\ \citenamefont
  {Winter}}]{Bonnet:2012kh}%
  \BibitemOpen
  \bibfield  {author} {\bibinfo {author} {\bibfnamefont {F.}~\bibnamefont
  {Bonnet}}, \bibinfo {author} {\bibfnamefont {M.}~\bibnamefont {Hirsch}},
  \bibinfo {author} {\bibfnamefont {T.}~\bibnamefont {Ota}}, \ and\ \bibinfo
  {author} {\bibfnamefont {W.}~\bibnamefont {Winter}},\ }\bibfield  {title}
  {\enquote {\bibinfo {title} {{Systematic decomposition of the neutrinoless
  double beta decay operator}},}\ }\href {\doibase 10.1007/JHEP03(2013)055}
  {\bibfield  {journal} {\bibinfo  {journal} {J.\ High Energy Phys.}\ }\textbf
  {\bibinfo {volume} {1303}},\ \bibinfo {pages} {055} (\bibinfo {year}
  {2013})},\ \bibinfo {note} {[Erratum:
  \href{http://link.springer.com/article/10.1007\%2FJHEP04\%282014\%29090} {J.\
  High Energy Phys. {\bf 1404}, 090 (2014)}]}\BibitemShut {NoStop}%
\bibitem [{\citenamefont {Atre}\ \emph {et~al.}(2009)\citenamefont {Atre},
  \citenamefont {Han}, \citenamefont {Pascoli},\ and\ \citenamefont
  {Zhang}}]{Atre:2009rg}%
  \BibitemOpen
  \bibfield  {author} {\bibinfo {author} {\bibfnamefont {A.}~\bibnamefont
  {Atre}}, \bibinfo {author} {\bibfnamefont {T.}~\bibnamefont {Han}}, \bibinfo
  {author} {\bibfnamefont {S.}~\bibnamefont {Pascoli}}, \ and\ \bibinfo
  {author} {\bibfnamefont {B.}~\bibnamefont {Zhang}},\ }\bibfield  {title}
  {\enquote {\bibinfo {title} {{The Search for Heavy Majorana Neutrinos}},}\
  }\href {\doibase 10.1088/1126-6708/2009/05/030} {\bibfield  {journal}
  {\bibinfo  {journal} {J.\ High Energy Phys.}\ }\textbf {\bibinfo {volume}
  {0905}},\ \bibinfo {pages} {030} (\bibinfo {year} {2009})}\BibitemShut
  {NoStop}%
%%CITATION = ARXIV:0901.3589;%%
\bibitem [{\citenamefont {Mitra}\ \emph {et~al.}(2012)\citenamefont {Mitra},
  \citenamefont {Senjanovic},\ and\ \citenamefont {Vissani}}]{Mitra:2011qr}%
  \BibitemOpen
  \bibfield  {author} {\bibinfo {author} {\bibfnamefont {M.}~\bibnamefont
  {Mitra}}, \bibinfo {author} {\bibfnamefont {G.}~\bibnamefont {Senjanovic}}, \
  and\ \bibinfo {author} {\bibfnamefont {F.}~\bibnamefont {Vissani}},\
  }\bibfield  {title} {\enquote {\bibinfo {title} {{Neutrinoless Double Beta
  Decay and Heavy Sterile Neutrinos}},}\ }\href {\doibase
  10.1016/j.nuclphysb.2011.10.035} {\bibfield  {journal} {\bibinfo  {journal}
  {Nucl.\ Phys.\ B}\ }\textbf {\bibinfo {volume} {856}},\ \bibinfo {pages}
  {26--73} (\bibinfo {year} {2012})}\BibitemShut {NoStop}%
%%CITATION = ARXIV:1108.0004;%%
\bibitem [{\citenamefont {Blennow}\ \emph {et~al.}(2010)\citenamefont
  {Blennow}, \citenamefont {Fernandez-Martinez}, \citenamefont {Lopez-Pavon},\
  and\ \citenamefont {Men\'endez}}]{Blennow:2010th}%
  \BibitemOpen
  \bibfield  {author} {\bibinfo {author} {\bibfnamefont {M.}~\bibnamefont
  {Blennow}}, \bibinfo {author} {\bibfnamefont {E.}~\bibnamefont
  {Fernandez-Martinez}}, \bibinfo {author} {\bibfnamefont {J.}~\bibnamefont
  {Lopez-Pavon}}, \ and\ \bibinfo {author} {\bibfnamefont {J.}~\bibnamefont
  {Men\'endez}},\ }\bibfield  {title} {\enquote {\bibinfo {title}
  {{Neutrinoless double beta decay in seesaw models}},}\ }\href {\doibase
  10.1007/JHEP07(2010)096} {\bibfield  {journal} {\bibinfo  {journal} {J.\ High
  Energy Phys.}\ }\textbf {\bibinfo {volume} {1007}},\ \bibinfo {pages} {096}
  (\bibinfo {year} {2010})}\BibitemShut {NoStop}%
%%CITATION = ARXIV:1005.3240;%%
\bibitem [{\citenamefont {Lopez-Pavon}\ \emph {et~al.}(2013)\citenamefont
  {Lopez-Pavon}, \citenamefont {Pascoli},\ and\ \citenamefont
  {Wong}}]{LopezPavon:2012zg}%
  \BibitemOpen
  \bibfield  {author} {\bibinfo {author} {\bibfnamefont {J.}~\bibnamefont
  {Lopez-Pavon}}, \bibinfo {author} {\bibfnamefont {S.}~\bibnamefont
  {Pascoli}}, \ and\ \bibinfo {author} {\bibfnamefont {C.}~\bibnamefont
  {Wong}},\ }\bibfield  {title} {\enquote {\bibinfo {title} {{Can heavy
  neutrinos dominate neutrinoless double beta decay?}}}\ }\href {\doibase
  10.1103/PhysRevD.87.093007} {\bibfield  {journal} {\bibinfo  {journal}
  {Phys.\ Rev.\ D}\ }\textbf {\bibinfo {volume} {87}},\ \bibinfo {pages}
  {093007} (\bibinfo {year} {2013})}\BibitemShut {NoStop}%
%%CITATION = ARXIV:1209.5342;%%
\bibitem [{\citenamefont {Dev}\ and\ \citenamefont
  {Mohapatra}(2015)}]{Dev:2015pga}%
  \BibitemOpen
  \bibfield  {author} {\bibinfo {author} {\bibfnamefont {P.~S.~B.}\
  \bibnamefont {Dev}}\ and\ \bibinfo {author} {\bibfnamefont {R.~N.}\
  \bibnamefont {Mohapatra}},\ }\bibfield  {title} {\enquote {\bibinfo {title}
  {{Unified explanation of the $eejj$, diboson and dijet resonances at the
  LHC}},}\ }\href {\doibase 10.1103/PhysRevLett.115.181803} {\bibfield
  {journal} {\bibinfo  {journal} {Phys.\ Rev.\ Lett.}\ }\textbf {\bibinfo
  {volume} {115}},\ \bibinfo {pages} {181803} (\bibinfo {year}
  {2015})}\BibitemShut {NoStop}%
%%CITATION = ARXIV:1508.02277;%%
\bibitem [{\citenamefont {Deppisch}\ \emph
  {et~al.}(2016{\natexlab{a}})\citenamefont {Deppisch}, \citenamefont {Graf},
  \citenamefont {Kulkarni}, \citenamefont {Patra}, \citenamefont {Rodejohann},
  \citenamefont {Sahu},\ and\ \citenamefont {Sarkar}}]{Deppisch:2015cua}%
  \BibitemOpen
  \bibfield  {author} {\bibinfo {author} {\bibfnamefont {F.~F.}\ \bibnamefont
  {Deppisch}}, \bibinfo {author} {\bibfnamefont {L.}~\bibnamefont {Graf}},
  \bibinfo {author} {\bibfnamefont {S.}~\bibnamefont {Kulkarni}}, \bibinfo
  {author} {\bibfnamefont {S.}~\bibnamefont {Patra}}, \bibinfo {author}
  {\bibfnamefont {W.}~\bibnamefont {Rodejohann}}, \bibinfo {author}
  {\bibfnamefont {N.}~\bibnamefont {Sahu}}, \ and\ \bibinfo {author}
  {\bibfnamefont {U.}~\bibnamefont {Sarkar}},\ }\bibfield  {title} {\enquote
  {\bibinfo {title} {{Reconciling the 2 TeV Excesses at the LHC in a Linear
  Seesaw Left-Right Model}},}\ }\href {\doibase 10.1103/PhysRevD.93.013011}
  {\bibfield  {journal} {\bibinfo  {journal} {Phys.\ Rev.\ D}\ }\textbf
  {\bibinfo {volume} {93}},\ \bibinfo {pages} {013011} (\bibinfo {year}
  {2016}{\natexlab{a}})}\BibitemShut {NoStop}%
%%CITATION = ARXIV:1508.05940;%%
\bibitem [{\citenamefont {Deppisch}\ \emph
  {et~al.}(2016{\natexlab{b}})\citenamefont {Deppisch}, \citenamefont {Hati},
  \citenamefont {Patra}, \citenamefont {Pritimita},\ and\ \citenamefont
  {Sarkar}}]{Deppisch:2016scs}%
  \BibitemOpen
  \bibfield  {author} {\bibinfo {author} {\bibfnamefont {F.~F.}\ \bibnamefont
  {Deppisch}}, \bibinfo {author} {\bibfnamefont {C.}~\bibnamefont {Hati}},
  \bibinfo {author} {\bibfnamefont {S.}~\bibnamefont {Patra}}, \bibinfo
  {author} {\bibfnamefont {P.}~\bibnamefont {Pritimita}}, \ and\ \bibinfo
  {author} {\bibfnamefont {U.}~\bibnamefont {Sarkar}},\ }\bibfield  {title}
  {\enquote {\bibinfo {title} {{Implications of the diphoton excess on
  Left-Right models and gauge unification}},}\ }\href {\doibase
  10.1016/j.physletb.2016.03.081} {\bibfield  {journal} {\bibinfo  {journal}
  {Phys.\ Lett.\ B}\ }\textbf {\bibinfo {volume} {757}},\ \bibinfo {pages}
  {223--230} (\bibinfo {year} {2016}{\natexlab{b}})}\BibitemShut {NoStop}%
%%CITATION = ARXIV:1601.00952;%%
\bibitem [{\citenamefont {Schechter}\ and\ \citenamefont
  {Valle}(1982)}]{Schechter:1981bd}%
  \BibitemOpen
  \bibfield  {author} {\bibinfo {author} {\bibfnamefont {J.}~\bibnamefont
  {Schechter}}\ and\ \bibinfo {author} {\bibfnamefont {J.~W.~F.}\ \bibnamefont
  {Valle}},\ }\bibfield  {title} {\enquote {\bibinfo {title} {{Neutrinoless
  Double beta Decay in $SU(2) \times U(1)$ Theories}},}\ }\href {\doibase
  10.1103/PhysRevD.25.2951} {\bibfield  {journal} {\bibinfo  {journal} {Phys.\
  Rev.\ D}\ }\textbf {\bibinfo {volume} {25}},\ \bibinfo {pages} {2951}
  (\bibinfo {year} {1982})}\BibitemShut {NoStop}%
%%CITATION = PHRVA,D25,2951;%%
\bibitem [{\citenamefont {Coleman}(1988)}]{Coleman}%
  \BibitemOpen
  \bibfield  {author} {\bibinfo {author} {\bibfnamefont {S.}~\bibnamefont
  {Coleman}},\ }\href
  {http://www.cambridge.org/us/academic/subjects/mathematics/%
  mathematical-physics/aspects-symmetry-selected-erice-lectures} {\emph
  {\bibinfo {title} {{Aspects of symmetry}}}}\ (\bibinfo  {publisher}
  {Cambridge University Press},\ \bibinfo {year} {1988})\ Chap.~\bibinfo
  {chapter} {4}\BibitemShut {NoStop}%
\bibitem [{\citenamefont {Duerr}\ \emph {et~al.}(2011)\citenamefont {Duerr},
  \citenamefont {Lindner},\ and\ \citenamefont {Merle}}]{Duerr:2011zd}%
  \BibitemOpen
  \bibfield  {author} {\bibinfo {author} {\bibfnamefont {M.}~\bibnamefont
  {Duerr}}, \bibinfo {author} {\bibfnamefont {M.}~\bibnamefont {Lindner}}, \
  and\ \bibinfo {author} {\bibfnamefont {A.}~\bibnamefont {Merle}},\ }\bibfield
   {title} {\enquote {\bibinfo {title} {{On the Quantitative Impact of the
  Schechter-Valle Theorem}},}\ }\href {\doibase 10.1007/JHEP06(2011)091}
  {\bibfield  {journal} {\bibinfo  {journal} {J.\ High Energy Phys.}\ }\textbf
  {\bibinfo {volume} {1106}},\ \bibinfo {pages} {091} (\bibinfo {year}
  {2011})}\BibitemShut {NoStop}%
%%CITATION = ARXIV:1105.0901;%%
\bibitem [{\citenamefont {Weinberg}(1982)}]{Weinberg:1981wj}%
  \BibitemOpen
  \bibfield  {author} {\bibinfo {author} {\bibfnamefont {S.}~\bibnamefont
  {Weinberg}},\ }\bibfield  {title} {\enquote {\bibinfo {title} {{Supersymmetry
  at Ordinary Energies. 1. Masses and Conservation Laws}},}\ }\href {\doibase
  10.1103/PhysRevD.26.287} {\bibfield  {journal} {\bibinfo  {journal} {Phys.\
  Rev.\ D}\ }\textbf {\bibinfo {volume} {26}},\ \bibinfo {pages} {287}
  (\bibinfo {year} {1982})}\BibitemShut {NoStop}%
%%CITATION = PHRVA,D26,287;%%
\bibitem [{\citenamefont {Mohapatra}\ and\ \citenamefont
  {Smirnov}(2006)}]{Mohapatra:2006gs}%
  \BibitemOpen
  \bibfield  {author} {\bibinfo {author} {\bibfnamefont {R.~N.}\ \bibnamefont
  {Mohapatra}}\ and\ \bibinfo {author} {\bibfnamefont {A.~Y.}\ \bibnamefont
  {Smirnov}},\ }\bibfield  {title} {\enquote {\bibinfo {title} {{Neutrino Mass
  and New Physics}},}\ }\href {\doibase 10.1146/annurev.nucl.56.080805.140534}
  {\bibfield  {journal} {\bibinfo  {journal} {Ann.\ Rev.\ Nucl.\ Part.\ Sci.}\
  }\textbf {\bibinfo {volume} {56}},\ \bibinfo {pages} {569--628} (\bibinfo
  {year} {2006})}\BibitemShut {NoStop}%
%%CITATION = HEP-PH/0603118;%%
\bibitem [{\citenamefont {Mohapatra}\ and\ \citenamefont
  {Pal}(1998)}]{Mohapatra:1998rq}%
  \BibitemOpen
  \bibfield  {author} {\bibinfo {author} {\bibfnamefont {R.~N.}\ \bibnamefont
  {Mohapatra}}\ and\ \bibinfo {author} {\bibfnamefont {P.~B.}\ \bibnamefont
  {Pal}},\ }\bibfield  {title} {\enquote {\bibinfo {title} {{Massive neutrinos
  in physics and astrophysics. (Second edition)}},}\ }\href@noop {} {\bibfield
  {journal} {\bibinfo  {journal} {World Sci.\ Lect.\ Notes Phys.}\ }\textbf
  {\bibinfo {volume} {60}},\ \bibinfo {pages} {1--397} (\bibinfo {year}
  {1998})}\BibitemShut {NoStop}%
%%CITATION = 00327,60,1;%%
\bibitem [{\citenamefont {Alekhin}\ \emph {et~al.}(2015)\citenamefont {Alekhin}
  \emph {et~al.}}]{Alekhin:2015byh}%
  \BibitemOpen
  \bibfield  {author} {\bibinfo {author} {\bibfnamefont {S.}~\bibnamefont
  {Alekhin}} \emph {et~al.},\ }\bibfield  {title} {\enquote {\bibinfo {title}
  {{A facility to Search for Hidden Particles at the CERN SPS: the SHiP physics
  case}},}\ }\href@noop {} {\  (\bibinfo {year} {2015})},\ \Eprint
  {http://arxiv.org/abs/1504.04855} {arXiv:1504.04855 [hep-ph]} \BibitemShut
  {NoStop}%
%%CITATION = ARXIV:1504.04855;%%
\bibitem [{\citenamefont {Anelli}\ \emph {et~al.}(2015)\citenamefont {Anelli}
  \emph {et~al.}}]{Anelli:2015pba}%
  \BibitemOpen
  \bibfield  {author} {\bibinfo {author} {\bibfnamefont {M.}~\bibnamefont
  {Anelli}} \emph {et~al.} (\bibinfo {collaboration} {SHiP Collaboration}),\
  }\bibfield  {title} {\enquote {\bibinfo {title} {{A facility to Search for
  Hidden Particles (SHiP) at the CERN SPS}},}\ }\href@noop {} {\  (\bibinfo
  {year} {2015})},\ \Eprint {http://arxiv.org/abs/1504.04956} {arXiv:1504.04956
  [physics.ins-det]} \BibitemShut {NoStop}%
%%CITATION = ARXIV:1504.04956;%%
\bibitem [{\citenamefont {Pati}\ and\ \citenamefont
  {Salam}(1974)}]{Pati:1974yy}%
  \BibitemOpen
  \bibfield  {author} {\bibinfo {author} {\bibfnamefont {J.~C.}\ \bibnamefont
  {Pati}}\ and\ \bibinfo {author} {\bibfnamefont {A.}~\bibnamefont {Salam}},\
  }\bibfield  {title} {\enquote {\bibinfo {title} {{Lepton Number as the Fourth
  Color}},}\ }\href {\doibase 10.1103/PhysRevD.10.275,
  10.1103/PhysRevD.11.703.2} {\bibfield  {journal} {\bibinfo  {journal} {Phys.\
  Rev.\ D}\ }\textbf {\bibinfo {volume} {10}},\ \bibinfo {pages} {275--289}
  (\bibinfo {year} {1974})}\BibitemShut {NoStop}%
%%CITATION = PHRVA,D10,275;%%
\bibitem [{\citenamefont {Keung}\ and\ \citenamefont
  {Senjanovic}(1983)}]{Keung:1983uu}%
  \BibitemOpen
  \bibfield  {author} {\bibinfo {author} {\bibfnamefont {W.~Y.}\ \bibnamefont
  {Keung}}\ and\ \bibinfo {author} {\bibfnamefont {G.}~\bibnamefont
  {Senjanovic}},\ }\bibfield  {title} {\enquote {\bibinfo {title} {{Majorana
  Neutrinos and the Production of the Right-handed Charged Gauge Boson}},}\
  }\href {\doibase 10.1103/PhysRevLett.50.1427} {\bibfield  {journal} {\bibinfo
   {journal} {Phys.\ Rev.\ Lett.}\ }\textbf {\bibinfo {volume} {50}},\ \bibinfo
  {pages} {1427} (\bibinfo {year} {1983})}\BibitemShut {NoStop}%
%%CITATION = PRLTA,50,1427;%%
\bibitem [{\citenamefont {Tello}\ \emph {et~al.}(2011)\citenamefont {Tello},
  \citenamefont {Nemevsek}, \citenamefont {Nesti}, \citenamefont {Senjanovic},\
  and\ \citenamefont {Vissani}}]{Tello:2010am}%
  \BibitemOpen
  \bibfield  {author} {\bibinfo {author} {\bibfnamefont {V.}~\bibnamefont
  {Tello}}, \bibinfo {author} {\bibfnamefont {M.}~\bibnamefont {Nemevsek}},
  \bibinfo {author} {\bibfnamefont {F.}~\bibnamefont {Nesti}}, \bibinfo
  {author} {\bibfnamefont {G.}~\bibnamefont {Senjanovic}}, \ and\ \bibinfo
  {author} {\bibfnamefont {F.}~\bibnamefont {Vissani}},\ }\bibfield  {title}
  {\enquote {\bibinfo {title} {{Left-Right Symmetry: from LHC to Neutrinoless
  Double Beta Decay}},}\ }\href {\doibase 10.1103/PhysRevLett.106.151801}
  {\bibfield  {journal} {\bibinfo  {journal} {Phys.\ Rev.\ Lett.}\ }\textbf
  {\bibinfo {volume} {106}},\ \bibinfo {pages} {151801} (\bibinfo {year}
  {2011})}\BibitemShut {NoStop}%
%%CITATION = ARXIV:1011.3522;%%
\bibitem [{\citenamefont {Nemevsek}\ \emph {et~al.}(2011)\citenamefont
  {Nemevsek}, \citenamefont {Nesti}, \citenamefont {Senjanovic},\ and\
  \citenamefont {Zhang}}]{Nemevsek:2011hz}%
  \BibitemOpen
  \bibfield  {author} {\bibinfo {author} {\bibfnamefont {M.}~\bibnamefont
  {Nemevsek}}, \bibinfo {author} {\bibfnamefont {F.}~\bibnamefont {Nesti}},
  \bibinfo {author} {\bibfnamefont {G.}~\bibnamefont {Senjanovic}}, \ and\
  \bibinfo {author} {\bibfnamefont {Y.}~\bibnamefont {Zhang}},\ }\bibfield
  {title} {\enquote {\bibinfo {title} {{First Limits on Left-Right Symmetry
  Scale from LHC Data}},}\ }\href {\doibase 10.1103/PhysRevD.83.115014}
  {\bibfield  {journal} {\bibinfo  {journal} {Phys.\ Rev.\ D}\ }\textbf
  {\bibinfo {volume} {83}},\ \bibinfo {pages} {115014} (\bibinfo {year}
  {2011})}\BibitemShut {NoStop}%
%%CITATION = ARXIV:1103.1627;%%
\bibitem [{\citenamefont {Dell'Oro}\ \emph {et~al.}(2014)\citenamefont
  {Dell'Oro}, \citenamefont {Marcocci},\ and\ \citenamefont
  {Vissani}}]{Dell'Oro:2014yca}%
  \BibitemOpen
  \bibfield  {author} {\bibinfo {author} {\bibfnamefont {S.}~\bibnamefont
  {Dell'Oro}}, \bibinfo {author} {\bibfnamefont {S.}~\bibnamefont {Marcocci}},
  \ and\ \bibinfo {author} {\bibfnamefont {F.}~\bibnamefont {Vissani}},\
  }\bibfield  {title} {\enquote {\bibinfo {title} {{New expectations and
  uncertainties on neutrinoless double beta decay}},}\ }\href {\doibase
  10.1103/PhysRevD.90.033005} {\bibfield  {journal} {\bibinfo  {journal}
  {Phys.\ Rev.\ D}\ }\textbf {\bibinfo {volume} {90}},\ \bibinfo {pages}
  {033005} (\bibinfo {year} {2014})}\BibitemShut {NoStop}%
%%CITATION = ARXIV:1404.2616;%%
\bibitem [{\citenamefont {Capozzi}\ \emph {et~al.}(2014)\citenamefont
  {Capozzi}, \citenamefont {Fogli}, \citenamefont {Lisi}, \citenamefont
  {Marrone}, \citenamefont {Montanino},\ and\ \citenamefont
  {Palazzo}}]{Capozzi:2013csa}%
  \BibitemOpen
  \bibfield  {author} {\bibinfo {author} {\bibfnamefont {F.}~\bibnamefont
  {Capozzi}}, \bibinfo {author} {\bibfnamefont {G.~L.}\ \bibnamefont {Fogli}},
  \bibinfo {author} {\bibfnamefont {E.}~\bibnamefont {Lisi}}, \bibinfo {author}
  {\bibfnamefont {A.}~\bibnamefont {Marrone}}, \bibinfo {author} {\bibfnamefont
  {D.}~\bibnamefont {Montanino}}, \ and\ \bibinfo {author} {\bibfnamefont
  {A.}~\bibnamefont {Palazzo}},\ }\bibfield  {title} {\enquote {\bibinfo
  {title} {{Status of three-neutrino oscillation parameters, circa 2013}},}\
  }\href {\doibase 10.1103/PhysRevD.89.093018} {\bibfield  {journal} {\bibinfo
  {journal} {Phys.\ Rev.\ D}\ }\textbf {\bibinfo {volume} {89}},\ \bibinfo
  {pages} {093018} (\bibinfo {year} {2014})}\BibitemShut {NoStop}%
%%CITATION = ARXIV:1312.2878;%%
\bibitem [{\citenamefont {Wolfenstein}(1978)}]{Wolfenstein:1977ue}%
  \BibitemOpen
  \bibfield  {author} {\bibinfo {author} {\bibfnamefont {L.}~\bibnamefont
  {Wolfenstein}},\ }\bibfield  {title} {\enquote {\bibinfo {title} {{Neutrino
  Oscillations in Matter}},}\ }\href {\doibase 10.1103/PhysRevD.17.2369}
  {\bibfield  {journal} {\bibinfo  {journal} {Phys.\ Rev.\ D}\ }\textbf
  {\bibinfo {volume} {17}},\ \bibinfo {pages} {2369--2374} (\bibinfo {year}
  {1978})}\BibitemShut {NoStop}%
%%CITATION = PHRVA,D17,2369;%%
\bibitem [{\citenamefont {Mikheev}\ and\ \citenamefont
  {Smirnov}(1985)}]{Mikheev:1986gs}%
  \BibitemOpen
  \bibfield  {author} {\bibinfo {author} {\bibfnamefont {S.P.}\ \bibnamefont
  {Mikheev}}\ and\ \bibinfo {author} {\bibfnamefont {A.~Y.}\ \bibnamefont
  {Smirnov}},\ }\bibfield  {title} {\enquote {\bibinfo {title} {{Resonance
  Amplification of Oscillations in Matter and Spectroscopy of Solar
  Neutrinos}},}\ }\href@noop {} {\bibfield  {journal} {\bibinfo  {journal}
  {Sov.\ J.\ Nucl.\ Phys.}\ }\textbf {\bibinfo {volume} {42}},\ \bibinfo
  {pages} {913--917} (\bibinfo {year} {1985})}\BibitemShut {NoStop}%
%%CITATION = SJNCA,42,913;%%
\bibitem [{\citenamefont {Abe}\ \emph {et~al.}(2008)\citenamefont {Abe} \emph
  {et~al.}}]{Abe:2008aa}%
  \BibitemOpen
  \bibfield  {author} {\bibinfo {author} {\bibfnamefont {S.}~\bibnamefont
  {Abe}} \emph {et~al.} (\bibinfo {collaboration} {KamLAND Collaboration}),\
  }\bibfield  {title} {\enquote {\bibinfo {title} {{Precision Measurement of
  Neutrino Oscillation Parameters with KamLAND}},}\ }\href {\doibase
  10.1103/PhysRevLett.100.221803} {\bibfield  {journal} {\bibinfo  {journal}
  {Phys.\ Rev.\ Lett.}\ }\textbf {\bibinfo {volume} {100}},\ \bibinfo {pages}
  {221803} (\bibinfo {year} {2008})}\BibitemShut {NoStop}%
%%CITATION = ARXIV:0801.4589;%%
\bibitem [{\citenamefont {Ghoshal}\ and\ \citenamefont
  {Petcov}(2011)}]{Ghoshal:2010wt}%
  \BibitemOpen
  \bibfield  {author} {\bibinfo {author} {\bibfnamefont {P.}~\bibnamefont
  {Ghoshal}}\ and\ \bibinfo {author} {\bibfnamefont {S.~T.}\ \bibnamefont
  {Petcov}},\ }\bibfield  {title} {\enquote {\bibinfo {title} {{Neutrino Mass
  Hierarchy Determination Using Reactor Antineutrinos}},}\ }\href {\doibase
  10.1007/JHEP03(2011)058} {\bibfield  {journal} {\bibinfo  {journal} {J.\ High
  Energy Phys.}\ }\textbf {\bibinfo {volume} {1103}},\ \bibinfo {pages} {058}
  (\bibinfo {year} {2011})}\BibitemShut {NoStop}%
%%CITATION = ARXIV:1011.1646;%%
\bibitem [{\citenamefont {Vissani}(1999)}]{Vissani:1999tu}%
  \BibitemOpen
  \bibfield  {author} {\bibinfo {author} {\bibfnamefont {F.}~\bibnamefont
  {Vissani}},\ }\bibfield  {title} {\enquote {\bibinfo {title} {{Signal of
  neutrinoless double beta decay, neutrino spectrum and oscillation
  scenarios}},}\ }\href {\doibase 10.1088/1126-6708/1999/06/022} {\bibfield
  {journal} {\bibinfo  {journal} {J.\ High Energy Phys.}\ }\textbf {\bibinfo
  {volume} {9906}},\ \bibinfo {pages} {022} (\bibinfo {year}
  {1999})}\BibitemShut {NoStop}%
%%CITATION = HEP-PH/9906525;%%
\bibitem [{\citenamefont {Feruglio}\ \emph {et~al.}(2002)\citenamefont
  {Feruglio}, \citenamefont {Strumia},\ and\ \citenamefont
  {Vissani}}]{Feruglio:2002af}%
  \BibitemOpen
  \bibfield  {author} {\bibinfo {author} {\bibfnamefont {F.}~\bibnamefont
  {Feruglio}}, \bibinfo {author} {\bibfnamefont {A.}~\bibnamefont {Strumia}}, \
  and\ \bibinfo {author} {\bibfnamefont {F.}~\bibnamefont {Vissani}},\
  }\bibfield  {title} {\enquote {\bibinfo {title} {{Neutrino oscillations and
  signals in $\beta$ and $0\nu2\beta$ experiments}},}\ }\href {\doibase
  10.1016/S0550-3213(02)00345-0} {\bibfield  {journal} {\bibinfo  {journal}
  {Nucl.\ Phys.\ B}\ }\textbf {\bibinfo {volume} {637}},\ \bibinfo {pages}
  {345--377} (\bibinfo {year} {2002})}\BibitemShut {NoStop}%
%%CITATION = HEP-PH/0201291;%%
\bibitem [{\citenamefont {Zel'dovich}\ and\ \citenamefont
  {Khlopov}(1981)}]{Zeldovich:1981wf}%
  \BibitemOpen
  \bibfield  {author} {\bibinfo {author} {\bibfnamefont {Y.~B.}\ \bibnamefont
  {Zel'dovich}}\ and\ \bibinfo {author} {\bibfnamefont {M.~Y.}\ \bibnamefont
  {Khlopov}},\ }\bibfield  {title} {\enquote {\bibinfo {title} {{The Neutrino
  Mass in Elementary Particle Physics and in Big Bang Cosmology}},}\ }\href
  {\doibase 10.1070/PU1981v024n09ABEH004816} {\bibfield  {journal} {\bibinfo
  {journal} {Sov.\ Phys.\ Usp.}\ }\textbf {\bibinfo {volume} {24}},\ \bibinfo
  {pages} {755--774} (\bibinfo {year} {1981})}\BibitemShut {NoStop}%
%%CITATION = SOPUA,24,755;%%
\bibitem [{\citenamefont {Fogli}\ \emph {et~al.}(2004)\citenamefont {Fogli},
  \citenamefont {Lisi}, \citenamefont {Marrone}, \citenamefont {Melchiorri},
  \citenamefont {Palazzo}, \citenamefont {Serra},\ and\ \citenamefont
  {Silk}}]{Fogli:2004as}%
  \BibitemOpen
  \bibfield  {author} {\bibinfo {author} {\bibfnamefont {G.~L.}\ \bibnamefont
  {Fogli}}, \bibinfo {author} {\bibfnamefont {E.}~\bibnamefont {Lisi}},
  \bibinfo {author} {\bibfnamefont {A.}~\bibnamefont {Marrone}}, \bibinfo
  {author} {\bibfnamefont {A.}~\bibnamefont {Melchiorri}}, \bibinfo {author}
  {\bibfnamefont {A.}~\bibnamefont {Palazzo}}, \bibinfo {author} {\bibfnamefont
  {P.}~\bibnamefont {Serra}}, \ and\ \bibinfo {author} {\bibfnamefont
  {J.}~\bibnamefont {Silk}},\ }\bibfield  {title} {\enquote {\bibinfo {title}
  {{Observables sensitive to absolute neutrino masses: Constraints and
  correlations from world neutrino data}},}\ }\href {\doibase
  10.1103/PhysRevD.70.113003} {\bibfield  {journal} {\bibinfo  {journal}
  {Phys.\ Rev.\ D}\ }\textbf {\bibinfo {volume} {70}},\ \bibinfo {pages}
  {113003} (\bibinfo {year} {2004})}\BibitemShut {NoStop}%
%%CITATION = HEP-PH/0408045;%%
\bibitem [{\citenamefont {Primack}\ \emph {et~al.}(1995)\citenamefont
  {Primack}, \citenamefont {Holtzman}, \citenamefont {Klypin},\ and\
  \citenamefont {Caldwell}}]{Primack:1994pe}%
  \BibitemOpen
  \bibfield  {author} {\bibinfo {author} {\bibfnamefont {J.~R.}\ \bibnamefont
  {Primack}}, \bibinfo {author} {\bibfnamefont {J.}~\bibnamefont {Holtzman}},
  \bibinfo {author} {\bibfnamefont {A.}~\bibnamefont {Klypin}}, \ and\ \bibinfo
  {author} {\bibfnamefont {D.~O.}\ \bibnamefont {Caldwell}},\ }\bibfield
  {title} {\enquote {\bibinfo {title} {{Cold + hot dark matter cosmology with
  $m(\nu_\mu) \approx m(\nu_\tau) \approx 2.4$\,eV}},}\ }\href {\doibase
  10.1103/PhysRevLett.74.2160} {\bibfield  {journal} {\bibinfo  {journal}
  {Phys.\ Rev.\ Lett.}\ }\textbf {\bibinfo {volume} {74}},\ \bibinfo {pages}
  {2160--2163} (\bibinfo {year} {1995})}\BibitemShut {NoStop}%
%%CITATION = ASTRO-PH/9411020;%%
\bibitem [{\citenamefont {Allen}\ \emph {et~al.}(2003)\citenamefont {Allen},
  \citenamefont {Schmidt},\ and\ \citenamefont {Bridle}}]{Allen:2003pta}%
  \BibitemOpen
  \bibfield  {author} {\bibinfo {author} {\bibfnamefont {S.~W.}\ \bibnamefont
  {Allen}}, \bibinfo {author} {\bibfnamefont {R.~W.}\ \bibnamefont {Schmidt}},
  \ and\ \bibinfo {author} {\bibfnamefont {S.~L.}\ \bibnamefont {Bridle}},\
  }\bibfield  {title} {\enquote {\bibinfo {title} {{A Preference for a non-zero
  neutrino mass from cosmological data}},}\ }\href {\doibase
  10.1046/j.1365-2966.2003.07022.x} {\bibfield  {journal} {\bibinfo  {journal}
  {Mon.\ Not.\ Roy.\ Astron.\ Soc.}\ }\textbf {\bibinfo {volume} {346}},\
  \bibinfo {pages} {593} (\bibinfo {year} {2003})}\BibitemShut {NoStop}%
%%CITATION = ASTRO-PH/0306386;%%
\bibitem [{\citenamefont {Battye}\ and\ \citenamefont
  {Moss}(2014)}]{Battye:2013xqa}%
  \BibitemOpen
  \bibfield  {author} {\bibinfo {author} {\bibfnamefont {R.~A.}\ \bibnamefont
  {Battye}}\ and\ \bibinfo {author} {\bibfnamefont {A.}~\bibnamefont {Moss}},\
  }\bibfield  {title} {\enquote {\bibinfo {title} {{Evidence for Massive
  Neutrinos from Cosmic Microwave Background and Lensing Observations}},}\
  }\href {\doibase 10.1103/PhysRevLett.112.051303} {\bibfield  {journal}
  {\bibinfo  {journal} {Phys.\ Rev.\ Lett.}\ }\textbf {\bibinfo {volume}
  {112}},\ \bibinfo {pages} {051303} (\bibinfo {year} {2014})}\BibitemShut
  {NoStop}%
%%CITATION = ARXIV:1308.5870;%%
\bibitem [{\citenamefont {Palanque-Delabrouille}\ \emph
  {et~al.}(2015{\natexlab{a}})\citenamefont {Palanque-Delabrouille} \emph
  {et~al.}}]{Palanque-Delabrouille:2014jca}%
  \BibitemOpen
  \bibfield  {author} {\bibinfo {author} {\bibfnamefont {N.}~\bibnamefont
  {Palanque-Delabrouille}} \emph {et~al.},\ }\bibfield  {title} {\enquote
  {\bibinfo {title} {{Constraint on neutrino masses from SDSS-III/BOSS
  Ly$\alpha$ forest and other cosmological probes}},}\ }\href {\doibase
  10.1088/1475-7516/2015/02/045} {\bibfield  {journal} {\bibinfo  {journal}
  {J.\ Cosm.\ Astropart.\ Phys.}\ }\textbf {\bibinfo {volume} {1502}},\
  \bibinfo {pages} {045} (\bibinfo {year} {2015}{\natexlab{a}})}\BibitemShut
  {NoStop}%
%%CITATION = ARXIV:1410.7244;%%
\bibitem [{\citenamefont {Lesgourgues}\ \emph {et~al.}(2013)\citenamefont
  {Lesgourgues}, \citenamefont {Mangano}, \citenamefont {Miele},\ and\
  \citenamefont {Pastor}}]{Lesgourgues&al:2013}%
  \BibitemOpen
  \bibfield  {author} {\bibinfo {author} {\bibfnamefont {J.}~\bibnamefont
  {Lesgourgues}}, \bibinfo {author} {\bibfnamefont {G.}~\bibnamefont
  {Mangano}}, \bibinfo {author} {\bibfnamefont {G.}~\bibnamefont {Miele}}, \
  and\ \bibinfo {author} {\bibfnamefont {S.}~\bibnamefont {Pastor}},\ }\href
  {http://www.cambridge.org/it/academic/subjects/physics/%
  particle-physics-and-nuclear-physics/neutrino-cosmology} {\emph {\bibinfo
  {title} {{Neutrino Cosmology}}}}\ (\bibinfo  {publisher} {Cambridge
  University Press},\ \bibinfo {year} {2013})\BibitemShut {NoStop}%
\bibitem [{\citenamefont {Wyman}\ \emph {et~al.}(2014)\citenamefont {Wyman},
  \citenamefont {Rudd}, \citenamefont {Vanderveld},\ and\ \citenamefont
  {Hu}}]{Wyman:2013lza}%
  \BibitemOpen
  \bibfield  {author} {\bibinfo {author} {\bibfnamefont {M.}~\bibnamefont
  {Wyman}}, \bibinfo {author} {\bibfnamefont {D.~H.}\ \bibnamefont {Rudd}},
  \bibinfo {author} {\bibfnamefont {R.~A.}\ \bibnamefont {Vanderveld}}, \ and\
  \bibinfo {author} {\bibfnamefont {W.}~\bibnamefont {Hu}},\ }\bibfield
  {title} {\enquote {\bibinfo {title} {{Neutrinos Help Reconcile Planck
  Measurements with the Local Universe}},}\ }\href {\doibase
  10.1103/PhysRevLett.112.051302} {\bibfield  {journal} {\bibinfo  {journal}
  {Phys.\ Rev.\ Lett.}\ }\textbf {\bibinfo {volume} {112}},\ \bibinfo {pages}
  {051302} (\bibinfo {year} {2014})}\BibitemShut {NoStop}%
%%CITATION = ARXIV:1307.7715;%%
\bibitem [{\citenamefont {Hamann}\ and\ \citenamefont
  {Hasenkamp}(2013)}]{Hamann:2013iba}%
  \BibitemOpen
  \bibfield  {author} {\bibinfo {author} {\bibfnamefont {J.}~\bibnamefont
  {Hamann}}\ and\ \bibinfo {author} {\bibfnamefont {J.}~\bibnamefont
  {Hasenkamp}},\ }\bibfield  {title} {\enquote {\bibinfo {title} {{A new life
  for sterile neutrinos: resolving inconsistencies using hot dark matter}},}\
  }\href {\doibase 10.1088/1475-7516/2013/10/044} {\bibfield  {journal}
  {\bibinfo  {journal} {J.\ Cosm.\ Astropart.\ Phys.}\ }\textbf {\bibinfo
  {volume} {1310}},\ \bibinfo {pages} {044} (\bibinfo {year}
  {2013})}\BibitemShut {NoStop}%
%%CITATION = ARXIV:1308.3255;%%
\bibitem [{\citenamefont {Ade}\ \emph {et~al.}(2014)\citenamefont {Ade} \emph
  {et~al.}}]{Ade:2013zuv}%
  \BibitemOpen
  \bibfield  {author} {\bibinfo {author} {\bibfnamefont {P.~A.~R.}\
  \bibnamefont {Ade}} \emph {et~al.} (\bibinfo {collaboration} {Planck
  Collaboration}),\ }\bibfield  {title} {\enquote {\bibinfo {title} {{Planck
  2013 results. XVI. Cosmological parameters}},}\ }\href {\doibase
  10.1051/0004-6361/201321591} {\bibfield  {journal} {\bibinfo  {journal}
  {Astron.\ Astrophys.}\ }\textbf {\bibinfo {volume} {571}},\ \bibinfo {pages}
  {A16} (\bibinfo {year} {2014})}\BibitemShut {NoStop}%
%%CITATION = ARXIV:1303.5076;%%
\bibitem [{\citenamefont {Kovalenko}\ \emph {et~al.}(2014)\citenamefont
  {Kovalenko}, \citenamefont {Krivoruchenko},\ and\ \citenamefont
  {\u{S}imkovic}}]{Kovalenko:2013eba}%
  \BibitemOpen
  \bibfield  {author} {\bibinfo {author} {\bibfnamefont {S.}~\bibnamefont
  {Kovalenko}}, \bibinfo {author} {\bibfnamefont {M.~I.}\ \bibnamefont
  {Krivoruchenko}}, \ and\ \bibinfo {author} {\bibfnamefont {F.}~\bibnamefont
  {\u{S}imkovic}},\ }\bibfield  {title} {\enquote {\bibinfo {title} {{Neutrino
  propagation in nuclear medium and neutrinoless double-beta decay}},}\ }\href
  {\doibase 10.1103/PhysRevLett.112.142503} {\bibfield  {journal} {\bibinfo
  {journal} {Phys.\ Rev.\ Lett.}\ }\textbf {\bibinfo {volume} {112}},\ \bibinfo
  {pages} {142503} (\bibinfo {year} {2014})}\BibitemShut {NoStop}%
%%CITATION = ARXIV:1311.4200;%%
\bibitem [{\citenamefont {Wong}(2011)}]{Wong:2011ip}%
  \BibitemOpen
  \bibfield  {author} {\bibinfo {author} {\bibfnamefont {Y.~Y.~Y.}\
  \bibnamefont {Wong}},\ }\bibfield  {title} {\enquote {\bibinfo {title}
  {{Neutrino mass in cosmology: status and prospects}},}\ }\href {\doibase
  10.1146/annurev-nucl-102010-130252} {\bibfield  {journal} {\bibinfo
  {journal} {Ann.\ Rev.\ Nucl.\ Part.\ Sci.}\ }\textbf {\bibinfo {volume}
  {61}},\ \bibinfo {pages} {69--98} (\bibinfo {year} {2011})}\BibitemShut
  {NoStop}%
%%CITATION = ARXIV:1111.1436;%%
\bibitem [{\citenamefont {Lesgourgues}\ and\ \citenamefont
  {Pastor}(2014)}]{Lesgourgues:2014zoa}%
  \BibitemOpen
  \bibfield  {author} {\bibinfo {author} {\bibfnamefont {J.}~\bibnamefont
  {Lesgourgues}}\ and\ \bibinfo {author} {\bibfnamefont {S.}~\bibnamefont
  {Pastor}},\ }\bibfield  {title} {\enquote {\bibinfo {title} {{Neutrino
  cosmology and Planck}},}\ }\href {\doibase 10.1088/1367-2630/16/6/065002}
  {\bibfield  {journal} {\bibinfo  {journal} {New J.\ Phys.}\ }\textbf
  {\bibinfo {volume} {16}},\ \bibinfo {pages} {065002} (\bibinfo {year}
  {2014})}\BibitemShut {NoStop}%
%%CITATION = ARXIV:1404.1740;%%
\bibitem [{\citenamefont {Palanque-Delabrouille}\ \emph
  {et~al.}(2015{\natexlab{b}})\citenamefont {Palanque-Delabrouille} \emph
  {et~al.}}]{Palanque-Delabrouille:2015pga}%
  \BibitemOpen
  \bibfield  {author} {\bibinfo {author} {\bibfnamefont {N.}~\bibnamefont
  {Palanque-Delabrouille}} \emph {et~al.},\ }\bibfield  {title} {\enquote
  {\bibinfo {title} {{Neutrino masses and cosmology with Lyman-alpha forest
  power spectrum}},}\ }\href {\doibase 10.1088/1475-7516/2015/11/011}
  {\bibfield  {journal} {\bibinfo  {journal} {J.\ Cosm.\ Astropart.\ Phys.}\
  }\textbf {\bibinfo {volume} {1511}},\ \bibinfo {pages} {011} (\bibinfo {year}
  {2015}{\natexlab{b}})}\BibitemShut {NoStop}%
\bibitem [{\citenamefont {Di~Valentino}\ \emph {et~al.}(2015)\citenamefont
  {Di~Valentino}, \citenamefont {Giusarma}, \citenamefont {Mena}, \citenamefont
  {Melchiorri},\ and\ \citenamefont {Silk}}]{DiValentino:2015sam}%
  \BibitemOpen
  \bibfield  {author} {\bibinfo {author} {\bibfnamefont {E.}~\bibnamefont
  {Di~Valentino}}, \bibinfo {author} {\bibfnamefont {E.}~\bibnamefont
  {Giusarma}}, \bibinfo {author} {\bibfnamefont {O.}~\bibnamefont {Mena}},
  \bibinfo {author} {\bibfnamefont {A.}~\bibnamefont {Melchiorri}}, \ and\
  \bibinfo {author} {\bibfnamefont {J.}~\bibnamefont {Silk}},\ }\bibfield
  {title} {\enquote {\bibinfo {title} {{Cosmological limits on neutrino
  unknowns versus low redshift priors}},}\ }\href@noop {} {\  (\bibinfo {year}
  {2015})},\ \Eprint {http://arxiv.org/abs/1511.00975} {arXiv:1511.00975
  [astro-ph.CO]} \BibitemShut {NoStop}%
%%CITATION = ARXIV:1511.00975;%%
\bibitem [{\citenamefont {Zhang}(2015)}]{Zhang:2015uhk}%
  \BibitemOpen
  \bibfield  {author} {\bibinfo {author} {\bibfnamefont {X.}~\bibnamefont
  {Zhang}},\ }\bibfield  {title} {\enquote {\bibinfo {title} {{Impacts of dark
  energy on weighing neutrinos after Planck 2015}},}\ }\href@noop {} {\
  (\bibinfo {year} {2015})},\ \Eprint {http://arxiv.org/abs/1511.02651}
  {arXiv:1511.02651 [astro-ph.CO]} \BibitemShut {NoStop}%
%%CITATION = ARXIV:1511.02651;%%
\bibitem [{\citenamefont {Cuesta}\ \emph {et~al.}(2015)\citenamefont {Cuesta},
  \citenamefont {Niro},\ and\ \citenamefont {Verde}}]{Cuesta:2015iho}%
  \BibitemOpen
  \bibfield  {author} {\bibinfo {author} {\bibfnamefont {A.~J.}\ \bibnamefont
  {Cuesta}}, \bibinfo {author} {\bibfnamefont {V.}~\bibnamefont {Niro}}, \ and\
  \bibinfo {author} {\bibfnamefont {L.}~\bibnamefont {Verde}},\ }\bibfield
  {title} {\enquote {\bibinfo {title} {{Neutrino mass limits: robust
  information from the power spectrum of galaxy surveys}},}\ }\href@noop {} {\
  (\bibinfo {year} {2015})},\ \Eprint {http://arxiv.org/abs/1511.05983}
  {arXiv:1511.05983 [astro-ph.CO]} \BibitemShut {NoStop}%
%%CITATION = ARXIV:1511.05983;%%
\bibitem [{\citenamefont {Loredo}\ and\ \citenamefont
  {Lamb}(2002)}]{Loredo:2001rx}%
  \BibitemOpen
  \bibfield  {author} {\bibinfo {author} {\bibfnamefont {T.~J.}\ \bibnamefont
  {Loredo}}\ and\ \bibinfo {author} {\bibfnamefont {D.~Q.}\ \bibnamefont
  {Lamb}},\ }\bibfield  {title} {\enquote {\bibinfo {title} {{Bayesian analysis
  of neutrinos observed from supernova SN-1987A}},}\ }\href {\doibase
  10.1103/PhysRevD.65.063002} {\bibfield  {journal} {\bibinfo  {journal}
  {Phys.\ Rev.\ D}\ }\textbf {\bibinfo {volume} {65}},\ \bibinfo {pages}
  {063002} (\bibinfo {year} {2002})}\BibitemShut {NoStop}%
%%CITATION = ASTRO-PH/0107260;%%
\bibitem [{\citenamefont {Pagliaroli}\ \emph {et~al.}(2010)\citenamefont
  {Pagliaroli}, \citenamefont {Rossi-Torres},\ and\ \citenamefont
  {Vissani}}]{Pagliaroli:2010ik}%
  \BibitemOpen
  \bibfield  {author} {\bibinfo {author} {\bibfnamefont {G.}~\bibnamefont
  {Pagliaroli}}, \bibinfo {author} {\bibfnamefont {F.}~\bibnamefont
  {Rossi-Torres}}, \ and\ \bibinfo {author} {\bibfnamefont {F.}~\bibnamefont
  {Vissani}},\ }\bibfield  {title} {\enquote {\bibinfo {title} {{Neutrino mass
  bound in the standard scenario for supernova electronic antineutrino
  emission}},}\ }\href {\doibase 10.1016/j.astropartphys.2010.02.007}
  {\bibfield  {journal} {\bibinfo  {journal} {Astropart.\ Phys.}\ }\textbf
  {\bibinfo {volume} {33}},\ \bibinfo {pages} {287--291} (\bibinfo {year}
  {2010})}\BibitemShut {NoStop}%
%%CITATION = ARXIV:1002.3349;%%
\bibitem [{\citenamefont {Kraus}\ \emph {et~al.}(2005)\citenamefont {Kraus}
  \emph {et~al.}}]{Kraus:2004zw}%
  \BibitemOpen
  \bibfield  {author} {\bibinfo {author} {\bibfnamefont {C.}~\bibnamefont
  {Kraus}} \emph {et~al.},\ }\bibfield  {title} {\enquote {\bibinfo {title}
  {{Final results from phase II of the Mainz neutrino mass search in tritium
  beta decay}},}\ }\href {\doibase 10.1140/epjc/s2005-02139-7} {\bibfield
  {journal} {\bibinfo  {journal} {Eur.\ Phys.\ J.\ C}\ }\textbf {\bibinfo
  {volume} {40}},\ \bibinfo {pages} {447--468} (\bibinfo {year}
  {2005})}\BibitemShut {NoStop}%
%%CITATION = HEP-EX/0412056;%%
\bibitem [{\citenamefont {Aseev}\ \emph {et~al.}(2011)\citenamefont {Aseev}
  \emph {et~al.}}]{Aseev:2011dq}%
  \BibitemOpen
  \bibfield  {author} {\bibinfo {author} {\bibfnamefont {V.~N.}\ \bibnamefont
  {Aseev}} \emph {et~al.},\ }\bibfield  {title} {\enquote {\bibinfo {title}
  {{An upper limit on electron antineutrino mass from Troitsk experiment}},}\
  }\href {\doibase 10.1103/PhysRevD.84.112003} {\bibfield  {journal} {\bibinfo
  {journal} {Phys.\ Rev.\ D}\ }\textbf {\bibinfo {volume} {84}},\ \bibinfo
  {pages} {112003} (\bibinfo {year} {2011})}\BibitemShut {NoStop}%
%%CITATION = ARXIV:1108.5034;%%
\bibitem [{\citenamefont {Osipowicz}\ \emph {et~al.}(2001)\citenamefont
  {Osipowicz} \emph {et~al.}}]{Osipowicz:2001sq}%
  \BibitemOpen
  \bibfield  {author} {\bibinfo {author} {\bibfnamefont {A.}~\bibnamefont
  {Osipowicz}} \emph {et~al.} (\bibinfo {collaboration} {KATRIN
  Collaboration}),\ }\bibfield  {title} {\enquote {\bibinfo {title} {{KATRIN: A
  Next generation tritium beta decay experiment with sub-eV sensitivity for the
  electron neutrino mass. Letter of intent}},}\ }\href@noop {} {\  (\bibinfo
  {year} {2001})},\ \Eprint {http://arxiv.org/abs/hep-ex/0109033}
  {arXiv:hep-ex/0109033 [hep-ex]} \BibitemShut {NoStop}%
%%CITATION = HEP-EX/0109033;%%
\bibitem [{\citenamefont {Alpert}\ \emph {et~al.}(2015)\citenamefont {Alpert}
  \emph {et~al.}}]{Alpert:2014lfa}%
  \BibitemOpen
  \bibfield  {author} {\bibinfo {author} {\bibfnamefont {B.}~\bibnamefont
  {Alpert}} \emph {et~al.},\ }\bibfield  {title} {\enquote {\bibinfo {title}
  {{HOLMES - The Electron Capture Decay of $^{163}$Ho to Measure the Electron
  Neutrino Mass with sub-eV sensitivity}},}\ }\href {\doibase
  10.1140/epjc/s10052-015-3329-5} {\bibfield  {journal} {\bibinfo  {journal}
  {Eur.\ Phys.\ J.\ C}\ }\textbf {\bibinfo {volume} {75}},\ \bibinfo {pages}
  {112} (\bibinfo {year} {2015})}\BibitemShut {NoStop}%
%%CITATION = ARXIV:1412.5060;%%
\bibitem [{\citenamefont {Doe}\ \emph {et~al.}(2013)\citenamefont {Doe} \emph
  {et~al.}}]{Doe:2013jfe}%
  \BibitemOpen
  \bibfield  {author} {\bibinfo {author} {\bibfnamefont {P.~J.}\ \bibnamefont
  {Doe}} \emph {et~al.} (\bibinfo {collaboration} {Project 8 Collaboration}),\
  }\bibfield  {title} {\enquote {\bibinfo {title} {{Project 8: Determining
  neutrino mass from tritium beta decay using a frequency-based method}},}\
  }\href@noop {} {\bibfield  {journal} {\bibinfo  {journal} {Conf.\ Proc.\
  C13-07-29.2}\ }\textbf {\bibinfo {volume} {\!\!}} (\bibinfo {year} {2013})},\
  \Eprint {http://arxiv.org/abs/1309.7093} {arXiv:1309.7093 [nucl-ex]}
  \BibitemShut {NoStop}%
%%CITATION = ARXIV:1309.7093;%%
\bibitem [{\citenamefont {Ranitzsch}\ \emph {et~al.}(2012)\citenamefont
  {Ranitzsch}, \citenamefont {Porst}, \citenamefont {Kempf}, \citenamefont
  {Pies}, \citenamefont {Schafer}, \citenamefont {Hengstler}, \citenamefont
  {Fleischmann}, \citenamefont {Enss},\ and\ \citenamefont
  {Gastaldo}}]{Ranitzsch:2012}%
  \BibitemOpen
  \bibfield  {author} {\bibinfo {author} {\bibfnamefont {P.~C.~O.}\
  \bibnamefont {Ranitzsch}}, \bibinfo {author} {\bibfnamefont {J.~P.}\
  \bibnamefont {Porst}}, \bibinfo {author} {\bibfnamefont {S.}~\bibnamefont
  {Kempf}}, \bibinfo {author} {\bibfnamefont {C.}~\bibnamefont {Pies}},
  \bibinfo {author} {\bibfnamefont {S.}~\bibnamefont {Schafer}}, \bibinfo
  {author} {\bibfnamefont {D.}~\bibnamefont {Hengstler}}, \bibinfo {author}
  {\bibfnamefont {A.}~\bibnamefont {Fleischmann}}, \bibinfo {author}
  {\bibfnamefont {C.}~\bibnamefont {Enss}}, \ and\ \bibinfo {author}
  {\bibfnamefont {L.}~\bibnamefont {Gastaldo}},\ }\bibfield  {title} {\enquote
  {\bibinfo {title} {{Development of Metallic Magnetic Calorimeters for High
  Precision Measurements of Calorimetric \ce{^{187}Re} and \ce{^{163}Ho}
  Spectra}},}\ }\href {\doibase 10.1007/s10909-012-0556-0} {\bibfield
  {journal} {\bibinfo  {journal} {J.\ Low Temp.\ Phys.}\ }\textbf {\bibinfo
  {volume} {167}},\ \bibinfo {pages} {1004--1014} (\bibinfo {year}
  {2012})}\BibitemShut {NoStop}%
\bibitem [{\citenamefont {Buccella}\ and\ \citenamefont
  {Falcone}(2004)}]{Buccella:2004cd}%
  \BibitemOpen
  \bibfield  {author} {\bibinfo {author} {\bibfnamefont {F.}~\bibnamefont
  {Buccella}}\ and\ \bibinfo {author} {\bibfnamefont {D.}~\bibnamefont
  {Falcone}},\ }\bibfield  {title} {\enquote {\bibinfo {title} {{Bounds for
  neutrinoless double beta decay in SO(10) inspired see-saw models}},}\ }\href
  {\doibase 10.1142/S0217732304016317} {\bibfield  {journal} {\bibinfo
  {journal} {Mod.\ Phys.\ Lett.\ A}\ }\textbf {\bibinfo {volume} {19}},\
  \bibinfo {pages} {2993} (\bibinfo {year} {2004})}\BibitemShut {NoStop}%
%%CITATION = HEP-PH/0404159;%%
\bibitem [{\citenamefont {Pascoli}\ and\ \citenamefont
  {Petcov}(2008)}]{Pascoli:2007qh}%
  \BibitemOpen
  \bibfield  {author} {\bibinfo {author} {\bibfnamefont {S.}~\bibnamefont
  {Pascoli}}\ and\ \bibinfo {author} {\bibfnamefont {S.~T.}\ \bibnamefont
  {Petcov}},\ }\bibfield  {title} {\enquote {\bibinfo {title} {{Majorana
  neutrinos, neutrino mass spectrum and the $|\left< m \right> | \sim
  10^{-3}$\,eV frontier in neutrinoless double beta decay}},}\ }\href {\doibase
  10.1103/PhysRevD.77.113003} {\bibfield  {journal} {\bibinfo  {journal}
  {Phys.\ Rev.\ D}\ }\textbf {\bibinfo {volume} {77}},\ \bibinfo {pages}
  {113003} (\bibinfo {year} {2008})}\BibitemShut {NoStop}%
%%CITATION = ARXIV:0711.4993;%%
\bibitem [{\citenamefont {Froggatt}\ and\ \citenamefont
  {Nielsen}(1979)}]{Froggatt:1978nt}%
  \BibitemOpen
  \bibfield  {author} {\bibinfo {author} {\bibfnamefont {C.~D.}\ \bibnamefont
  {Froggatt}}\ and\ \bibinfo {author} {\bibfnamefont {H.~B.}\ \bibnamefont
  {Nielsen}},\ }\bibfield  {title} {\enquote {\bibinfo {title} {{Hierarchy of
  Quark Masses, Cabibbo Angles and CP Violation}},}\ }\href {\doibase
  10.1016/0550-3213(79)90316-X} {\bibfield  {journal} {\bibinfo  {journal}
  {Nucl.\ Phys.\ B}\ }\textbf {\bibinfo {volume} {147}},\ \bibinfo {pages}
  {277} (\bibinfo {year} {1979})}\BibitemShut {NoStop}%
%%CITATION = NUPHA,B147,277;%%
\bibitem [{\citenamefont {Vissani}(2001)}]{Vissani:2001im}%
  \BibitemOpen
  \bibfield  {author} {\bibinfo {author} {\bibfnamefont {F.}~\bibnamefont
  {Vissani}},\ }\bibfield  {title} {\enquote {\bibinfo {title} {{Expected
  properties of massive neutrinos for mass matrices with a dominant block and
  random coefficients order unity}},}\ }\href {\doibase
  10.1016/S0370-2693(01)00485-3} {\bibfield  {journal} {\bibinfo  {journal}
  {Phys.\ Lett.\ B}\ }\textbf {\bibinfo {volume} {508}},\ \bibinfo {pages}
  {79--84} (\bibinfo {year} {2001})}\BibitemShut {NoStop}%
%%CITATION = HEP-PH/0102236;%%
\bibitem [{\citenamefont {Vissani}\ \emph {et~al.}(2003)\citenamefont
  {Vissani}, \citenamefont {Narayan},\ and\ \citenamefont
  {Berezinsky}}]{Vissani:2003aj}%
  \BibitemOpen
  \bibfield  {author} {\bibinfo {author} {\bibfnamefont {F.}~\bibnamefont
  {Vissani}}, \bibinfo {author} {\bibfnamefont {M.}~\bibnamefont {Narayan}}, \
  and\ \bibinfo {author} {\bibfnamefont {V.}~\bibnamefont {Berezinsky}},\
  }\bibfield  {title} {\enquote {\bibinfo {title} {{$U_{e3}$ from physics above
  the GUT scale}},}\ }\href {\doibase 10.1016/j.physletb.2003.07.002}
  {\bibfield  {journal} {\bibinfo  {journal} {Phys.\ Lett.\ B}\ }\textbf
  {\bibinfo {volume} {571}},\ \bibinfo {pages} {209--216} (\bibinfo {year}
  {2003})}\BibitemShut {NoStop}%
%%CITATION = HEP-PH/0305233;%%
\bibitem [{\citenamefont {Primakoff}\ and\ \citenamefont
  {Rosen}(1959)}]{Primakoff:1959}%
  \BibitemOpen
  \bibfield  {author} {\bibinfo {author} {\bibfnamefont {H.}~\bibnamefont
  {Primakoff}}\ and\ \bibinfo {author} {\bibfnamefont {S.~P.}\ \bibnamefont
  {Rosen}},\ }\bibfield  {title} {\enquote {\bibinfo {title} {{Double beta
  decay}},}\ }\href {http://stacks.iop.org/0034-4885/22/i=1/a=305} {\bibfield
  {journal} {\bibinfo  {journal} {Rep.\ Prog.\ Phys.}\ }\textbf {\bibinfo
  {volume} {22}},\ \bibinfo {pages} {121} (\bibinfo {year} {1959})}\BibitemShut
  {NoStop}%
\bibitem [{\citenamefont {Doi}\ \emph {et~al.}(1981)\citenamefont {Doi},
  \citenamefont {Kotani}, \citenamefont {Nishiura}, \citenamefont {Okuda},\
  and\ \citenamefont {Takasugi}}]{Doi:1981mj}%
  \BibitemOpen
  \bibfield  {author} {\bibinfo {author} {\bibfnamefont {M.}~\bibnamefont
  {Doi}}, \bibinfo {author} {\bibfnamefont {T.}~\bibnamefont {Kotani}},
  \bibinfo {author} {\bibfnamefont {H.}~\bibnamefont {Nishiura}}, \bibinfo
  {author} {\bibfnamefont {K.}~\bibnamefont {Okuda}}, \ and\ \bibinfo {author}
  {\bibfnamefont {E.}~\bibnamefont {Takasugi}},\ }\bibfield  {title} {\enquote
  {\bibinfo {title} {{Neutrino Mass, the Right-handed Interaction and the
  Double Beta Decay. 2. General Properties and Data Analysis}},}\ }\href
  {\doibase 10.1143/PTP.66.1765} {\bibfield  {journal} {\bibinfo  {journal}
  {Prog.\ Theor.\ Phys.}\ }\textbf {\bibinfo {volume} {66}},\ \bibinfo {pages}
  {1765} (\bibinfo {year} {1981})},\ \bibinfo {note} {[Erratum:
  \href{http://ptp.oxfordjournals.org/content/68/1/348} {Prog.\ Theor.\ Phys.
  {\bf 68}, 348 (1982)}]}\BibitemShut {NoStop}%
%%CITATION = PTPKA,66,1765;%%
\bibitem [{\citenamefont {Doi}\ \emph {et~al.}(1983)\citenamefont {Doi},
  \citenamefont {Kotani}, \citenamefont {Nishiura},\ and\ \citenamefont
  {Takasugi}}]{Doi:1982dn}%
  \BibitemOpen
  \bibfield  {author} {\bibinfo {author} {\bibfnamefont {M.}~\bibnamefont
  {Doi}}, \bibinfo {author} {\bibfnamefont {T.}~\bibnamefont {Kotani}},
  \bibinfo {author} {\bibfnamefont {H.}~\bibnamefont {Nishiura}}, \ and\
  \bibinfo {author} {\bibfnamefont {E.}~\bibnamefont {Takasugi}},\ }\bibfield
  {title} {\enquote {\bibinfo {title} {{Double Beta Decay}},}\ }\href {\doibase
  10.1143/PTP.69.602} {\bibfield  {journal} {\bibinfo  {journal} {Prog.\
  Theor.\ Phys.}\ }\textbf {\bibinfo {volume} {69}},\ \bibinfo {pages} {602}
  (\bibinfo {year} {1983})}\BibitemShut {NoStop}%
%%CITATION = PTPKA,69,602;%%
\bibitem [{\citenamefont {Tomoda}(1991)}]{Tomoda:1990rs}%
  \BibitemOpen
  \bibfield  {author} {\bibinfo {author} {\bibfnamefont {T.}~\bibnamefont
  {Tomoda}},\ }\bibfield  {title} {\enquote {\bibinfo {title} {{Double beta
  decay}},}\ }\href {\doibase 10.1088/0034-4885/54/1/002} {\bibfield  {journal}
  {\bibinfo  {journal} {Rep.\ Prog.\ Phys.}\ }\textbf {\bibinfo {volume}
  {54}},\ \bibinfo {pages} {53--126} (\bibinfo {year} {1991})}\BibitemShut
  {NoStop}%
%%CITATION = RPPHA,54,53;%%
\bibitem [{\citenamefont {Kotila}\ and\ \citenamefont
  {Iachello}(2012)}]{Kotila:2012zza}%
  \BibitemOpen
  \bibfield  {author} {\bibinfo {author} {\bibfnamefont {J.}~\bibnamefont
  {Kotila}}\ and\ \bibinfo {author} {\bibfnamefont {F.}~\bibnamefont
  {Iachello}},\ }\bibfield  {title} {\enquote {\bibinfo {title} {{Phase space
  factors for double-$\beta$ decay}},}\ }\href {\doibase
  10.1103/PhysRevC.85.034316} {\bibfield  {journal} {\bibinfo  {journal}
  {Phys.\ Rev.\ C}\ }\textbf {\bibinfo {volume} {85}},\ \bibinfo {pages}
  {034316} (\bibinfo {year} {2012})}\BibitemShut {NoStop}%
%%CITATION = ARXIV:1209.5722;%%
\bibitem [{\citenamefont {Stoica}\ and\ \citenamefont
  {Mirea}(2013)}]{Stoica:2013lka}%
  \BibitemOpen
  \bibfield  {author} {\bibinfo {author} {\bibfnamefont {S.}~\bibnamefont
  {Stoica}}\ and\ \bibinfo {author} {\bibfnamefont {M.}~\bibnamefont {Mirea}},\
  }\bibfield  {title} {\enquote {\bibinfo {title} {{New calculations for phase
  space factors involved in double-$\beta$ decay}},}\ }\href {\doibase
  10.1103/PhysRevC.88.037303} {\bibfield  {journal} {\bibinfo  {journal}
  {Phys.\ Rev.\ C}\ }\textbf {\bibinfo {volume} {88}},\ \bibinfo {pages}
  {037303} (\bibinfo {year} {2013})}\BibitemShut {NoStop}%
%%CITATION = ARXIV:1307.0290;%%
\bibitem [{\citenamefont {\u{S}tef\'anik}\ \emph {et~al.}(2015)\citenamefont
  {\u{S}tef\'anik}, \citenamefont {Dvornick\'y}, \citenamefont {\u{S}imkovic},\
  and\ \citenamefont {Vogel}}]{Stefanik:2015twa}%
  \BibitemOpen
  \bibfield  {author} {\bibinfo {author} {\bibfnamefont {D.}~\bibnamefont
  {\u{S}tef\'anik}}, \bibinfo {author} {\bibfnamefont {R.}~\bibnamefont
  {Dvornick\'y}}, \bibinfo {author} {\bibfnamefont {F.}~\bibnamefont
  {\u{S}imkovic}}, \ and\ \bibinfo {author} {\bibfnamefont {P.}~\bibnamefont
  {Vogel}},\ }\bibfield  {title} {\enquote {\bibinfo {title} {{Reexamining the
  light neutrino exchange mechanism of the $0\nu\beta\beta$ decay with left-
  and right-handed leptonic and hadronic currents}},}\ }\href {\doibase
  10.1103/PhysRevC.92.055502} {\bibfield  {journal} {\bibinfo  {journal}
  {Phys.\ Rev.\ C}\ }\textbf {\bibinfo {volume} {92}},\ \bibinfo {pages}
  {055502} (\bibinfo {year} {2015})}\BibitemShut {NoStop}%
%%CITATION = ARXIV:1506.07145;%%
\bibitem [{\citenamefont {Barea}\ \emph {et~al.}(2015)\citenamefont {Barea},
  \citenamefont {Kotila},\ and\ \citenamefont {Iachello}}]{Barea:2015kwa}%
  \BibitemOpen
  \bibfield  {author} {\bibinfo {author} {\bibfnamefont {J.}~\bibnamefont
  {Barea}}, \bibinfo {author} {\bibfnamefont {J.}~\bibnamefont {Kotila}}, \
  and\ \bibinfo {author} {\bibfnamefont {F.}~\bibnamefont {Iachello}},\
  }\bibfield  {title} {\enquote {\bibinfo {title} {{$0\nu\beta\beta$ and
  $2\nu\beta\beta$ nuclear matrix elements in the interacting boson model with
  isospin restoration}},}\ }\href {\doibase 10.1103/PhysRevC.91.034304}
  {\bibfield  {journal} {\bibinfo  {journal} {Phys.\ Rev.\ C}\ }\textbf
  {\bibinfo {volume} {91}},\ \bibinfo {pages} {034304} (\bibinfo {year}
  {2015})}\BibitemShut {NoStop}%
%%CITATION = PHRVA,C91,034304;%%
\bibitem [{\citenamefont {\u{S}imkovic}\ \emph
  {et~al.}(2013{\natexlab{a}})\citenamefont {\u{S}imkovic}, \citenamefont
  {Rodin}, \citenamefont {Faessler},\ and\ \citenamefont
  {Vogel}}]{Simkovic:2013qiy}%
  \BibitemOpen
  \bibfield  {author} {\bibinfo {author} {\bibfnamefont {F.}~\bibnamefont
  {\u{S}imkovic}}, \bibinfo {author} {\bibfnamefont {V.}~\bibnamefont {Rodin}},
  \bibinfo {author} {\bibfnamefont {A.}~\bibnamefont {Faessler}}, \ and\
  \bibinfo {author} {\bibfnamefont {P.}~\bibnamefont {Vogel}},\ }\bibfield
  {title} {\enquote {\bibinfo {title} {{$0\nu\beta\beta$ and $2\nu\beta\beta$
  nuclear matrix elements, quasiparticle random-phase approximation, and
  isospin symmetry restoration}},}\ }\href {\doibase
  10.1103/PhysRevC.87.045501} {\bibfield  {journal} {\bibinfo  {journal}
  {Phys.\ Rev.\ C}\ }\textbf {\bibinfo {volume} {87}},\ \bibinfo {pages}
  {045501} (\bibinfo {year} {2013}{\natexlab{a}})}\BibitemShut {NoStop}%
%%CITATION = ARXIV:1302.1509;%%
\bibitem [{\citenamefont {Men\'{e}ndez}\ \emph {et~al.}(2009)\citenamefont
  {Men\'{e}ndez}, \citenamefont {Poves}, \citenamefont {Caurier},\ and\
  \citenamefont {Nowacki}}]{Menendez:2008jp}%
  \BibitemOpen
  \bibfield  {author} {\bibinfo {author} {\bibfnamefont {J.}~\bibnamefont
  {Men\'{e}ndez}}, \bibinfo {author} {\bibfnamefont {A.}~\bibnamefont {Poves}},
  \bibinfo {author} {\bibfnamefont {E.}~\bibnamefont {Caurier}}, \ and\
  \bibinfo {author} {\bibfnamefont {F.}~\bibnamefont {Nowacki}},\ }\bibfield
  {title} {\enquote {\bibinfo {title} {{Disassembling the Nuclear Matrix
  Elements of the Neutrinoless beta beta Decay}},}\ }\href {\doibase
  10.1016/j.nuclphysa.2008.12.005} {\bibfield  {journal} {\bibinfo  {journal}
  {Nucl.\ Phys.\ A}\ }\textbf {\bibinfo {volume} {818}},\ \bibinfo {pages}
  {139--151} (\bibinfo {year} {2009})}\BibitemShut {NoStop}%
%%CITATION = ARXIV:0801.3760;%%
\bibitem [{\citenamefont {Caurier}\ \emph {et~al.}(2005)\citenamefont
  {Caurier}, \citenamefont {Martinez-Pinedo}, \citenamefont {Nowacki},
  \citenamefont {Poves},\ and\ \citenamefont {Zuker}}]{Caurier:2004gf}%
  \BibitemOpen
  \bibfield  {author} {\bibinfo {author} {\bibfnamefont {E.}~\bibnamefont
  {Caurier}}, \bibinfo {author} {\bibfnamefont {G.}~\bibnamefont
  {Martinez-Pinedo}}, \bibinfo {author} {\bibfnamefont {F.}~\bibnamefont
  {Nowacki}}, \bibinfo {author} {\bibfnamefont {A.}~\bibnamefont {Poves}}, \
  and\ \bibinfo {author} {\bibfnamefont {A.~P.}\ \bibnamefont {Zuker}},\
  }\bibfield  {title} {\enquote {\bibinfo {title} {{The Shell model as unified
  view of nuclear structure}},}\ }\href {\doibase 10.1103/RevModPhys.77.427}
  {\bibfield  {journal} {\bibinfo  {journal} {Rev.\ Mod.\ Phys.}\ }\textbf
  {\bibinfo {volume} {77}},\ \bibinfo {pages} {427--488} (\bibinfo {year}
  {2005})}\BibitemShut {NoStop}%
%%CITATION = NUCL-TH/0402046;%%
\bibitem [{\citenamefont {Hyvarinen}\ and\ \citenamefont
  {Suhonen}(2015)}]{Hyvarinen:2015bda}%
  \BibitemOpen
  \bibfield  {author} {\bibinfo {author} {\bibfnamefont {J.}~\bibnamefont
  {Hyvarinen}}\ and\ \bibinfo {author} {\bibfnamefont {J.}~\bibnamefont
  {Suhonen}},\ }\bibfield  {title} {\enquote {\bibinfo {title} {{Nuclear matrix
  elements for $0\nu\beta\beta$ decays with light or heavy Majorana-neutrino
  exchange}},}\ }\href {\doibase 10.1103/PhysRevC.91.024613} {\bibfield
  {journal} {\bibinfo  {journal} {Phys.\ Rev.\ C}\ }\textbf {\bibinfo {volume}
  {91}},\ \bibinfo {pages} {024613} (\bibinfo {year} {2015})}\BibitemShut
  {NoStop}%
%%CITATION = PHRVA,C91,024613;%%
\bibitem [{\citenamefont {Rath}\ \emph {et~al.}(2013)\citenamefont {Rath},
  \citenamefont {Chandra}, \citenamefont {Chaturvedi}, \citenamefont {Lohani},
  \citenamefont {Raina},\ and\ \citenamefont {Hirsch}}]{Rath:2013fma}%
  \BibitemOpen
  \bibfield  {author} {\bibinfo {author} {\bibfnamefont {P.~K.}\ \bibnamefont
  {Rath}}, \bibinfo {author} {\bibfnamefont {R.}~\bibnamefont {Chandra}},
  \bibinfo {author} {\bibfnamefont {K.}~\bibnamefont {Chaturvedi}}, \bibinfo
  {author} {\bibfnamefont {P.}~\bibnamefont {Lohani}}, \bibinfo {author}
  {\bibfnamefont {P.~K.}\ \bibnamefont {Raina}}, \ and\ \bibinfo {author}
  {\bibfnamefont {J.~G.}\ \bibnamefont {Hirsch}},\ }\bibfield  {title}
  {\enquote {\bibinfo {title} {{Neutrinoless $\beta\beta$ decay transition
  matrix elements within mechanisms involving light Majorana neutrinos,
  classical Majorons, and sterile neutrinos}},}\ }\href {\doibase
  10.1103/PhysRevC.88.064322} {\bibfield  {journal} {\bibinfo  {journal}
  {Phys.\ Rev.\ C}\ }\textbf {\bibinfo {volume} {88}},\ \bibinfo {pages}
  {064322} (\bibinfo {year} {2013})}\BibitemShut {NoStop}%
%%CITATION = ARXIV:1308.0460;%%
\bibitem [{\citenamefont {Rodriguez}\ and\ \citenamefont
  {Martinez-Pinedo}(2010)}]{Rodriguez:2010mn}%
  \BibitemOpen
  \bibfield  {author} {\bibinfo {author} {\bibfnamefont {T.~R.}\ \bibnamefont
  {Rodriguez}}\ and\ \bibinfo {author} {\bibfnamefont {G.}~\bibnamefont
  {Martinez-Pinedo}},\ }\bibfield  {title} {\enquote {\bibinfo {title} {{Energy
  density functional study of nuclear matrix elements for neutrinoless
  $\beta\beta$ decay}},}\ }\href {\doibase 10.1103/PhysRevLett.105.252503}
  {\bibfield  {journal} {\bibinfo  {journal} {Phys.\ Rev.\ Lett.}\ }\textbf
  {\bibinfo {volume} {105}},\ \bibinfo {pages} {252503} (\bibinfo {year}
  {2010})}\BibitemShut {NoStop}%
%%CITATION = ARXIV:1008.5260;%%
\bibitem [{\citenamefont {\u{S}imkovic}\ \emph {et~al.}(1999)\citenamefont
  {\u{S}imkovic}, \citenamefont {Pantis}, \citenamefont {Vergados},\ and\
  \citenamefont {Faessler}}]{Simkovic:1999re}%
  \BibitemOpen
  \bibfield  {author} {\bibinfo {author} {\bibfnamefont {F.}~\bibnamefont
  {\u{S}imkovic}}, \bibinfo {author} {\bibfnamefont {G.}~\bibnamefont
  {Pantis}}, \bibinfo {author} {\bibfnamefont {J.~D.}\ \bibnamefont
  {Vergados}}, \ and\ \bibinfo {author} {\bibfnamefont {A.}~\bibnamefont
  {Faessler}},\ }\bibfield  {title} {\enquote {\bibinfo {title} {{Additional
  nucleon current contributions to neutrinoless double beta decay}},}\ }\href
  {\doibase 10.1103/PhysRevC.60.055502} {\bibfield  {journal} {\bibinfo
  {journal} {Phys.\ Rev.\ C}\ }\textbf {\bibinfo {volume} {60}},\ \bibinfo
  {pages} {055502} (\bibinfo {year} {1999})}\BibitemShut {NoStop}%
%%CITATION = HEP-PH/9905509;%%
\bibitem [{\citenamefont {Barea}\ \emph {et~al.}(2013)\citenamefont {Barea},
  \citenamefont {Kotila},\ and\ \citenamefont {Iachello}}]{Barea:2013bz}%
  \BibitemOpen
  \bibfield  {author} {\bibinfo {author} {\bibfnamefont {J.}~\bibnamefont
  {Barea}}, \bibinfo {author} {\bibfnamefont {J.}~\bibnamefont {Kotila}}, \
  and\ \bibinfo {author} {\bibfnamefont {F.}~\bibnamefont {Iachello}},\
  }\bibfield  {title} {\enquote {\bibinfo {title} {{Nuclear matrix elements for
  double-$\beta$ decay}},}\ }\href {\doibase 10.1103/PhysRevC.87.014315}
  {\bibfield  {journal} {\bibinfo  {journal} {Phys.\ Rev.\ C}\ }\textbf
  {\bibinfo {volume} {87}},\ \bibinfo {pages} {014315} (\bibinfo {year}
  {2013})}\BibitemShut {NoStop}%
%%CITATION = ARXIV:1301.4203;%%
\bibitem [{\citenamefont {Bahcall}\ \emph {et~al.}(2004)\citenamefont
  {Bahcall}, \citenamefont {Murayama},\ and\ \citenamefont
  {Pena-Garay}}]{Bahcall:2004ip}%
  \BibitemOpen
  \bibfield  {author} {\bibinfo {author} {\bibfnamefont {J.N.}\ \bibnamefont
  {Bahcall}}, \bibinfo {author} {\bibfnamefont {H.}~\bibnamefont {Murayama}}, \
  and\ \bibinfo {author} {\bibfnamefont {C.}~\bibnamefont {Pena-Garay}},\
  }\bibfield  {title} {\enquote {\bibinfo {title} {{What can we learn from
  neutrinoless double beta decay experiments?}}}\ }\href {\doibase
  10.1103/PhysRevD.70.033012} {\bibfield  {journal} {\bibinfo  {journal}
  {Phys.\ Rev.\ D}\ }\textbf {\bibinfo {volume} {70}},\ \bibinfo {pages}
  {033012} (\bibinfo {year} {2004})}\BibitemShut {NoStop}%
%%CITATION = HEP-PH/0403167;%%
\bibitem [{\citenamefont {Rodin}\ \emph {et~al.}(2003)\citenamefont {Rodin},
  \citenamefont {Faessler}, \citenamefont {\u{S}imkovic},\ and\ \citenamefont
  {Vogel}}]{Rodin:2003eb}%
  \BibitemOpen
  \bibfield  {author} {\bibinfo {author} {\bibfnamefont {V.~A.}\ \bibnamefont
  {Rodin}}, \bibinfo {author} {\bibfnamefont {A.}~\bibnamefont {Faessler}},
  \bibinfo {author} {\bibfnamefont {F.}~\bibnamefont {\u{S}imkovic}}, \ and\
  \bibinfo {author} {\bibfnamefont {P.}~\bibnamefont {Vogel}},\ }\bibfield
  {title} {\enquote {\bibinfo {title} {{On the uncertainty in the
  $0\nu\beta\beta$ decay nuclear matrix elements}},}\ }\href {\doibase
  10.1103/PhysRevC.68.044302} {\bibfield  {journal} {\bibinfo  {journal}
  {Phys.\ Rev.\ C}\ }\textbf {\bibinfo {volume} {68}},\ \bibinfo {pages}
  {044302} (\bibinfo {year} {2003})}\BibitemShut {NoStop}%
%%CITATION = NUCL-TH/0305005;%%
\bibitem [{\citenamefont {Rodin}\ \emph {et~al.}(2006)\citenamefont {Rodin},
  \citenamefont {Faessler}, \citenamefont {\u{S}imkovic},\ and\ \citenamefont
  {Vogel}}]{Rodin:2006yk}%
  \BibitemOpen
  \bibfield  {author} {\bibinfo {author} {\bibfnamefont {V.~A.}\ \bibnamefont
  {Rodin}}, \bibinfo {author} {\bibfnamefont {A.}~\bibnamefont {Faessler}},
  \bibinfo {author} {\bibfnamefont {F.}~\bibnamefont {\u{S}imkovic}}, \ and\
  \bibinfo {author} {\bibfnamefont {P.}~\bibnamefont {Vogel}},\ }\bibfield
  {title} {\enquote {\bibinfo {title} {{Assessment of uncertainties in QRPA
  $0\nu\beta\beta$ nuclear matrix elements}},}\ }\href {\doibase
  10.1016/j.nuclphysa.2005.12.004} {\bibfield  {journal} {\bibinfo  {journal}
  {Nucl.\ Phys.\ A}\ }\textbf {\bibinfo {volume} {766}},\ \bibinfo {pages}
  {107--131} (\bibinfo {year} {2006})}\BibitemShut {NoStop}%
%%CITATION = ARXIV:0706.4304;%%
\bibitem [{\citenamefont {Faessler}\ \emph {et~al.}(2014)\citenamefont
  {Faessler}, \citenamefont {Gonz\'alez}, \citenamefont {Kovalenko},\ and\
  \citenamefont {\u{S}imkovic}}]{Faessler:2014kka}%
  \BibitemOpen
  \bibfield  {author} {\bibinfo {author} {\bibfnamefont {A.}~\bibnamefont
  {Faessler}}, \bibinfo {author} {\bibfnamefont {M.}~\bibnamefont
  {Gonz\'alez}}, \bibinfo {author} {\bibfnamefont {S.}~\bibnamefont
  {Kovalenko}}, \ and\ \bibinfo {author} {\bibfnamefont {F.}~\bibnamefont
  {\u{S}imkovic}},\ }\bibfield  {title} {\enquote {\bibinfo {title} {{Arbitrary
  mass Majorana neutrinos in neutrinoless double beta decay}},}\ }\href
  {\doibase 10.1103/PhysRevD.90.096010} {\bibfield  {journal} {\bibinfo
  {journal} {Phys.\ Rev.\ D}\ }\textbf {\bibinfo {volume} {90}},\ \bibinfo
  {pages} {096010} (\bibinfo {year} {2014})}\BibitemShut {NoStop}%
%%CITATION = ARXIV:1408.6077;%%
\bibitem [{\citenamefont {\u{S}imkovic}\ \emph
  {et~al.}(2013{\natexlab{b}})\citenamefont {\u{S}imkovic}, \citenamefont
  {Bilenky}, \citenamefont {Faessler},\ and\ \citenamefont
  {Gutsche}}]{Simkovic:2012hq}%
  \BibitemOpen
  \bibfield  {author} {\bibinfo {author} {\bibfnamefont {F.}~\bibnamefont
  {\u{S}imkovic}}, \bibinfo {author} {\bibfnamefont {S.~M.}\ \bibnamefont
  {Bilenky}}, \bibinfo {author} {\bibfnamefont {A.}~\bibnamefont {Faessler}}, \
  and\ \bibinfo {author} {\bibfnamefont {T.}~\bibnamefont {Gutsche}},\
  }\bibfield  {title} {\enquote {\bibinfo {title} {{Possibility of measuring
  the CP Majorana phases in $0\nu\beta\beta$ decay}},}\ }\href {\doibase
  10.1103/PhysRevD.87.073002} {\bibfield  {journal} {\bibinfo  {journal}
  {Phys.\ Rev.\ D}\ }\textbf {\bibinfo {volume} {87}},\ \bibinfo {pages}
  {073002} (\bibinfo {year} {2013}{\natexlab{b}})}\BibitemShut {NoStop}%
%%CITATION = ARXIV:1210.1306;%%
\bibitem [{\citenamefont {Faessler}\ \emph {et~al.}(2008)\citenamefont
  {Faessler}, \citenamefont {Fogli}, \citenamefont {Lisi}, \citenamefont
  {Rodin}, \citenamefont {Rotunno},\ and\ \citenamefont
  {\u{S}imkovic}}]{Faessler:2007hu}%
  \BibitemOpen
  \bibfield  {author} {\bibinfo {author} {\bibfnamefont {A.}~\bibnamefont
  {Faessler}}, \bibinfo {author} {\bibfnamefont {G.~L.}\ \bibnamefont {Fogli}},
  \bibinfo {author} {\bibfnamefont {E.}~\bibnamefont {Lisi}}, \bibinfo {author}
  {\bibfnamefont {V.}~\bibnamefont {Rodin}}, \bibinfo {author} {\bibfnamefont
  {A.~M.}\ \bibnamefont {Rotunno}}, \ and\ \bibinfo {author} {\bibfnamefont
  {F.}~\bibnamefont {\u{S}imkovic}},\ }\bibfield  {title} {\enquote {\bibinfo
  {title} {{Overconstrained estimates of neutrinoless double beta decay within
  the QRPA}},}\ }\href {\doibase 10.1088/0954-3899/35/7/075104} {\bibfield
  {journal} {\bibinfo  {journal} {J.\ Phys.\ G}\ }\textbf {\bibinfo {volume}
  {35}},\ \bibinfo {pages} {075104} (\bibinfo {year} {2008})}\BibitemShut
  {NoStop}%
%%CITATION = ARXIV:0711.3996;%%
\bibitem [{\citenamefont {Suhonen}\ and\ \citenamefont
  {Civitarese}(2013)}]{Suhonen:2013laa}%
  \BibitemOpen
  \bibfield  {author} {\bibinfo {author} {\bibfnamefont {J.}~\bibnamefont
  {Suhonen}}\ and\ \bibinfo {author} {\bibfnamefont {O.}~\bibnamefont
  {Civitarese}},\ }\bibfield  {title} {\enquote {\bibinfo {title} {{Probing the
  quenching of $g_A$ by single and double beta decays}},}\ }\href {\doibase
  10.1016/j.physletb.2013.06.042} {\bibfield  {journal} {\bibinfo  {journal}
  {Phys.\ Lett.\ B}\ }\textbf {\bibinfo {volume} {725}},\ \bibinfo {pages}
  {153--157} (\bibinfo {year} {2013})}\BibitemShut {NoStop}%
%%CITATION = PHLTA,B725,153;%%
\bibitem [{\citenamefont {Lisi}\ \emph {et~al.}(2015)\citenamefont {Lisi},
  \citenamefont {Rotunno},\ and\ \citenamefont {\u{S}imkovic}}]{Lisi:2015yma}%
  \BibitemOpen
  \bibfield  {author} {\bibinfo {author} {\bibfnamefont {E.}~\bibnamefont
  {Lisi}}, \bibinfo {author} {\bibfnamefont {A.}~\bibnamefont {Rotunno}}, \
  and\ \bibinfo {author} {\bibfnamefont {F.}~\bibnamefont {\u{S}imkovic}},\
  }\bibfield  {title} {\enquote {\bibinfo {title} {{Degeneracies of particle
  and nuclear physics uncertainties in neutrinoless double beta decay}},}\
  }\href {\doibase 10.1103/PhysRevD.92.093004} {\bibfield  {journal} {\bibinfo
  {journal} {Phys.\ Rev.\ D}\ }\textbf {\bibinfo {volume} {92}},\ \bibinfo
  {pages} {093004} (\bibinfo {year} {2015})}\BibitemShut {NoStop}%
%%CITATION = ARXIV:1506.04058;%%
\bibitem [{\citenamefont {Konieczka}\ \emph {et~al.}(2016)\citenamefont
  {Konieczka}, \citenamefont {Baczyk},\ and\ \citenamefont
  {Satula}}]{Konieczka:2015ela}%
  \BibitemOpen
  \bibfield  {author} {\bibinfo {author} {\bibfnamefont {M.}~\bibnamefont
  {Konieczka}}, \bibinfo {author} {\bibfnamefont {P.}~\bibnamefont {Baczyk}}, \
  and\ \bibinfo {author} {\bibfnamefont {W.}~\bibnamefont {Satula}},\
  }\bibfield  {title} {\enquote {\bibinfo {title} {{Beta-decay study within
  multi-reference density functional theory and beyond}},}\ }\href {\doibase
  10.1103/PhysRevC.93.042501} {\bibfield  {journal} {\bibinfo  {journal} {Phys.
  Rev.}\ }\textbf {\bibinfo {volume} {C93}},\ \bibinfo {pages} {042501}
  (\bibinfo {year} {2016})}\BibitemShut {NoStop}%
%%CITATION = ARXIV:1509.04480;%%
\bibitem [{\citenamefont {Fujita}\ and\ \citenamefont
  {Ikeda}(1965)}]{Fujita:1965}%
  \BibitemOpen
  \bibfield  {author} {\bibinfo {author} {\bibfnamefont {J.~I.}\ \bibnamefont
  {Fujita}}\ and\ \bibinfo {author} {\bibfnamefont {K.}~\bibnamefont {Ikeda}},\
  }\bibfield  {title} {\enquote {\bibinfo {title} {{Existence of isobaric
  states and beta decay of heavier nuclei}},}\ }\href {\doibase
  10.1016/0029-5582(65)90119-7} {\bibfield  {journal} {\bibinfo  {journal}
  {Nucl.\ Phys.}\ }\textbf {\bibinfo {volume} {67}},\ \bibinfo {pages}
  {145--177} (\bibinfo {year} {1965})}\BibitemShut {NoStop}%
\bibitem [{\citenamefont {Wilkinson}(1973{\natexlab{a}})}]{Wilkinson:1973zz}%
  \BibitemOpen
  \bibfield  {author} {\bibinfo {author} {\bibfnamefont {D.~H.}\ \bibnamefont
  {Wilkinson}},\ }\bibfield  {title} {\enquote {\bibinfo {title}
  {{Renormalization of the Axial-Vector Coupling Constant in Nuclear beta
  Decay}},}\ }\href {\doibase 10.1103/PhysRevC.7.930} {\bibfield  {journal}
  {\bibinfo  {journal} {Phys.\ Rev.\ C}\ }\textbf {\bibinfo {volume} {7}},\
  \bibinfo {pages} {930--936} (\bibinfo {year}
  {1973}{\natexlab{a}})}\BibitemShut {NoStop}%
%%CITATION = PHRVA,C7,930;%%
\bibitem [{\citenamefont {Wilkinson}(1973{\natexlab{b}})}]{Wilkinson:1973_2}%
  \BibitemOpen
  \bibfield  {author} {\bibinfo {author} {\bibfnamefont {D.~H.}\ \bibnamefont
  {Wilkinson}},\ }\bibfield  {title} {\enquote {\bibinfo {title}
  {{Renormalization of the axial-vector coupling constant in nuclear
  $\beta$-decay (II)}},}\ }\href {\doibase 10.1016/0375-9474(73)90840-3}
  {\bibfield  {journal} {\bibinfo  {journal} {Nucl.\ Phys.\ A}\ }\textbf
  {\bibinfo {volume} {209}},\ \bibinfo {pages} {470--484} (\bibinfo {year}
  {1973}{\natexlab{b}})}\BibitemShut {NoStop}%
\bibitem [{\citenamefont {Wilkinson}(1974)}]{Wilkinson:1974}%
  \BibitemOpen
  \bibfield  {author} {\bibinfo {author} {\bibfnamefont {D.~H.}\ \bibnamefont
  {Wilkinson}},\ }\bibfield  {title} {\enquote {\bibinfo {title}
  {{Renormalization of the axial-vector coupling constant in nuclear
  $\beta$-decay (III)}},}\ }\href {\doibase 10.1016/0375-9474(74)90347-9}
  {\bibfield  {journal} {\bibinfo  {journal} {Nucl.\ Phys.\ A}\ }\textbf
  {\bibinfo {volume} {225}},\ \bibinfo {pages} {365--381} (\bibinfo {year}
  {1974})}\BibitemShut {NoStop}%
\bibitem [{\citenamefont {Ericson}(1971)}]{Ericson:1971}%
  \BibitemOpen
  \bibfield  {author} {\bibinfo {author} {\bibfnamefont {M.}~\bibnamefont
  {Ericson}},\ }\bibfield  {title} {\enquote {\bibinfo {title} {{Axial vector
  nuclear sum rules and exchange effects}},}\ }\href {\doibase
  10.1016/0003-4916(71)90029-7} {\bibfield  {journal} {\bibinfo  {journal}
  {Ann.\ Phys.}\ }\textbf {\bibinfo {volume} {63}},\ \bibinfo {pages}
  {562--576} (\bibinfo {year} {1971})}\BibitemShut {NoStop}%
\bibitem [{\citenamefont {Barshay}\ \emph {et~al.}(1974)\citenamefont
  {Barshay}, \citenamefont {Brown},\ and\ \citenamefont
  {Rho}}]{Barshay:1974fh}%
  \BibitemOpen
  \bibfield  {author} {\bibinfo {author} {\bibfnamefont {S.}~\bibnamefont
  {Barshay}}, \bibinfo {author} {\bibfnamefont {G.E.}\ \bibnamefont {Brown}}, \
  and\ \bibinfo {author} {\bibfnamefont {M.}~\bibnamefont {Rho}},\ }\bibfield
  {title} {\enquote {\bibinfo {title} {{A New Interpretation of the
  Pion-Nucleus Optical Potential for Pionic Atoms}},}\ }\href {\doibase
  10.1103/PhysRevLett.32.787} {\bibfield  {journal} {\bibinfo  {journal}
  {Phys.\ Rev.\ Lett.}\ }\textbf {\bibinfo {volume} {32}},\ \bibinfo {pages}
  {787} (\bibinfo {year} {1974})},\ \bibinfo {note} {[Erratum:
  \href{http://journals.aps.org/prl/abstract/10.1103/PhysRevLett.32.1149.3}
  {Phys.\ Rev.\ Lett.\ {\bf 32}, 1149 (1974)}]}\BibitemShut {NoStop}%
%%CITATION = PRLTA,32,787;%%
\bibitem [{\citenamefont {Shimizu}\ \emph {et~al.}(1974)\citenamefont
  {Shimizu}, \citenamefont {Ichimura},\ and\ \citenamefont
  {Arima}}]{Shimizu:1974}%
  \BibitemOpen
  \bibfield  {author} {\bibinfo {author} {\bibfnamefont {K.}~\bibnamefont
  {Shimizu}}, \bibinfo {author} {\bibfnamefont {M.}~\bibnamefont {Ichimura}}, \
  and\ \bibinfo {author} {\bibfnamefont {A.}~\bibnamefont {Arima}},\ }\bibfield
   {title} {\enquote {\bibinfo {title} {{Magnetic moments and GT-type β-decay
  matrix elements in nuclei with a $LS$ doubly closed shell plus or minus one
  nucleon}},}\ }\href {\doibase 10.1016/0375-9474(74)90407-2} {\bibfield
  {journal} {\bibinfo  {journal} {Nucl.\ Phys.\ A}\ }\textbf {\bibinfo {volume}
  {226}},\ \bibinfo {pages} {282--318} (\bibinfo {year} {1974})}\BibitemShut
  {NoStop}%
\bibitem [{\citenamefont {Men\'{e}ndez}\ \emph {et~al.}(2011)\citenamefont
  {Men\'{e}ndez}, \citenamefont {Gazit},\ and\ \citenamefont
  {Schwenk}}]{Menendez:2011qq}%
  \BibitemOpen
  \bibfield  {author} {\bibinfo {author} {\bibfnamefont {J.}~\bibnamefont
  {Men\'{e}ndez}}, \bibinfo {author} {\bibfnamefont {D.}~\bibnamefont {Gazit}},
  \ and\ \bibinfo {author} {\bibfnamefont {A.}~\bibnamefont {Schwenk}},\
  }\bibfield  {title} {\enquote {\bibinfo {title} {{Chiral two-body currents in
  nuclei: Gamow-Teller transitions and neutrinoless double-beta decay}},}\
  }\href {\doibase 10.1103/PhysRevLett.107.062501} {\bibfield  {journal}
  {\bibinfo  {journal} {Phys.\ Rev.\ Lett.}\ }\textbf {\bibinfo {volume}
  {107}},\ \bibinfo {pages} {062501} (\bibinfo {year} {2011})}\BibitemShut
  {NoStop}%
%%CITATION = ARXIV:1103.3622;%%
\bibitem [{\citenamefont {Ekstr$\ddot{\text o}$m}\ \emph
  {et~al.}(2014)\citenamefont {Ekstr$\ddot{\text o}$m}, \citenamefont {Jansen},
  \citenamefont {Wendt}, \citenamefont {Hagen}, \citenamefont {Papenbrock},
  \citenamefont {Bacca}, \citenamefont {Carlsson},\ and\ \citenamefont
  {Gazit}}]{Ekstrom:2014iya}%
  \BibitemOpen
  \bibfield  {author} {\bibinfo {author} {\bibfnamefont {A.}~\bibnamefont
  {Ekstr$\ddot{\text o}$m}}, \bibinfo {author} {\bibfnamefont {G.~R.}\
  \bibnamefont {Jansen}}, \bibinfo {author} {\bibfnamefont {K.~A.}\
  \bibnamefont {Wendt}}, \bibinfo {author} {\bibfnamefont {G.}~\bibnamefont
  {Hagen}}, \bibinfo {author} {\bibfnamefont {T.}~\bibnamefont {Papenbrock}},
  \bibinfo {author} {\bibfnamefont {S.}~\bibnamefont {Bacca}}, \bibinfo
  {author} {\bibfnamefont {B.}~\bibnamefont {Carlsson}}, \ and\ \bibinfo
  {author} {\bibfnamefont {D.}~\bibnamefont {Gazit}},\ }\bibfield  {title}
  {\enquote {\bibinfo {title} {{Effects of three-nucleon forces and two-body
  currents on Gamow-Teller strengths}},}\ }\href {\doibase
  10.1103/PhysRevLett.113.262504} {\bibfield  {journal} {\bibinfo  {journal}
  {Phys.\ Rev.\ Lett.}\ }\textbf {\bibinfo {volume} {113}},\ \bibinfo {pages}
  {262504} (\bibinfo {year} {2014})}\BibitemShut {NoStop}%
%%CITATION = ARXIV:1406.4696;%%
\bibitem [{\citenamefont {Engel}\ \emph {et~al.}(2014)\citenamefont {Engel},
  \citenamefont {\u{S}imkovic},\ and\ \citenamefont {Vogel}}]{Engel:2014pha}%
  \BibitemOpen
  \bibfield  {author} {\bibinfo {author} {\bibfnamefont {J.}~\bibnamefont
  {Engel}}, \bibinfo {author} {\bibfnamefont {F.}~\bibnamefont {\u{S}imkovic}},
  \ and\ \bibinfo {author} {\bibfnamefont {P.}~\bibnamefont {Vogel}},\
  }\bibfield  {title} {\enquote {\bibinfo {title} {{Chiral Two-Body Currents
  and Neutrinoless Double-Beta Decay in the QRPA}},}\ }\href {\doibase
  10.1103/PhysRevC.89.064308} {\bibfield  {journal} {\bibinfo  {journal}
  {Phys.\ Rev.\ C}\ }\textbf {\bibinfo {volume} {89}},\ \bibinfo {pages}
  {064308} (\bibinfo {year} {2014})}\BibitemShut {NoStop}%
%%CITATION = ARXIV:1403.7860;%%
\bibitem [{\citenamefont {Cappuzzello}\ \emph {et~al.}(2015)\citenamefont
  {Cappuzzello}, \citenamefont {Agodi}, \citenamefont {Bond\`i}, \citenamefont
  {Carbone}, \citenamefont {Cavallaro},\ and\ \citenamefont
  {Foti}}]{Cappuzzello:2015oza}%
  \BibitemOpen
  \bibfield  {author} {\bibinfo {author} {\bibfnamefont {F.}~\bibnamefont
  {Cappuzzello}}, \bibinfo {author} {\bibfnamefont {C.}~\bibnamefont {Agodi}},
  \bibinfo {author} {\bibfnamefont {M.}~\bibnamefont {Bond\`i}}, \bibinfo
  {author} {\bibfnamefont {D.}~\bibnamefont {Carbone}}, \bibinfo {author}
  {\bibfnamefont {M.}~\bibnamefont {Cavallaro}}, \ and\ \bibinfo {author}
  {\bibfnamefont {A.}~\bibnamefont {Foti}},\ }\bibfield  {title} {\enquote
  {\bibinfo {title} {{The role of nuclear reactions in the problem of
  $0\nu\beta\beta$ decay and the NUMEN project at INFN-LNS}},}\ }\href
  {\doibase 10.1088/1742-6596/630/1/012018} {\bibfield  {journal} {\bibinfo
  {journal} {J.\ Phys.\ Conf.\ Ser.}\ }\textbf {\bibinfo {volume} {630}},\
  \bibinfo {pages} {012018} (\bibinfo {year} {2015})}\BibitemShut {NoStop}%
%%CITATION = 00462,630,012018;%%
\bibitem [{\citenamefont {Agostini}\ \emph {et~al.}(2013)\citenamefont
  {Agostini} \emph {et~al.}}]{Agostini:2013mzu}%
  \BibitemOpen
  \bibfield  {author} {\bibinfo {author} {\bibfnamefont {M.}~\bibnamefont
  {Agostini}} \emph {et~al.} (\bibinfo {collaboration} {GERDA Collaboration}),\
  }\bibfield  {title} {\enquote {\bibinfo {title} {{Results on Neutrinoless
  Double-$\beta$ Decay of \ce{^{76}Ge} from Phase I of the GERDA
  Experiment}},}\ }\href {\doibase 10.1103/PhysRevLett.111.122503} {\bibfield
  {journal} {\bibinfo  {journal} {Phys.\ Rev.\ Lett.}\ }\textbf {\bibinfo
  {volume} {111}},\ \bibinfo {pages} {122503} (\bibinfo {year}
  {2013})}\BibitemShut {NoStop}%
%%CITATION = ARXIV:1307.4720;%%
\bibitem [{\citenamefont {Kovalenko}\ \emph {et~al.}(2009)\citenamefont
  {Kovalenko}, \citenamefont {Lu},\ and\ \citenamefont
  {Schmidt}}]{Kovalenko:2009td}%
  \BibitemOpen
  \bibfield  {author} {\bibinfo {author} {\bibfnamefont {S.}~\bibnamefont
  {Kovalenko}}, \bibinfo {author} {\bibfnamefont {Z.}~\bibnamefont {Lu}}, \
  and\ \bibinfo {author} {\bibfnamefont {I.}~\bibnamefont {Schmidt}},\
  }\bibfield  {title} {\enquote {\bibinfo {title} {{Lepton Number Violating
  Processes Mediated by Majorana Neutrinos at Hadron Colliders}},}\ }\href
  {\doibase 10.1103/PhysRevD.80.073014} {\bibfield  {journal} {\bibinfo
  {journal} {Phys.\ Rev.\ D}\ }\textbf {\bibinfo {volume} {80}},\ \bibinfo
  {pages} {073014} (\bibinfo {year} {2009})}\BibitemShut {NoStop}%
%%CITATION = ARXIV:0907.2533;%%
\bibitem [{\citenamefont {Bergsma}\ \emph {et~al.}(1986)\citenamefont {Bergsma}
  \emph {et~al.}}]{Bergsma:1985is}%
  \BibitemOpen
  \bibfield  {author} {\bibinfo {author} {\bibfnamefont {F.}~\bibnamefont
  {Bergsma}} \emph {et~al.} (\bibinfo {collaboration} {CHARM Collaboration}),\
  }\bibfield  {title} {\enquote {\bibinfo {title} {{A Search for Decays of
  Heavy Neutrinos in the Mass Range (0.5--2.8)\,GeV}},}\ }\href {\doibase
  10.1016/0370-2693(86)91601-1} {\bibfield  {journal} {\bibinfo  {journal}
  {Phys.\ Lett.}\ }\textbf {\bibinfo {volume} {B\,166}},\ \bibinfo {pages}
  {473} (\bibinfo {year} {1986})}\BibitemShut {NoStop}%
%%CITATION = PHLTA,B166,473;%%
\bibitem [{\citenamefont {Abreu}\ \emph {et~al.}(1997)\citenamefont {Abreu}
  \emph {et~al.}}]{Abreu:1996pa}%
  \BibitemOpen
  \bibfield  {author} {\bibinfo {author} {\bibfnamefont {P.}~\bibnamefont
  {Abreu}} \emph {et~al.} (\bibinfo {collaboration} {DELPHI Collaboration}),\
  }\bibfield  {title} {\enquote {\bibinfo {title} {{Search for neutral heavy
  leptons produced in Z decays}},}\ }\href {\doibase 10.1007/s002880050370}
  {\bibfield  {journal} {\bibinfo  {journal} {Z.\ Phys.}\ }\textbf {\bibinfo
  {volume} {C\,74}},\ \bibinfo {pages} {57--71} (\bibinfo {year}
  {1997})}\BibitemShut {NoStop}%
%%CITATION = ZEPYA,C74,57;%%
\bibitem [{\citenamefont {Bernardi}\ \emph {et~al.}(1988)\citenamefont
  {Bernardi} \emph {et~al.}}]{Bernardi:1987ek}%
  \BibitemOpen
  \bibfield  {author} {\bibinfo {author} {\bibfnamefont {G.}~\bibnamefont
  {Bernardi}} \emph {et~al.},\ }\bibfield  {title} {\enquote {\bibinfo {title}
  {{Further Limits on Heavy Neutrino Couplings}},}\ }\href {\doibase
  10.1016/0370-2693(88)90563-1} {\bibfield  {journal} {\bibinfo  {journal}
  {Phys.\ Lett.}\ }\textbf {\bibinfo {volume} {B\,203}},\ \bibinfo {pages}
  {332} (\bibinfo {year} {1988})}\BibitemShut {NoStop}%
%%CITATION = PHLTA,B203,332;%%
\bibitem [{\citenamefont {Britton}\ \emph
  {et~al.}(1992{\natexlab{a}})\citenamefont {Britton} \emph
  {et~al.}}]{Britton:1992pg}%
  \BibitemOpen
  \bibfield  {author} {\bibinfo {author} {\bibfnamefont {D.I.}\ \bibnamefont
  {Britton}} \emph {et~al.},\ }\bibfield  {title} {\enquote {\bibinfo {title}
  {{Measurement of the $\pi^+ \rightarrow e^+\,\nu$ branching ratio}},}\ }\href
  {\doibase 10.1103/PhysRevLett.68.3000} {\bibfield  {journal} {\bibinfo
  {journal} {Phys.\ Rev.\ Lett.}\ }\textbf {\bibinfo {volume} {68}},\ \bibinfo
  {pages} {3000--3003} (\bibinfo {year} {1992}{\natexlab{a}})}\BibitemShut
  {NoStop}%
%%CITATION = PRLTA,68,3000;%%
\bibitem [{\citenamefont {Britton}\ \emph
  {et~al.}(1992{\natexlab{b}})\citenamefont {Britton} \emph
  {et~al.}}]{Britton:1992xv}%
  \BibitemOpen
  \bibfield  {author} {\bibinfo {author} {\bibfnamefont {D.~I.}\ \bibnamefont
  {Britton}} \emph {et~al.},\ }\bibfield  {title} {\enquote {\bibinfo {title}
  {{Improved search for massive neutrinos in $\pi^+ \rightarrow e^+\,\nu$
  decay}},}\ }\href {\doibase 10.1103/PhysRevD.46.R885} {\bibfield  {journal}
  {\bibinfo  {journal} {Phys.\ Rev.\ D}\ }\textbf {\bibinfo {volume} {46}},\
  \bibinfo {pages} {885--887} (\bibinfo {year}
  {1992}{\natexlab{b}})}\BibitemShut {NoStop}%
%%CITATION = PHRVA,D46,885;%%
\bibitem [{\citenamefont {Brugnera}\ and\ \citenamefont
  {Garfagnini}(2013)}]{Brugnera:2013xma}%
  \BibitemOpen
  \bibfield  {author} {\bibinfo {author} {\bibfnamefont {R.}~\bibnamefont
  {Brugnera}}\ and\ \bibinfo {author} {\bibfnamefont {A.}~\bibnamefont
  {Garfagnini}},\ }\bibfield  {title} {\enquote {\bibinfo {title} {{Status of
  the $GERDA$ Experiment at the Laboratori Nazionali del Gran Sasso}},}\ }\href
  {\doibase 10.1155/2013/506186} {\bibfield  {journal} {\bibinfo  {journal}
  {Adv.\ High Energy Phys.}\ }\textbf {\bibinfo {volume} {2013}},\ \bibinfo
  {pages} {506186} (\bibinfo {year} {2013})}\BibitemShut {NoStop}%
%%CITATION = 00642,2013,506186;%%
\bibitem [{\citenamefont {Saakyan}(2013)}]{Saakyan:2013yna}%
  \BibitemOpen
  \bibfield  {author} {\bibinfo {author} {\bibfnamefont {R.}~\bibnamefont
  {Saakyan}},\ }\bibfield  {title} {\enquote {\bibinfo {title} {{Two-Neutrino
  Double-Beta Decay}},}\ }\href {\doibase 10.1146/annurev-nucl-102711-094904}
  {\bibfield  {journal} {\bibinfo  {journal} {Ann.\ Rev.\ Nucl.\ Part.\ Sci.}\
  }\textbf {\bibinfo {volume} {63}},\ \bibinfo {pages} {503--529} (\bibinfo
  {year} {2013})}\BibitemShut {NoStop}%
%%CITATION = ARNUA,63,503;%%
\bibitem [{\citenamefont {Robertson}(2013)}]{Robertson:2013cy}%
  \BibitemOpen
  \bibfield  {author} {\bibinfo {author} {\bibfnamefont {R.~G.~H.}\
  \bibnamefont {Robertson}},\ }\bibfield  {title} {\enquote {\bibinfo {title}
  {{Empirical Survey of Neutrinoless Double Beta Decay Matrix Elements}},}\
  }\href {\doibase 10.1142/S0217732313500211} {\bibfield  {journal} {\bibinfo
  {journal} {Mod.\ Phys.\ Lett.\ A}\ }\textbf {\bibinfo {volume} {28}},\
  \bibinfo {pages} {1350021} (\bibinfo {year} {2013})}\BibitemShut {NoStop}%
%%CITATION = ARXIV:1301.1323;%%
\bibitem [{\citenamefont {G\'{o}mez-Cadenas}\ and\ \citenamefont
  {Mart\`in-Albo}(2015)}]{Gomez-Cadenas:2015twa}%
  \BibitemOpen
  \bibfield  {author} {\bibinfo {author} {\bibfnamefont {J.~J.}\ \bibnamefont
  {G\'{o}mez-Cadenas}}\ and\ \bibinfo {author} {\bibfnamefont {J.}~\bibnamefont
  {Mart\`in-Albo}},\ }\bibfield  {title} {\enquote {\bibinfo {title}
  {{Phenomenology of neutrinoless double beta decay}},}\ }\href
  {http://pos.sissa.it/cgi-bin/reader/conf.cgi?confid=229} {\bibfield
  {journal} {\bibinfo  {journal} {PoS}\ }\textbf {\bibinfo {volume}
  {(GSSI14)}},\ \bibinfo {pages} {004} (\bibinfo {year} {2015})}\BibitemShut
  {NoStop}%
%%CITATION = ARXIV:1502.00581;%%
\bibitem [{\citenamefont {Elliott}\ and\ \citenamefont
  {Vogel}(2002)}]{Elliott:2002xe}%
  \BibitemOpen
  \bibfield  {author} {\bibinfo {author} {\bibfnamefont {S.~R.}\ \bibnamefont
  {Elliott}}\ and\ \bibinfo {author} {\bibfnamefont {P.}~\bibnamefont
  {Vogel}},\ }\bibfield  {title} {\enquote {\bibinfo {title} {{Double beta
  decay}},}\ }\href {\doibase 10.1146/annurev.nucl.52.050102.090641} {\bibfield
   {journal} {\bibinfo  {journal} {Ann.\ Rev.\ Nucl.\ Part.\ Sci.}\ }\textbf
  {\bibinfo {volume} {52}},\ \bibinfo {pages} {115--151} (\bibinfo {year}
  {2002})}\BibitemShut {NoStop}%
%%CITATION = HEP-PH/0202264;%%
\bibitem [{\citenamefont {Andreotti}\ \emph {et~al.}(2011)\citenamefont
  {Andreotti} \emph {et~al.}}]{Andreotti:2010vj}%
  \BibitemOpen
  \bibfield  {author} {\bibinfo {author} {\bibfnamefont {E.}~\bibnamefont
  {Andreotti}} \emph {et~al.},\ }\bibfield  {title} {\enquote {\bibinfo {title}
  {{\ce{^{130}Te} Neutrinoless Double-Beta Decay with CUORICINO}},}\ }\href
  {\doibase 10.1016/j.astropartphys.2011.02.002} {\bibfield  {journal}
  {\bibinfo  {journal} {Astropart.\ Phys.}\ }\textbf {\bibinfo {volume} {34}},\
  \bibinfo {pages} {822--831} (\bibinfo {year} {2011})}\BibitemShut {NoStop}%
%%CITATION = ARXIV:1012.3266;%%
\bibitem [{\citenamefont {Alfonso}\ \emph {et~al.}(2015)\citenamefont {Alfonso}
  \emph {et~al.}}]{Alfonso:2015wka}%
  \BibitemOpen
  \bibfield  {author} {\bibinfo {author} {\bibfnamefont {K.}~\bibnamefont
  {Alfonso}} \emph {et~al.} (\bibinfo {collaboration} {CUORE Collaboration}),\
  }\bibfield  {title} {\enquote {\bibinfo {title} {{Search for Neutrinoless
  Double-Beta Decay of $^{130}$Te with CUORE-0}},}\ }\href {\doibase
  10.1103/PhysRevLett.115.102502} {\bibfield  {journal} {\bibinfo  {journal}
  {Phys.\ Rev.\ Lett.}\ }\textbf {\bibinfo {volume} {115}},\ \bibinfo {pages}
  {102502} (\bibinfo {year} {2015})}\BibitemShut {NoStop}%
%%CITATION = ARXIV:1504.02454;%%
\bibitem [{\citenamefont {Klapdor-Kleingrothaus}\ \emph
  {et~al.}(2001{\natexlab{a}})\citenamefont {Klapdor-Kleingrothaus} \emph
  {et~al.}}]{KlapdorKleingrothaus:2000sn}%
  \BibitemOpen
  \bibfield  {author} {\bibinfo {author} {\bibfnamefont {H.~V.}\ \bibnamefont
  {Klapdor-Kleingrothaus}} \emph {et~al.},\ }\bibfield  {title} {\enquote
  {\bibinfo {title} {{Latest results from the Heidelberg-Moscow double beta
  decay experiment}},}\ }\href {\doibase 10.1007/s100500170022} {\bibfield
  {journal} {\bibinfo  {journal} {Eur.\ Phys.\ J.\ A}\ }\textbf {\bibinfo
  {volume} {12}},\ \bibinfo {pages} {147--154} (\bibinfo {year}
  {2001}{\natexlab{a}})}\BibitemShut {NoStop}%
%%CITATION = HEP-PH/0103062;%%
\bibitem [{\citenamefont {Aalseth}\ \emph {et~al.}(1999)\citenamefont {Aalseth}
  \emph {et~al.}}]{Aalseth:1999ji}%
  \BibitemOpen
  \bibfield  {author} {\bibinfo {author} {\bibfnamefont {C.~E.}\ \bibnamefont
  {Aalseth}} \emph {et~al.} (\bibinfo {collaboration} {IGEX Collaboration}),\
  }\bibfield  {title} {\enquote {\bibinfo {title} {{Neutrinoless double-beta
  decay of \ce{^{76}Ge}: First results from the International Germanium
  Experiment (IGEX) with six isotopically enriched detectors}},}\ }\href
  {\doibase 10.1103/PhysRevC.59.2108} {\bibfield  {journal} {\bibinfo
  {journal} {Phys.\ Rev.\ C}\ }\textbf {\bibinfo {volume} {59}},\ \bibinfo
  {pages} {2108--2113} (\bibinfo {year} {1999})}\BibitemShut {NoStop}%
%%CITATION = PHRVA,C59,2108;%%
\bibitem [{\citenamefont {Aalseth}\ \emph
  {et~al.}(2002{\natexlab{a}})\citenamefont {Aalseth} \emph
  {et~al.}}]{Aalseth:2002rf}%
  \BibitemOpen
  \bibfield  {author} {\bibinfo {author} {\bibfnamefont {C.~E.}\ \bibnamefont
  {Aalseth}} \emph {et~al.} (\bibinfo {collaboration} {IGEX Collaboration}),\
  }\bibfield  {title} {\enquote {\bibinfo {title} {{The IGEX \ce{^{76}Ge}
  neutrinoless double beta decay experiment: Prospects for next generation
  experiments}},}\ }\href {\doibase 10.1103/PhysRevD.65.092007} {\bibfield
  {journal} {\bibinfo  {journal} {Phys.\ Rev.\ D}\ }\textbf {\bibinfo {volume}
  {65}},\ \bibinfo {pages} {092007} (\bibinfo {year}
  {2002}{\natexlab{a}})}\BibitemShut {NoStop}%
%%CITATION = HEP-EX/0202026;%%
\bibitem [{\citenamefont {Ackermann}\ \emph {et~al.}(2013)\citenamefont
  {Ackermann} \emph {et~al.}}]{Ackermann:2012xja}%
  \BibitemOpen
  \bibfield  {author} {\bibinfo {author} {\bibfnamefont {K.~H.}\ \bibnamefont
  {Ackermann}} \emph {et~al.} (\bibinfo {collaboration} {GERDA
  Collaboration}),\ }\bibfield  {title} {\enquote {\bibinfo {title} {{The GERDA
  experiment for the search of $0\nu\beta\beta$ decay in \ce{^{76}Ge}}},}\
  }\href {\doibase 10.1140/epjc/s10052-013-2330-0} {\bibfield  {journal}
  {\bibinfo  {journal} {Eur.\ Phys.\ J.\ C}\ }\textbf {\bibinfo {volume}
  {73}},\ \bibinfo {pages} {2330} (\bibinfo {year} {2013})}\BibitemShut
  {NoStop}%
%%CITATION = ARXIV:1212.4067;%%
\bibitem [{\citenamefont {Arnold}\ \emph {et~al.}(2014)\citenamefont {Arnold}
  \emph {et~al.}}]{Arnold:2013dha}%
  \BibitemOpen
  \bibfield  {author} {\bibinfo {author} {\bibfnamefont {R.}~\bibnamefont
  {Arnold}} \emph {et~al.} (\bibinfo {collaboration} {NEMO Collaboration}),\
  }\bibfield  {title} {\enquote {\bibinfo {title} {{Search for neutrinoless
  double-beta decay of \ce{^{100}Mo} with the NEMO-3 detector}},}\ }\href
  {\doibase 10.1103/PhysRevD.89.111101} {\bibfield  {journal} {\bibinfo
  {journal} {Phys.\ Rev.\ D}\ }\textbf {\bibinfo {volume} {89}},\ \bibinfo
  {pages} {111101} (\bibinfo {year} {2014})}\BibitemShut {NoStop}%
%%CITATION = ARXIV:1311.5695;%%
\bibitem [{\citenamefont {Albert}\ \emph {et~al.}(2014)\citenamefont {Albert}
  \emph {et~al.}}]{Albert:2014awa}%
  \BibitemOpen
  \bibfield  {author} {\bibinfo {author} {\bibfnamefont {J.~B.}\ \bibnamefont
  {Albert}} \emph {et~al.} (\bibinfo {collaboration} {EXO Collaboration}),\
  }\bibfield  {title} {\enquote {\bibinfo {title} {{Search for Majorana
  neutrinos with the first two years of EXO-200 data}},}\ }\href {\doibase
  10.1038/nature13432} {\bibfield  {journal} {\bibinfo  {journal} {Nature}\
  }\textbf {\bibinfo {volume} {510}},\ \bibinfo {pages} {229--234} (\bibinfo
  {year} {2014})}\BibitemShut {NoStop}%
%%CITATION = ARXIV:1402.6956;%%
\bibitem [{\citenamefont {Gando}\ \emph {et~al.}(2013)\citenamefont {Gando}
  \emph {et~al.}}]{Gando:2012zm}%
  \BibitemOpen
  \bibfield  {author} {\bibinfo {author} {\bibfnamefont {A.}~\bibnamefont
  {Gando}} \emph {et~al.} (\bibinfo {collaboration} {KamLAND-Zen
  Collaboration}),\ }\bibfield  {title} {\enquote {\bibinfo {title} {{Limit on
  Neutrinoless $\beta\beta$ Decay of \ce{^{136}Xe} from the First Phase of
  KamLAND-Zen and Comparison with the Positive Claim in \ce{^{76}Ge}}},}\
  }\href {\doibase 10.1103/PhysRevLett.110.062502} {\bibfield  {journal}
  {\bibinfo  {journal} {Phys.\ Rev.\ Lett.}\ }\textbf {\bibinfo {volume}
  {110}},\ \bibinfo {pages} {062502} (\bibinfo {year} {2013})}\BibitemShut
  {NoStop}%
%%CITATION = ARXIV:1211.3863;%%
\bibitem [{\citenamefont {Asakura}\ \emph {et~al.}(2015)\citenamefont {Asakura}
  \emph {et~al.}}]{Asakura:2014lma}%
  \BibitemOpen
  \bibfield  {author} {\bibinfo {author} {\bibfnamefont {K.}~\bibnamefont
  {Asakura}} \emph {et~al.} (\bibinfo {collaboration} {KamLAND-Zen
  Collaboration}),\ }\bibfield  {title} {\enquote {\bibinfo {title} {{Results
  from KamLAND-Zen}},}\ }\href {\doibase 10.1063/1.4915593} {\bibfield
  {journal} {\bibinfo  {journal} {AIP Conf.\ Proc.}\ }\textbf {\bibinfo
  {volume} {1666}},\ \bibinfo {pages} {170003} (\bibinfo {year}
  {2015})}\BibitemShut {NoStop}%
%%CITATION = ARXIV:1409.0077;%%
\bibitem [{\citenamefont {Artusa}\ \emph {et~al.}(2015)\citenamefont {Artusa}
  \emph {et~al.}}]{Artusa:2014lgv}%
  \BibitemOpen
  \bibfield  {author} {\bibinfo {author} {\bibfnamefont {D.~R.}\ \bibnamefont
  {Artusa}} \emph {et~al.} (\bibinfo {collaboration} {CUORE Collaboration}),\
  }\bibfield  {title} {\enquote {\bibinfo {title} {{Searching for neutrinoless
  double-beta decay of $^{130}$Te with CUORE}},}\ }\href {\doibase
  10.1155/2015/879871} {\bibfield  {journal} {\bibinfo  {journal} {Adv.\ High
  Energy Phys.}\ }\textbf {\bibinfo {volume} {2015}},\ \bibinfo {pages}
  {879871} (\bibinfo {year} {2015})}\BibitemShut {NoStop}%
%%CITATION = ARXIV:1402.6072;%%
\bibitem [{\citenamefont {Pattavina}(2015)}]{Pattavina:2015_LUCIFER}%
  \BibitemOpen
  \bibfield  {author} {\bibinfo {author} {\bibfnamefont {L.}~\bibnamefont
  {Pattavina}},\ }\bibfield  {title} {\enquote {\bibinfo {title}
  {{Scintillating bolometers for the LUCIFER project}},}\ }\href@noop {} {\
  (\bibinfo {year} {2015})},\ \bibinfo {note} {[To appear in the proceedings of
  the XIV International Conference on Topics in Astroparticle and Underground
  Physics
  (\href{http://taup2015.to.infn.it/scientific-program/parallel-sessions/}%
  {TAUP 2015}), Torino, Italy, September 2015]}\BibitemShut {NoStop}%
\bibitem [{\citenamefont {Abgrall}\ \emph {et~al.}(2014)\citenamefont {Abgrall}
  \emph {et~al.}}]{Abgrall:2013rze}%
  \BibitemOpen
  \bibfield  {author} {\bibinfo {author} {\bibfnamefont {N.}~\bibnamefont
  {Abgrall}} \emph {et~al.} (\bibinfo {collaboration} {MAJORANA
  Collaboration}),\ }\bibfield  {title} {\enquote {\bibinfo {title} {{The
  MAJORANA DEMONSTRATOR Neutrinoless Double-Beta Decay Experiment}},}\ }\href
  {\doibase 10.1155/2014/365432} {\bibfield  {journal} {\bibinfo  {journal}
  {Adv.\ High Energy Phys.}\ }\textbf {\bibinfo {volume} {2014}},\ \bibinfo
  {pages} {365432} (\bibinfo {year} {2014})}\BibitemShut {NoStop}%
%%CITATION = ARXIV:1308.1633;%%
\bibitem [{\citenamefont {G\'{o}mez-Cadenas}\ \emph {et~al.}(2014)\citenamefont
  {G\'{o}mez-Cadenas} \emph {et~al.}}]{Gomez-Cadenas:2013lta}%
  \BibitemOpen
  \bibfield  {author} {\bibinfo {author} {\bibfnamefont {J.~J.}\ \bibnamefont
  {G\'{o}mez-Cadenas}} \emph {et~al.} (\bibinfo {collaboration} {NEXT
  Collaboration}),\ }\bibfield  {title} {\enquote {\bibinfo {title} {{Present
  status and future perspectives of the NEXT experiment}},}\ }\href {\doibase
  10.1155/2014/907067} {\bibfield  {journal} {\bibinfo  {journal} {Adv.\ High
  Energy Phys.}\ }\textbf {\bibinfo {volume} {2014}},\ \bibinfo {pages}
  {907067} (\bibinfo {year} {2014})}\BibitemShut {NoStop}%
%%CITATION = ARXIV:1307.3914;%%
\bibitem [{\citenamefont {Laing}(2015)}]{Laing:2015_NEXT}%
  \BibitemOpen
  \bibfield  {author} {\bibinfo {author} {\bibfnamefont {A.}~\bibnamefont
  {Laing}},\ }\bibfield  {title} {\enquote {\bibinfo {title} {{The NEXT double
  beta decay experiment}},}\ }\href@noop {} {\  (\bibinfo {year} {2015})},\
  \bibinfo {note} {[To appear in the proceedings of the XIV International
  Conference on Topics in Astroparticle and Underground Physics
  (\href{http://taup2015.to.infn.it/scientific-program/parallel-sessions/}%
  {TAUP 2015}), Torino, Italy, September 2015]}\BibitemShut {NoStop}%
\bibitem [{\citenamefont {Kim}(2015)}]{Kim:2015_AMoRE}%
  \BibitemOpen
  \bibfield  {author} {\bibinfo {author} {\bibfnamefont {Y.~H.}\ \bibnamefont
  {Kim}},\ }\bibfield  {title} {\enquote {\bibinfo {title} {{The AMoRE
  project}},}\ }\href@noop {} {\  (\bibinfo {year} {2015})},\ \bibinfo {note}
  {[To appear in the proceedings of the XIV International Conference on Topics
  in Astroparticle and Underground Physics
  (\href{http://taup2015.to.infn.it/scientific-program/parallel-sessions/}%
  {TAUP 2015}), Torino, Italy, September 2015]}\BibitemShut {NoStop}%
\bibitem [{\citenamefont {Ostrovskiy}(2015)}]{Ostrovskiy:2015_nEXO}%
  \BibitemOpen
  \bibfield  {author} {\bibinfo {author} {\bibfnamefont {I.}~\bibnamefont
  {Ostrovskiy}},\ }\bibfield  {title} {\enquote {\bibinfo {title} {{nEXO: The
  next generation \ce{^{136}Xe} neutrinoless double beta decay search}},}\
  }\href@noop {} {\  (\bibinfo {year} {2015})},\ \bibinfo {note} {[To appear in
  the proceedings of the XIV International Conference on Topics in
  Astroparticle and Underground Physics
  (\href{http://taup2015.to.infn.it/scientific-program/parallel-sessions/}%
  {TAUP 2015}), Torino, Italy, September 2015]}\BibitemShut {NoStop}%
\bibitem [{\citenamefont {Ji}(2015)}]{Ji:2015_PandaX}%
  \BibitemOpen
  \bibfield  {author} {\bibinfo {author} {\bibfnamefont {X.}~\bibnamefont
  {Ji}},\ }\bibfield  {title} {\enquote {\bibinfo {title} {{PandaX and
  0nuDBD}},}\ }\href@noop {} {\  (\bibinfo {year} {2015})},\ \bibinfo {note}
  {[Presentation at the International Workshop on Baryon and Lepton Number
  Violation
  (\href{http://www.physics.umass.edu/blv2015/workshop/program-description/thursday}{BLV
  2015}), Amherst, MA, USA, April 2015]}\BibitemShut {NoStop}%
\bibitem [{\citenamefont {Andringa}\ \emph {et~al.}(2015)\citenamefont
  {Andringa} \emph {et~al.}}]{Andringa:2015tza}%
  \BibitemOpen
  \bibfield  {author} {\bibinfo {author} {\bibfnamefont {S.}~\bibnamefont
  {Andringa}} \emph {et~al.} (\bibinfo {collaboration} {SNO+ Collaboration}),\
  }\bibfield  {title} {\enquote {\bibinfo {title} {{Current Status and Future
  Prospects of the SNO+ Experiment}},}\ }\href {\doibase 10.1155/2016/6194250}
  {\bibfield  {journal} {\bibinfo  {journal} {Adv.\ High Energy Phys.}\
  }\textbf {\bibinfo {volume} {2016}},\ \bibinfo {pages} {6194250} (\bibinfo
  {year} {2015})}\BibitemShut {NoStop}%
%%CITATION = ARXIV:1508.05759;%%
\bibitem [{\citenamefont {Arnold}\ \emph {et~al.}(2015)\citenamefont {Arnold}
  \emph {et~al.}}]{Arnold:2015wpy}%
  \BibitemOpen
  \bibfield  {author} {\bibinfo {author} {\bibfnamefont {R.}~\bibnamefont
  {Arnold}} \emph {et~al.} (\bibinfo {collaboration} {NEMO-3 Collaboration}),\
  }\bibfield  {title} {\enquote {\bibinfo {title} {{Result of the search for
  neutrinoless double-$\beta$ decay in $^{100}$Mo with the NEMO-3
  experiment}},}\ }\href {\doibase 10.1103/PhysRevD.92.072011} {\bibfield
  {journal} {\bibinfo  {journal} {Phys.\ Rev.\ D}\ }\textbf {\bibinfo {volume}
  {92}},\ \bibinfo {pages} {072011} (\bibinfo {year} {2015})}\BibitemShut
  {NoStop}%
%%CITATION = ARXIV:1506.05825;%%
\bibitem [{\citenamefont {Blot}(2015)}]{Blot:2015_SuperNEMO}%
  \BibitemOpen
  \bibfield  {author} {\bibinfo {author} {\bibfnamefont {S.}~\bibnamefont
  {Blot}},\ }\bibfield  {title} {\enquote {\bibinfo {title} {{Recent results
  from the NEMO-3 experiment and the status of SuperNEMO}},}\ }\href@noop {} {\
   (\bibinfo {year} {2015})},\ \bibinfo {note} {[To appear in the proceedings
  of the XIV International Conference on Topics in Astroparticle and
  Underground Physics
  (\href{http://taup2015.to.infn.it/scientific-program/parallel-sessions/}%
  {TAUP 2015}), Torino, Italy, September 2015]}\BibitemShut {NoStop}%
\bibitem [{\citenamefont {Fireman}(1948)}]{Fireman:1948}%
  \BibitemOpen
  \bibfield  {author} {\bibinfo {author} {\bibfnamefont {E.~L.}\ \bibnamefont
  {Fireman}},\ }\bibfield  {title} {\enquote {\bibinfo {title} {{Artificial
  Radioactive Substances}},}\ }\href {\doibase 10.1103/PhysRev.74.1201}
  {\bibfield  {journal} {\bibinfo  {journal} {Phys.\ Rev.}\ }\textbf {\bibinfo
  {volume} {74}},\ \bibinfo {pages} {1238} (\bibinfo {year}
  {1948})}\BibitemShut {NoStop}%
\bibitem [{\citenamefont {Fireman}(1949)}]{Fireman:1949qs}%
  \BibitemOpen
  \bibfield  {author} {\bibinfo {author} {\bibfnamefont {E.~L.}\ \bibnamefont
  {Fireman}},\ }\bibfield  {title} {\enquote {\bibinfo {title} {{A measurement
  of the half-life of double beta-decay from \ce{_{50}Sn^{124}}}},}\ }\href
  {\doibase 10.1103/PhysRev.75.323} {\bibfield  {journal} {\bibinfo  {journal}
  {Phys.\ Rev.}\ }\textbf {\bibinfo {volume} {75}},\ \bibinfo {pages}
  {323--324} (\bibinfo {year} {1949})}\BibitemShut {NoStop}%
%%CITATION = PHRVA,75,323;%%
\bibitem [{\citenamefont {Elliott}\ \emph {et~al.}(1987)\citenamefont
  {Elliott}, \citenamefont {Hahn},\ and\ \citenamefont {Moe}}]{Elliott:1987kp}%
  \BibitemOpen
  \bibfield  {author} {\bibinfo {author} {\bibfnamefont {S.~R.}\ \bibnamefont
  {Elliott}}, \bibinfo {author} {\bibfnamefont {A.~A.}\ \bibnamefont {Hahn}}, \
  and\ \bibinfo {author} {\bibfnamefont {M.~K.}\ \bibnamefont {Moe}},\
  }\bibfield  {title} {\enquote {\bibinfo {title} {{Direct Evidence for Two
  Neutrino Double Beta Decay in $^{82}$Se}},}\ }\href {\doibase
  10.1103/PhysRevLett.59.2020} {\bibfield  {journal} {\bibinfo  {journal}
  {Phys.\ Rev.\ Lett.}\ }\textbf {\bibinfo {volume} {59}},\ \bibinfo {pages}
  {2020--2023} (\bibinfo {year} {1987})}\BibitemShut {NoStop}%
%%CITATION = PRLTA,59,2020;%%
\bibitem [{\citenamefont
  {Klapdor-Kleingrothaus}(2010)}]{KlapdorKleingrothaus:2010zza}%
  \BibitemOpen
  \bibfield  {author} {\bibinfo {author} {\bibfnamefont {H.~V.}\ \bibnamefont
  {Klapdor-Kleingrothaus}},\ }\href
  {http://www.worldscibooks.com/physics/6921.html} {\emph {\bibinfo {title}
  {{Seventy years of double beta decay: From nuclear physics to
  beyond-standard-model particle physics}}}}\ (\bibinfo  {publisher} {World
  Scientific},\ \bibinfo {year} {2010})\BibitemShut {NoStop}%
%%CITATION = INSPIRE-861595;%%
\bibitem [{\citenamefont {Tretyak}\ and\ \citenamefont
  {Zdesenko}(2002)}]{Tretyak:2002dx}%
  \BibitemOpen
  \bibfield  {author} {\bibinfo {author} {\bibfnamefont {V.~I.}\ \bibnamefont
  {Tretyak}}\ and\ \bibinfo {author} {\bibfnamefont {Y.~G.}\ \bibnamefont
  {Zdesenko}},\ }\bibfield  {title} {\enquote {\bibinfo {title} {{Tables of
  double beta decay data: An update}},}\ }\href {\doibase
  10.1006/adnd.2001.0873} {\bibfield  {journal} {\bibinfo  {journal} {Atom.\
  Data Nucl.\ Data Tabl.}\ }\textbf {\bibinfo {volume} {80}},\ \bibinfo {pages}
  {83--116} (\bibinfo {year} {2002})}\BibitemShut {NoStop}%
%%CITATION = ADNDA,80,83;%%
\bibitem [{\citenamefont {Barabash}(2011)}]{Barabash:2011mf}%
  \BibitemOpen
  \bibfield  {author} {\bibinfo {author} {\bibfnamefont {A.~S.}\ \bibnamefont
  {Barabash}},\ }\bibfield  {title} {\enquote {\bibinfo {title} {{Double Beta
  Decay: Historical Review of 75 Years of Research}},}\ }\href {\doibase
  10.1134/S1063778811030070} {\bibfield  {journal} {\bibinfo  {journal} {Phys.\
  Atom.\ Nucl.}\ }\textbf {\bibinfo {volume} {74}},\ \bibinfo {pages}
  {603--613} (\bibinfo {year} {2011})}\BibitemShut {NoStop}%
%%CITATION = ARXIV:1104.2714;%%
\bibitem [{\citenamefont {Klapdor-Kleingrothaus}\ \emph
  {et~al.}(2001{\natexlab{b}})\citenamefont {Klapdor-Kleingrothaus},
  \citenamefont {Dietz}, \citenamefont {Harney},\ and\ \citenamefont
  {Krivosheina}}]{KlapdorKleingrothaus:2001ke}%
  \BibitemOpen
  \bibfield  {author} {\bibinfo {author} {\bibfnamefont {H.~V.}\ \bibnamefont
  {Klapdor-Kleingrothaus}}, \bibinfo {author} {\bibfnamefont {A.}~\bibnamefont
  {Dietz}}, \bibinfo {author} {\bibfnamefont {H.~L.}\ \bibnamefont {Harney}}, \
  and\ \bibinfo {author} {\bibfnamefont {I.~V.}\ \bibnamefont {Krivosheina}},\
  }\bibfield  {title} {\enquote {\bibinfo {title} {{Evidence for neutrinoless
  double beta decay}},}\ }\href {\doibase 10.1142/S0217732301005825} {\bibfield
   {journal} {\bibinfo  {journal} {Mod.\ Phys.\ Lett.\ A}\ }\textbf {\bibinfo
  {volume} {16}},\ \bibinfo {pages} {2409--2420} (\bibinfo {year}
  {2001}{\natexlab{b}})}\BibitemShut {NoStop}%
%%CITATION = HEP-PH/0201231;%%
\bibitem [{\citenamefont {Klapdor-Kleingrothaus}\ and\ \citenamefont
  {Krivosheina}(2006)}]{KlapdorKleingrothaus:2006ff}%
  \BibitemOpen
  \bibfield  {author} {\bibinfo {author} {\bibfnamefont {H.~V.}\ \bibnamefont
  {Klapdor-Kleingrothaus}}\ and\ \bibinfo {author} {\bibfnamefont {I.~V.}\
  \bibnamefont {Krivosheina}},\ }\bibfield  {title} {\enquote {\bibinfo {title}
  {{The evidence for the observation of 0nu beta beta decay: The identification
  of 0nu beta beta events from the full spectra}},}\ }\href {\doibase
  10.1142/S0217732306020937} {\bibfield  {journal} {\bibinfo  {journal} {Mod.\
  Phys.\ Lett.\ A}\ }\textbf {\bibinfo {volume} {21}},\ \bibinfo {pages}
  {1547--1566} (\bibinfo {year} {2006})}\BibitemShut {NoStop}%
%%CITATION = MPLAE,A21,1547;%%
\bibitem [{\citenamefont {Aalseth}\ \emph
  {et~al.}(2002{\natexlab{b}})\citenamefont {Aalseth} \emph
  {et~al.}}]{Aalseth:2002dt}%
  \BibitemOpen
  \bibfield  {author} {\bibinfo {author} {\bibfnamefont {C.~E.}\ \bibnamefont
  {Aalseth}} \emph {et~al.},\ }\bibfield  {title} {\enquote {\bibinfo {title}
  {{Comment on `Evidence for neutrinoless double beta decay'}},}\ }\href
  {\doibase 10.1142/S0217732302007715} {\bibfield  {journal} {\bibinfo
  {journal} {Mod.\ Phys.\ Lett.\ A}\ }\textbf {\bibinfo {volume} {17}},\
  \bibinfo {pages} {1475--1478} (\bibinfo {year}
  {2002}{\natexlab{b}})}\BibitemShut {NoStop}%
%%CITATION = HEP-EX/0202018;%%
\bibitem [{\citenamefont {Zdesenko}\ \emph {et~al.}(2002)\citenamefont
  {Zdesenko}, \citenamefont {Danevich},\ and\ \citenamefont
  {Tretyak}}]{Zdesenko:2002kz}%
  \BibitemOpen
  \bibfield  {author} {\bibinfo {author} {\bibfnamefont {Y.~G.}\ \bibnamefont
  {Zdesenko}}, \bibinfo {author} {\bibfnamefont {F.~A.}\ \bibnamefont
  {Danevich}}, \ and\ \bibinfo {author} {\bibfnamefont {V.~I.}\ \bibnamefont
  {Tretyak}},\ }\bibfield  {title} {\enquote {\bibinfo {title} {{Has
  neutrinoless double beta decay of Ge-76 been really observed?}}}\ }\href
  {\doibase 10.1016/S0370-2693(02)02705-3} {\bibfield  {journal} {\bibinfo
  {journal} {Phys.\ Lett.\ B}\ }\textbf {\bibinfo {volume} {546}},\ \bibinfo
  {pages} {206--215} (\bibinfo {year} {2002})}\BibitemShut {NoStop}%
%%CITATION = PHLTA,B546,206;%%
\bibitem [{\citenamefont {Artusa}\ \emph {et~al.}(2014)\citenamefont {Artusa}
  \emph {et~al.}}]{Artusa:2014wnl}%
  \BibitemOpen
  \bibfield  {author} {\bibinfo {author} {\bibfnamefont {D.~R.}\ \bibnamefont
  {Artusa}} \emph {et~al.} (\bibinfo {collaboration} {CUORE Collaboration}),\
  }\bibfield  {title} {\enquote {\bibinfo {title} {{Exploring the Neutrinoless
  Double Beta Decay in the Inverted Neutrino Hierarchy with Bolometric
  Detectors}},}\ }\href {\doibase 10.1140/epjc/s10052-014-3096-8} {\bibfield
  {journal} {\bibinfo  {journal} {Eur.\ Phys.\ J.\ C}\ }\textbf {\bibinfo
  {volume} {74}},\ \bibinfo {pages} {3096} (\bibinfo {year}
  {2014})}\BibitemShut {NoStop}%
%%CITATION = ARXIV:1404.4469;%%
\bibitem [{\citenamefont {Dell'Oro}\ \emph {et~al.}(2015)\citenamefont
  {Dell'Oro}, \citenamefont {Marcocci}, \citenamefont {Viel},\ and\
  \citenamefont {Vissani}}]{Dell'Oro:2015tia}%
  \BibitemOpen
  \bibfield  {author} {\bibinfo {author} {\bibfnamefont {S.}~\bibnamefont
  {Dell'Oro}}, \bibinfo {author} {\bibfnamefont {S.}~\bibnamefont {Marcocci}},
  \bibinfo {author} {\bibfnamefont {M.}~\bibnamefont {Viel}}, \ and\ \bibinfo
  {author} {\bibfnamefont {F.}~\bibnamefont {Vissani}},\ }\bibfield  {title}
  {\enquote {\bibinfo {title} {{The contribution of light Majorana neutrinos to
  neutrinoless double beta decay and cosmology}},}\ }\href {\doibase
  10.1088/1475-7516/2015/12/023} {\bibfield  {journal} {\bibinfo  {journal}
  {J.\ Cosm.\ Astropart.\ Phys.}\ }\textbf {\bibinfo {volume} {1512}},\
  \bibinfo {pages} {023} (\bibinfo {year} {2015})}\BibitemShut {NoStop}%
%%CITATION = ARXIV:1505.02722;%%
\bibitem [{\citenamefont {Palanque-Delabrouille}\ \emph
  {et~al.}(2013)\citenamefont {Palanque-Delabrouille} \emph
  {et~al.}}]{Palanque-Delabrouille:2013gaa}%
  \BibitemOpen
  \bibfield  {author} {\bibinfo {author} {\bibfnamefont {N.}~\bibnamefont
  {Palanque-Delabrouille}} \emph {et~al.},\ }\bibfield  {title} {\enquote
  {\bibinfo {title} {{The one-dimensional Ly-$\alpha$ forest power spectrum
  from BOSS}},}\ }\href {\doibase 10.1051/0004-6361/201322130} {\bibfield
  {journal} {\bibinfo  {journal} {Astron.\ Astrophys.}\ }\textbf {\bibinfo
  {volume} {559}},\ \bibinfo {pages} {A85} (\bibinfo {year}
  {2013})}\BibitemShut {NoStop}%
%%CITATION = ARXIV:1306.5896;%%
\bibitem [{\citenamefont {Cirigliano}\ \emph {et~al.}(2004)\citenamefont
  {Cirigliano}, \citenamefont {Kurylov}, \citenamefont {Ramsey-Musolf},\ and\
  \citenamefont {Vogel}}]{Cirigliano:2004tc}%
  \BibitemOpen
  \bibfield  {author} {\bibinfo {author} {\bibfnamefont {V.}~\bibnamefont
  {Cirigliano}}, \bibinfo {author} {\bibfnamefont {A.}~\bibnamefont {Kurylov}},
  \bibinfo {author} {\bibfnamefont {M.~J.}\ \bibnamefont {Ramsey-Musolf}}, \
  and\ \bibinfo {author} {\bibfnamefont {P.}~\bibnamefont {Vogel}},\ }\bibfield
   {title} {\enquote {\bibinfo {title} {{Neutrinoless double beta decay and
  lepton flavor violation}},}\ }\href {\doibase 10.1103/PhysRevLett.93.231802}
  {\bibfield  {journal} {\bibinfo  {journal} {Phys.\ Rev.\ Lett.}\ }\textbf
  {\bibinfo {volume} {93}},\ \bibinfo {pages} {231802} (\bibinfo {year}
  {2004})}\BibitemShut {NoStop}%
%%CITATION = HEP-PH/0406199;%%
\end{thebibliography}%

%\onecolumngrid	
\clearpage
\vfill\eject	
%\small

\tableofcontents

\end{document}